\documentclass[a4paper,11pt,twocolumn,notitlepage]{article}

\usepackage{tkz-berge} 
\usepackage{amsmath}
\usepackage{amssymb}
\usepackage{cite}
\usepackage{slashed}
\usepackage{multirow}
\usepackage{lscape} 
\usepackage{longtable} 
\usepackage{graphicx}
\usepackage{array}
\usepackage[top=3cm, bottom=3.2cm, left=2.2cm, right=2.1cm]{geometry}

\usepackage{amsmath}
\makeatletter
\g@addto@macro\normalsize{
  \setlength\abovedisplayskip{5pt}
  \setlength\belowdisplayskip{5pt}
  \setlength\abovedisplayshortskip{5pt}
  \setlength\belowdisplayshortskip{5pt}
}
\makeatother

\newcommand{\ket}[1]{\left| {#1} \right\rangle}
\newcommand{\bra}[1]{\left\langle {#1} \right|}

\newcolumntype{V}{>{\centering\arraybackslash} m{.21\linewidth} }

\begin{document} 
\title{Brane Tilings as On-shell Diagrams}
\author{Alvaro M. Alhambra - Supervisor: Yang-Hui He \\ Department of Physics, University of Oxford} 
\maketitle

\begin{abstract}
A new way of computing scattering amplitudes in a certain very important QFT ($\mathcal{N}=4$ SYM) has recently been developed, in which an algebraic structure called the positive Grassmannian plays a very important role. The mathematics of the positive Grassmannian involve, among other things, bipartite graphs, which also appear in the formulation of a certain class of conformal field theories that are currently being generalized into \emph{Bipartite Field Theories} (BFT). The fact that the same structures appear in two such different realms of physics suggests a deeper connection between the two that is yet to be fully unveiled. Here we explore that potential connection by looking at the graphs of a certain class of BFTs, the brane tilings, in terms of the new mathematics developed for the computation of the amplitudes. This way we produce a set of data that will hopefully be useful in the development of that connection. 
\end{abstract}

\section{Introduction}
In particle physics, the most important processes we are usually able to compute are \emph{scattering} ones,  in which a set of incoming \emph{free} particles (meaning particles with definite momentum, represented by eigenstates of a quadratic Hamiltonian, with no interaction terms) interact in a region of spacetime and form a possibly different set of outgoing free particles.
Understanding these amounts to calculating the so-called $S$-matrix, which maps the initial and final states by giving the amplitudes of the transitions
\begin{equation}
\mathcal{A}({i\to f})=\bra{f}S\ket{i}
.\end{equation}
There is a very well-understood procedure for calculating these amplitudes, with origin in the work of Feynman, Schwinger, Dyson and others during the 60s, based on the famous \emph{Feynman diagrams}, which is now part of any introductory course in QFT \cite{Peskin}. It enables the computation of the amplitudes to all orders in perturbation theory, although in practice only the first few leading orders are calculated. 

Still, this method, although originally aimed towards simplifying the calculations, does get very complicated quickly for an increasing number of particles, even if the final answers are usually very simple. 
This fact suggests that there may be more straightforward ways of computing the perturbative $S$ matrix, in which aspects not obvious in the standard description are manifest, and viceversa. It is even thought that such method will bring a different way of looking at QFT that might shed some light into some of the big open problems in physics, by creating an alternative method of computing local physical observables, the amplitudes \cite{SimplestQFT,Amplituhedron,SUSY2013}.

A research program for finding such a potential new formulation has been very actively developed in the past decade with very rapid success. A new and ``very unexpected''\footnote{In the words of Edward Witten.} picture \cite{NAH,Amplituhedron} has very recently been developed for the scattering amplitudes of a very important gauge theory, $\mathcal{N}=4$ Super Yang-Mills, widely present in the context of string theory, and in particular in the AdS/CFT correspondence \cite{GreenLimit,ADS/CFT,Gubser,WittenADS}, and it is hoped to be generalized to more complicated theories in the future. Remarkably, this new way of computing amplitudes moves its main focus away from locality and unitarity, ubiquitous in the Feyman diagram formalism and quantum mechanics in general, in exchange for making a different symmetry of the $S$ matrix manifest, the so-called Yangian symmetry. A brief account of how this line of work has progressed in the past times is given in Section \ref{ScatteringAmplitudes}.

A key in the development of the understanding of this new formulation was finding that a certain algebraic object plays a fundamental role in this story, the \emph{positive Grassmannian} \cite{NAH}. This object was being studied by mathematicians \cite{Postnikov,Postnikov2,MarshRietsch} almost at the same time as physicists realized of its importance, making it a wonderful instance of an unexpected encounter between physics and mathematics. Section \ref{positroid} this report will be devoted to explaining some of these new mathematics and the way in which they are relevant to physics.

The mathematics of the positive Grassmannian naturally involve \emph{bipartite graphs}, which are  a kind of graphs with black and white vertices which are only joined to those of the opposite color. It turns out that these objects also play a crucial role in the classification of some field theories with certain symmetry properties, the \emph{bipartite field theories} (BTFs) \cite{FrancoBipartite} that, as much as the already mentioned $\mathcal{N}=4$ SYM does, make an appearance in the context of string theory and the AdS/CFT correspondence \cite{Bipartita}. These are explained in Section \ref{Bipartites}. 

Here, in Section \ref{graphs}, and with the help of a software package specific for working with positive geometry and the positive Grassmannian \cite{Positroids}, we compute the list of bipartite graphs on a 2-dimensional torus that produce consistent BFTs (so-called \emph{brane tilings}) \cite{GraphClassification} in terms of the mathematics of the positive Grassmannian (the full list is shown in Appendix \ref{list}). The aim of this is to help develop a link between scattering amplitudes and BFTs that will hopefully be expanded and better understood in the future. Some effort is currently being made in sheding some light in this so far speculative connection \cite{FrancoDirections,FrancoBipartite,FrancoGeometry}, and the work shown in this report intends to be a small contribution in this direction.

\section{Scattering amplitudes} \label{ScatteringAmplitudes}
The details of $\mathcal{N}=4$ SYM are beyond the scope of this report, but it is worth giving some details of the amplitudes of the theory. The sort of amplitudes we will be talking about are contributions to some order of the so-called ${T}$ matrix $\bra{f}T\ket{i}=\bra{f}S-1\ket{i}$ in $\mathcal{N}=4$ SYM, which is a supersymmetric  $SU(N)$ gauge theory of massless particles. The leading-order amplitudes will be the tree-level amplitudes, while higher order ones will be loop amplitudes. In particular, the objects to calculate are the \emph{color-stripped} amplitudes $\mathcal{A}_n(1,...,n)$ of $n$ outgoing particles (given QFT's crossing symmetry we can make them all outgoing) defined as
\begin{equation}
\mathcal{A}_n=\sum_\sigma \text{Tr}(T_{\sigma(1)}...T_{\sigma(n)}) \mathcal{A}({\sigma(1)},....,{\sigma(n)}),
\end{equation}
where $T_i$ are the generators of the $SU(N)$ algebra corresponding to certain colors of the particles and $\sigma$ are the $n$-permutations with different cyclic ordering (so that the sum includes all different permutations up to cyclic rotations - note that this is the way to sum over all possible traces of $n$ generators). Because we are in a supersymmetric theory it is more natural to talk about \emph{superamplitudes}, in which the helicities of the outgoing particles (excitations of \emph{superfields}) do not need to be specified during calculations \cite{ScattAmpli}. Instead, our amplitude will contain $\delta$-functions in the so-called $\tilde{\eta}_I$ Grassmann variables (despite the name, these are not related to Grassmannians) which we can later integrate out, and depending in they way in which we decide to do this we obtain amplitudes corresponding to different combinations of helicities. For example, from the same superamplitude we can obtain the amplitude of scattering of two $+\frac{1}{2}$ fermions and two $+1$ gluons or the amplitude of three $+1$ gluons and a scalar. We give more details about supersymmetry and superamplitudes in $\mathcal{N}=4$ in Appendix \ref{SUSY}.
\subsection{Towards the positive Grassmannian}
Some great progress was already being made in amplitudes from the early 90s, mainly by Bern, Dixon, Kosower and others \cite{Bern1994,BernDixonKosower2}, but what triggered this particular line of study of amplitudes, in terms of twistors and recursion relations, was Witten's 2003 paper \cite{WittenTwistor}. There, he noted that the scattering amplitudes of $\mathcal{N}=4$ SYM are dual to  solutions of a certain string theory when the kinematical data is represented in twistor space \cite{Penrose,Ferber}. 
This inspired a further finding of crucial importance: the BCFW recursion relations \cite{BCFRecursion,BCFW}. In 2005, Britto, Cachazo, Feng and Witten himself found a recursive way of building scattering amplitudes in gauge theories at \emph{tree} level (in processes with no loops, or internal particles over which we need to integrate) from a sum of combinations of lower-point amplitudes (BCFW terms, these findings are explained further in Appendix \ref{Recursion}). Soon after, the language of the BCFW was translated into to twistor diagrams \cite{Hodges2005}, forecasting how important twistor geometry and twistor diagrams (later to be identified as on-shell diagrams) would be.

In the BCFW construction of amplitudes, the various terms added are separately non-local (due to some so-called ``spurious poles"), but when adding them these poles magically cancel into some manifestly local amplitude, and hence we see the first sign of locality being an emergent feature, as opposed to a manifest one as it occurs in the Feynman diagrams language.

The next breakthrough came during 2007, when a new remarkable symmetry \cite{Alday,DrummondFirst} was found for amplitudes of $\mathcal{N}=4$ in the \emph{planar} limit (which we explain in Section \ref{OnShell}). This \emph{dual superconformal symmetry} \cite{DrummondReview} was seen to merge with another well-known symmetry of the theory, the superconformal symmetry, into an infinite-dimensional symmetry (meaning an infinite number of generators can be built out of the generators of the two symmetries together) named a \emph{Yangian} symmetry \cite{DrummondYangT}, which is better understood in twistor space \cite{Mason2009}. Remarkably, the BCFW terms that make up an amplitude were found to be separately Yangian invariant.

Meanwhile, an idea was conjectured that involved calculating amplitudes using the Grassmannian for the first time (not yet the \emph{positive} Grassmannian) \cite{SMatrixTwistor,DualityS}. 
It was found that the residues of an integrand using the Grassmannian naturally yields terms that are invariant under Yangian symmetry \cite{GrassmannianOrigin,DrummondYang}, the way BCFW terms are. In particular, this integrand generates the so-called \emph{leading singularities}, which are algebraic functions of external momenta with the right symmetry properties that, when grouped together in certain ways, give different representations of scattering amplitudes (such as, for instance, the BCFW representation) \cite{Residues}. Still, this was far from well understood, as it was not known how it should produce the right combination of leading singularities. 

It was soon found \cite{Hodges2009} that, when writing the right BCFW terms of certain amplitudes with a new set of kinematical variables (the \emph{momentum twistor} variables), and in such a way that we get a local answer, the result looks like a \emph{polytope}\footnote{A generalization of a polygon to $n$ dimensions.} in that twistor space, hinting towards a possible geometrical understanding of the Grassmannian integral \cite{NotePolytopes}.
This new space of variables allowed for the natural generalization of BCFW relations from tree-level to all-loop orders \cite{AllLoop,BoelsLoop}, from which it was found that the integrand for any amplitude could be written in terms of a set of variables $\alpha_i$ such that they appear in the integration measure as 
\begin{equation}
\prod_i \text{d}(\text{log}(\alpha_i)) = \prod_i \frac{\text{d}\alpha_i}{\alpha_i}.
\end{equation}
This important integration measure and some work preceding it is further explained in Appendix \ref{Recursion}.
Hence, what was known by 2010/11 was that the amplitudes could be thought of as a sum of residues of an integrand over that $\text{d}(\text{log}(\alpha_i))$ form that generates leading singularities with the right symmetry properties (the most important being the Yangian symmetry), and that the result will probably look like ``some volume of some polytope'' made of a sum of such leading singularities that has to yield a local, unitary answer. 

The turning point ocurred with the publication of \cite{NAH} in December of 2012, where the positive Grassmannian comes into scene. Before then there were very good hints as to how the integrand should be written and how the leading singularities produced should be added \cite{Hodges2009,AllLoop,Residues,NotePolytopes}, but suddenly physicists realized that what they were really doing was an integration over subsets or \emph{cells} of the \emph{positive Grassmannian}, not the Grassmannian itself, and that within the mathematics of the positive Grassmannian there is a well-defined procedure to write the integrand as a function of the already mentioned $\alpha_i$ variables. These variables have a natural interpretation in this framework when thinking of the cells in terms of permutations and graphs, which represent the way in which small amplitudes are glued together into higher-order ones. We now proceed to define the Grassmannian, its positive stratification into cells and the way in which this mysterious integrand over them is generated through the gluing of simple 3-particle amplitudes.

\section{The positive Grassmannian and its \emph{positroid stratification}}\label{positroid}

The Grassmanian $\text{Gr}(k,n)$ is the set of $k$-planes through the origin in $\mathbb{R}^n$, represented as a $k \times n$ matrix where the rows are a basis of the $k$-plane. This means that the action of the general linear group $GL(k)$ has no effect on the particular $k$-plane, so we can gauge-fix the matrix to have the identity as one of its minors. In this way, any one element of $\text{Gr}(2,4)$ can be, for example 
\begin{equation}
C=
\begin{pmatrix}
1 & 0 & c_1 & c_2 \\
0 & 1 & c_3 & c_4
\end{pmatrix},~~~~~c_i \in \mathbb{R}.
\end{equation}
We shall also define an orthogonal complement $C^\perp$ (an $(n-k)\times n$ matrix), such that $C \cdot C^\perp=0$. \\
The positive Grassmannian $\text{Gr}^+(k,n)$ is then simply the subset of the Grassmannian whose elements have all their $k \times k$ minors with positive determinant, and, as we discuss in detail in Appendix \ref{Positive}, constitutes a generalization of the space of the \emph{inside} of a polygon.
It turns out that it is possible to map the so-called \emph{positroid cells} of $\text{Gr}^+(k,n)$ (a certain kind of subsets) to permutations and to bipartite graphs (as will be discussed in detail below), and that in such map we produce $C$ matrices representing those cells, which contain some $\alpha_i$ variables. These will produce the integrand we are looking for \cite{NAH}, which is 
\begin{align} \label{eq:Integrand}
\prod_i \frac{\text{d}\alpha_i}{\alpha_i} \delta^{k\times4}(C \cdot \tilde{\eta}) \delta^{k \times 2} (C \cdot \tilde{\lambda}) \delta^{2 \times (n-k)} (\lambda \cdot C^\perp).
\end{align}
Here, $C$ is the $k \times n$ matrix of a cell (in the amplitude, $n$ will be the total number of particles, and $k$ is the number of negative helicity ones), $\lambda$ and $\tilde{\lambda}$ are $2 \times n$ matrices $(\lambda_1....\lambda_n)$ representing, respectively, the left-handed and right-handed spinors that describe the momenta of the outgoing particles (remember that we do not need to specify helicities, the only input are the 4-momenta), and $\tilde{\eta}$ is a $4\times n$ matrix with the superspace variables described in Section \ref{ScatteringAmplitudes} and Appendix \ref{SUSY}. The last of these $\delta$-functions is found via the auxiliary integral
\begin{equation}
\delta^{2(n-k)}(\lambda \cdot C^\perp)=\int \text{d}^{2\times k}\rho\, \delta^{2\times n}(\rho \cdot C - \lambda).
\end{equation}
The twistorial representation of momenta is ubiquitous in this line of work, but here for simplicity we write the integrand in terms of the more familiar spinors.

The form \eqref{eq:Integrand}, once properly written and integrated, and with potential IR and UV divergences taken care of (when possible), will produce any leading singularity, but we first need to \emph{i)} be able to write $C$ from the corresponding \emph{on-shell diagrams} (to be explained in \ref{OnShell}) and \emph{ii)} know the right contour of integration, which amounts to identifying the leading singularities we want with the right on-shell diagrams. 

Following \cite{NAH}, we will proceed to explain how the first of these points is done, involving the mathematics that will later be connected to the bipartite field theories. We will not explain the second point here, which has only been fully developed as late as December 2013 \cite{Amplituhedron,IntoAmplituhedron}. It is worth mentioning, however, that, at least at tree-level, knowing how to do this entails a geometrical understanding of amplitudes as polytopes in a space where our kinematical variables are, instead of spinors, \emph{momentum twistor} variables \cite{ScattAmpli}. Such polytopes are the so-called \emph{amplituhedrons} \cite{Amplituhedron,IntoAmplituhedron}, and their inside is mapped by their corresponding positroid cells (this is explained further in Appendix \ref{Positive}). 

\subsection{On-shell diagrams, graphs and permutations}\label{OnShell}
With the BCFW recursion method for tree-level and its generalization to all loops \cite{AllLoop} we see that we can express the amplitudes as sums of combinations of smaller amplitudes (the leading singularities), and such combinations are represented by objects called \emph{on-shell diagrams}. These on-shell diagrams will be bicolored graphs and, because we are only dealing with amplitudes in the planar limit, these graphs will in principle be planar, meaning that they can be drawn on a plane, with no edges intersecting\footnote{Outside of the context of on-shell diagrams, the planar limit is also defined as the approximation of amplitudes to first order in $1/N$, with $N$ the number of colors of the theory.}. A detailed description of the correspondence between on-shell diagrams and the full amplitudes is beyond the scope of this report, but for our purposes it will be enough to keep in mind that a tree amplitude is built as a sum of one or more leading singularities generated from certain diagrams, in a way consistent with locality and unitarity, and that loop amplitudes are generally just written in terms of the integrand, as their final form presents a number of additional difficulties.

The smallest amplitudes we can have, and the ones out of which we build our on-shell diagrams, are 3-point amplitudes, which we will represent as trivalent white vertices when the total helicity of the particles is  +1 and trivalent black vertices when it is -1, as these are the only two possible sums of helicities that give a non-zero amplitude. They are:
\begin{figure}[!h]
~~~~~~~~~\includegraphics[width=0.12\textwidth]{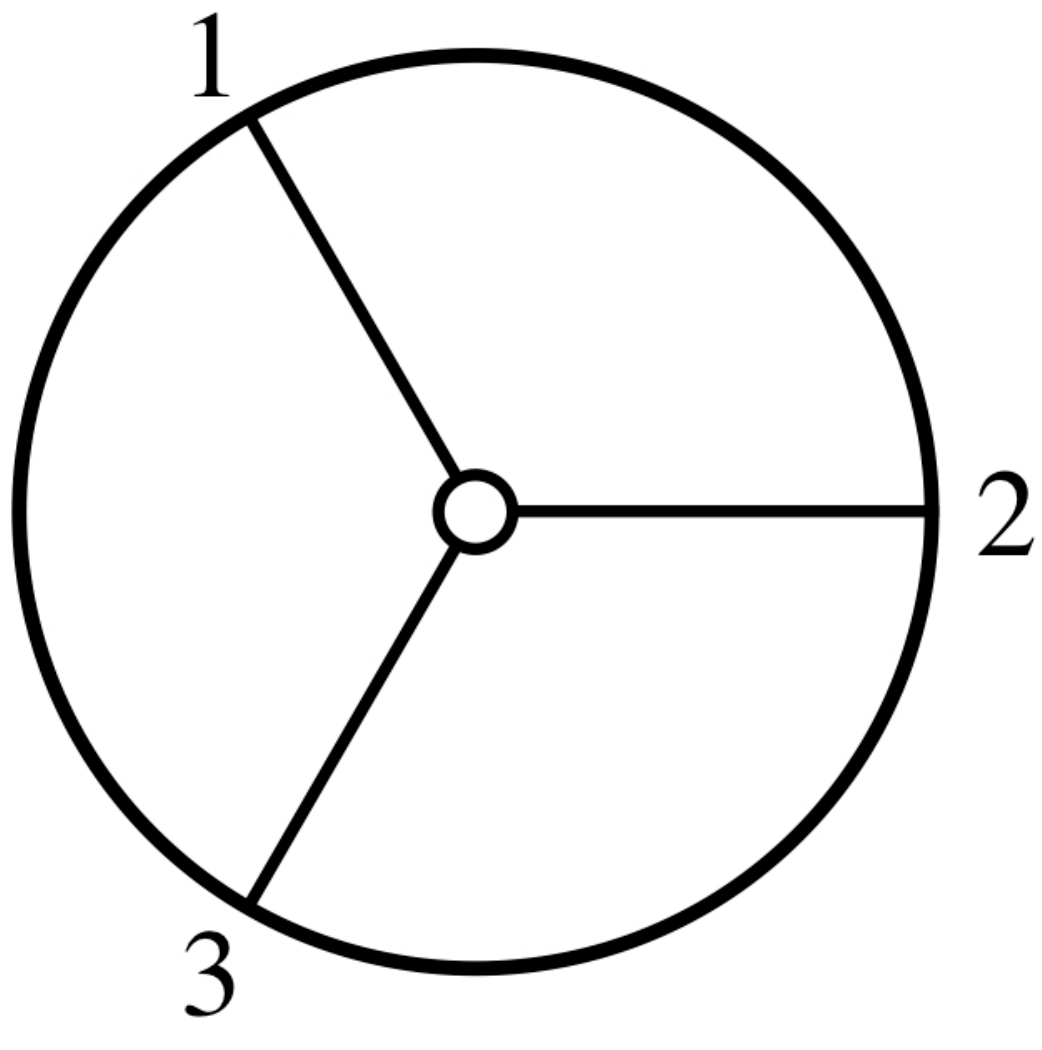} ~~~~~~~~\includegraphics[width=0.12\textwidth]{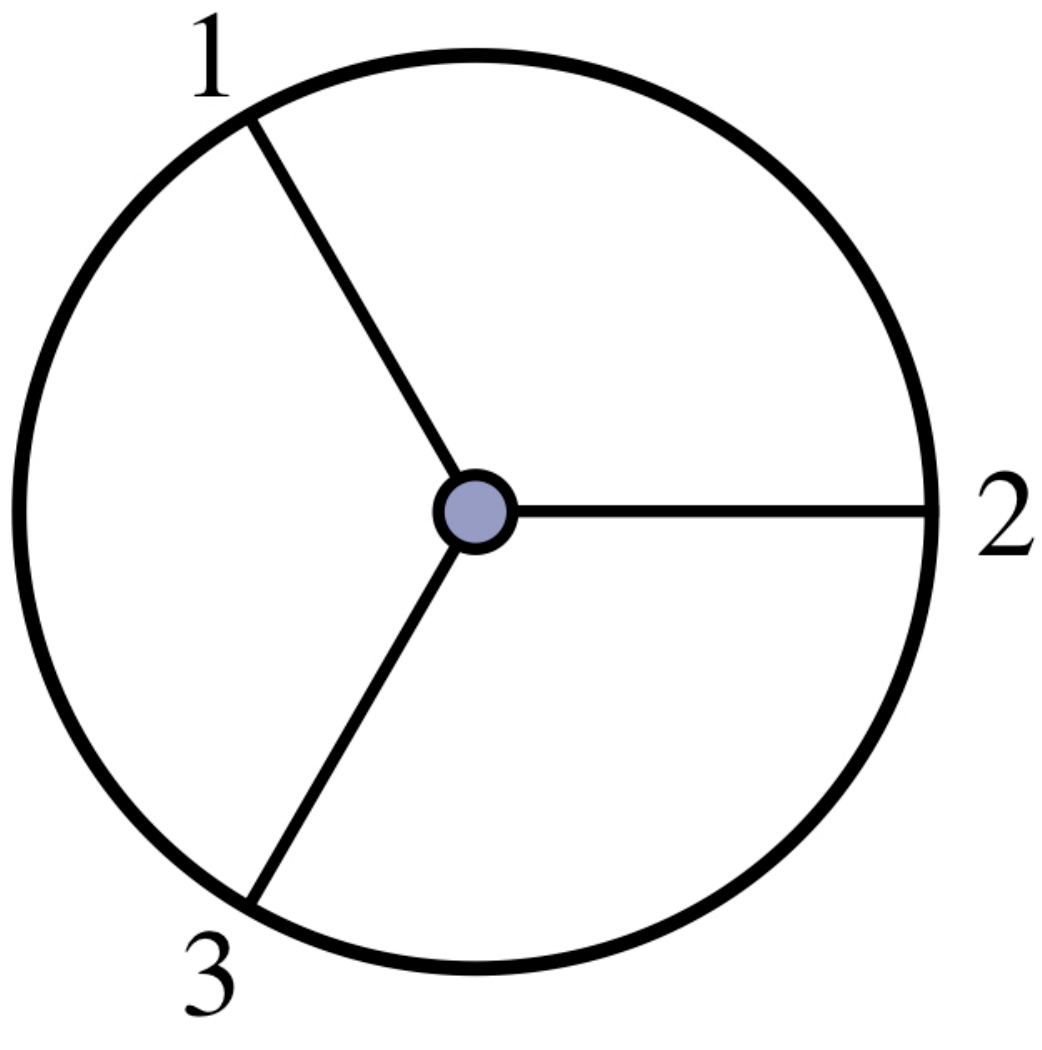} .
\end{figure}

And the corresponding amplitudes are
\begin{equation}
A_3^{(1)}=\frac{\delta^{1\times4}([23] \tilde{\eta_1} + [31] \tilde{\eta_2}  + [12] \tilde{\eta_3} )}{[12][23][31]} \delta^{2\times2}(\lambda \cdot  \tilde{\lambda})\nonumber
\end{equation}
for the white vertex and
\begin{equation}
A_3^{(2)}=\frac{\delta^{2\times4}(\lambda_1 \tilde{\eta_1} + \lambda_2 \tilde{\eta_2}  + \lambda_3 \tilde{\eta_3} )}{\langle12\rangle \langle23\rangle \langle31\rangle} \delta^{2\times2}(\lambda \cdot  \tilde{\lambda})\nonumber
\end{equation}
for the black vertex. The $[ij]$ and $\langle ij\rangle$ are Lorentz invariant products of, respectively, left handed $\lambda_i$ and right-handed $\tilde{\lambda}_i$ spinors.
It turns out that the form of these amplitudes is fixed by Poincar\'e invariance only \cite{NAH,ScattAmpli}, and will hence be our ideal building blocks. Because we are dealing with superamplitudes, no helicities need to be specified in the external legs of the graph.

\begin{figure}
~~~~~~~~~~~~\includegraphics[width=0.3\textwidth]{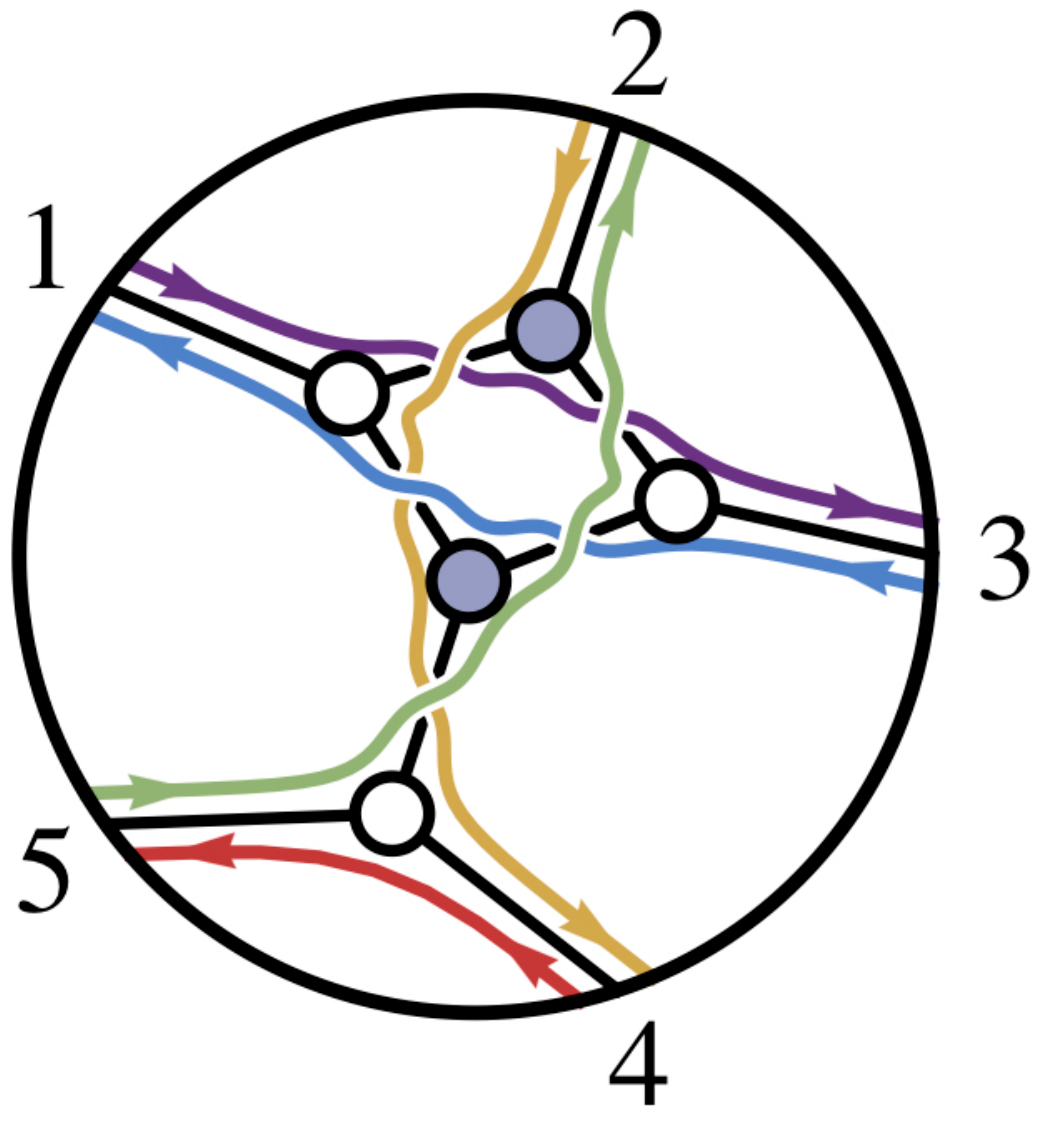} 
\caption  {A graph and its path prescription corresponding to the \emph{decorated} permutation $\sigma=\{3,4,6,5,7\}$, or the \emph{undecorated} permutation $\sigma^*=\{3,4,1,5,2\}$, hence the corresponding $C \in \text{Gr}^+(2,5)$. Here 1, 2 and 4 are \emph{sinks}, and 3 and 5 are \emph{sources}.}  \label{fig:paths} \end{figure}

To build more complicated on-shell diagrams we need to join many of these vertices together into a graph. For example, the $k=2$ 4-particle amplitude (denoted as $\mathcal{A}_4^\text{MHV}$) is generated from the diagram \cite{NAH}

\begin{equation}
\begin{array}{V c c}
\includegraphics[width=0.14\textwidth]{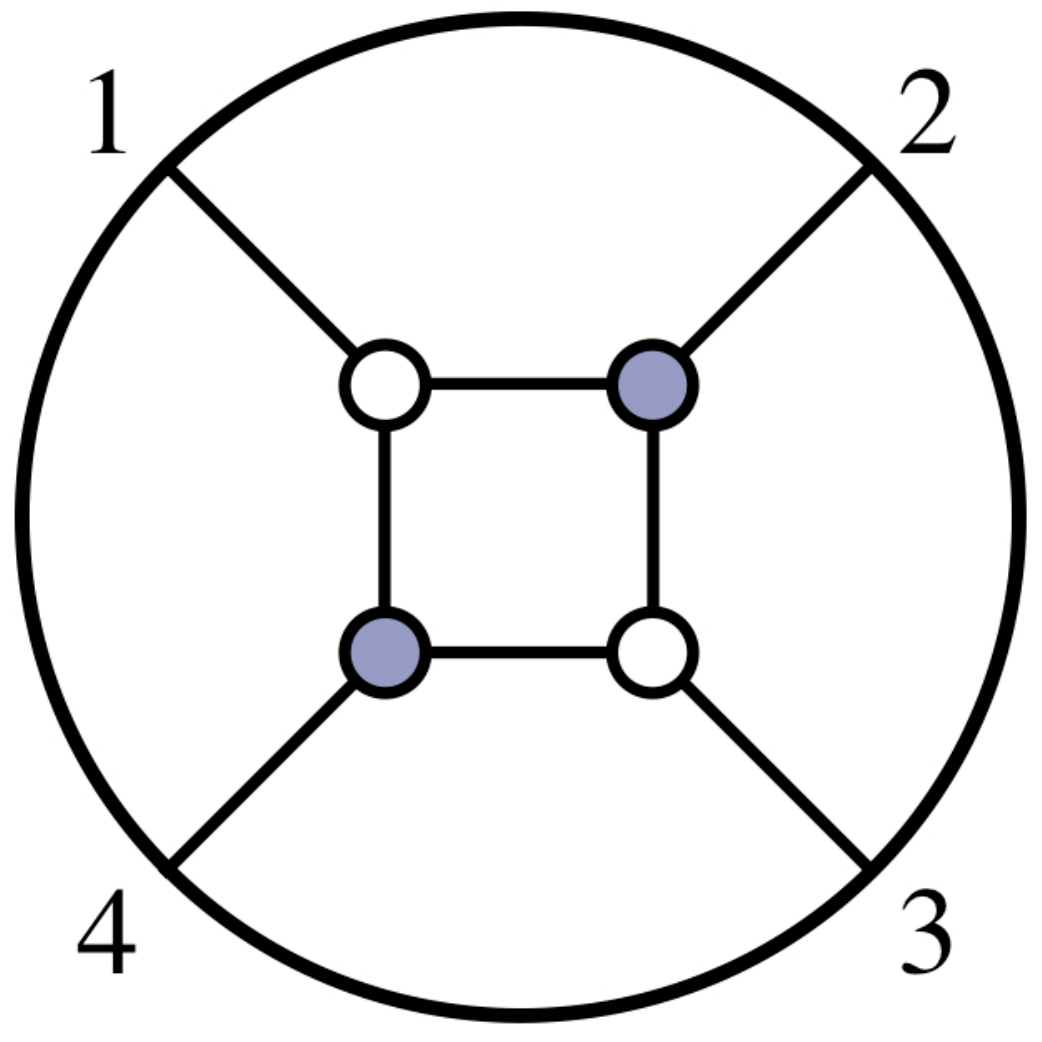}  
&\,\,\,\,\,\,\,\,\, \rightarrow &\mathcal{A}_4^\text{MHV}=\frac{\delta^8(\sum_i^n \tilde{\lambda_i} \tilde{\eta}_i) \delta^4(\sum_i^n P^\mu_i)}{\langle12\rangle \langle23\rangle \langle34\rangle \langle41\rangle}.
\end{array}
\end{equation}
It turns out that, in an on-shell diagram, there are certain internal operations that leave its physical information unchanged. These are
\begin{itemize}
\item \underline{Merge}: we may put two adjacent vertices of the same color together. This is the condition that enables us to make the graphs always bipartite. 
\begin{figure}[!h]
~~~~~~~~\includegraphics[width=0.33\textwidth]{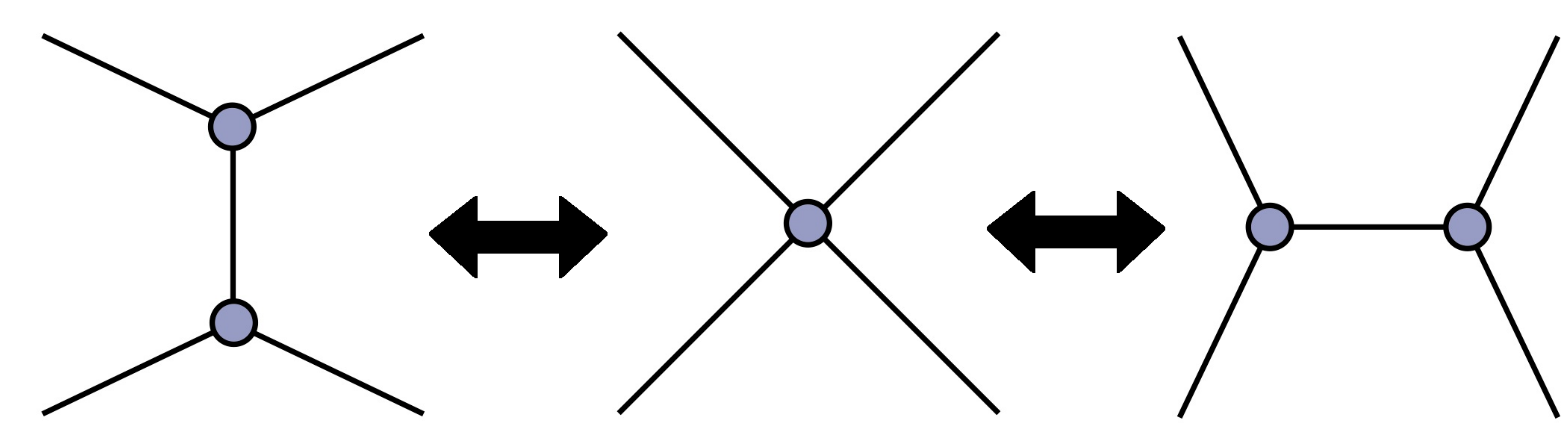} 
\end{figure}
\item \underline{Square move}: we can see that the 4-particle amplitude graph is cyclically invariant, which means that it can be rotated with no effect on the final permutation. This is of course still the case if such 4-particle graph is embedded within a bigger one. The rotated graph can then always be made bipartite again via merging. 
\begin{figure}[!h]
~~~~~~~~~~~~\includegraphics[width=0.3\textwidth]{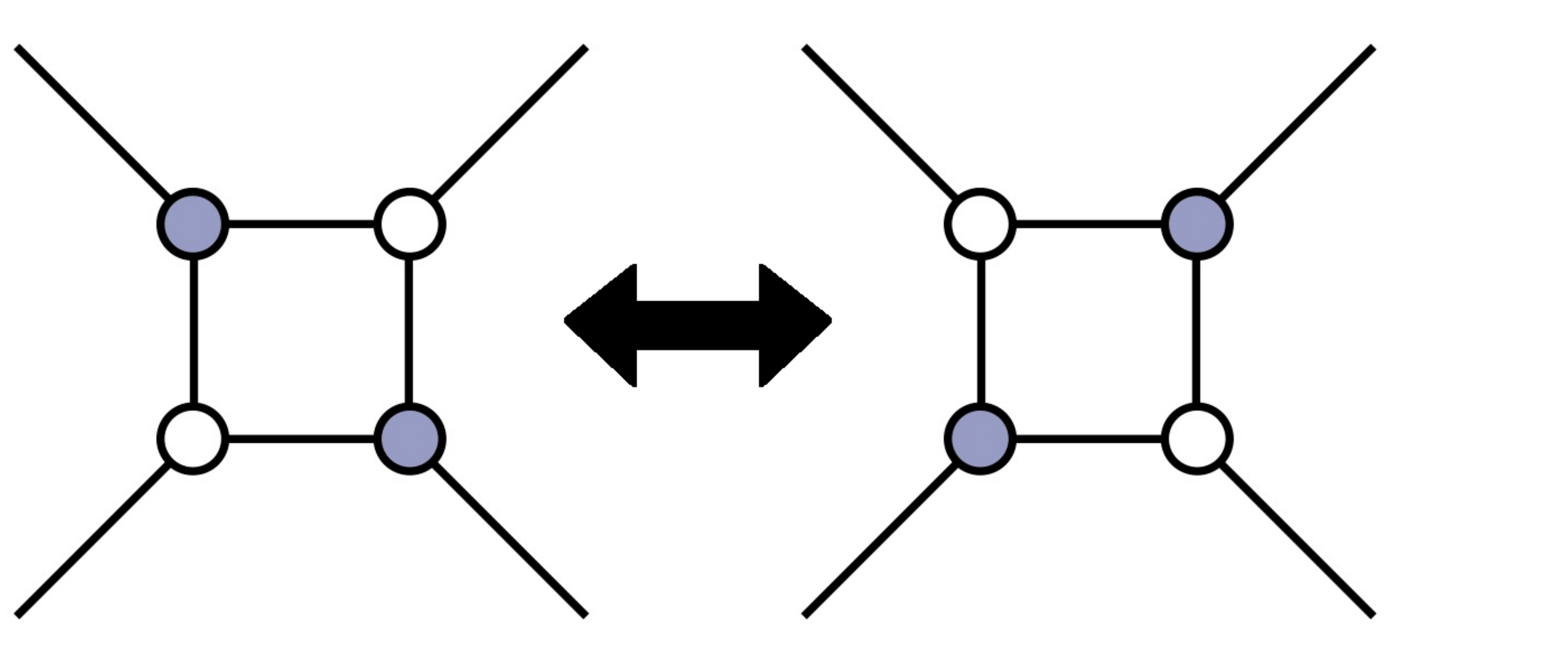} 
\end{figure}
\item \underline{Bubble deletion}: in which we eliminate two vertices joined together by two edges. As opposed to the other two, it reduces the number of faces by one, meaning (as we will see later) that the number of $\alpha_i$ is also decreased. In any case, this is not a problem in terms of the integrand; a bubble deletion means that a $\text{d}(\text{log}\alpha_i)$ is factored out of the differential form, and then upon taking a residue around $\alpha_i=0$  we get the form of the reduced diagram. A graph in which not all possible bubbles have been deleted is a \emph{reducible} graph, and they are needed in the integrands of loop amplitudes (this can be seen from the discussion in Appendix \ref{Recursion}).
\begin{figure}[!h]
~~~~~~~~\includegraphics[width=0.45\textwidth]{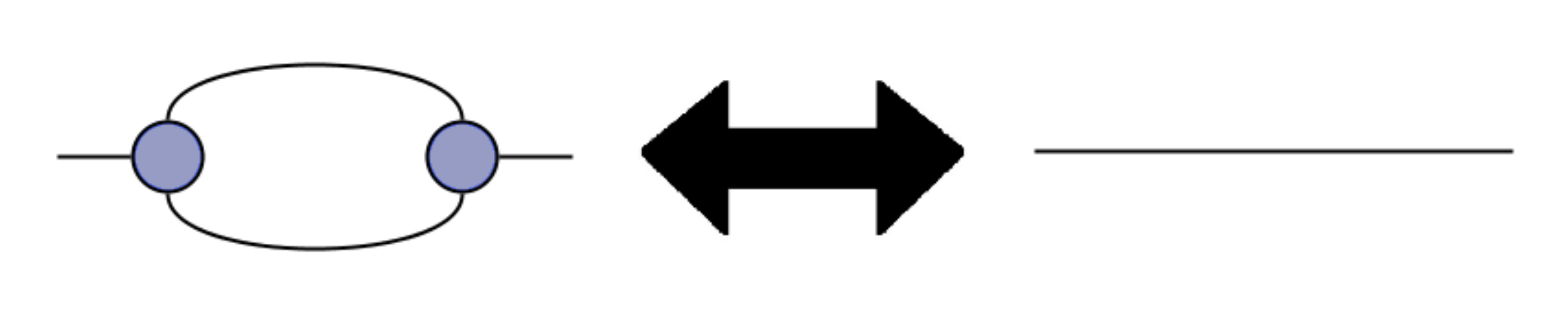} 
\end{figure}
\end{itemize}
There is a way to characterize the set of graphs with the same physical information, which is via a permutation $\sigma(a)$. This is a bijective map of the integers $1...n$ to itselfs which can be represented as $\{\sigma(1),...,\sigma(n)\}$. In any on-shell diagram/bi-colored graph, we choose any outer node $a$ (from $1$ to $n$) and start a path in which we turn \emph{left} in any white node and \emph{right} in any black node. This path will end in a possibly different outer node from $1$ to $n$, and that number will be $\sigma(a)$. It is guaranteed that if two paths have different starting point they will also have different finishing points, and viceversa, so our permutation is always well-defined for planar bicolored graphs. It is easy to see that none of the operations described above change the associated permutation. An example of the path prescription is shown in Fig.\ref{fig:paths}.

The permutation already tells us the total number of negative helicity particles $k$ of that on-shell diagram, which will correspond to the number of rows in the Grassmannian cell. If we \emph{decorate} a permutation $\sigma^*(a)$ by assigning $\sigma(a)=\sigma^*(a)+n$ when $\sigma^*(a)<a$ (as shown in the example in Fig. \ref{fig:paths}), then the number of elements bigger than $n$ in that permutation (the ``decorated" ones) will be $k$. This way, the map is not from $1...n$ into itself but to $1....2n$. The outer nodes that are decorated in the permutation will be called \emph{sources}, while the others are \emph{sinks}. Again, an example is given in Fig.\ref{fig:paths}.

\subsection{The BCFW bridge}\label{BCFWbridge}
The canonical operation from which we construct our on-shell diagrams from simpler ones is the addition of a BCFW bridge. If we have two adjacent external legs, then such a bridge corresponds to the operation
\begin{figure}[!h]
\hskip-0.7cm\includegraphics[width=0.55\textwidth]{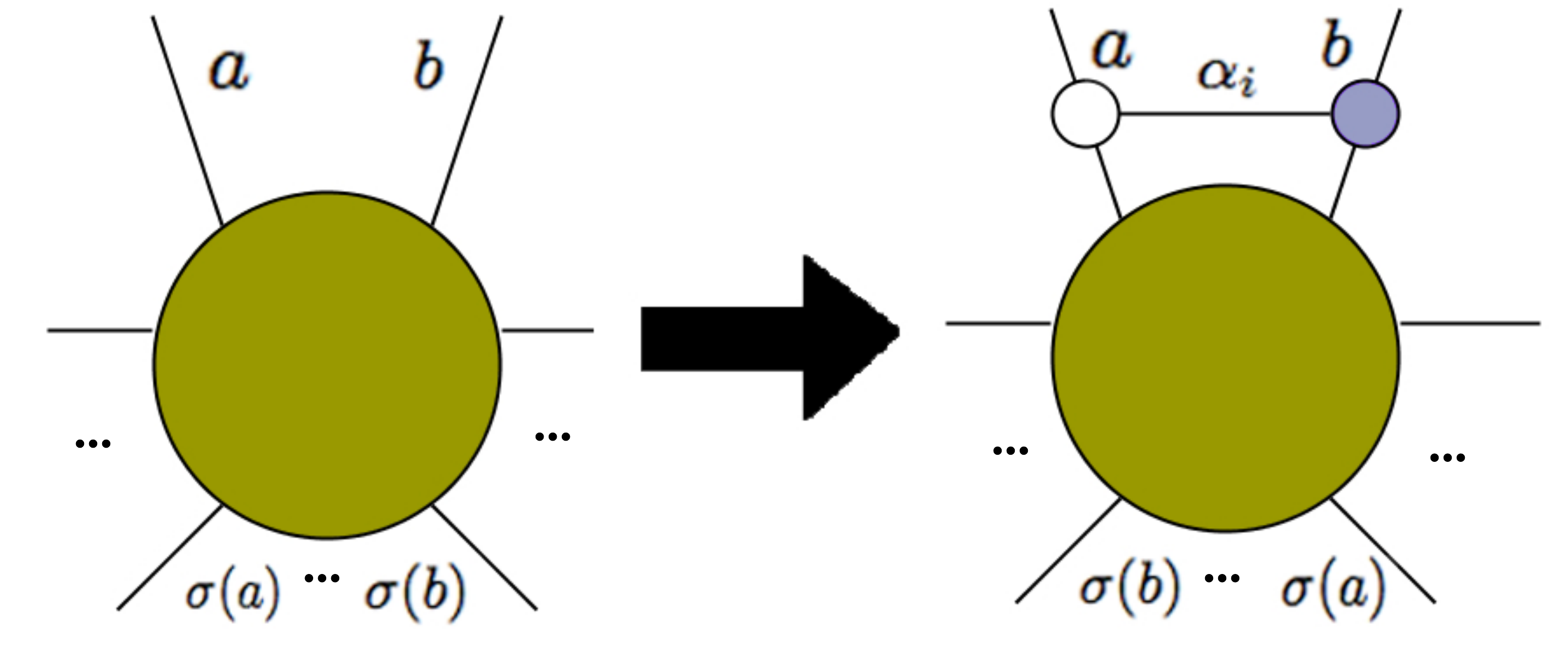}
\end{figure}
\\ \\ \\ \\ \\
which changes $\sigma(b)$ by an adjacent permutation of the two nodes involved in the bridge (we can also allow the two nodes of the bridge to have some self-identified outer nodes inbetween).
We assign a variable $\alpha_i$ to the new edge, such that the momenta entering the rest of the graphs through those two legs is changed to 
\begin{align}\label{BCFWDeform}
&\lambda_a \rightarrow \lambda_{\hat{a}}=\lambda_a  ~~~~~~\text{and}~~~~~~  \lambda_b \rightarrow \lambda_{\hat{b}}=\lambda_b+\alpha_i \lambda_a \nonumber \\ &
\tilde{\lambda}_a \rightarrow \tilde{\lambda}_{\hat{a}}=\tilde{\lambda}_a-\alpha_i \tilde{\lambda}_b ~~~~~~~~ \tilde{\lambda_b} \rightarrow \tilde{\lambda}_{\hat{b}}=\tilde{\lambda_b}
\nonumber \\
&
\tilde{\eta}_a \rightarrow \tilde{\eta}_{\hat{a}}=\tilde{\eta}_a-\alpha_i \tilde{\eta}_b ~~~~~~~~~ \tilde{\eta_b} \rightarrow \tilde{\eta}_{\hat{b}}=\tilde{\eta_b}
\end{align}
The addition of a BCFW bridge to the diagram changes the corresponding integrand $f_0$ to 

\begin{align}
&f(\ldots;\lambda_a,\widetilde{\lambda}_a,\widetilde{\eta}_a;\lambda_{b},\widetilde{\lambda}_{b},\widetilde{\eta}_{b};\ldots) \\& \nonumber
=\frac{d\alpha_i}{\alpha_i}f_0(\ldots;\lambda_{\hat{a}},\widetilde{\lambda}_{\hat{a}},\widetilde{\eta}_{\hat{a}};\lambda_{\hat{b}},\widetilde{\lambda}_{\hat{b}},\widetilde{\eta}_{\hat{b}};\ldots).
\end{align}
Which, given the form of \eqref{eq:Integrand}, corresponds to an additional $\text{d}(\text{log}\alpha_i)$ factor and a change in the \emph{columns} $c_i$ of the matrix $C$ of the form
\begin{equation} \label{eq:AdjTrans}
c_{b} \rightarrow c_{b}+(-1)^q \alpha_i c_a\,.
 \end{equation} 
 Where $q$ in the number of columns $c_s$ between the two involved in the bridge such that $\sigma(s)=s+n$ (the number of sources that are self-identified and that are between the two outer nodes involved in the construction of the bridge). We can see that if $C\in \text{Gr}^+(k,n)$ before the bridge, the operation \eqref{eq:AdjTrans} leaves this positivity intact as long as $\alpha_i>0$. For this to occur we need the extra $(-1)^q$ factor.
 \\
The fact that the bridge corresponds to an adjacent permutation in $\sigma(a)$ means we can decompose any on-shell diagram into the bridges that built it, by looking at how undoing such adjacent permutations leads to a decoration of the identity $\{1....n\}$. There is not a unique way of doing this, but for convenience here we will be using the lexicographic BCFW decomposition, which is:
\\
\begin{quote}
\textbf{Lexicographic BCFW-Bridge Decomposition:}
Given a permutation $\sigma$ (which is not a decoration of the identity, in which case the decomposition is already done) we decompose it as $(ac)\circ \sigma'$ where $1\leq a < c \leq n$ is the \emph{lexicographically-first pair}\footnote{The lexicographically first order is analogous to alphabetical order in which the numbers are the letters of the alphabet. This way, the adjacent decomposition of (23) will precede (34) or (24), but not (12).} (as chosen by our decomposition convention) separated only by self-identified legs $b$ (such that $\sigma (b)=b \, \text{mod}(n)$) and for which $\sigma(a)<\sigma(c)$, and where $\sigma '$ is the permutation with the adjacent transposition of the $(ac)$ pair ``undone''. The process is repeated until $\sigma'$ is a decoration of the identity. 
\end{quote}
This essentially amounts to an ordered way of decomposing $\sigma(a)$ into a well-defined series of adjacent permutations that lead to the identity.
Examples of this are given in Section \ref{cell} and Appendix \ref{BCFWDecomposition}, where we also show the graph construction explicitly.

A caveat is in order here. We mentioned that the bubble-deletion procedure takes loop-integrands to lower dimensional integrands by the removal of $\alpha_i$, and that this leaves the permutation invariant. This means the permutation characterizes a whole family of non-reduced graphs which have a reduced graph in common, and hence the BCFW construction procedure only produces the reduced graph given by that permutation. The main consequence for the computation of amplitudes is that, currently, loop integrands cannot be written by combinatorial means only (as they require non-reduced integrands). In any case, this will not be important for our means, as the graphs we will be dealing with are found to be always fully reduced.

\subsection{From the bipartite graph to the positroid cell}\label{cell}
We are now ready to explain the full process of how to map an on-shell diagram (a bipartite graph) to a cell of $\text{Gr}^+(k,n)$, along with the example corresponding to graph 2.3b in our list of bipartite field theories (explained in Section \ref{graphs}). The steps are:
\begin{itemize} 
\item Identify the permutation $\sigma$ associated with the graph via the zig-zag paths, and decorate it such that $k$ of its elements are bigger than $n$. In our case the graph shown below has a decorated permutation $\sigma(a)=\{2,5,6,9,7,10\}$, from an undecorated $\sigma(a)=\{2,5,6,3,1,4\}$.
\begin{figure}[!h]
~~~~~~~~~~~~~~~~~~~~\includegraphics[width=0.18\textwidth]{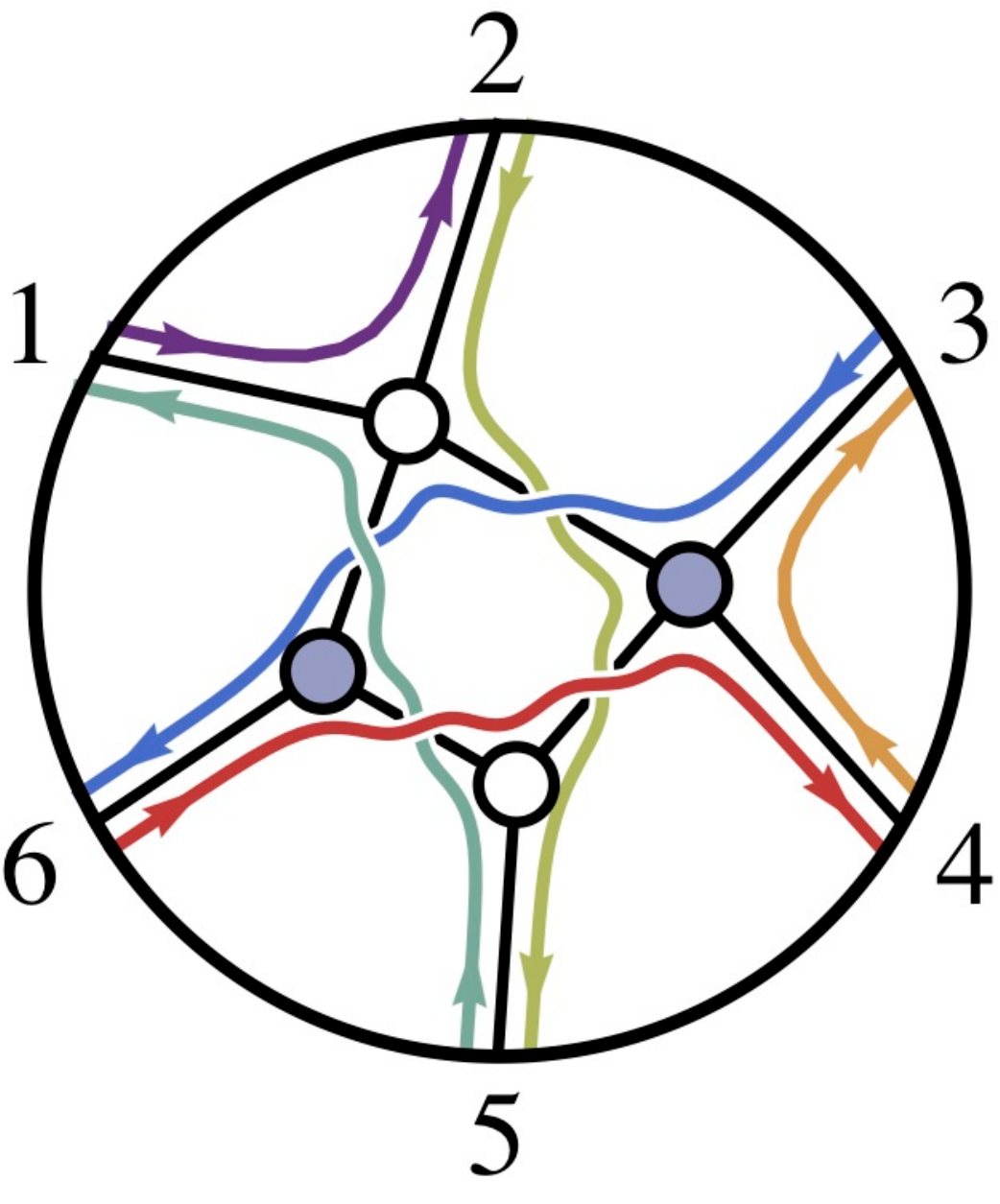}
\label{fig:example} \end{figure}
\item Decompose the permutation via the BCFW procedure to find an ordered path of adjacent permutations. These correspond to operations of the form of \eqref{eq:AdjTrans}. The result of this is a decoration of the identity $\sigma^*$ with $k$ decorated terms, each of which will be labeled by $j\in \{1...k\}$ starting from the left. In our example:
\begin{equation} \nonumber
\hskip-0.3cm\begin{array}{|c|cccccc|c|}
\hline
\text{Pair} &1&2&3&4&5&6& {} \\ 
(ac)&\downarrow&\downarrow&\downarrow&\downarrow&\downarrow&\downarrow&\text{BCFW shift} \\ \hline
\multirow{2}{*}{(1\,2)} &{\color{blue}2}&{\color{blue}5}&6&9&7&10&\multirow{2}{*}{$c_{2}\mapsto c_{2}+\alpha_{6}c_{1}$} \\
\multirow{2}{*}{(1\,3)} &{\color{blue}5}&{\color{gray}2}&{\color{blue}6}&9&7&10&\multirow{2}{*}{$c_{3}\mapsto c_{3}+\alpha_{5}c_{1}$} \\
\multirow{2}{*}{(3\,4)} &6&{\color{gray}2}&{\color{blue}5}&{\color{blue}9}&7&10&\multirow{2}{*}{$c_{3}\mapsto c_{3}+\alpha_{4}c_{3}$} \\
\multirow{2}{*}{(4\,5)}&6&{\color{gray}2}&{\color{gray}9}&{\color{blue}5}&{\color{blue}7}&10&\multirow{2}{*}{$c_{5}\mapsto c_{5}+\alpha_{3}c_{4}$} \\
\multirow{2}{*}{(1\,4)} &{\color{blue}6}&{\color{gray}2}&{\color{gray}9}&{\color{blue}7}&{\color{gray}5}&10&\multirow{2}{*}{$c_{4}\mapsto c_{4}-\alpha_{2}c_{1}$} \\
\multirow{2}{*}{(4\,6)} &{\color{gray}7}&{\color{gray}2}&{\color{gray}9}&{\color{blue}6}&{\color{gray}5}&{\color{blue}10}&\multirow{2}{*}{$c_{6}\mapsto c_{6}+\alpha_{1}c_{4}$} \\
{} &{\color{gray}7}&{\color{gray}2}&{\color{gray}9}&{\color{gray}10}&{\color{gray}5}&{\color{gray}6}&{} \\
\hline
\end{array}
\end{equation}

Here the permutated pairs are highlighted in blue and the elements of the decorated identity are in gray. Our decoration of the identity is $\sigma^*=\{7,2,9,10,5,6\}$.

\item Given $\sigma^*$, we write a $k \times n$ matrix whose only non-zero entries are ones at the $i$-th column and $j$-th row, where $i$ is such that $\sigma^*(i)=n+i$ and $j$ is as was just defined. This is
\small{\begin{equation}\left(
\begin{array}{cccccc}
 1 & 0 & 0 & 0 & 0 & 0 \\
 0 & 0 & 1 & 0 & 0 & 0 \\
 0 & 0 & 0 & 1 & 0 & 0 \\ 
\end{array}
\right).\nonumber
\end{equation}}

\item Starting with this matrix, we perform the list of operations of the BCFW decomposition in reverse order, such that our first one would be $c_{6}\mapsto c_{6}+\alpha_{1}c_{4}$ and the last one $c_{2}\mapsto c_{2}+\alpha_{6}c_{1}$. This yields the final $C$ matrix we use in our integrand \eqref{eq:Integrand}, which is
\small{\begin{equation}\left(
\begin{array}{cccccc}
 1 & \alpha_6 & \alpha_5 & -\alpha_2 & -\alpha_2 \alpha_3 & 0 \\
 0 & 0 & 1 & \alpha_4 & 0 & 0 \\
 0 & 0 & 0 & 1 & \alpha_3 & \alpha_1 \\
\end{array}
\right) \nonumber.
\end{equation}}
\end{itemize}
It must be said, however, that the relation between the matrix and the permutation is much deeper than this procedure suggests. As we explain in Appendix \ref{permutation}, the permutation encodes the mutual dependence of the adjacent columns $c_i$. 

From this construction we can see that our integrand will have $d$ $\alpha_i$ variables, where $d+1$ is the number of faces of the graph. This number $d$ will be the \emph{dimension} of the cell $C$. But the $\delta$-functions of the integrand \eqref{eq:Integrand} only impose $2n-4$ constraints, given that we have $2n$ $\delta$'s minus the 4 we need for conservation of momentum (amplitudes always have a factor of the form $\delta^4(\sum P_i^\mu)$). This means that our integrand only yields a leading singularity of a tree amplitude in cases where $d=2n-4$. If  $d<2n-4$ we have that there are constraints on the external momenta of particles, and if $d>2n-4$ not all of the $\alpha_i$ are specified. In the latter case these extra degrees of freedom may also correspond to the loop momenta in the amplitudes, which need to later be integrated over.

Hence, in general, the correct statement is that on-shell diagrams map not to leading singularities but to integrands only, all of which are in any case part of the general picture of positive geometry. The integral from such integrands will then in most cases be a purely formal object, much more work is needed to compute obsevable amplitudes (some of which is explained in, among others, \cite{AllLoop,NAH,ScattAmpli,IntoAmplituhedron}). In our case, when computing our graphs from BFTs we find that only exceptionally do these map to leading singularities.

With this construction method and the rest of the explanations above we have only scratched the surface of the depths of positive geometry and planar bicolored graphs. Appendices \ref{Positive}, \ref{permutation}, and \ref{BCFWDecomposition} expand on these issues, and the more thorough explanation of positive geometry and plabic graphs can be found in \cite{NAH} and the references therein. In any case, we now have enough tools to compute the bipartite graphs from the BFTs; in the next section we proceed to explain where the ones considered here are taken from.
\section{Bipartite field theories}\label{Bipartites}
Following the AdS/CFT conjecture \cite{ADS/CFT}, the quest for conformal field theories with well-defined AdS duals led to the study of \emph{quiver gauge theories}, with $\mathcal{N}=1$ \cite{Quiver1}. These were found to have some remarkable combinatorial properties \cite{BraneTiling1} that were better undestood in terms of \emph{brane tilings}, which are bipartite graphs defined on a 2D torus \cite{BraneTiling2,BraneTilingReview} (unlike the planar ones we have so far been dealing with). This class is now understood as a subset of the \emph{bipartite field theories} \cite{FrancoBipartite,Franco2014}, in which the study of CFTs defined through a graph has been extended to Riemann surfaces beyond a torus \cite{FrancoDirections,FrancoGeometry,DBranes}. In these theories, the graph defines the Lagrangian interaction terms of the field theory by the following rules:
\begin{itemize}
\item{Internal faces:} each face is associated a $SU(N)$ gauge group.
\item{Internal edges: each edge is assigned a certain \emph{chiral multiplet} (a supersymmetric field representing all possible helicities). These will have a local gauge symmetry given by the product of the groups associated to the two faces it separates.}
\item{Nodes: each node will be a superpotential term in the Lagrangian given by the product of all the fields whose edges have an endpoint in that node. We assign plus signs to white nodes and minus signs to black nodes.}
\end{itemize}
Graphs with boundaries, such as planar ones in a disk, need extra care and have additional procedures to account for the external faces and nodes \cite{FrancoDirections}. The full development of such rules for boundaries will be a key in understanding the relevance of positive geometry (where currently only planar graphs with boundaries are understood) to the general BFTs. In any case, the rules above are all that is needed for a brane tiling to define a conformal field theory.

An algorithm to generate all possible such brane tilings was presented in \cite{GraphClassification}, where a full list of theories with at most 6 superpotential terms is produced (in the language of graphs, a total of 6 nodes). Here we take all the consistent graphs from that list  and look at them in terms of the positive geometry explained in Section \ref{positroid} by constructing their corresponding positroid cells. The full list of brane tilings and their cells is shown in Appendix \ref{list}.
For example, one of the brane tilings directly taken from that work, listed as 2.3, is
\begin{figure}[!h]
~~~~~~~~\includegraphics[width=0.25\textwidth]{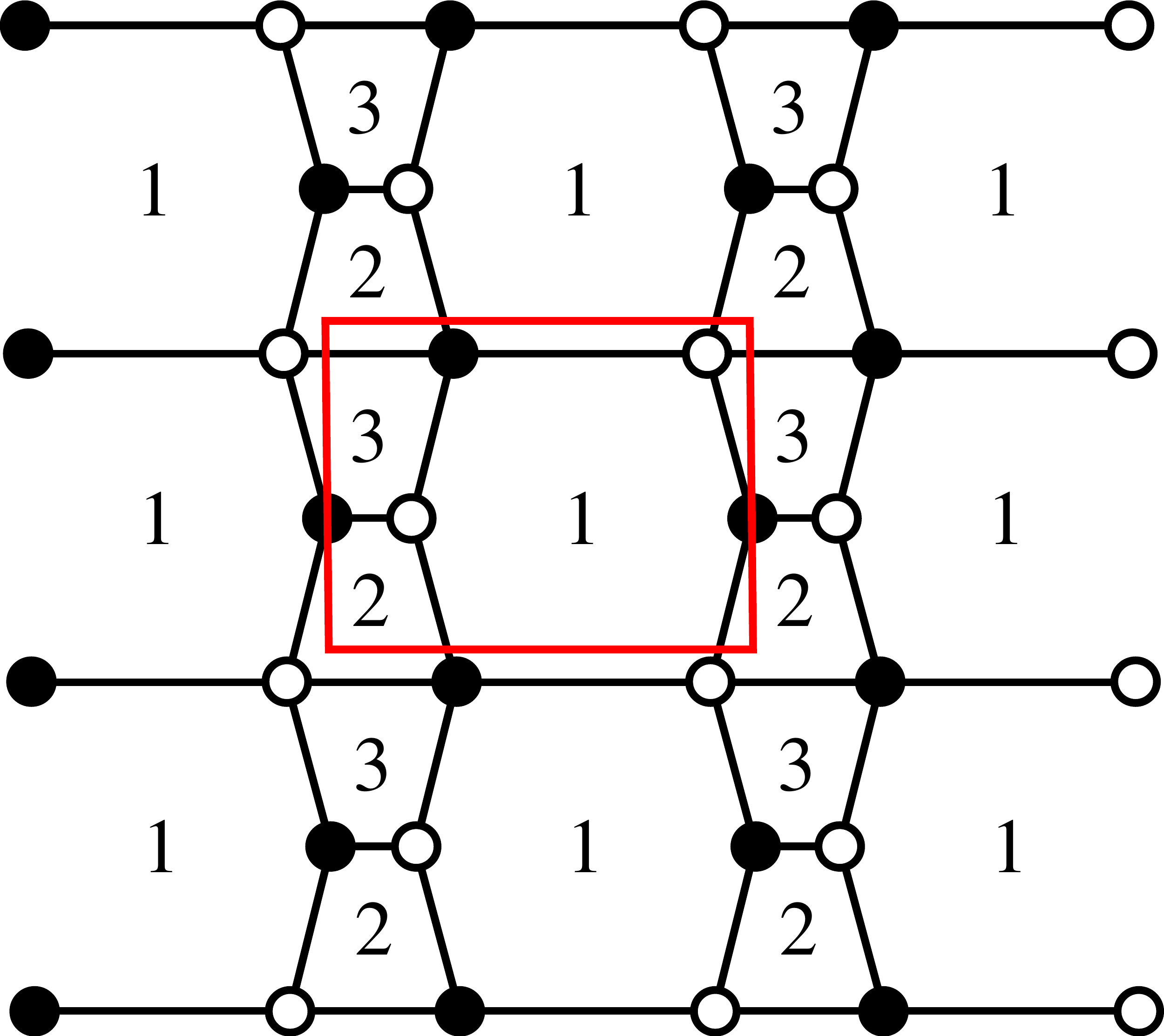},
\end{figure}

\noindent with a superpotential given by 
\begin{align}\label{Superpotential}
H_I=-X_{12}^1 . X_{23}^2 . X_{31}^1+X_{12}^2 .X_{23}^2.X_{31}^2 \\ \nonumber
+\phi_1.X_{12}^1.X_{23}^1.X_{31}^1-\phi_1.X_{12}^2.X_{23}^1.X_{31}^2.
\end{align}
Where the $X$ and $\phi$ are the 4 different superfields (each with its particular gauge symmetries) and each term is given by one of the nodes in the fundamental region.
We see that the relevant part of the graph is defined as the fundamental region (the red square), and the toric embedding is represented by writing it within a lattice of such regions. Going to the end of a fundamental region in a 2D torus takes you directly to the opposite end. If we write it as a lattice, instead of going to that opposite end we reach the opposite end of the next fundamental region; the two spaces (lattice and torus) are hence mathematically equivalent.
 
The main reason why it is believed that the positive Grassmannian framework may also be important here is because the operations described in \ref{OnShell} also have a clear meaning in the context of these theories. Namely, a square move links a BFT with its Seiberg dual\footnote{Seiberg duality is a remarkable property of $\mathcal{N}=1$ theories which makes pairs of them equivalent at low energies.} \cite{Seiberg}, and the merging of adjacent vertices is related to integrating out mass terms of the Lagrangian (essentially the opposite procedure of a Higgs symmetry breaking). The bridge removal operation also has a natural understanding in the BFTs, as it is briefly described in Appendix \ref{Positive}.
All these coincidences may indicate that, at least for planar BFTs with boundaries, a form of an associated permutation may be useful in their classification, and that positive geometry may play an important role in their future understanding. Also, the procedures described above for generating Grassmannian cells are currently being extended to non-planar graphs \cite{FrancoGeometry,ToricGeometry}, and it is hoped that such new mathematics will generate progress in the BFTs side of the story.

A natural thing to do from here is hence to map these brane tilings into on-shell diagrams and see how much information can we extract from their positive geometry. This is the subject of the next section. 
\section{The positive geometry of brane tilings}\label{graphs}
\subsection{Generation of the diagrams from the tilings}\label{Generation}
The present difficulty in mapping brane tilings and on-shell diagrams is that while the former (at least, the ones considered here) are defined on a torus, the relevant graphs in the context of the positive Grassmannian are planar and with boundaries. This means there is an ambiguity in translating the fundamental region from ones to others, and the map from brane tilings to corresponding on-shell diagrams is not well-defined. Indeed, on-shell diagrams must at least have a boundary to have any physical meaning at all, which we need to artificially assign to brane tilings. For the example above, one can place the fundamental region such that it contains either
\begin{figure}[!h]
\includegraphics[width=0.25\textwidth]{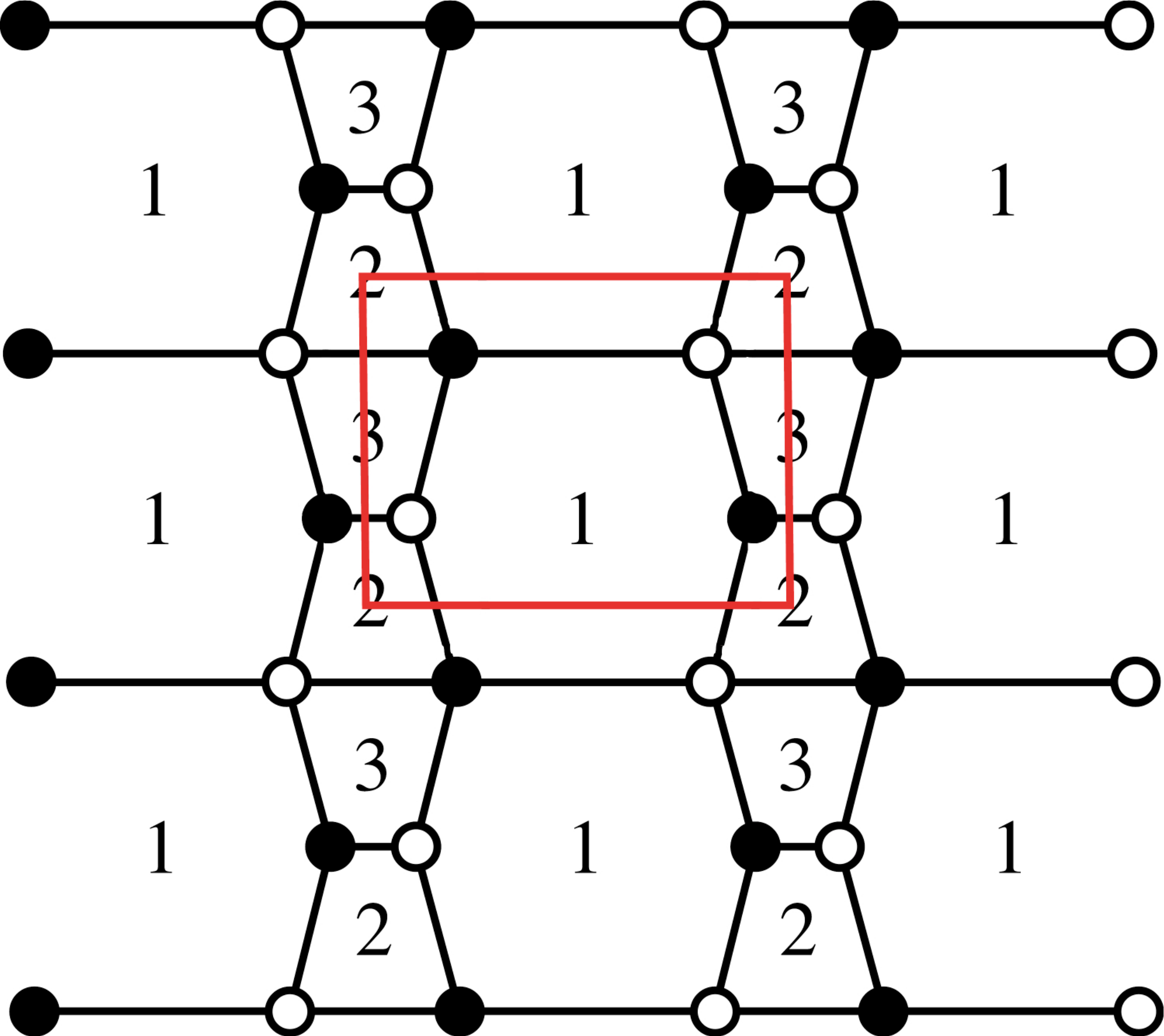}~~
\includegraphics[width=0.2\textwidth]{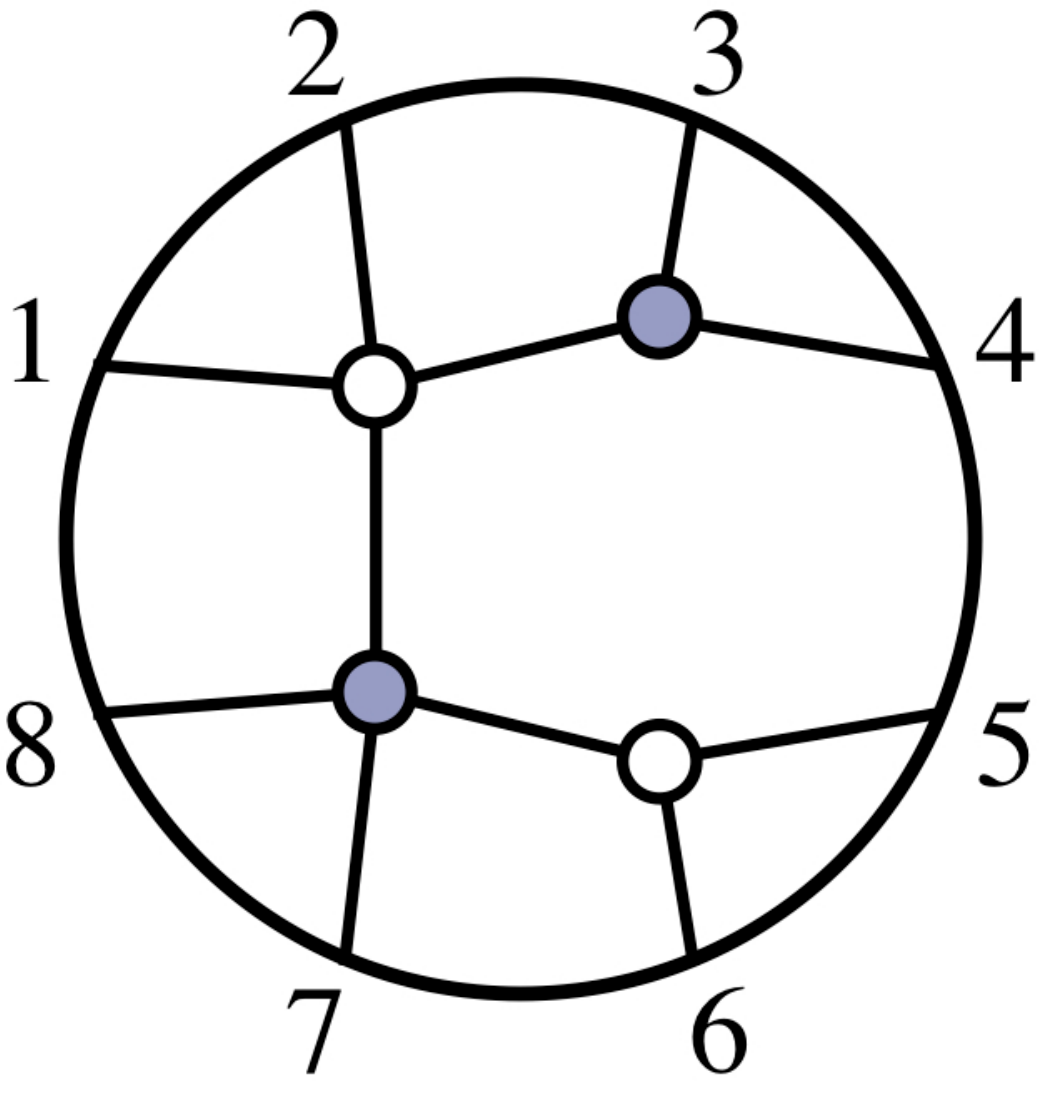}
\end{figure}
\\ \\
With permutation $\{2,4,8,11,6,9,13,15\}$ and positroid cell
\begin{equation}
\tiny{\left(
\begin{array}{cccccccc}
 1 & \alpha_7 & \alpha_6 & 0 & -\alpha_3 & -\alpha_3 \alpha_4 & 0 & 0 \\
 0 & 0 & 1 & \alpha_5 & 0 & 0 & 0 & 0 \\
 0 & 0 & 0 & 0 & 1 & \alpha_4 & \alpha_2 & 0 \\
 0 & 0 & 0 & 0 & 0 & 0 & 1 & \alpha_1 \\
\end{array}
\right)},
\end{equation}
or, by moving the fundamental region in a different way
\begin{figure}[!h]
\includegraphics[width=0.25\textwidth]{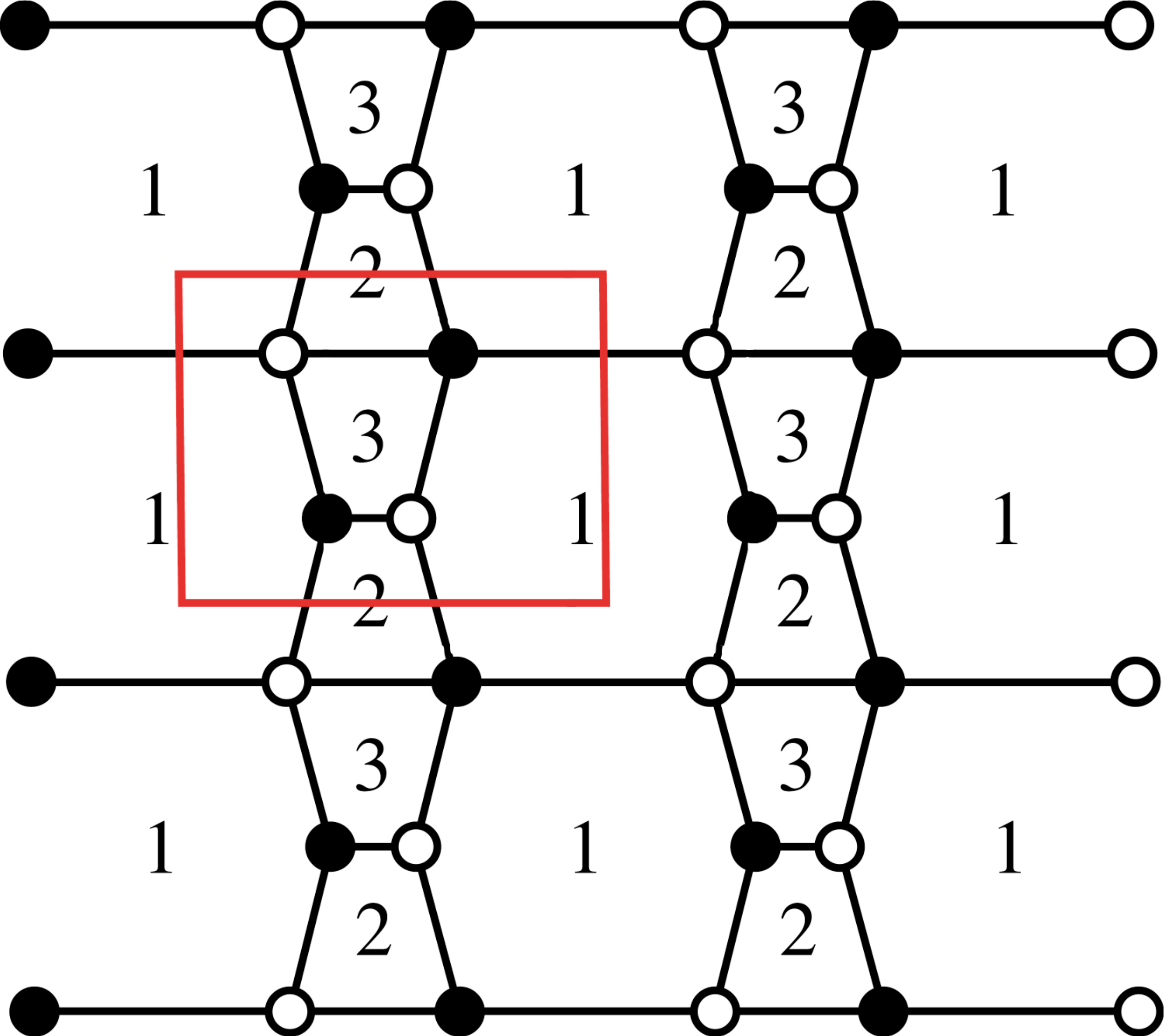}~~
\includegraphics[width=0.2\textwidth]{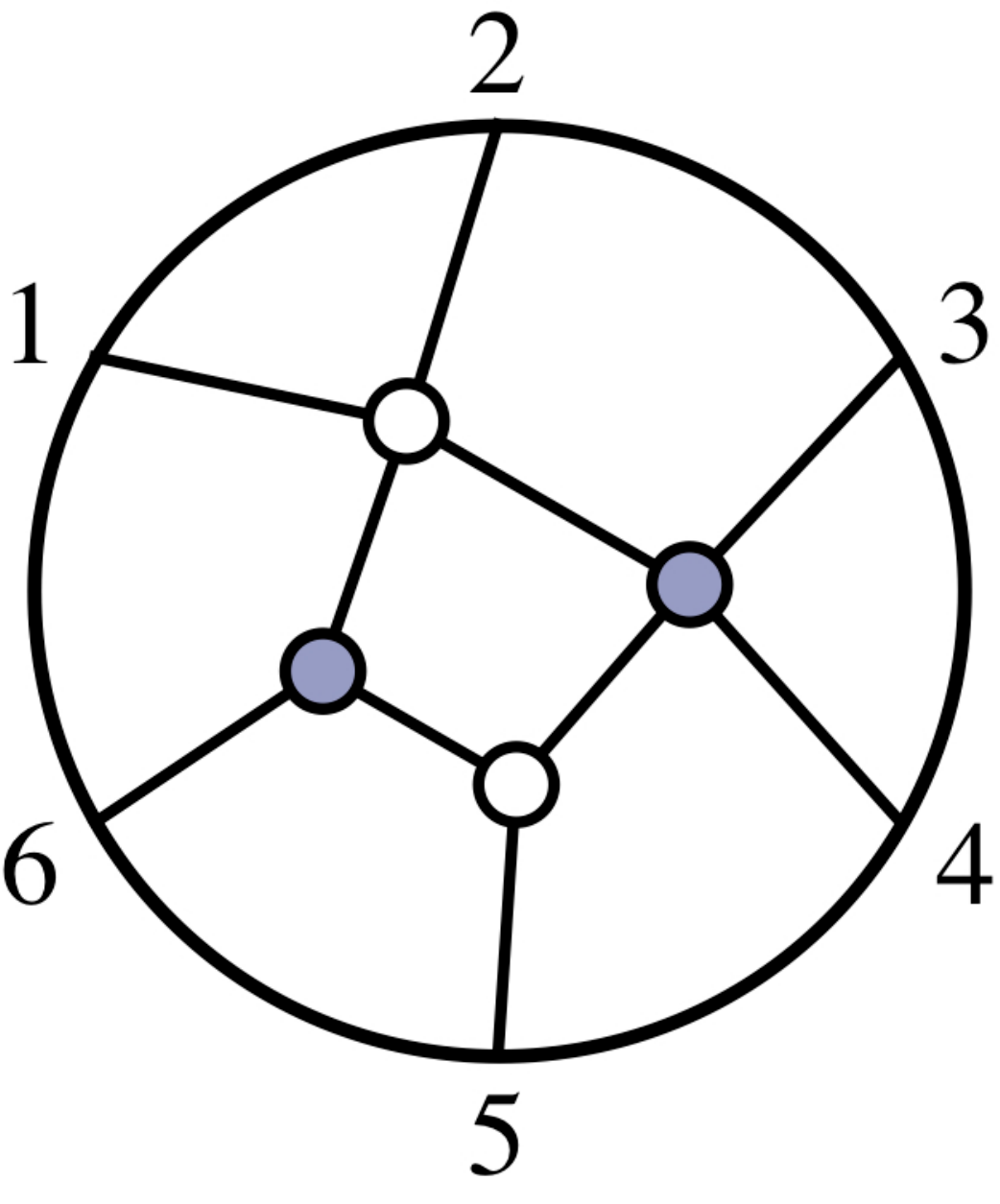},
\end{figure}

with permutation $\{2,5,6,9,7,10\}$ and cell
\begin{equation}
\small{\left.\left(
\begin{array}{cccccc}
 1 & \alpha_6 & \alpha_5 & -\alpha_2 & -\alpha_2 \alpha_3 & 0 \\
 0 & 0 & 1 & \alpha_4 & 0 & 0 \\
 0 & 0 & 0 & 1 & \alpha_3 & \alpha_1 \\
\end{array}
\right)\right. .}
\end{equation}
This is a particularly illustrative example because, even if these two on-shell diagrams come from the same brane tiling, they appear to be completely different. They do not have the same $n$, nor the same $k$, nor the same number of dimensions. One caveat we must mention is that there is not a fixed way of assigning the labels $\{1....n\}$ to the outer edges (there is a cyclic redundancy), but for consistency we will choose them such that the final permutation is the lexicographically minimal.

For this case it is easy to see there are no more possibilities, but, as the graphs complicate, the number of different cells we can find increases significantly. Beyond the method of moving the cell boundary around the graph, spanning all the ways for the nodes to combine within it (as we have shown in this example), we have so far been unable to find an algorithm to produce all the possible ways of making this translation. 

In any case, a list with all the consistent graphs in \cite{GraphClassification} with the different on-shell diagrams it can contain (if not all, at least the ones we have been able to find), and with their associated permutations and cells, has been produced and can be found in Appendix \ref{list}. Although the list is not exhaustive, we hope the data presented here will help to develop the link between BFTs and scattering amplitudes.

\subsection{Features of the diagrams}\label{Features}
The different graphs from each brane tiling have some common features. It is easy to see from the ways in which the boundary of our diagrams are defined that, if $n_I$ is the number of internal edges and $n$ the external ones (also, the number of particles and number of columns of $C$), $n+2n_I$ is the same for all diagrams coming from one given tiling. We also have Euler's identity for planar graphs, 
\begin{equation}
n_F+n_V=1+n_I+n,
\end{equation}
where $n_V$ and $n_F$ are the number of vertices and of faces, and the fact that the number of faces  is related to the dimension of the cell, ($n_F=d+1$). Putting them together we get that the number 
\begin{equation}
d+n_V+n_I=n+2n_I
\end{equation}
is also the same for all diagrams generated from the same tiling. All these, by construction, will have the same number of vertices, and hence, the number $d+n_I$ is also fixed. This can be easily checked in Appendix \ref{list}. For instance, in our example 2.3 one graph has
\begin{equation}
\begin{array}{V c}
\includegraphics[width=0.15\textwidth]{23}
& \,\,\,\,\,\,\,\,\,\,\,\,\, \begin{array}{c} n=8\\n_I=3\\n_V=4\\d=7\\d+n_I=10 \end{array}
\end{array},
\end{equation}
and the other ones has 
\begin{equation}
\begin{array}{V c}
\includegraphics[width=0.15\textwidth]{23b}
& \,\,\,\,\,\,\,\,\,\,\,\,\, \begin{array}{c} n=6\\n_I=4\\n_V=4\\d=6\\d+n_I=10 \end{array}
\end{array}.
\end{equation}
These similarities, however, are only at the level of the graphs. The permutations represented by the different diagrams do not in general have any apparent common features. The arbitrariness in the generation of the boundary means the zig-zag paths also have arbitrary endpoints, and this is reflected in the fact that their positive geometry is quite different (different structure and dimensions of $C$). 

As a way through this, one could think of defining such zig-zag paths to live in the 2D torus of the tilings in a way in which they will have no endpoints, just as loops within the torus. It is currently known that such paths are related to the \emph{moduli} space of the corresponding field theories \cite{FrancoBipartite} (a space related to the string theory embedding of such theories, which is in this case a Calabi-Yau 3-fold), but beyond this it is unclear what further lessons can be learnt from them. Along these lines, an idea to explore could be whether these paths on the torus can be mapped to some algebraic structure like planar graphs map to cells in $\text{Gr}^+(k,n)$. This would perhaps require the development of new mathematics beyond the positive Grassmannian.

Another general feature of all the diagrams listed is that they do not in general have enough loops to generate just a leading singularity (meaning that the dimension of the cell $d<2n-4$), so they represent singularities that impose constraints in the external momenta of the particles. For example, the simplest diagram (1.1), given by 

\begin{equation}
\begin{array}{V c}
\includegraphics[width=0.15\textwidth]{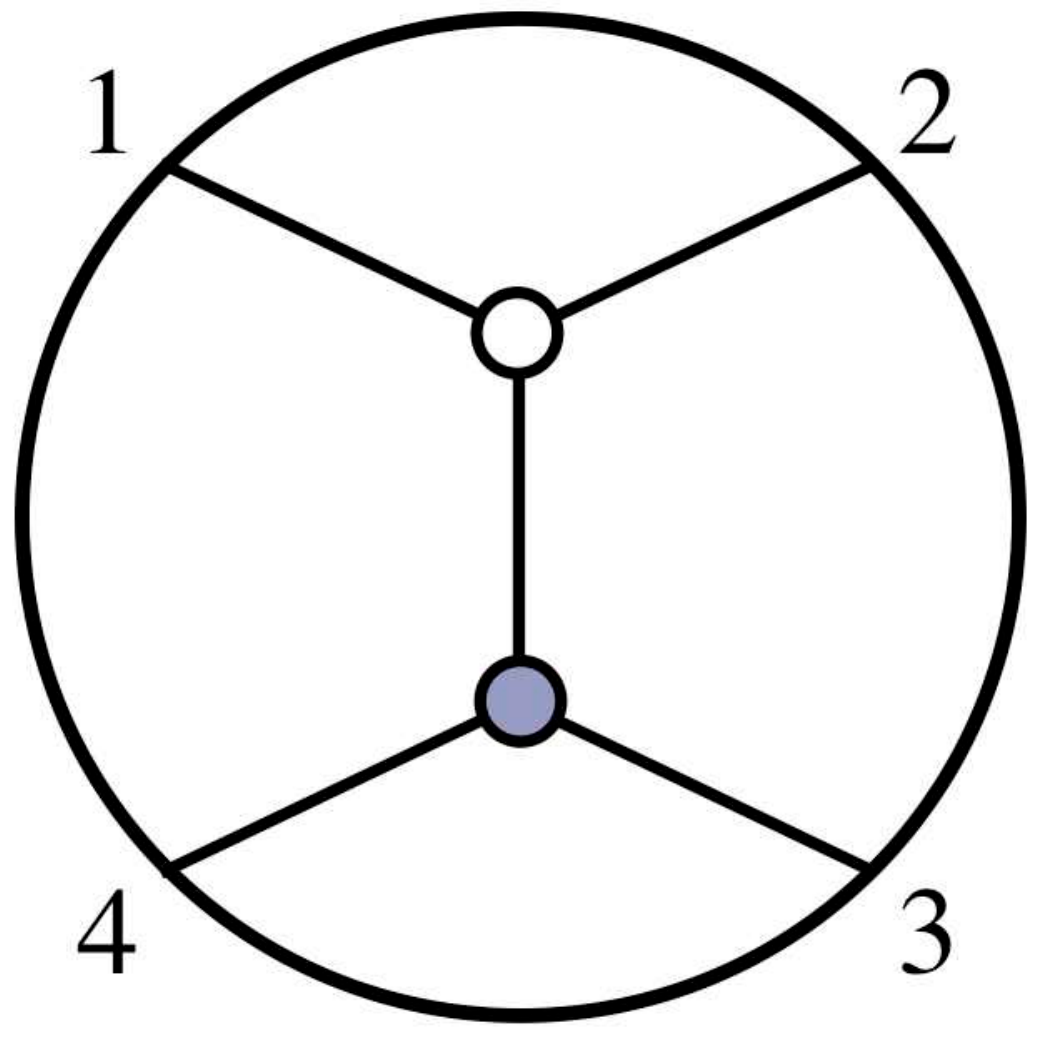} &\,\,\,\,\,\,\,\,\,\,\,\, \small{\left(
\begin{array}{cccc}
 1 & \alpha_3 & \alpha_2 & 0 \\
 0 & 0 & 1 & \alpha_1 \\
\end{array}
\right)},
\end{array}
\end{equation}

\noindent represents a leading singularity with $k=2$ and $n=4$ that, because it over-determines the $\alpha_i$, produces a constraint on external momenta. As mentioned in Section \ref{OnShell}, to get the actual 4-particle tree amplitude one needs the leading singularity generated from
\begin{equation}
\begin{array}{V c}
\includegraphics[width=0.15\textwidth]{n4k2} &\,\,\,\,\,\,\,\,\,\,\,\, \small{\left(
\begin{array}{cccc}
 1 & \alpha_2+\alpha_4 & \alpha_2 \alpha_3 & 0 \\
 0 & 1 & \alpha_3 & \alpha_1 \\
\end{array}
\right)},
\end{array}
\end{equation}
\noindent which has one face (hence one more dimension), meaning $2n-4=d=4$. This adds a further difficulty in our identification of brane tilings with loop and tree amplitudes: in general, they are not diagrams relevant to actual physical processes. 

Related to this there is a further common feature of brane tilings: all the graphs are fully reduced, in the sense that no bubbles can be deleted from them. This can be seen for our diagrams in Appendix \ref{list} by checking that in those with more than one loop (graphs with one or no loops are straightforwardly reduced) cannot be made to have any bubbles via square moves and merges. We are able to state this because the BCFW construction procedure outlined here allows only for reduced graphs, and we observe that none of the diagrams listed appear to be any different (in particular, with less loops) than their brane tiling counterparts from \cite{GraphClassification}. Heuristically, the main reason for this feature is the low number of loops our diagrams have.

\section{Conclusions and outlook}
In the report we showed how to map bipartite graphs to cells of $\text{Gr}^+(k,n)$, and we explained how these cells appear in the integrand of the amplitudes of $\mathcal{N}=4$ SYM to any loop order. Here, we use this map to associate cells to the fundamental regions of brane tilings, for which we tabulate the possible moves of each fundamental region. We have seen that the on-shell diagrams produced, when coming from the same tiling, have some common features such as the same number of dimensions and internal lines, and that these are generally too low-dimensional to produce a leading singularity. We hope the data in Appendix \ref{list} will be useful in establishing a meaningful correspondence between BFTs and amplitudes.

In this report we have attempted to explore this connection through the direct identification of theories with on-shell diagrams, but in order to proceed further we first need to understand:
\begin{itemize}
\item \emph{i)} how to properly generate boundaries for BFTs without it, as it is our case, and how to make physical sense of such a procedure. 
\item \emph{ii)} what does it mean, physically, to have a correspondence between a field theory and a scattering amplitude in a different theory. For example, it is obvious that arbitrarily complicated graphs are of relevance in scattering amplitudes of many-particle collisions \cite{NAH}, but it is less clear of what relevance arbitrarily complicated gauge theories are. 
\end{itemize}

We mentioned that non-reduced graphs are necessary in the loop integrands (indeed, the removed $\alpha_i$ variables can represent the loop momenta), but our identification of graphs in terms of permutations makes no distinction between reduced and non-reduced diagrams (given that bubble deletion leaves it invariant). This shows there is still work to do in the classification of non-reduced diagrams, as this limitation currently means the all-loop integrand cannot yet be determined by purely combinatorial methods \cite{NAH}. In connection to BFTs, it would be interesting to see whether more complicated theories than the ones listed here, potentially with non-reduced graphs, may find the combinatorial description of non-reduced graphs useful \cite{FrancoBipartite}.

The latest developments have led to a purely geometrical understanding of amplitudes in planar $\mathcal{N}=4$ SYM \cite{Amplituhedron,IntoAmplituhedron}. There are a number of ways to keep advancing in this direction. We have already mentioned the extension of this framework beyond the planar limit, for which, motivated by BFTs, some small progress has been made \cite{FrancoGeometry}, despite the fact that the non-planar sector does not have the very important dual superconformal symmetry. An indication on how this may proceed is the fact that it has been shown that toric geometry is a very natural tool to study the topology of the positive Grassmannian \cite{Postnikov2,ToricGeometry,FrancoGeometry}, as the positroid cells can be mapped to certain toric varieties.
Further lines of work, as suggested in \cite{NAH}, may aim to implement positive geometry in amplitudes beyond $\mathcal{N}=4$. In particular, amplitudes of theories with less to no supersymmetry can be represented by on-shell diagrams in which we define a \emph{perfect ordering}, where the edges are assigned a direction representing the flow of helicities between the 3-point amplitudes of the diagram.

In the end, the aim is to obtain a general way of computing scattering amplitudes that makes no reference to locality or unitarity, but makes other symmetries manifest (in the present case, the Yangian symmetry). The ultimate motivation of this program \cite{Amplituhedron,SimplestQFT} is that we may be able to calculate local observables (amplitudes) without making any reference to space-time, emergent through the final structure of such observables. Such a dual formulation would make no reference to concepts like Hilbert space, Lagrangians, path integrals,...and may be a hint towards a deeper theory of nature in a way similar to how the variational principle formulation of classical mechanics is closer to path-integral quantum mechanics than Newton's laws are \cite{Amplituhedron,SUSY2013}.

The BFTs may play an important role in this story. The fact that the the same combinatorics appear in both places through bipartite graphs, and, specially, the fact that BFTs have a degree of SUSY ($\mathcal{N}=1$) that may be described through on-shell diagrams \cite{NAH} make them candidates for being next in the line for their amplitudes to be understood geometrically. However, much progress is needed in elucidating how realistic this possibility is.

In any case, one does not have to go that far into the connection. It may be that there is not a strong physical link between amplitudes in $\mathcal{N}=4$ SYM and BFTs, but the new mathematics brought up by the study of those amplitudes, starting with the positive Grassmannian, may still be relevant in the analysis and classification of the BFTs. This is the line we mainly want to push with this work, where we explored a possible way of identifying brane tilings and cells of the Grassmannian. 

The main difficulty of the analysis here is that the graphs relevant for scattering amplitudes always have a boundary, while the brane tilings do not. This is one of the motivations that has led to the development of the more general BFTs \cite{FrancoBipartite,Franco2014} (for graphs with boundaries), but the opposite line of development, the mathematics for graphs with no boundaries to help the understanding of brane tilings, are yet to be envisioned (were they possible at all). As mentioned in the previous section, one could hope, for example, that some algebraic structure similar to the positive Grassmannian may be linked with these, the way in which the positive Grassmannian links to BFTs on a disk. The coincidences explained at the end of Section \ref{Bipartites}, that square moves, merges, and edge deletion, have a meaning in both realms, may by themselves justify the pursuit of such aims.

There is still a lot of work to do if we fully want to unveil the role that BFTs play in the amplitudes research program and viceversa, but there are enough reasons to believe that insightful results will come with it. 

\appendix

\section{Supersymmetry and superamplitudes}\label{SUSY}
For supersymmetric theories with maximal supersymmetry, we can define a \emph{superspace} with coordinates $\tilde{\eta}_A$ with the $A$ going from 1 to $\mathcal{N}$ (4 in this case). These variables are said to be \emph{fermionic} as opposed to standard \emph{bosonic} spacetime coordinates because the generators of the Lie group follow anti-commutation relations instead of commutation ones. If we label the generators $Q^A$ and their Hermitian conjugates $\tilde{Q}_A$, these are
\begin{equation}
\{Q_\alpha^A,\tilde{Q}_{\dot{\beta}B}\}=2 \sigma_{\alpha \dot{\beta}}^\mu P_\mu \delta^A_B.
\end{equation}
Where $\sigma_{\alpha \dot{\beta}}^\mu$ are Pauli matrices, $P_\mu$ is an associated 4-momentum and $\delta_B^A$ is a Kronecker delta. Given this, the superspace transformations act on states as $U=\text{exp}(i \tilde{\eta}_I Q_I)$, and supersymmetric Lagrangians are invariant under these transformations. 
The effect of the SUSY generators on the different fields and one-particle states is that of transforming them into a lower-helicity one-particle state. Given that we have 4 of them in $\mathcal{N}=4$, if we start with a fundamental positive helicity gluon $g^+$ denoted as $\ket{\Omega}$, then our theory will have four gluinos $\lambda^A$ of helicity $(+1/2)$ (for each of the $A$ indices), 6 scalars $S^{AB}$ (for each combination $AB$), four $(-1/2)$ gluinos $\lambda^{ABC}$ and 1 negative helicity gluon $g^-$.

These are hence all the possible particles that can arise in our scattering processes. It turns out that SUSY enables us to write all different 16 states into a single \emph{chiral superfield} as an expansion of the fermionic variables $\tilde{\eta}_A$. This is
\begin{align}
\Omega=&g^++\tilde{\eta}_A \lambda^A-\frac{1}{2!}\tilde{\eta}_A\tilde{\eta}_BS^{AB} \nonumber \\ &-\frac{1}{3!}\tilde{\eta}_A\tilde{\eta}_B\tilde{\eta}_C \lambda^{ABC} + \tilde{\eta}_1\tilde{\eta}_2\tilde{\eta}_3\tilde{\eta}_4 g^-.
\end{align}
These are the superparticles that will be outgoing in our amplitudes, and the fact that we can write it like this means we do not have to specify helicities when writing our on-shell diagrams. These $\tilde{\eta}_A$ variables then show up for the amplitudes in $\delta$-functions as shown in \eqref{eq:Integrand}, and the number of these delta functions ($4\times k$) determines the total number of negative helicity states $k$, which is also the number of rows in the Grassmannian. For example, $k=2$ corresponds to the lowest non-zero amplitude, the MHV or \emph{maximally helicity violating}. From that amplitude we can extract the amplitude with all $g^+$ and 2 $g^-$, but we can also extract the amplitude for all $g^+$ except 4 scalars, or any other combination, provided that the total sum of the helicities is the same. The way this proceeds is via an integral. If we have the $n$-particle MHV amplitude $\mathcal{A}_n^{\text{MHV}}(\lambda,\tilde{\lambda},\tilde{\eta})$, then, for example, to get the amplitude where particles $i$ and $j$ are negative helicity gluons we integrate the following way
\begin{align}
\mathcal{A}&(g_1^+...g^-_i...g^-_j...g^+_n)=\\ & \nonumber \int \prod_{a=1,\,a \neq i,j}^n \text{d}^4 \tilde{\eta}_a \tilde{\eta}^4_a \int  \text{d}^4 \tilde{\eta}_i \text{d}^4 \tilde{\eta}_j \mathcal{A}_n^{\text{MHV}}
\end{align}
where the $\text{d}^4 \tilde{\eta}_a$ are the differentials over the 4-dimensional superspace of each of the $n$ particles. Any other combination of helicities will then include a different product of $\tilde{\eta}_a$ in the integrand.

\section{Recursion relations, the tree-level BCFW scheme and the loop integrand}\label{Recursion}
As mentioned in Section \ref{ScatteringAmplitudes}, a number of methods have been developed that help compute amplitudes from lower point ones. In these on-shell approaches, the idea is to explote the analytic properties of the amplitudes with the external momenta deformed and made complex. The most famous of them are the BCFW recursion relations at tree level \cite{BCFRecursion,BCFW}, which we proceed to outline, following \cite{ScattAmpli}.
The deformation of momenta entails the introduction of a set of $n$ complex-valued vectors $r_i^\mu$ added to the external momenta
\begin{equation}\label{Deformation}
\hat{p}_i^\mu \equiv p_i^\mu + z r_i^\mu,
\end{equation}

\noindent where $z \in \mathbb{C}$ and $r_i^\mu$ is a set of complex-valued vectors  such that

\begin{itemize}
\item $\sum_{i=1}^n r_i^\mu =0 $
\item $r_i \cdot r_j=0$ for all $i,j$, including $i=j$
\item $p_i \cdot r_i=0$ for each $i$
\end{itemize}

Given this definition, this set has three important properties
\begin{enumerate}
\item Momentum conservation still holds: $\sum_{i=1}^n\hat{p}_i^\mu=0$.
\item Shifted momenta are on-shell: $\hat{p}_i^2=0$.
\item If we define a non-trivial subset of momenta $\{p^\mu_i\}_{i \in I}$, and $P_I^\mu=\sum_{i \in I} p_i^\mu $ then (defining $R_I=\sum_I r^\mu_i$)
\begin{equation}
\hat{P}^2_I=P_I^2+z2P_I \cdot R_I
\end{equation}
And hence, if $z_I=-\frac{P_I^2}{2P_I\cdot R_I}$
\begin{equation}\label{Poles}
\hat{P}_I^2=-\frac{P_I^2}{z_I}(z-z_I).
\end{equation}
\end{enumerate}
The claim now is that the deformed amplitude will have a set of poles at some $z_I$, and that the residues at those poles can be written from lower-point amplitudes. If we consider the Feyman diagrams of the amplitude, the only place where such poles can appear is in the internal shifted propagators, which by conservation of momentum must be of the form $1/\hat{P}_I^2$ for some set $I$, and, from \eqref{Poles}, these must be simple poles. It can be seen from the definition that the residue of that pole $z_I$ corresponds to that propagator going on-shell.

If we then consider the function $\mathcal{A}_n(z)/z$, then by Cauchy's theorem (the sum of the residues must be zero) we have that
\begin{equation}
\mathcal{A}_n=-\sum_{z_I}\text{Res}_{z=z_I}\frac{A_n(z)}{z}+B_n,
\end{equation}
where $\mathcal{A}_n$ is both the residue at $z=0$ and the tree amplitude, the sum is over all different poles, and $B_n$ is the pole at $z\rightarrow\infty$ (which will in most cases be zero). From our explanation above, the residues at $z_I$ correspond to progators going on-shell, meaning that the amplitude factorizes such that
\begin{equation}
\text{Res}_{z=z_I}\frac{A_n(z)}{z}=-A_L(z_I)\frac{1}{P_I^2}A_R(z_I).
\end{equation}
And finally the whole amplitude can be expressed as the sum of such poles, made out of lower-point amplitudes ($A_L$ and $A_R$) with one of its external particles having momentum $P_I$
\begin{equation}\label{BCFWSum}
\mathcal{A}_n=\sum_{z_I} \frac{A_n(z)}{z}=\sum_{z_I}-A_L(z_I)\frac{1}{P_I^2}A_R(z_I).
\end{equation}

The BCFW recursion relations are a type of recursion relation in which the deformation goes through only two adjacent external legs in such a way that the spinor associated to two external particles are deformed by (exactly like in \eqref{BCFWDeform})

\begin{align}
&\lambda_a \rightarrow \lambda_{\hat{a}}=\lambda_a  ~~~~~~\text{and}~~~~~~  \lambda_b \rightarrow \lambda_{\hat{b}}=\lambda_b+\alpha_i \lambda_a \nonumber \\ &
\tilde{\lambda}_a \rightarrow \tilde{\lambda}_{\hat{a}}=\tilde{\lambda}_a-\alpha_i \tilde{\lambda}_b ~~~~~~~~ \tilde{\lambda_b} \rightarrow \tilde{\lambda}_{\hat{b}}=\tilde{\lambda_b}
\end{align}
which can be proved to be equivalent to a deformation such as that given in \eqref{Deformation}.
But this is only one method, there are alternative recursive procedures with different deformation prescriptions, such as the CSW construction \cite{CSW}. It turns out that the different methods give different decompositions of the tree amplitudes which, in the language of the tree amplituhedron, correspond to different ways of triangulating it, to later add up the volume of all those triangles together \cite{Amplituhedron}. The integrand described in this work \eqref{eq:Integrand} has a singularity structure such that the residues of the $\alpha_i$ match the terms in the BCFW expansion \eqref{BCFWSum}.

For specific details and examples of how this works in practice we refer the reader to \cite{ScattAmpli,FengReview}, but in general this is enough to see how amplitudes are built from smaller ones via recursion relations, an idea that has culminated in the recursion relations for general planar on-shell diagrams outlined in Section \ref{positroid}. 

We see that what we just described only works at tree level, where the poles are simple and well defined, and there is no room for either IR or UV divergences. This gets much more complicated for loop amplitudes (essentially, higher order terms of the perturbative $S$-matrix), for which some techniques started to be developed in the early 90's with the work of Bern, Dixon, Kosower and others \cite{Bern1994,BernDixonKosower2,BernDixonKosower,BernReview}. These \emph{generalized unitarity} methods showed, among other things, that the analytic structure (the poles of the integrand associated to it) of loop amplitudes is much more rich and complex than at tree level. In particular, as mentioned in \cite{NAH}, the 1-loop amplitude for $n=4$ is related to the corresponding tree amplitudes by (as seen first in \cite{GreenLimit})
\begin{align}\label{1loop}
&\mathcal{A}^{1-loop}_4=\mathcal{A}^{tree}_4 \times \\ & \nonumber \int \frac{d^4(l)(p^\mu_1+p^\mu_2)^2(p^\mu_1+p^\mu_3)^2}{l^2 (l+p^\mu_1)^2(l+p^\mu_1+p^\mu_2)^2(l-p^\mu_4)^2},
\end{align}
where $l$ is the 4-momentum of the loop and $p_i$ the 4-momenta of the particles. We can see that this is IR divergent when integrated (given the amount and degree of the poles), and hence the focus is placed on the integrand rather than the integral itself (as it is done in general for loop amplitudes in $\mathcal{N}=4$). 
The merit of the work \cite{AllLoop} is that it was found that, in general, such loop integrands can be described as a $\d(\text{log}(\alpha_i))$ if we make a certain change of variables (which is motivated by the description of kinematical data in momentum twistor space, following \cite{Hodges2005,Hodges2009}) and that, together with the analytic structure of the amplitudes at tree level, any amplitude at any loop-order has a corresponding integrand with a singularity structure given by some product of $d(\text{log}(\alpha_i))$. The final integrand is, of course, over the cells of the positive Grassmannian, as given by \eqref{eq:Integrand}. In this particular case, the loop part of the integrand \eqref{1loop} can be written as
\begin{align}
\nonumber \int \frac{d^4l(p^\mu_1+p^\mu_2)^2(p^\mu_1+p^\mu_3)^2}{l^2 (l+p^\mu_1)^2(l+p^\mu_1+p^\mu_2)^2(l-p^\mu_4)^2}=\nonumber \\ \nonumber 
\int d(\text{log}(\frac{l^2}{(l-l^*)^2})) d(\text{log}(\frac{(l+p^\mu_1)^2}{(l-l^*)^2})) \\d(\text{log}(\frac{(l+p^\mu_1+p^\mu_2)^2}{(l-l^*)^2})) d(\text{log}(\frac{(l-p^\mu_4)^2}{(l-l^*)^2})),
\end{align}
where $l^*$ is defined as one of the two points null-separated from all the external momenta (such that $(l^*-p^\mu_i)^2=0$. This, in the context of the positive Grassmannian, is written as the simple form
\begin{equation}
\int \frac{d \alpha_1}{\alpha_1}\frac{d \alpha_2}{\alpha_2}\frac{d \alpha_3}{\alpha_3}\frac{d \alpha_4}{\alpha_4}.
\end{equation}
This finding was one of the first indicators that a simple way of writing the integrand of the loop amplitudes existed, and ultimately led to the positive Grassmannian framework \cite{NAH} described throughout this work.

\section{Positive geometry and the inside of polytopes}\label{Positive}
If we have 3 masses $c_i$ in $\mathbb{R}^2$ such that they form a triangle, the ``centre of mass" 
\begin{equation}
\vec{y}=c_1\vec{r}_1+c_2\vec{r}_2+c_3\vec{r}_3,
\end{equation}
will be a point in the inside of that triangle as long as the $c_i>0$, and by varying these we map the whole inside of the triangle. We can see this means $(c_1,c_2,c_3) \in \text{Gr}^+(1,3)$. We also note that the points where any $c_i=0$ correspond to the boundaries (the edges) of the triangle.

This can be extended easily to $\text{Gr}^+(1,n)$ if we instead of a triangle have $n$ points forming a polytope, and the vectors $r_i$ do not have to be in any particular space for the inside of the polytope to make mathematical sense. Indeed, the physically relevant space (\emph{momentum twistor space}) will be the projective space $\mathbb{CP}^3$. It turns out that for the external points to have a well defined inside (meaning that they form a well-defined convex polytope) the matrix of the coordinates of those points must meet a condition of having all its cyclic minors positive \cite{Amplituhedron}.

Furthermore, such notion of the inside of a polytope can be generalized if we enlarge our notion of positivity with the positive Grassmannian. We generalize the condition $c_i>0$ to the condition that the minors of a $k\times n$ matrix $C$ need to be positive. Indeed the polytopes whose volume yields the amplitude (the amplituhedrons) have an inside mapped by $\text{Gr}^+(k,n)$. We define the $\alpha_i$ variables such that they span the different sectors or cells of $\text{Gr}^+(k,n)$, and, in analogy with the case of a triangle, we have that setting one of these to zero corresponds to moving along the boundary of such a polytope. \\
The volume of any of these objects can in general be obtained from an integrand with logarithmic singularities at its $\alpha_i=0$ boundaries \cite{Amplituhedron,ScattAmpli}, which means that an integrand such as \eqref{eq:Integrand} may indeed yield a volume that in the right cases will correspond to a scattering amplitude. \\
For the case of scattering amplitudes at tree level, this means that, given an integrand \eqref{eq:Integrand} with a certain $n$ and $k$ of dimension $2n-3$, its various possible residues give $2n-4$-dimensional forms that can be integrated into leading singularities, and the right combination of such leading singularities (such that the sum is local and unitary) gives an amplitude $\mathcal{A}_n^k$, which can be seen as a volume in momentum twistor space.

This boundary operation amounts to removing an edge in the graph (note that not all edges can be removed this way, only those coming from a BCFW decomposition). This operation also has a definite meaning in the context of BFTs, where it is identified as a \emph{higgsing}. This means that the scalar field of the chiral multiplet represented by the edge (the superfield) acquires a non-zero vacuum expectation value \cite{FrancoBipartite}.
\section{The permutation as the label of the cells}\label{permutation}
The ``boundary" operation of setting $\alpha_i=0$ explained in \ref{Positive} is the opposite to the addition of a BCFW bridge that we explain in Section \ref{BCFWbridge}. Indeed, we see that the BCFW operation essentially complicates the dependence between adjacent columns of $C$. Given the construction procedure and the fact that the permutation includes all the information about a cell, one question we can ask is: how does the permutation encode the mutual dependence between the columns of the $C$ matrix? It turns out that, if we define a rank function 
\begin{equation}
r[a;b]=\text{rank}(c_a .....c_{b}),
\end{equation}

Then, $\sigma(a)$ is such that it is the smallest number with the property that
\begin{equation}
r[a;\sigma(a)]=r[a+1;\sigma(a)].
\end{equation}
This also implies that $r[a;\sigma(a)-1]>r[a+1;\sigma(a)-1]$ and that, reciprocally, $a$ is the maximal column such that $r[a;\sigma(a)]=r[a;\sigma(a)-1]$. 
A way to easily compute this permutation from C is via writing a table of ranks of the adjacent columns of $C$. For example, for the permutation
\begin{equation}
\{ 3,7,6,10,9,8,13,12 \}\nonumber
\end{equation}
which has a corresponding matrix (in the lexicographic scheme, as defined in Section \ref{BCFWbridge})
\begin{equation}
\small{ \left(
\begin{array}{cccccccc}
 1 & \alpha_9 & 0 & -\alpha_5 & -\alpha_5 \alpha_6 & 0 & 0 & 0 \\
 0 & 1 & \alpha_8 & \alpha_7 & 0 & 0 & 0 & 0 \\
 0 & 0 & 0 & 1 & \alpha_3+\alpha_6 & \alpha_3 \alpha_4 & 0 & -\alpha_1 \\
 0 & 0 & 0 & 0 & 1 & \alpha_4 & \alpha_2 & 0 \\
\end{array}
\right)} \nonumber
\end{equation}

We can write a chart listing the combinations of adjacent vectors of different ranks
\begin{equation}
\begin{array}{c | c}
\text{consec. chains of colums} & \text{span} \text{(rank)} \\
\hline
(1){\color{blue}(2)}(3)(4)(5)(6){\color{blue}(7)}(8) & \mathbb{P}^0 \,(1) \\
\hline
{\color{blue}(123)}(34){\color{blue}(45)}(56){\color{blue}(678)}(81) & \mathbb{P}^1 \,(2)\\
\hline
{\color{blue}(3456) (56781) (81234)} & \mathbb{P}^2 \,(3)
\end{array}\nonumber
\end{equation}
In rank $r-1$ there are some sets $\{c_{a-1}...c_{\sigma(a)-1}\}$ for which there is a corresponding set of rank $r$ of the form $\{c_{a}...c_{\sigma(a)}\} $. These $r-1$ subsets will be called the maximal planes', and are such that they are the smallest for which addition of one column to the right and to the left has the same effect in terms of rank. These are hence the sets of columns which we need to read off the permutation. In the table above these maximal planes are highlighted in blue. We have left out many aspects of positive geometry and the combinatorial description of the positive Grassmannian due to space constraints, but the full discussion of these issues can be found in \cite{NAH} and mathematical literature such as \cite{Postnikov,Postnikov2}.
\onecolumn
\section{A further example of the lexicographic BCFW decomposition scheme}\label{BCFWDecomposition}
Here we show another example of the decomposition of the permutation, together with a careful description of the building of the graph step by step through BCFW bridges which will hopefully make the procedure clearer. 

If we choose the decorated permutation $\sigma=\{ 3,7,6,10,9,8,13,12 \}$, the decomposition, as defined in Section \ref{BCFWbridge} is given by
\begin{equation} \nonumber
\hskip0.1cm\begin{array}{|c|cccccccc|c|}
\hline
{} &1&2&3&4&5&6&7&8& {} \\ 
\tau&\downarrow&\downarrow&\downarrow&\downarrow&\downarrow&\downarrow&\downarrow&\downarrow&\text{BCFW shift} \\ \hline
\multirow{2}{*}{(1\,2)} &{\color{blue}3}&{\color{blue}7}&6&10&9&8&13&12&\multirow{2}{*}{$c_{2}\mapsto c_{2}+\alpha_{9}c_{1}$} \\
\multirow{2}{*}{(2\,3)} &7&{\color{blue}3}&{\color{blue}6}&10&9&8&13&12&\multirow{2}{*}{$c_{3}\mapsto c_{3}+\alpha_{8}c_{2}$} \\
\multirow{2}{*}{(2\,4)} &7&{\color{blue}6}&{\color{gray}3}&{\color{blue}10}&9&8&13&12&\multirow{2}{*}{$c_{4}\mapsto c_{4}+\alpha_{7}c_{2}$} \\
\multirow{2}{*}{(4\,5)} &7&{\color{gray}10}&{\color{gray}3}&{\color{blue}6}&{\color{blue}9}&8&13&12&\multirow{2}{*}{$c_{5}\mapsto c_{5}+\alpha_{6}c_{4}$} \\
\multirow{2}{*}{(1\,4)} &{\color{blue}7}&{\color{gray}10}&{\color{gray}3}&{\color{blue}9}&6&8&13&12&\multirow{2}{*}{$c_{4}\mapsto c_{4}-\alpha_{5}c_{1}$} \\
\multirow{2}{*}{(5\,6)} &{\color{gray}9}&{\color{gray}10}&{\color{gray}3}&7&{\color{blue}6}&{\color{blue}8}&13&12&\multirow{2}{*}{$c_{6}\mapsto c_{6}+\alpha_{4}c_{5}$} \\
\multirow{2}{*}{(4\,5)} &{\color{gray}9}&{\color{gray}10}&{\color{gray}3}&{\color{blue}7}&{\color{blue}8}&{\color{gray}6}&13&12&\multirow{2}{*}{$c_{5}\mapsto c_{5}+\alpha_{3}c_{4}$} \\
\multirow{2}{*}{(5\,7)} &{\color{gray}9}&{\color{gray}10}&{\color{gray}3}&8&{\color{blue}7}&{\color{gray}6}&{\color{blue}13}&12&\multirow{2}{*}{$c_{7}\mapsto c_{7}+\alpha_{2}c_{5}$} \\
\multirow{2}{*}{(4\,8)} &{\color{gray}9}&{\color{gray}10}&{\color{gray}3}&{\color{blue}8}&{\color{gray}13}&{\color{gray}6}&{\color{gray}7}&{\color{blue}12}&\multirow{2}{*}{$c_{8}\mapsto c_{8}-\alpha_{1}c_{4}$} \\
{} &{\color{gray}9}&{\color{gray}10}&{\color{gray}3}&{\color{gray}12}&{\color{gray}13}&{\color{gray}6}&{\color{gray}7}&{\color{gray}8}&{} \\
\hline
\end{array}
\end{equation}
In this table, the numbers permuted by the bridge are highlighted, and the ones corresponding to a decorated identity are in light grey.

If we construct this graph bridge by bridge via taking this decomposition backwards, the steps are the following (these correspond to all the adjacent permutations in the table above)
\begin{figure}[!h]
\hskip0.5cm\includegraphics[width=0.95\textwidth]{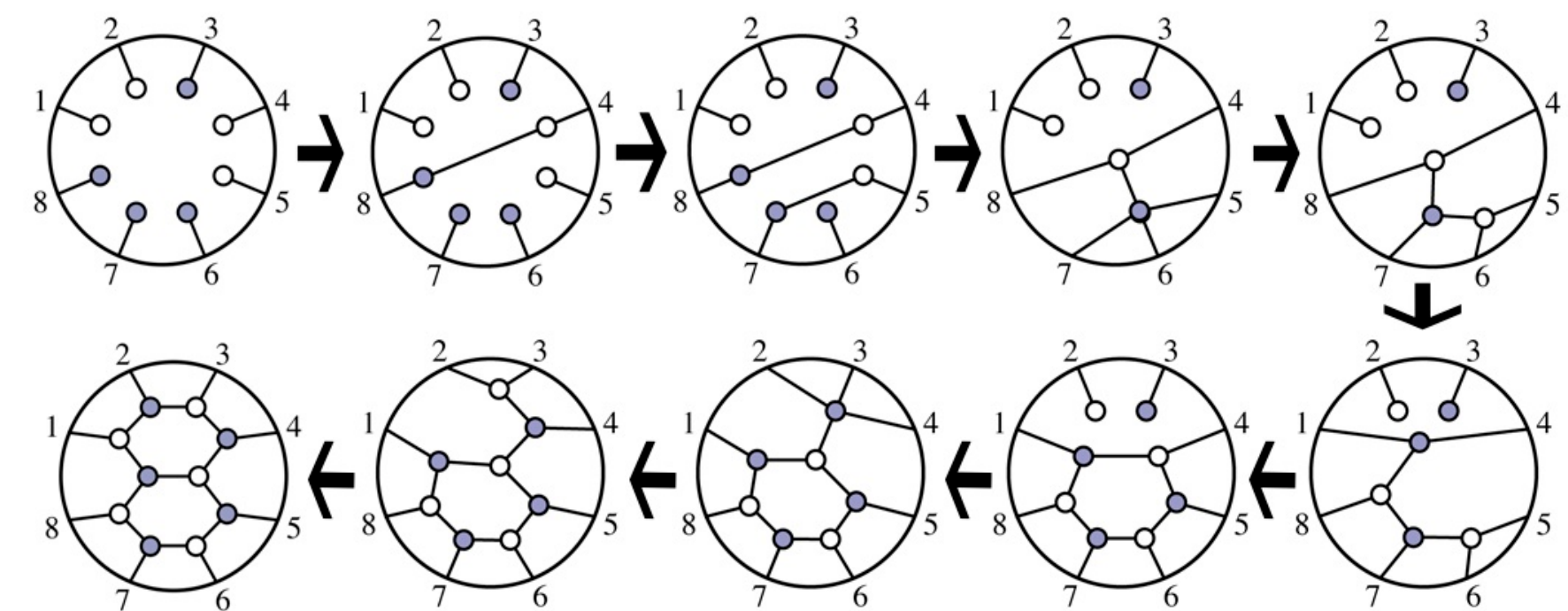}.
\label{fig:decomp} \end{figure} 

As we can see, each of these steps entails the creation of a bridge of the form shown in the first graph of Section \ref{BCFWbridge}. The final cell, created the way shown in Section \ref{cell}, has a matrix $C$
\begin{equation}
{ \left(
\begin{array}{cccccccc}
 1 & \alpha_9 & 0 & -\alpha_5 & -\alpha_5 \alpha_6 & 0 & 0 & 0 \\
 0 & 1 & \alpha_8 & \alpha_7 & 0 & 0 & 0 & 0 \\
 0 & 0 & 0 & 1 & \alpha_3+\alpha_6 & \alpha_3 \alpha_4 & 0 & -\alpha_1 \\
 0 & 0 & 0 & 0 & 1 & \alpha_4 & \alpha_2 & 0 \\
\end{array}
\right)}.
\end{equation}

Another more carefully worked example of how to construct the matrix is given in Section \ref{cell}.

\section{The positroid cells of the list of brane tilings}\label{list}
In the next few pages we show the on-shell diagrams that we have taken from the list of brane tilings (bipartite graphs on a 2D torus) from \cite{GraphClassification}. In that work, all the consistent and inconsistent brane tilings for up to 6 superpotential terms (with 3 white and black vertices) are produced and listed. Here, we show the on-shell diagrams that have been produced from each of those tilings through the procedure explained in Section \ref{Generation}, and the corresponding positroid cells calculated as it is explained in Section \ref{cell}. 

To facilitate the comparison, the diagrams are listed under the same label as the brane tiling they come from in the original list such that all diagrams coming from tiling $3.2$ in \cite{GraphClassification} are labeled $3.2a$, $3.2b$,... We show the different cells to which a boundary was assigned, with the dimensions of the cell ($n$, $k$ and $d$), the permutation of the diagram and the representation of their positroid cells $C$. There is an arbitrariness as to how to assign the labels $1.....n$ to the external particles, but for practical convenience and consistency we always choose the permutation with the lowest lexicographic order (analogous to the alphabetical order, but changing numbers for letters).

The first things we can see is that in general the diagrams do not have enough loops to reach the $2n-4$ dimensions necessary to produce just a leading singularity, and instead they produce a leading singularity with additional constraints on the external momenta. We also see that diagrams coming from the same tiling do have different parameters, complicating the analysis of these in terms of the positive geometry presented here. The rest of the discussion on the features of these graphs can be found in Section \ref{Features}.


\begin{landscape}
\begin{longtable}{| c | V | l |}

 \hline \# & Graph/Perm & Cell \\
\hline
\endhead

$\begin{array}{c} (1.1a) \\n=4\\k=2\\d=3 \end{array}$&
 \includegraphics[width=0.1\textwidth]{11}  
\newline\tiny\{2,4,5,7\} & 
\tiny{ $\left(
\begin{array}{cccc}
 1 & \alpha_3 & \alpha_2 & 0 \\
 0 & 0 & 1 & \alpha_1 \\
\end{array}
\right)$}
\\
\hline

$\begin{array}{c} (1.2a) \\n=6\\k=3\\d=5 \end{array}$ & 
\includegraphics[width=0.1\textwidth]{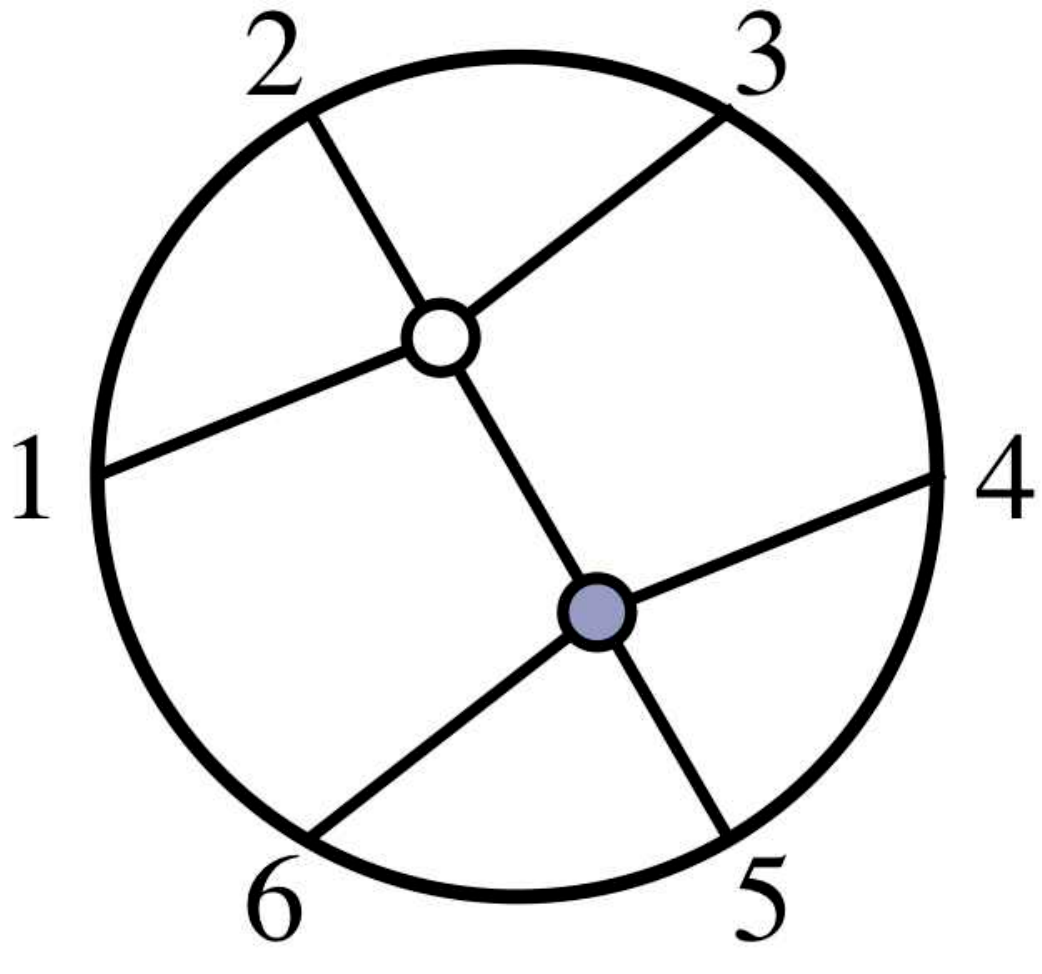}  
\newline\tiny\{2,3,6,7,10,11\} &
\tiny{$\left(
\begin{array}{cccccc}
 1 & \alpha_5 & \alpha_4 & \alpha_3 & 0 & 0 \\
 0 & 0 & 0 & 1 & \alpha_2 & 0 \\
 0 & 0 & 0 & 0 & 1 & \alpha_1 \\
\end{array}
\right)$}
\\
\hline




$\begin{array}{c} (2.1a) \\n=8\\k=4\\d=6 \end{array}$ &
\includegraphics[width=0.1\textwidth]{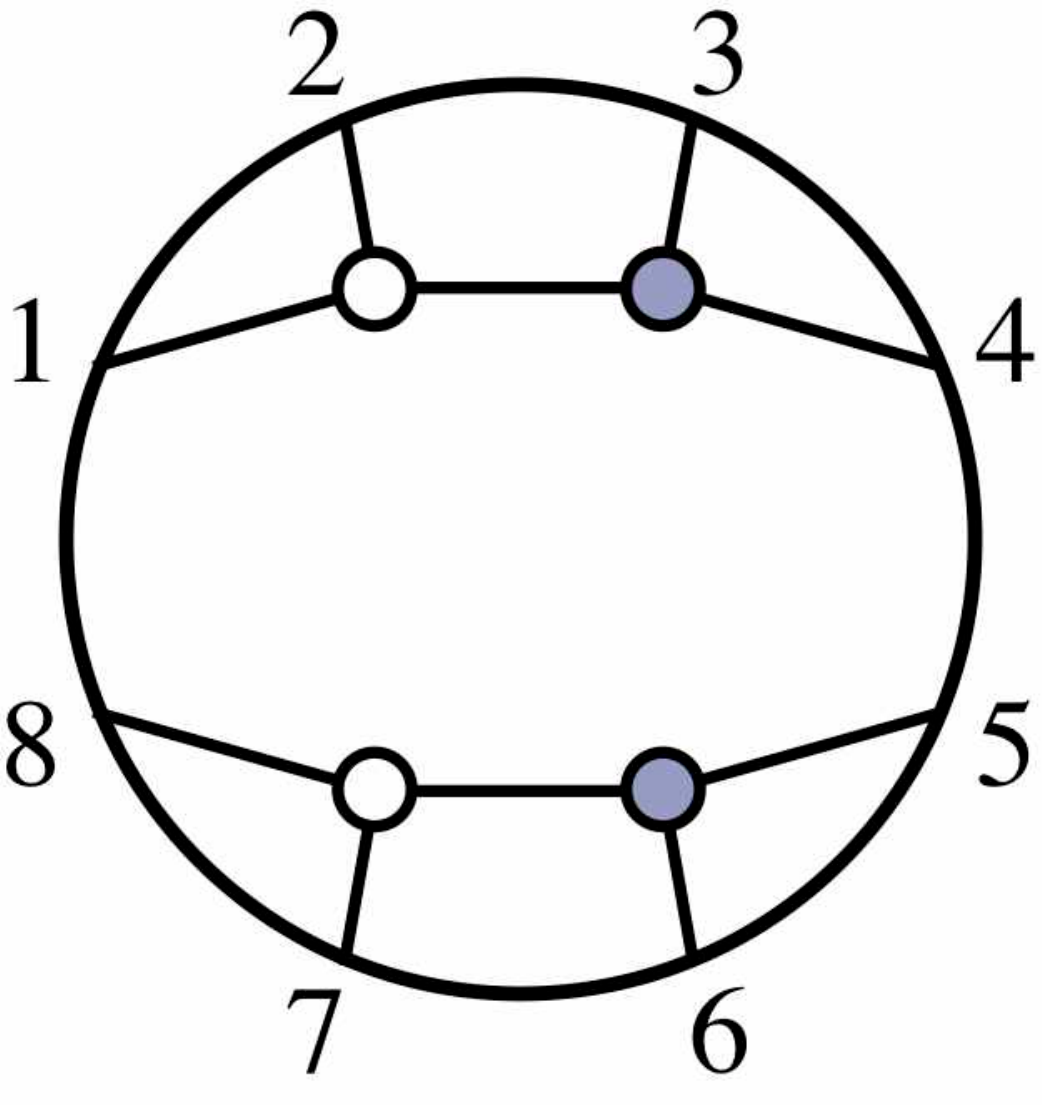}  
\newline\tiny\{2,4,9,11,7,13,8,14\} &
\tiny{$\left(
\begin{array}{cccccccc}
 1 & \alpha_6 & \alpha_5 & 0 & 0 & 0 & 0 & 0 \\
 0 & 0 & 1 & \alpha_4 & 0 & 0 & 0 & 0 \\
 0 & 0 & 0 & 0 & 1 & \alpha_3 & 0 & 0 \\
 0 & 0 & 0 & 0 & 0 & 1 & \alpha_2 & \alpha_1 \\
\end{array}
\right)$}
\\
\hline

$\begin{array}{c} (2.1b) \\n=6\\k=3\\d=5 \end{array}$ &
\includegraphics[width=0.1\textwidth]{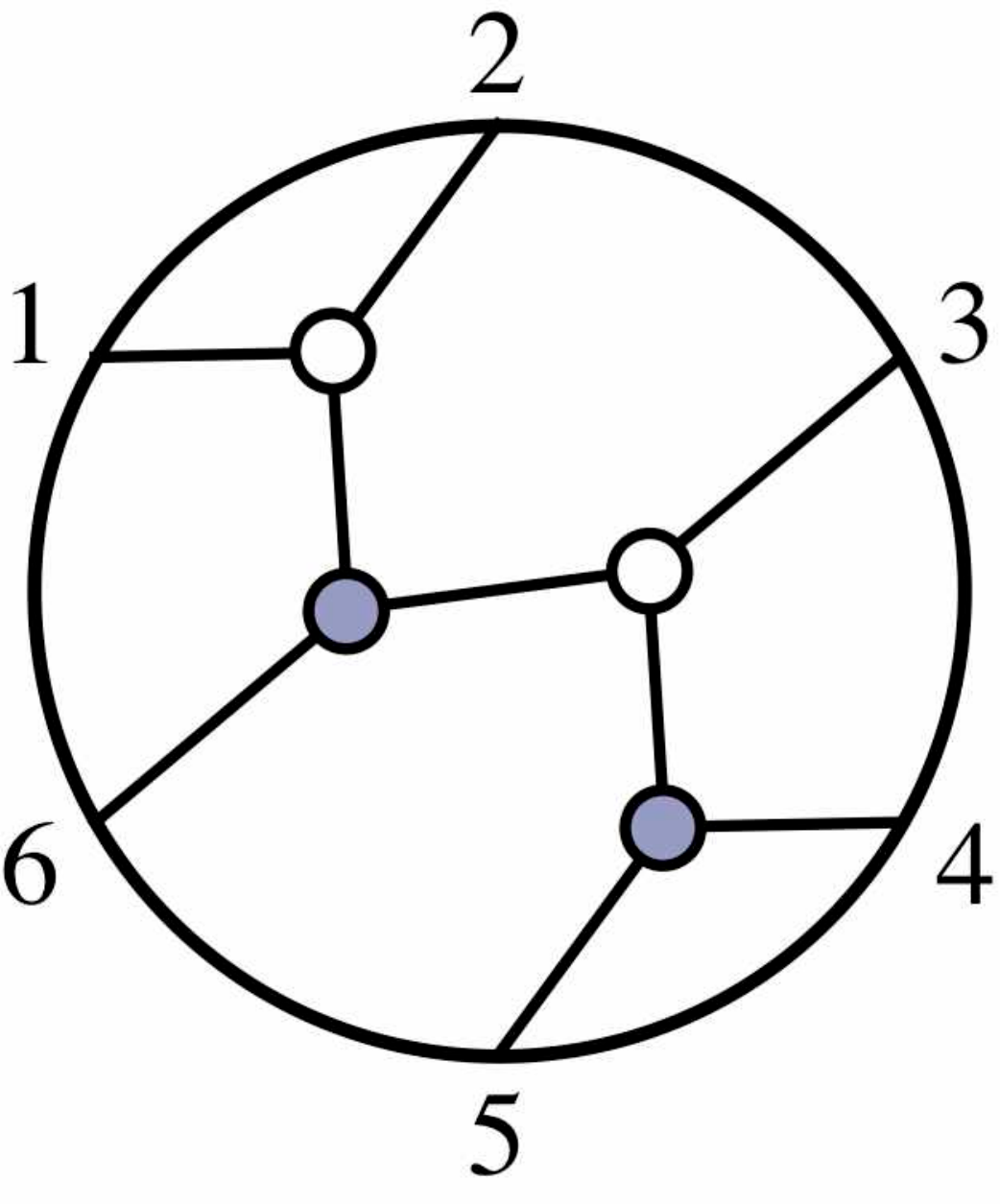}  
\newline \tiny\{2, 6, 5, 7, 10, 9\}&
\tiny{$\left(
\begin{array}{cccccc}
 1 & \alpha_5 & \alpha_3 & \alpha_3 \alpha_4 & 0 & 0 \\
 0 & 0 & 1 & \alpha_4 & 0 & -\alpha_1 \\
 0 & 0 & 0 & 1 & \alpha_2 & 0 \\
\end{array}
\right)$}
\\
\hline

$\begin{array}{c} (2.2a) \\n=8\\k=4\\d=7 \end{array}$ &
\includegraphics[width=0.1\textwidth]{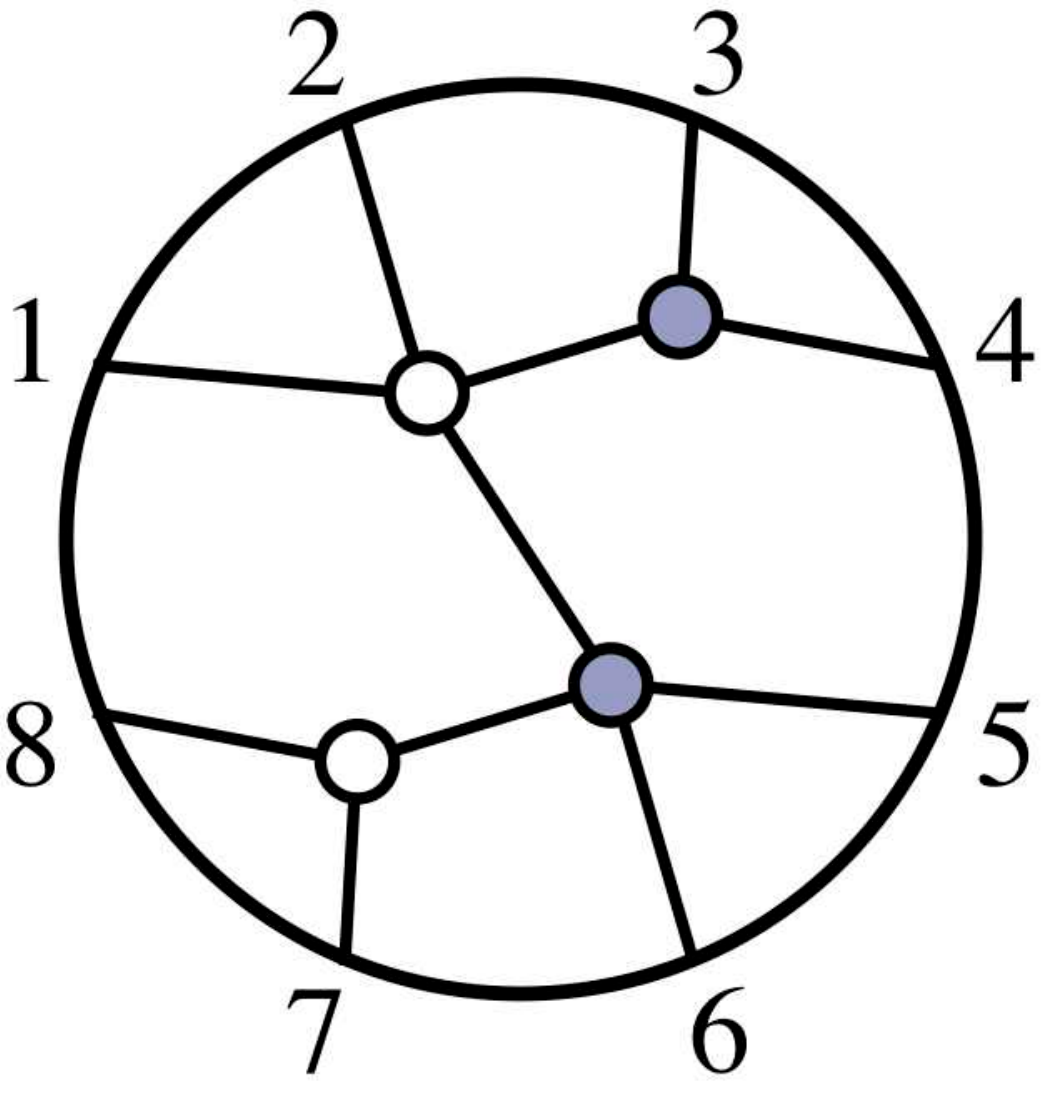}  
\newline\tiny\{2,4,7,11,9,13,8,14\} &
{\tiny
$\left(
\begin{array}{cccccccc}
 1 & \alpha_7 & \alpha_6 & 0 & -\alpha_4 & 0 & 0 & 0 \\
 0 & 0 & 1 & \alpha_5 & 0 & 0 & 0 & 0 \\
 0 & 0 & 0 & 0 & 1 & \alpha_3 & 0 & 0 \\
 0 & 0 & 0 & 0 & 0 & 1 & \alpha_2 & \alpha_1 \\
\end{array}
\right)$}
\\
\hline

$\begin{array}{c} (2.2b) \\n=8\\k=4\\d=7 \end{array}$&
\includegraphics[width=0.1\textwidth]{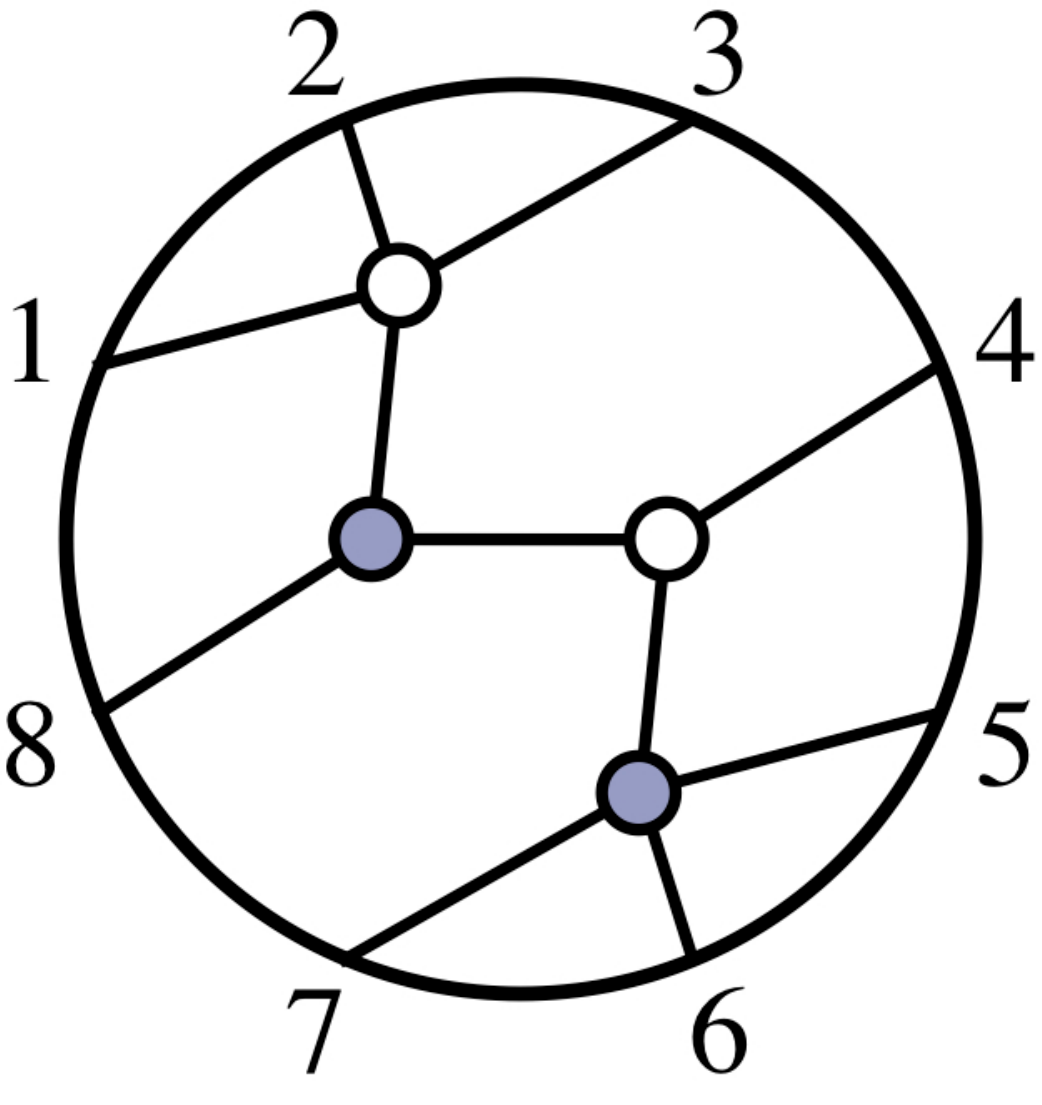}  
\newline\tiny\{2,3,8,7,9,13,14,12\} &
{\tiny
$\left(
\begin{array}{cccccccc}
 1 & \alpha_7 & \alpha_6 & \alpha_4 & \alpha_4 \alpha_5 & 0 & 0 & 0 \\
 0 & 0 & 0 & 1 & \alpha_5 & 0 & 0 & \alpha_1 \\
 0 & 0 & 0 & 0 & 1 & \alpha_3 & 0 & 0 \\
 0 & 0 & 0 & 0 & 0 & 1 & \alpha_2 & 0 \\
\end{array}
\right)$}
\\
\hline

$\begin{array}{c} (2.2c) \\n=8\\k=4\\d=7 \end{array}$&
\includegraphics[width=0.1\textwidth]{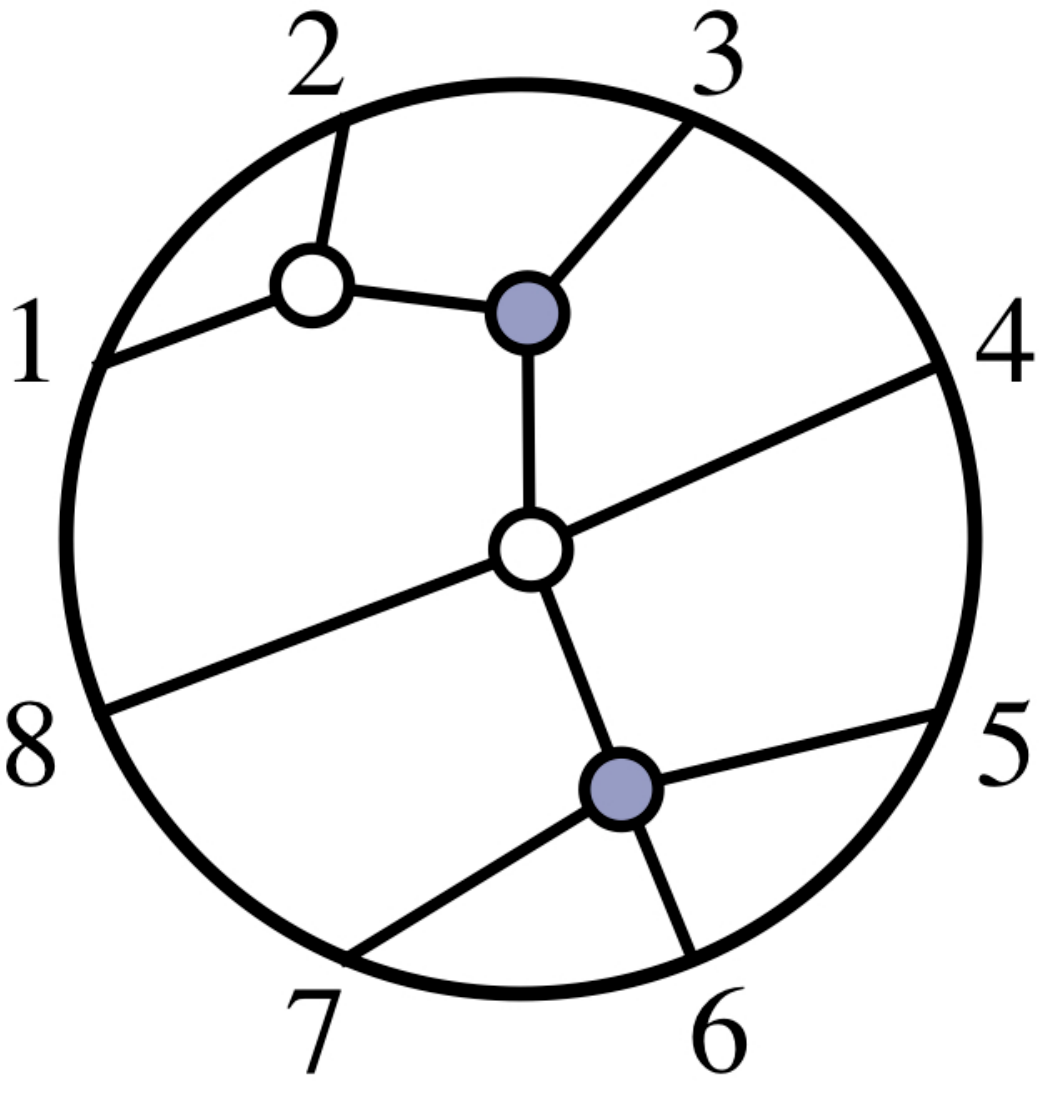}  
\newline\tiny\{2,4,9,7,8,13,14,11\} &
{\tiny
$\left(
\begin{array}{cccccccc}
 1 & \alpha_7 & \alpha_6 & 0 & 0 & 0 & 0 & 0 \\
 0 & 0 & 1 & \alpha_5 & \alpha_4 & 0 & 0 & \alpha_1 \\
 0 & 0 & 0 & 0 & 1 & \alpha_3 & 0 & 0 \\
 0 & 0 & 0 & 0 & 0 & 1 & \alpha_2 & 0 \\
\end{array}
\right)$}
\\
\hline

$\begin{array}{c} (2.2d) \\n=8\\k=4\\d=7 \end{array}$ &
\includegraphics[width=0.1\textwidth]{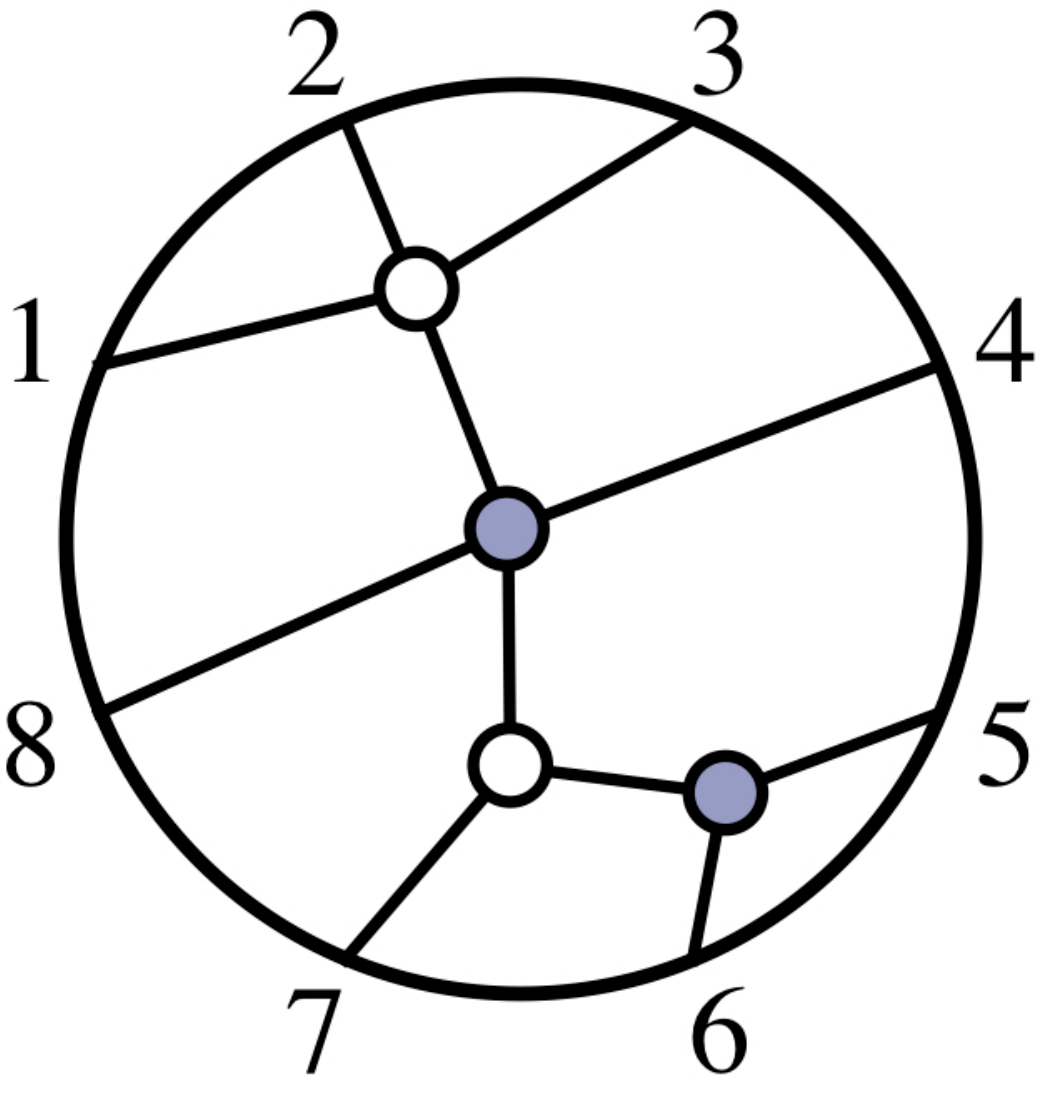}  
\newline \tiny\{2,3,8,9,7,13,12,14\}&
{\tiny
$\left(
\begin{array}{cccccccc}
 1 & \alpha_7 & \alpha_6 & \alpha_5 & 0 & 0 & 0 & 0 \\
 0 & 0 & 0 & 1 & 0 & -\alpha_2 & -\alpha_2 \alpha_3 & 0 \\
 0 & 0 & 0 & 0 & 1 & \alpha_4 & 0 & 0 \\
 0 & 0 & 0 & 0 & 0 & 1 & \alpha_3 & \alpha_1 \\
\end{array}
\right)$}
\\
\hline

$\begin{array}{c} (2.3a) \\n=8\\k=4\\d=7 \end{array}$ &
\includegraphics[width=0.1\textwidth]{23}  
\newline\tiny\{2, 4, 7, 6, 9, 11\} &
{\tiny
$\left(
\begin{array}{cccccccc}
 1 & \alpha_7 & \alpha_6 & 0 & -\alpha_3 & -\alpha_3 \alpha_4 & 0 & 0 \\
 0 & 0 & 1 & \alpha_5 & 0 & 0 & 0 & 0 \\
 0 & 0 & 0 & 0 & 1 & \alpha_4 & \alpha_2 & 0 \\
 0 & 0 & 0 & 0 & 0 & 0 & 1 & \alpha_1 \\
\end{array}
\right)$}
\\
\hline

$\begin{array}{c} (2.3b) \\n=6\\k=3\\d=6 \end{array}$ &
\includegraphics[width=0.1\textwidth]{23b}  
\newline\tiny\{2,5,6,9,7,10\} &
{\tiny
$\left(
\begin{array}{cccccc}
 1 & \alpha_6 & \alpha_5 & -\alpha_2 & -\alpha_2 \alpha_3 & 0 \\
 0 & 0 & 1 & \alpha_4 & 0 & 0 \\
 0 & 0 & 0 & 1 & \alpha_3 & \alpha_1 \\
\end{array}
\right)$}
\\
\hline

$\begin{array}{c} (2.4a) \\n=10\\k=5\\d=9 \end{array}$ &
\includegraphics[width=0.1\textwidth]{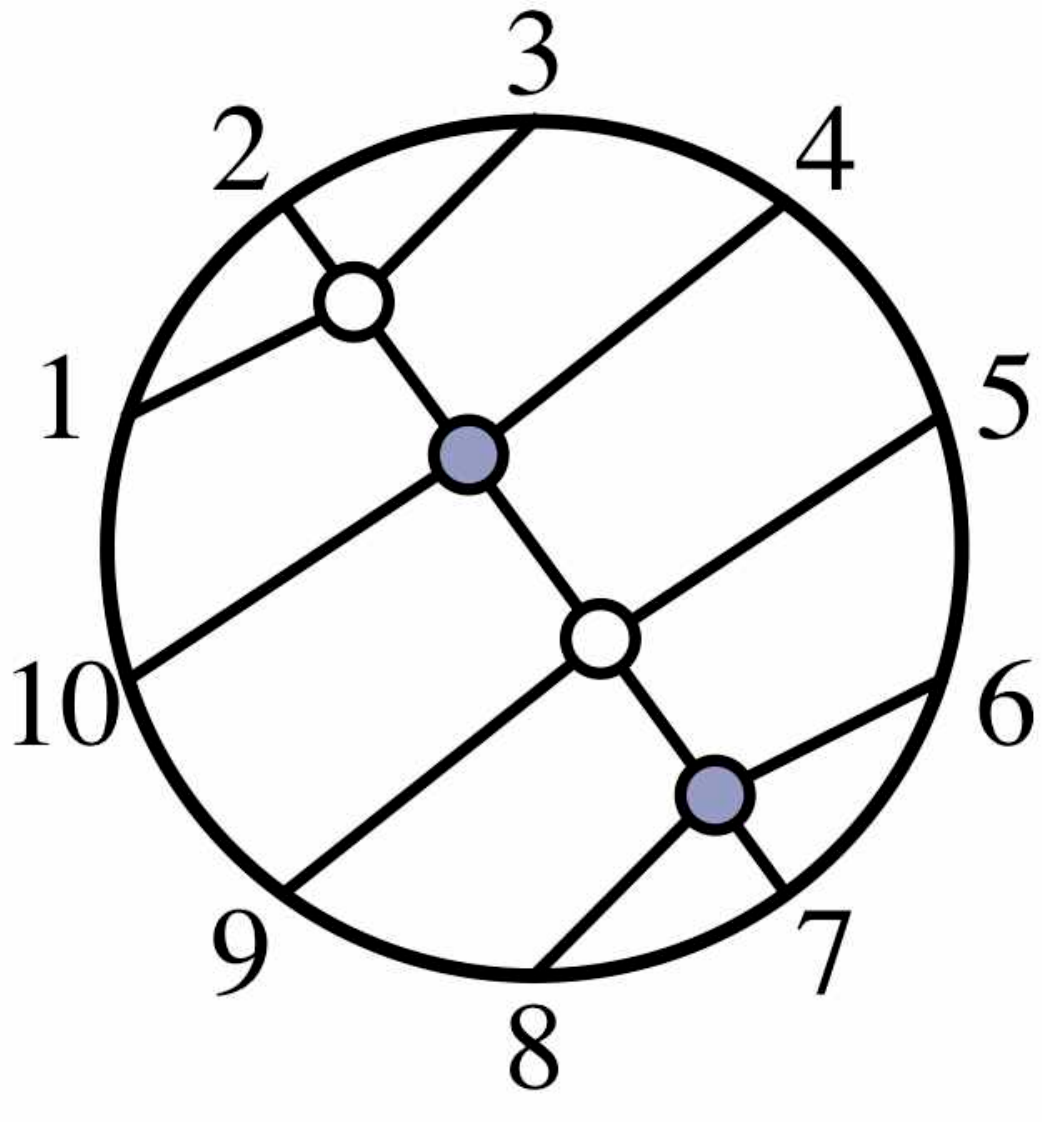}  
\newline\tiny \{2,3,10,11,8,9,16,17,14,15\}&
{\tiny
$\left(
\begin{array}{cccccccccc}
 1 & \alpha_9 & \alpha_8 & \alpha_7 & 0 & 0 & 0 & 0 & 0 & 0 \\
 0 & 0 & 0 & 1 & \alpha_2 & \alpha_2 \alpha_6 & 0 & 0 & \alpha_2 \alpha_3 & 0 \\
 0 & 0 & 0 & 0 & 1 & \alpha_6 & 0 & 0 & \alpha_3 & \alpha_1 \\
 0 & 0 & 0 & 0 & 0 & 1 & \alpha_5 & 0 & 0 & 0 \\
 0 & 0 & 0 & 0 & 0 & 0 & 1 & \alpha_4 & 0 & 0 \\
\end{array}
\right)$}
\\
\hline

$\begin{array}{c} (2.5a) \\n=10\\k=5\\d=9 \end{array}$ &
\includegraphics[width=0.1\textwidth]{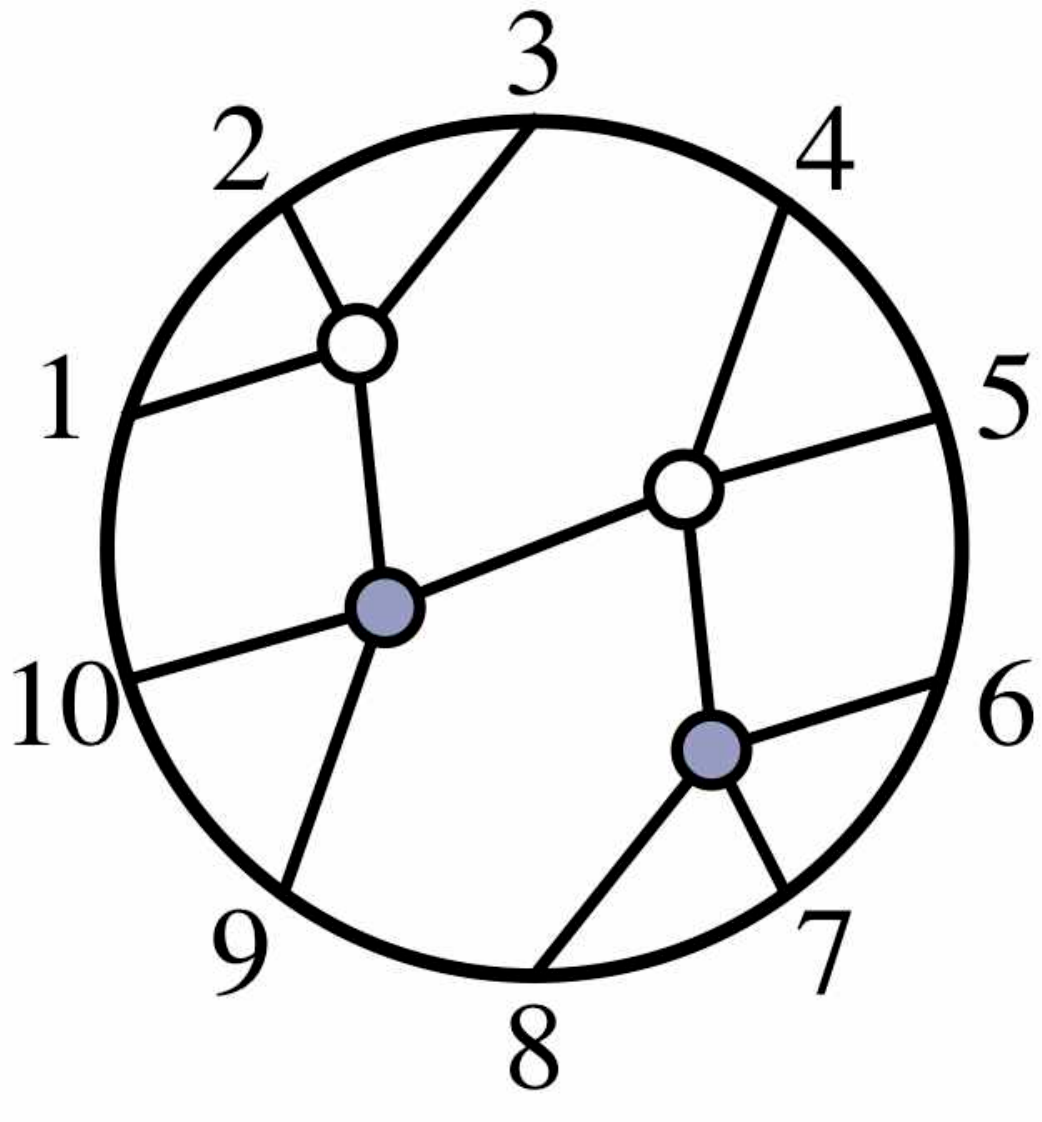}  
\newline\tiny \{2,3,10,5,8,11,16,17,14,19\}&
{\tiny
$\left(
\begin{array}{cccccccccc}
 1 & \alpha_9 & \alpha_8 & \alpha_5 & \alpha_5 \alpha_7 & \alpha_5 \alpha_6 & 0 & 0 & 0 & 0 \\
 0 & 0 & 0 & 1 & \alpha_7 & \alpha_6 & 0 & 0 & \alpha_2 & 0 \\
 0 & 0 & 0 & 0 & 0 & 1 & \alpha_4 & 0 & 0 & 0 \\
 0 & 0 & 0 & 0 & 0 & 0 & 1 & \alpha_3 & 0 & 0 \\
 0 & 0 & 0 & 0 & 0 & 0 & 0 & 0 & 1 & \alpha_1 \\
\end{array}
\right)$}
\\
\hline

$\begin{array}{c} (2.5b) \\n=8\\k=4\\d=8 \end{array}$ &
\includegraphics[width=0.1\textwidth]{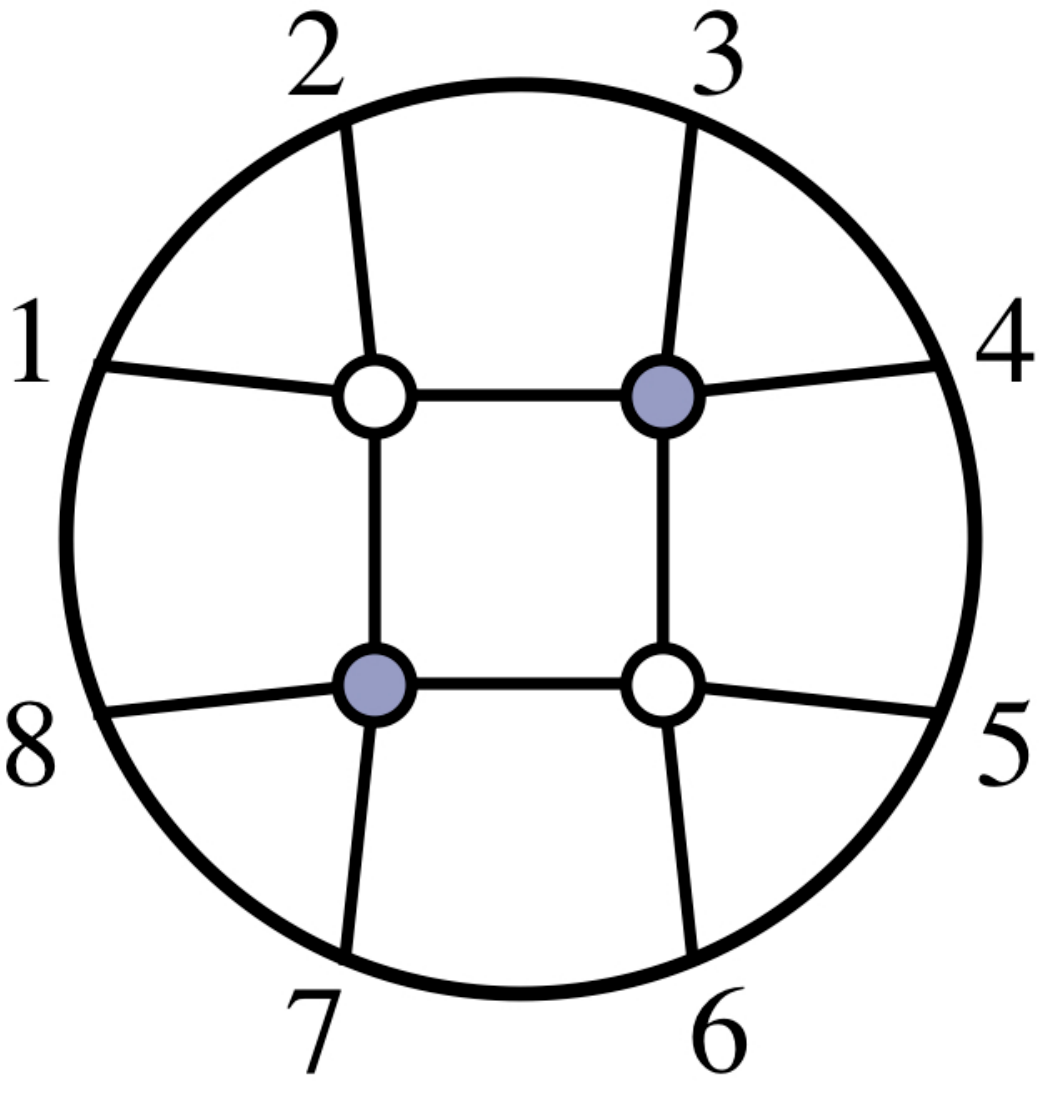}  
\newline\tiny \{2,5,8,11,6,9,12,15\}&
{\tiny
$\left(
\begin{array}{cccccccc}
 1 & \alpha_8 & \alpha_7 & -\alpha_3 & -\alpha_3 \alpha_5 & -\alpha_3 \alpha_4 & 0 & 0 \\
 0 & 0 & 1 & \alpha_6 & 0 & 0 & 0 & 0 \\
 0 & 0 & 0 & 1 & \alpha_5 & \alpha_4 & \alpha_2 & 0 \\
 0 & 0 & 0 & 0 & 0 & 0 & 1 & \alpha_1 \\
\end{array}
\right)$}
\\
\hline

$\begin{array}{c} (3.1a) \\n=12\\k=6\\d=9 \end{array}$ &
\includegraphics[width=0.1\textwidth]{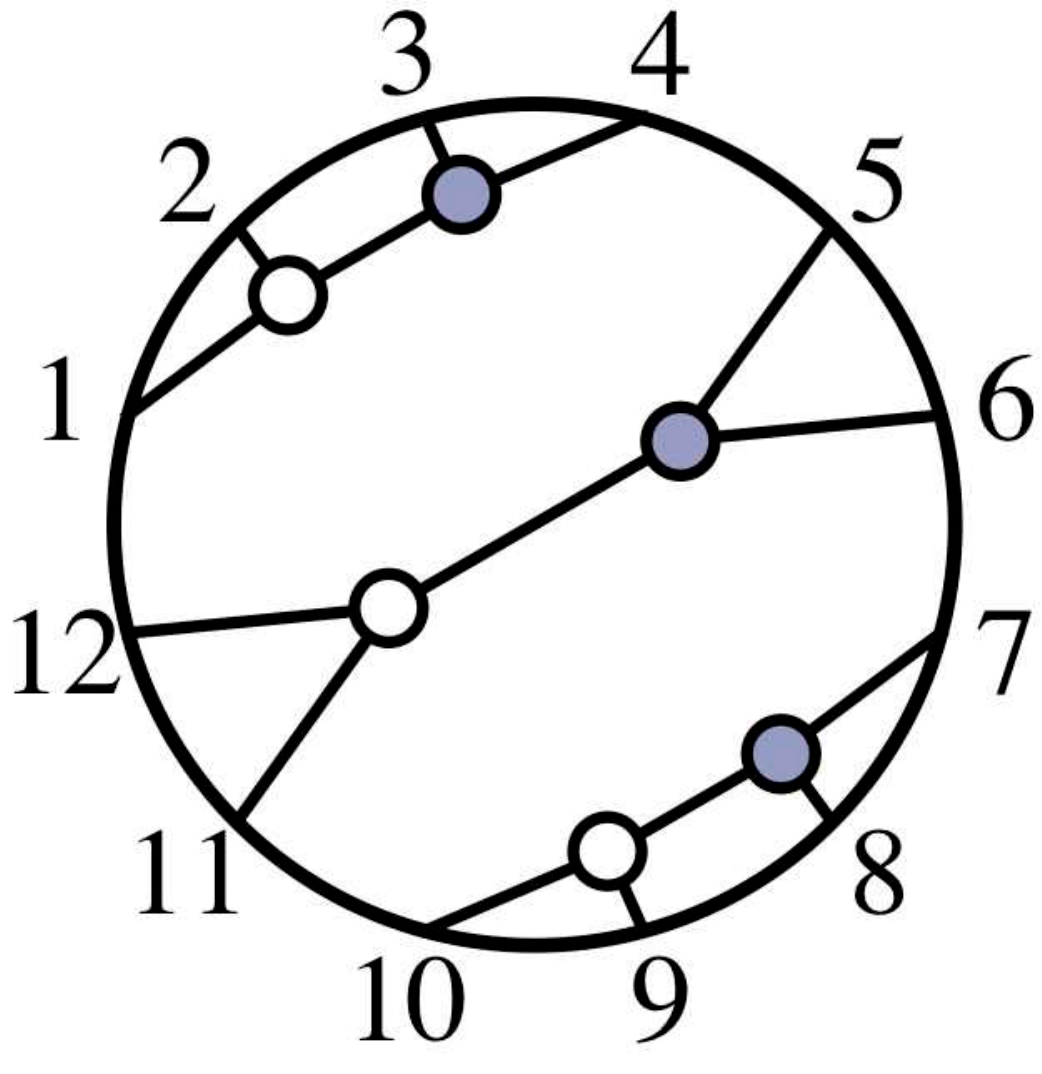}  
\newline\tiny \{2,4,13,15,11,17,9,19,10,20,12,18\} &
{\tiny
$\left(
\begin{array}{cccccccccccc}
 1 & \alpha_9 & \alpha_8 & 0 & 0 & 0 & 0 & 0 & 0 & 0 & 0 & 0 \\
 0 & 0 & 1 & \alpha_7 & 0 & 0 & 0 & 0 & 0 & 0 & 0 & 0 \\
 0 & 0 & 0 & 0 & 1 & \alpha_6 & 0 & 0 & 0 & 0 & 0 & 0 \\
 0 & 0 & 0 & 0 & 0 & 1 & 0 & 0 & 0 & 0 & \alpha_2 & \alpha_1 \\
 0 & 0 & 0 & 0 & 0 & 0 & 1 & \alpha_5 & 0 & 0 & 0 & 0 \\
 0 & 0 & 0 & 0 & 0 & 0 & 0 & 1 & \alpha_4 & \alpha_3 & 0 & 0 \\
\end{array}
\right)$}
\\
\hline

$\begin{array}{c} (3.1b) \\n=8\\k=4\\d=7 \end{array}$ &
\includegraphics[width=0.1\textwidth]{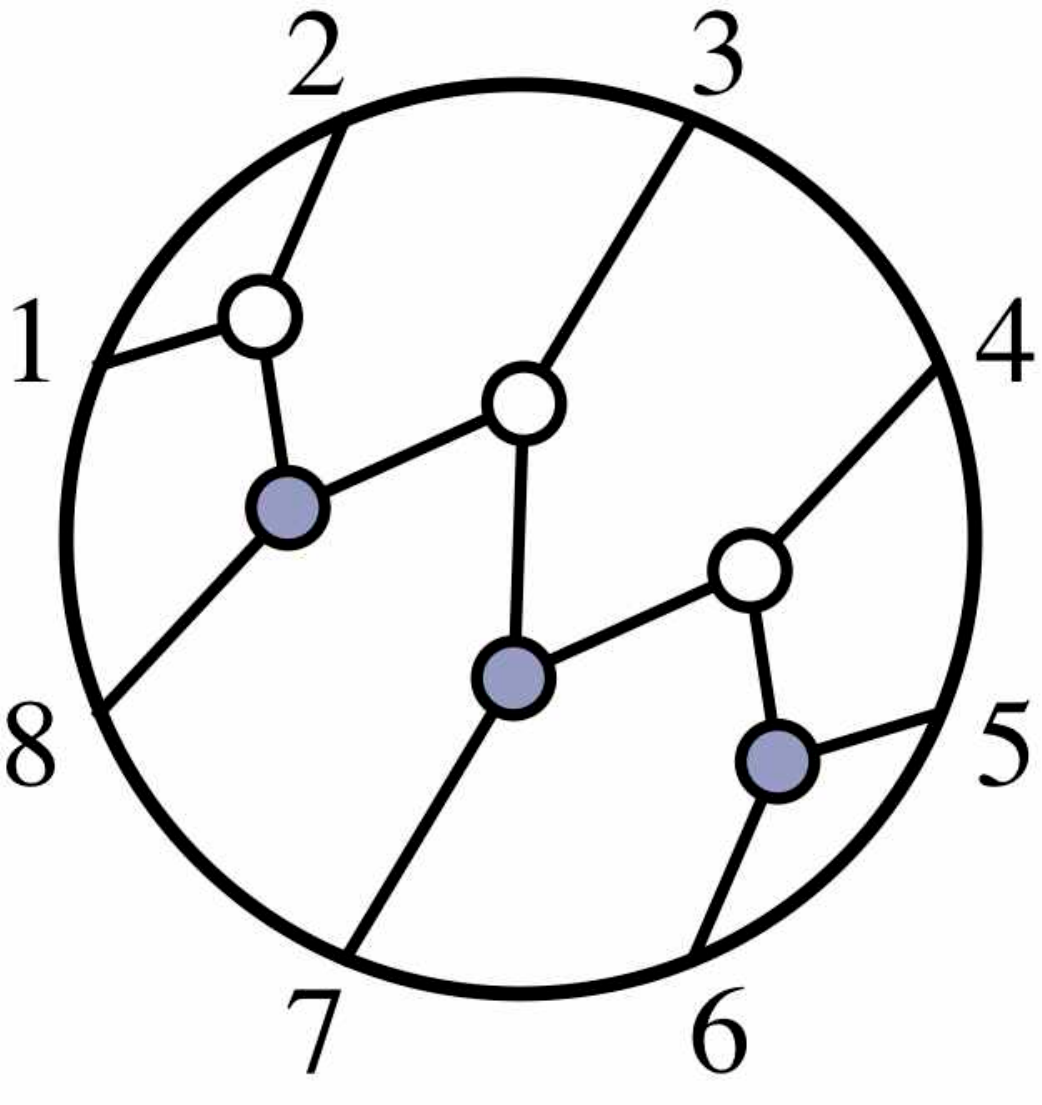}  
\newline\tiny\{2,8,7,6,9,13,12,11\}&
{\tiny
$\left(
\begin{array}{cccccccc}
 1 & \alpha_7 & \alpha_4 & \alpha_4 \alpha_5 & \alpha_4 \alpha_5 \alpha_6 & 0 & 0 & 0 \\
 0 & 0 & 1 & \alpha_5 & \alpha_5 \alpha_6 & 0 & 0 & \alpha_1 \\
 0 & 0 & 0 & 1 & \alpha_6 & 0 & -\alpha_2 & 0 \\
 0 & 0 & 0 & 0 & 1 & \alpha_3 & 0 & 0 \\
\end{array}
\right)$}
\\
\hline

$\begin{array}{c} (3.2a) \\n=6\\k=3\\d=6 \end{array}$ &
\includegraphics[width=0.1\textwidth]{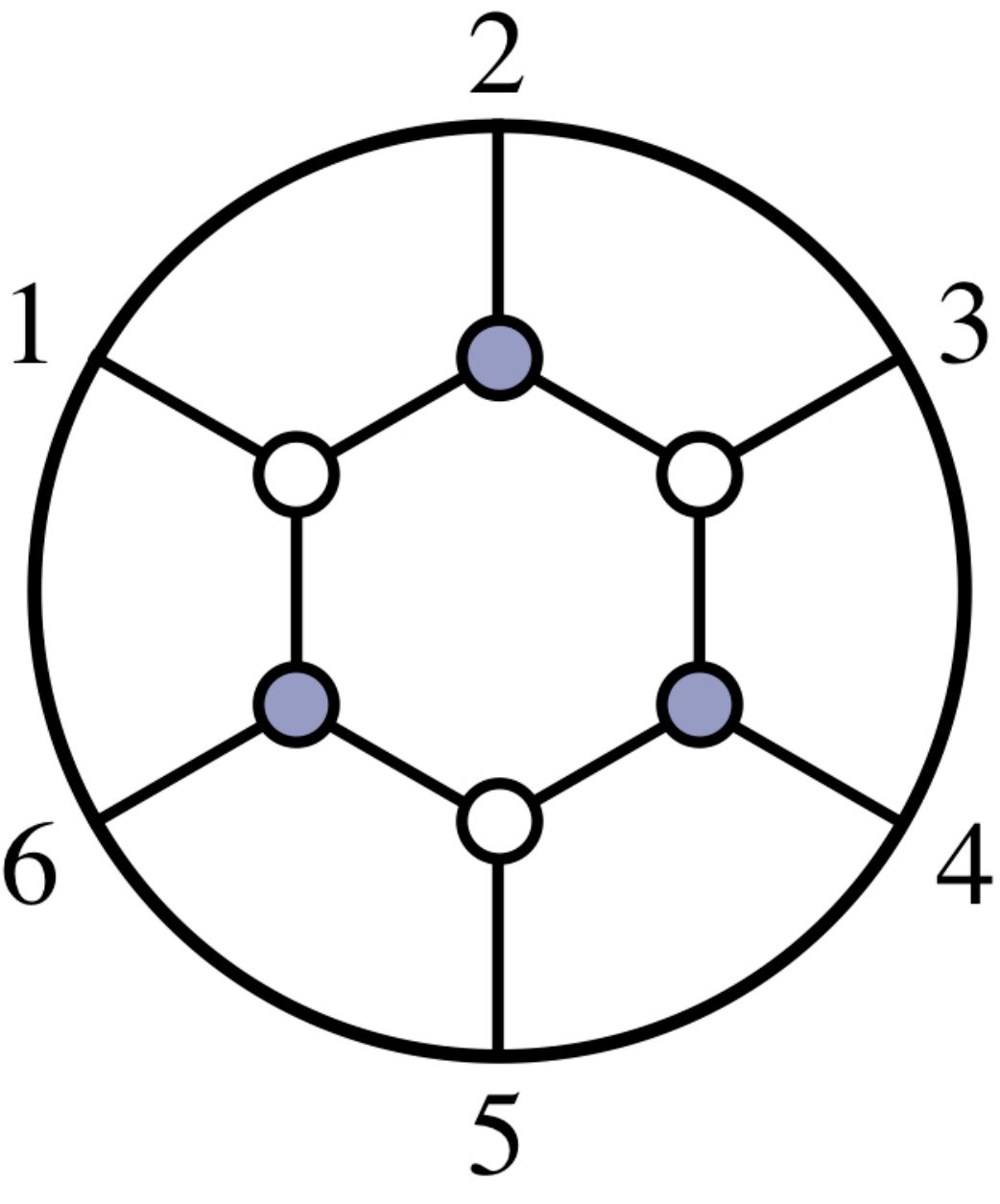}  
\newline\tiny\{3,6,5,8,7,10\}&
{\tiny
$\left(
\begin{array}{cccccc}
 1 & \alpha_6 & 0 & -\alpha_2 & -\alpha_2 \alpha_3 & 0 \\
 0 & 1 & \alpha_5 & \alpha_4 & 0 & 0 \\
 0 & 0 & 0 & 1 & \alpha_3 & \alpha_1 \\
\end{array}
\right)$}
\\
\hline

$\begin{array}{c} (3.2b) \\n=8\\k=4\\d=7 \end{array}$ &
\includegraphics[width=0.1\textwidth]{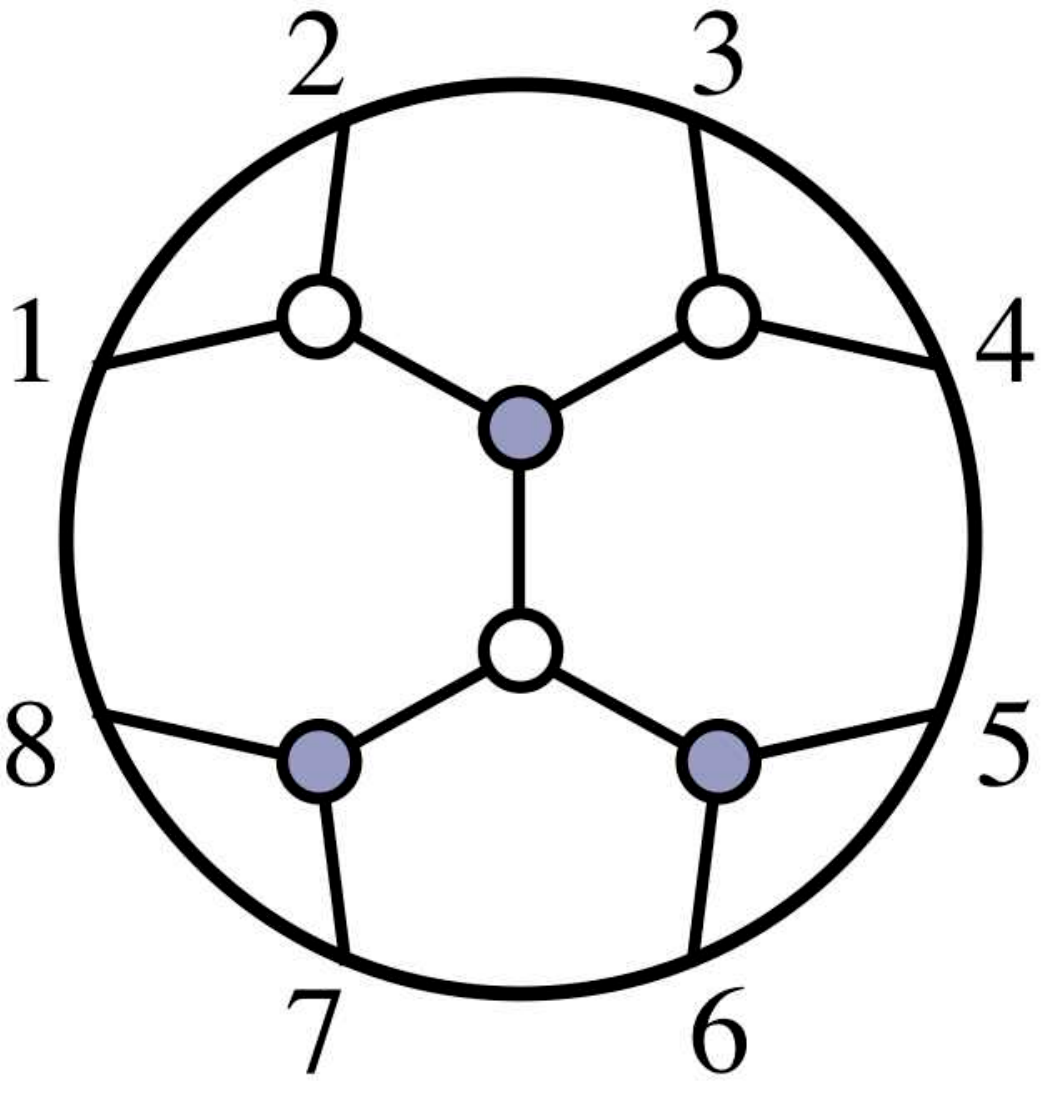}  
\newline\tiny\{2,6,4,9,8,13,11,15\}&
{\tiny
$\left(
\begin{array}{cccccccc}
 1 & \alpha_7 & \alpha_5 & \alpha_5 \alpha_6 & 0 & 0 & 0 & 0 \\
 0 & 0 & 1 & \alpha_6 & \alpha_4 & 0 & -\alpha_2 & 0 \\
 0 & 0 & 0 & 0 & 1 & \alpha_3 & 0 & 0 \\
 0 & 0 & 0 & 0 & 0 & 0 & 1 & \alpha_1 \\
\end{array}
\right)$}
\\
\hline

$\begin{array}{c} (3.2c) \\n=8\\k=4\\d=7 \end{array}$ &
\includegraphics[width=0.1\textwidth]{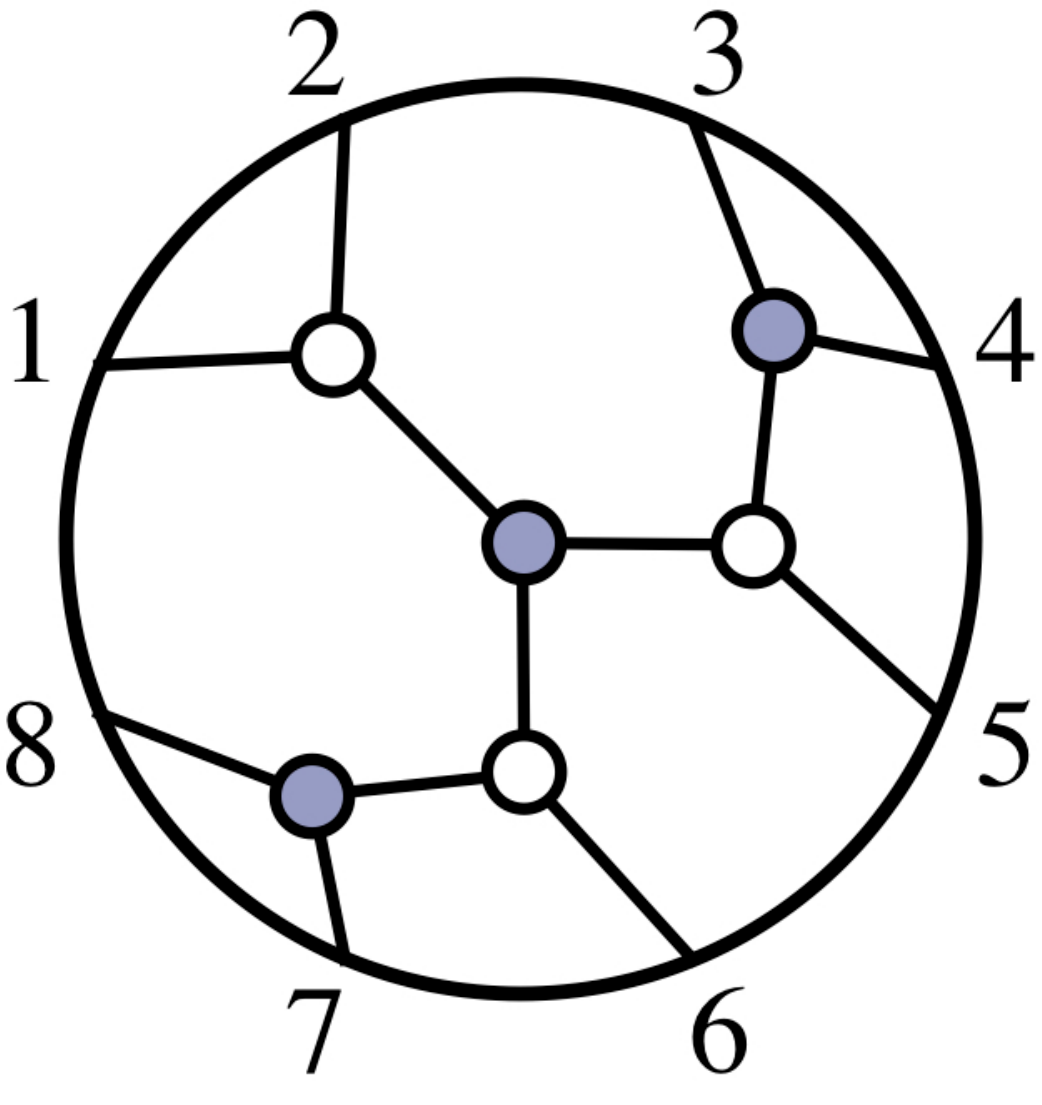}  
\newline \tiny\{2,6,5,11,9,8,12,15\}&
{\tiny
$\left(
\begin{array}{cccccccc}
 1 & \alpha_7 & 0 & -\alpha_4 & -\alpha_4 \alpha_5 & 0 & 0 & 0 \\
 0 & 0 & 1 & \alpha_6 & 0 & 0 & 0 & 0 \\
 0 & 0 & 0 & 1 & \alpha_5 & \alpha_3 & \alpha_2 & 0 \\
 0 & 0 & 0 & 0 & 0 & 0 & 1 & \alpha_1 \\
\end{array}
\right)$}
\\
\hline

$\begin{array}{c} (3.2d) \\n=8\\k=4\\d=7 \end{array}$ &
\includegraphics[width=0.1\textwidth]{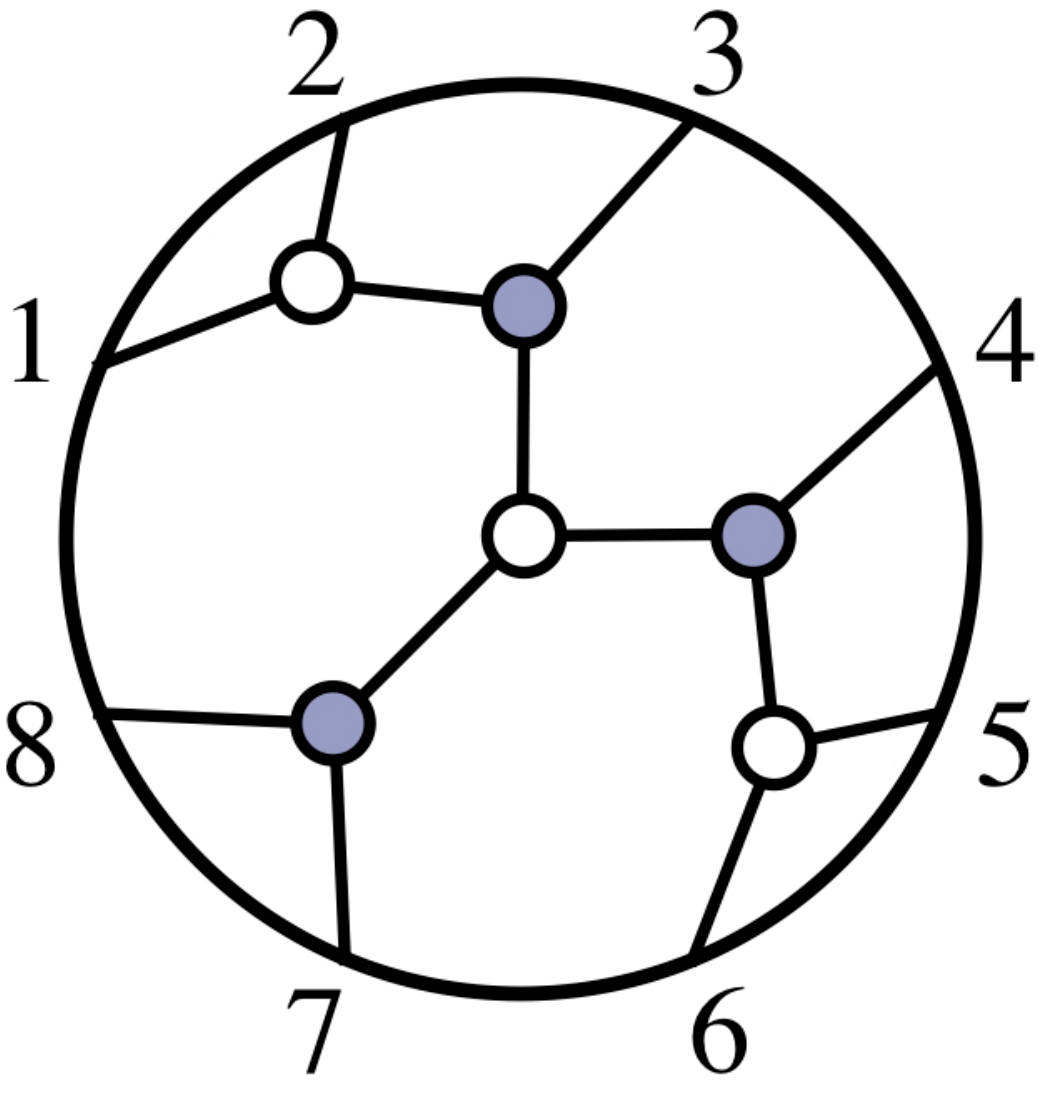}  
\newline \tiny\{2,5,9,8,6,12,11,15\}&
{\tiny
$\left(
\begin{array}{cccccccc}
 1 & \alpha_7 & \alpha_6 & 0 & 0 & 0 & 0 & 0 \\
 0 & 0 & 1 & \alpha_5 & 0 & 0 & -\alpha_2 & 0 \\
 0 & 0 & 0 & 1 & \alpha_4 & \alpha_3 & 0 & 0 \\
 0 & 0 & 0 & 0 & 0 & 0 & 1 & \alpha_1 \\
\end{array}
\right)$}
\\
\hline

$\begin{array}{c} (3.4a) \\n=10\\k=5\\d=9 \end{array}$ &
\includegraphics[width=0.1\textwidth]{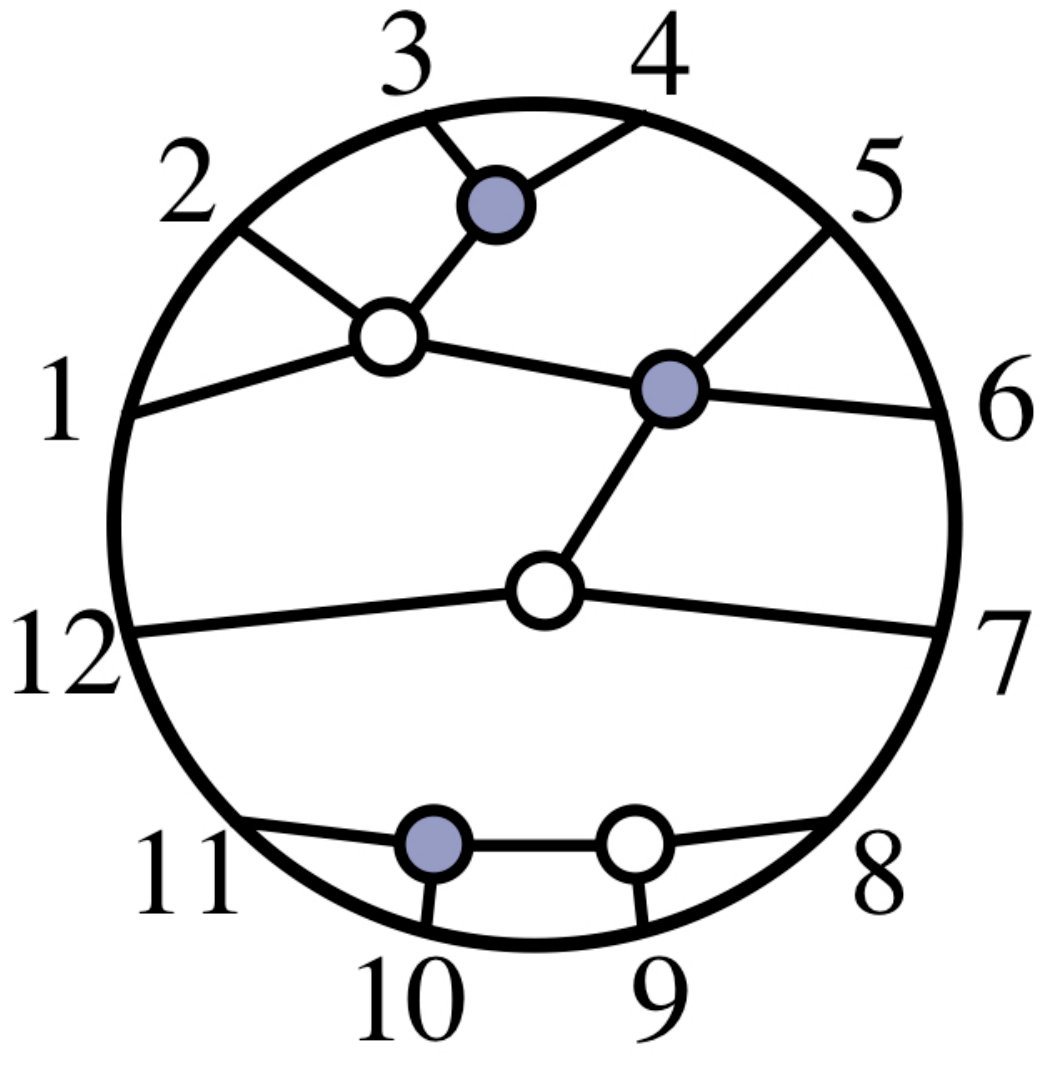}  
\newline{\tiny \{2,4,7,15,13,17,12,9,11,20,22,18\}}&
{\tiny
$\left(
\begin{array}{cccccccccccc}
 1 & \alpha_{10} & \alpha_9 & 0 & -\alpha_7 & 0 & 0 & 0 & 0 & 0 & 0 & 0 \\
 0 & 0 & 1 & \alpha_8 & 0 & 0 & 0 & 0 & 0 & 0 & 0 & 0 \\
 0 & 0 & 0 & 0 & 1 & \alpha_6 & 0 & 0 & 0 & 0 & 0 & 0 \\
 0 & 0 & 0 & 0 & 0 & 1 & \alpha_5 & 0 & 0 & 0 & 0 & \alpha_1 \\
 0 & 0 & 0 & 0 & 0 & 0 & 0 & 1 & \alpha_4 & \alpha_3 & 0 & 0 \\
 0 & 0 & 0 & 0 & 0 & 0 & 0 & 0 & 0 & 1 & \alpha_2 & 0 \\
\end{array}
\right)$}
\\
\hline

$\begin{array}{c} (3.4b) \\n=10\\k=5\\d=9 \end{array}$ &
\includegraphics[width=0.1\textwidth]{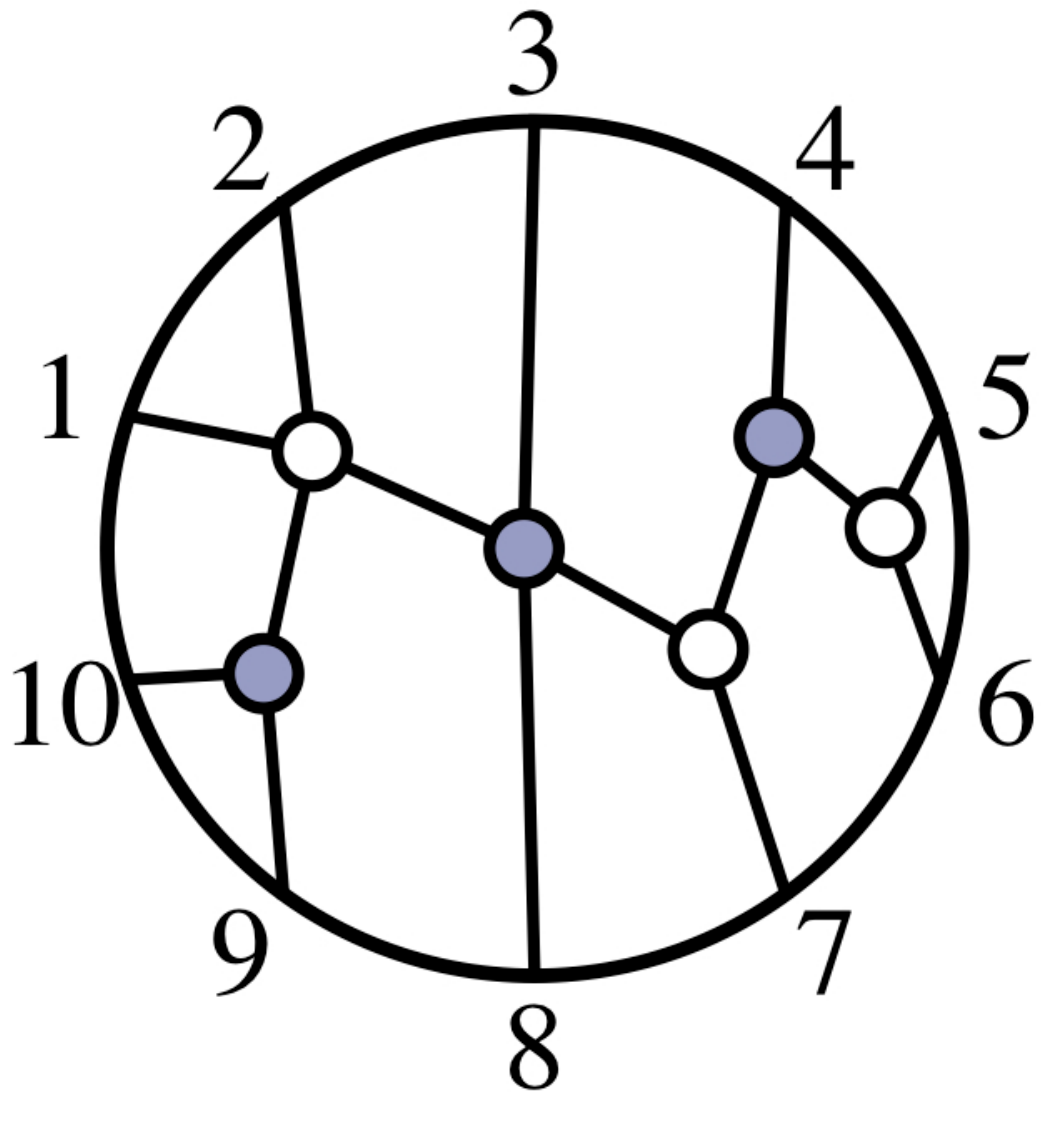}  
\newline{\tiny \{2,8,10,7,6,14,13,15,11,19\}}&
{\tiny
$\left(
\begin{array}{cccccccccc}
 1 & \alpha_9 & \alpha_8 & 0 & 0 & 0 & 0 & 0 & -\alpha_2 & 0 \\
 0 & 0 & 1 & 0 & -\alpha_4 & -\alpha_4 \alpha_7 & -\alpha_4 \alpha_5 & 0 & 0 & 0 \\
 0 & 0 & 0 & 1 & \alpha_6 & \alpha_6 \alpha_7 & 0 & 0 & 0 & 0 \\
 0 & 0 & 0 & 0 & 1 & \alpha_7 & \alpha_5 & \alpha_3 & 0 & 0 \\
 0 & 0 & 0 & 0 & 0 & 0 & 0 & 0 & 1 & \alpha_1 \\
\end{array}
\right)$}
\\
\hline

$\begin{array}{c} (3.4c) \\n=10\\k=5\\d=9 \end{array}$ &
\includegraphics[width=0.1\textwidth]{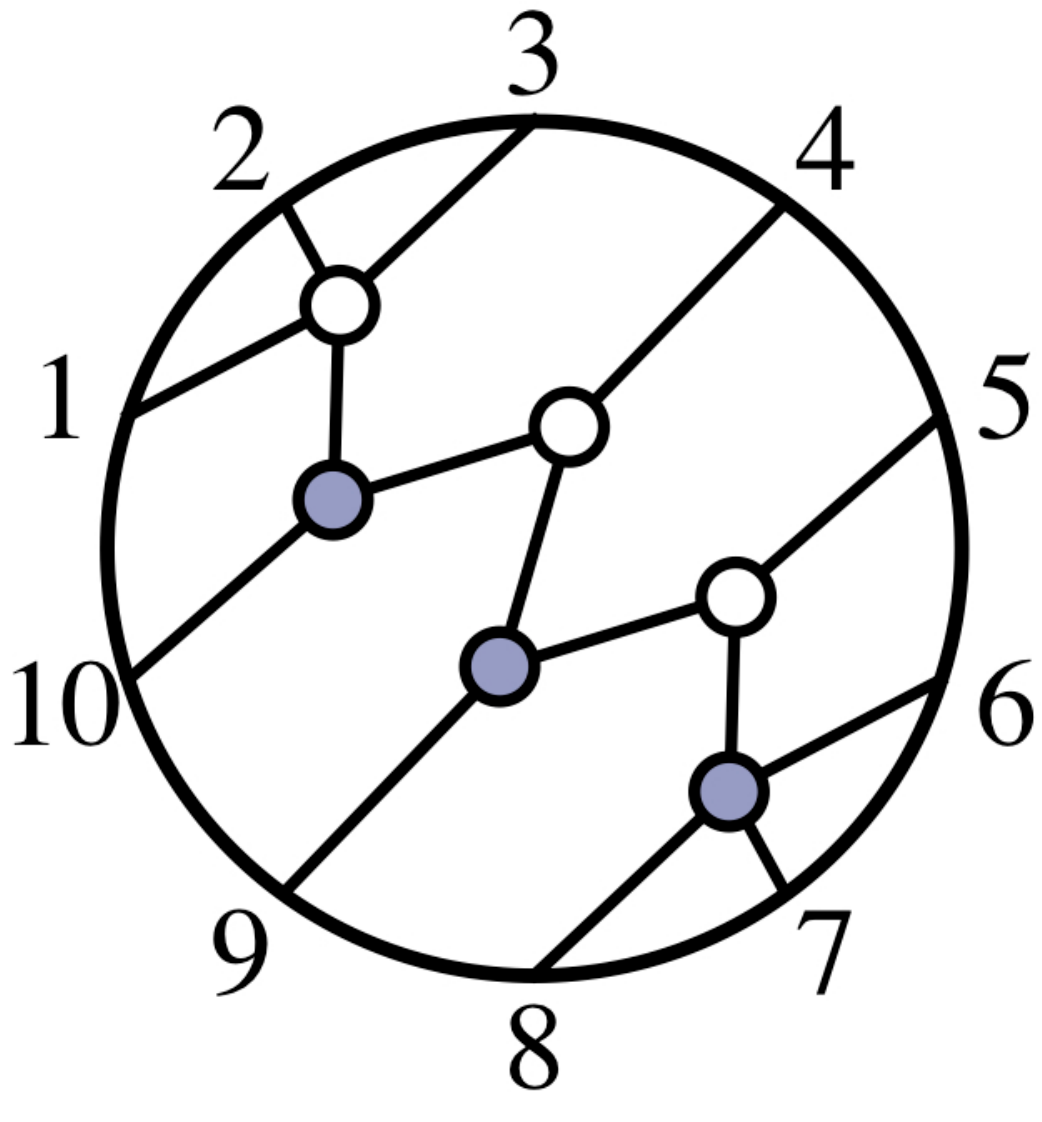}  
\newline\tiny\{2,3,10,9,8,11,16,17,15,14\} &
{\tiny
$\left(
\begin{array}{cccccccccc}
 1 & \alpha_9 & \alpha_8 & \alpha_5 & \alpha_5 \alpha_6 & \alpha_5 \alpha_6 \alpha_7 & 0 & 0 & 0 & 0 \\
 0 & 0 & 0 & 1 & \alpha_6 & \alpha_6 \alpha_7 & 0 & 0 & 0 & -\alpha_1 \\
 0 & 0 & 0 & 0 & 1 & \alpha_7 & 0 & 0 & \alpha_2 & 0 \\
 0 & 0 & 0 & 0 & 0 & 1 & \alpha_4 & 0 & 0 & 0 \\
 0 & 0 & 0 & 0 & 0 & 0 & 1 & \alpha_3 & 0 & 0 \\
\end{array}
\right)$}
\\
\hline

$\begin{array}{c} (3.4d) \\n=10\\k=5\\d=9 \end{array}$ &
\includegraphics[width=0.1\textwidth]{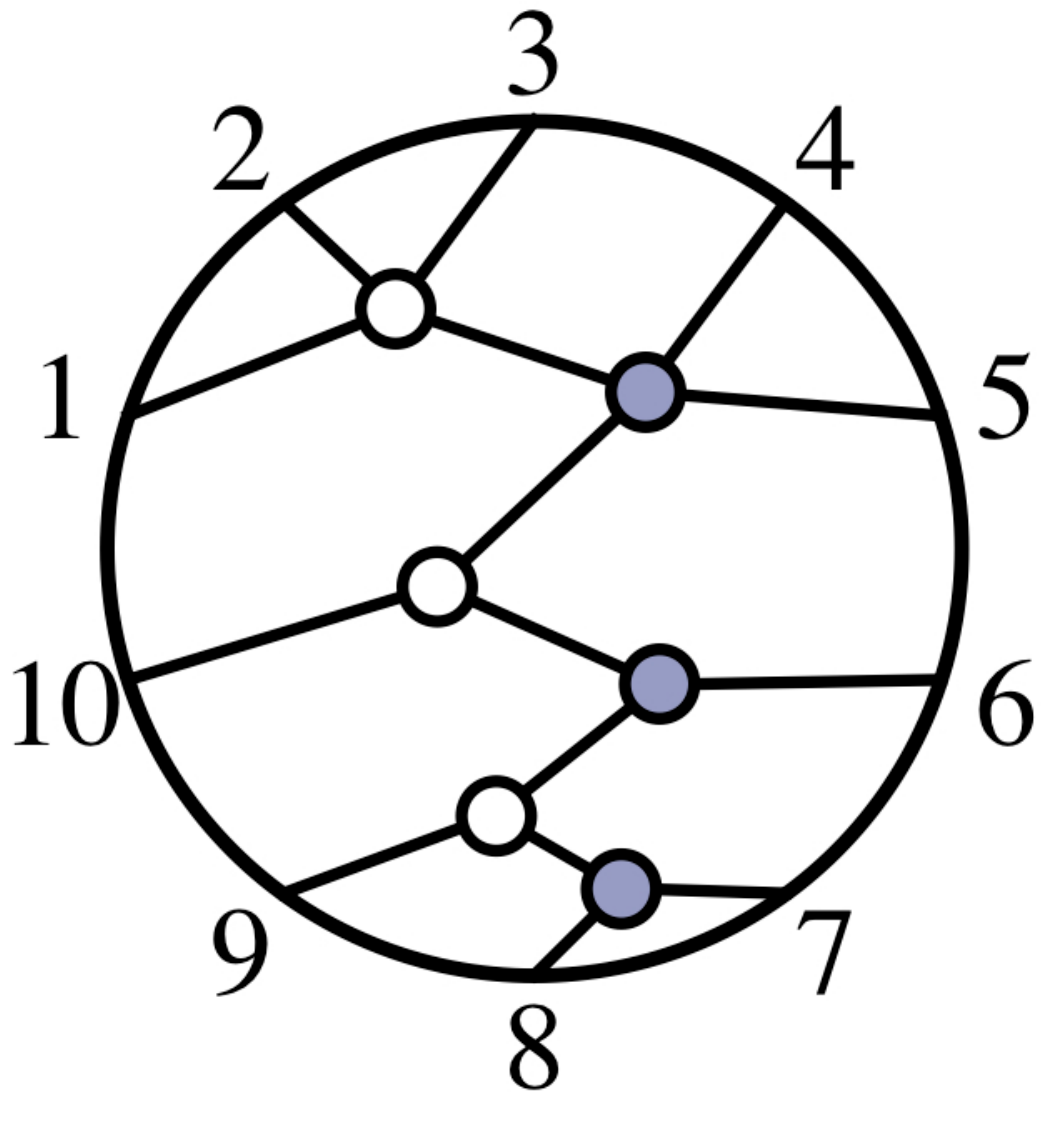}  
\newline\tiny\{2,3,8,11,14,10,9,17,16,15\} &
{\tiny
$\left(
\begin{array}{cccccccccc}
 1 & \alpha_9 & \alpha_8 & \alpha_7 & 0 & 0 & 0 & 0 & 0 & 0 \\
 0 & 0 & 0 & 1 & \alpha_6 & 0 & 0 & 0 & 0 & 0 \\
 0 & 0 & 0 & 0 & 1 & \alpha_5 & 0 & 0 & 0 & \alpha_1 \\
 0 & 0 & 0 & 0 & 0 & 1 & \alpha_4 & 0 & -\alpha_2 & 0 \\
 0 & 0 & 0 & 0 & 0 & 0 & 1 & \alpha_3 & 0 & 0 \\
\end{array}
\right)$}
\\
\hline

$\begin{array}{c} (3.5a) \\n=10\\k=5\\d=9 \end{array}$ &
\includegraphics[width=0.1\textwidth]{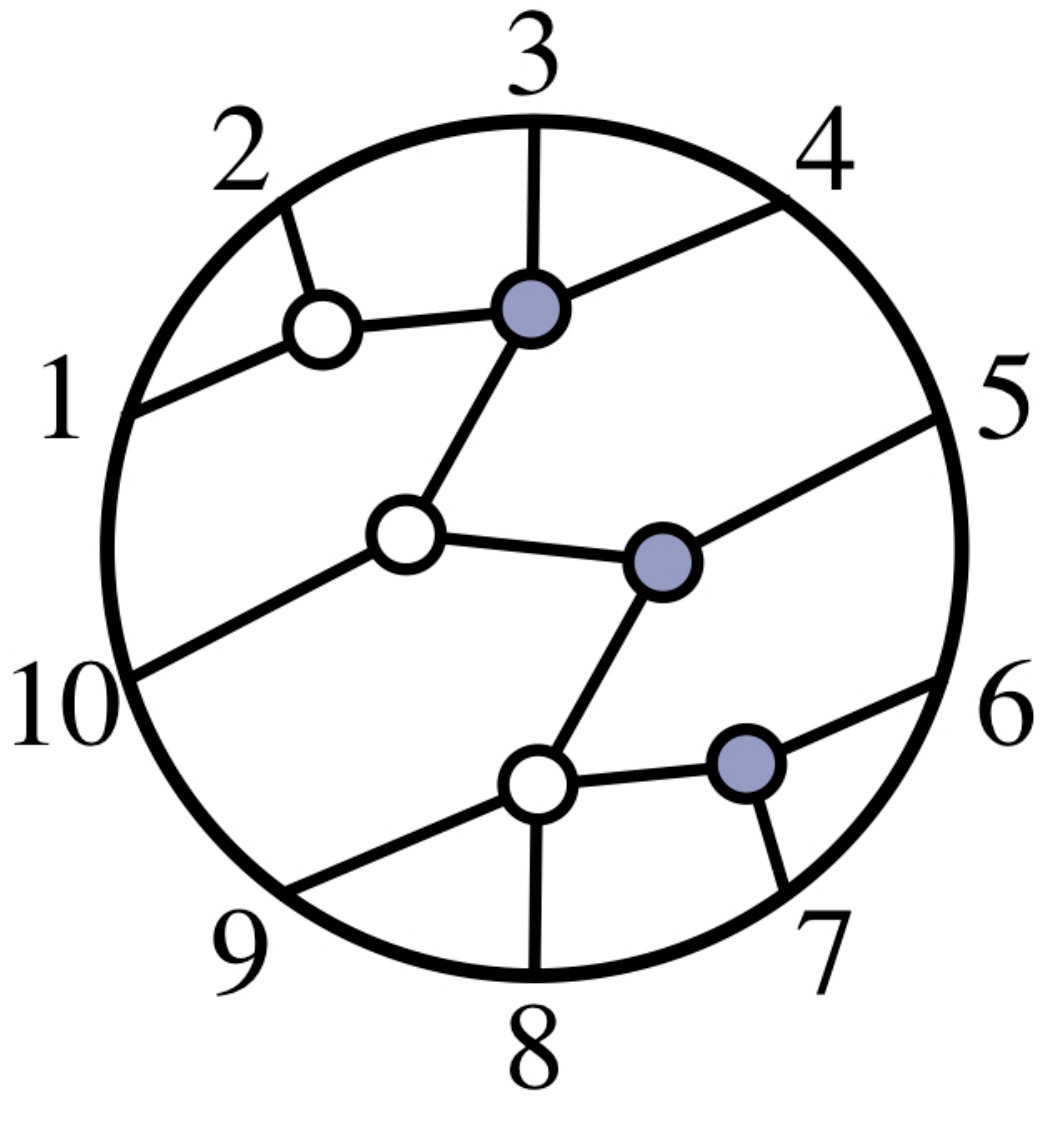}  
\newline{\tiny \{2,7,11,13,10,8,16,9,15,14\}}&
{\tiny
$\left(
\begin{array}{cccccccccc}
 1 & \alpha_9 & \alpha_8 & 0 & 0 & 0 & 0 & 0 & 0 & 0 \\
 0 & 0 & 1 & \alpha_7 & 0 & 0 & 0 & 0 & 0 & 0 \\
 0 & 0 & 0 & 1 & \alpha_6 & 0 & 0 & 0 & 0 & \alpha_1 \\
 0 & 0 & 0 & 0 & 1 & \alpha_5 & 0 & -\alpha_3 & -\alpha_2 & 0 \\
 0 & 0 & 0 & 0 & 0 & 1 & \alpha_4 & 0 & 0 & 0 \\
\end{array}
\right)$}
\\
\hline

$\begin{array}{c} (3.5b) \\n=12\\k=6\\d=10 \end{array}$ &
\includegraphics[width=0.1\textwidth]{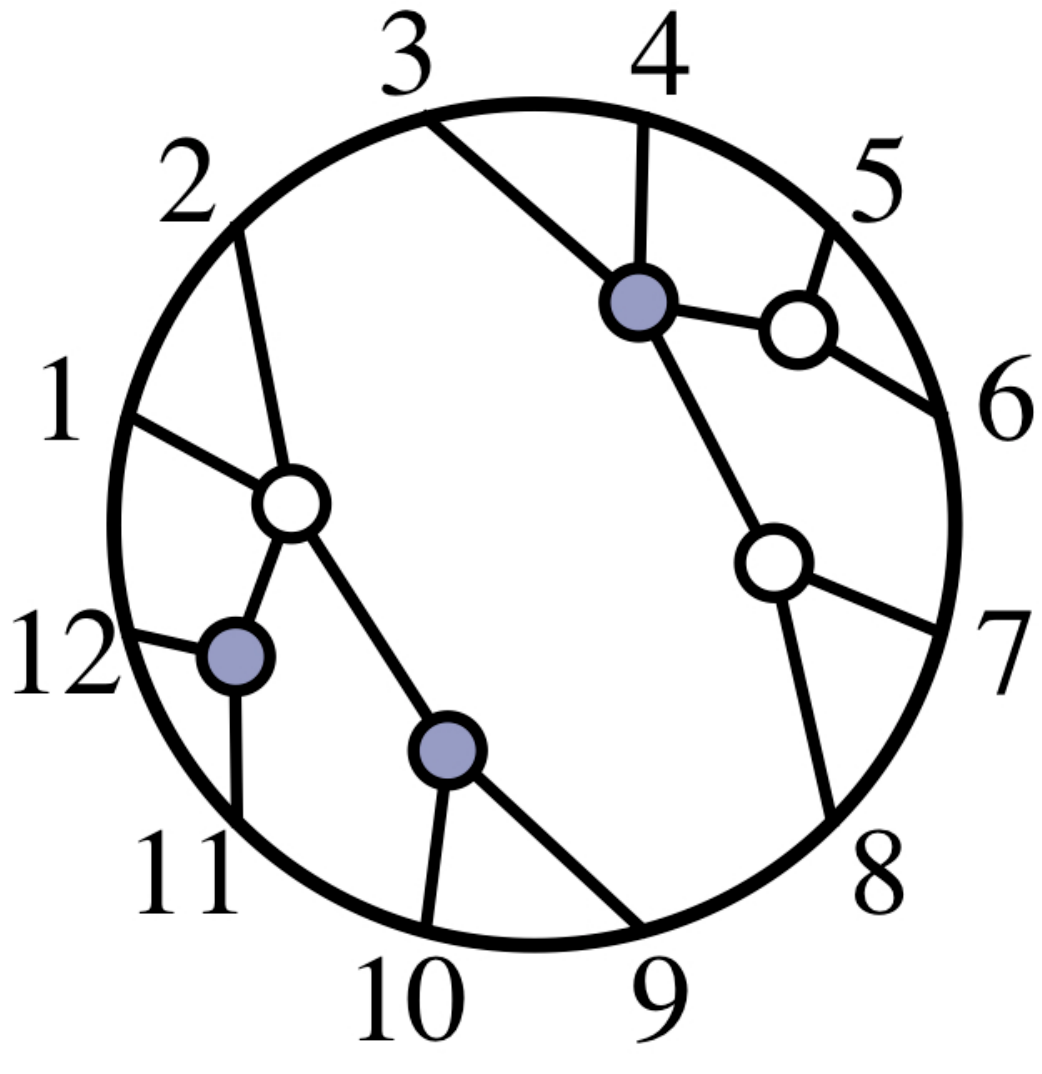}  
\newline{\tiny \{2,10,7,15,6,16,8,17,12,21,13,23\}}&
{\tiny
$\left(
\begin{array}{cccccccccccc}
 1 & \alpha_{10} & 0 & 0 & 0 & 0 & 0 & 0 & -\alpha_4 & 0 & \alpha_2 & 0 \\
 0 & 0 & 1 & \alpha_9 & 0 & 0 & 0 & 0 & 0 & 0 & 0 & 0 \\
 0 & 0 & 0 & 1 & \alpha_7 & \alpha_7 \alpha_8 & 0 & 0 & 0 & 0 & 0 & 0 \\
 0 & 0 & 0 & 0 & 1 & \alpha_8 & \alpha_6 & \alpha_5 & 0 & 0 & 0 & 0 \\
 0 & 0 & 0 & 0 & 0 & 0 & 0 & 0 & 1 & \alpha_3 & 0 & 0 \\
 0 & 0 & 0 & 0 & 0 & 0 & 0 & 0 & 0 & 0 & 1 & \alpha_1 \\
\end{array}
\right)$}
\\
\hline

$\begin{array}{c} (3.5c) \\n=10\\k=5\\d=9 \end{array}$ &
\includegraphics[width=0.1\textwidth]{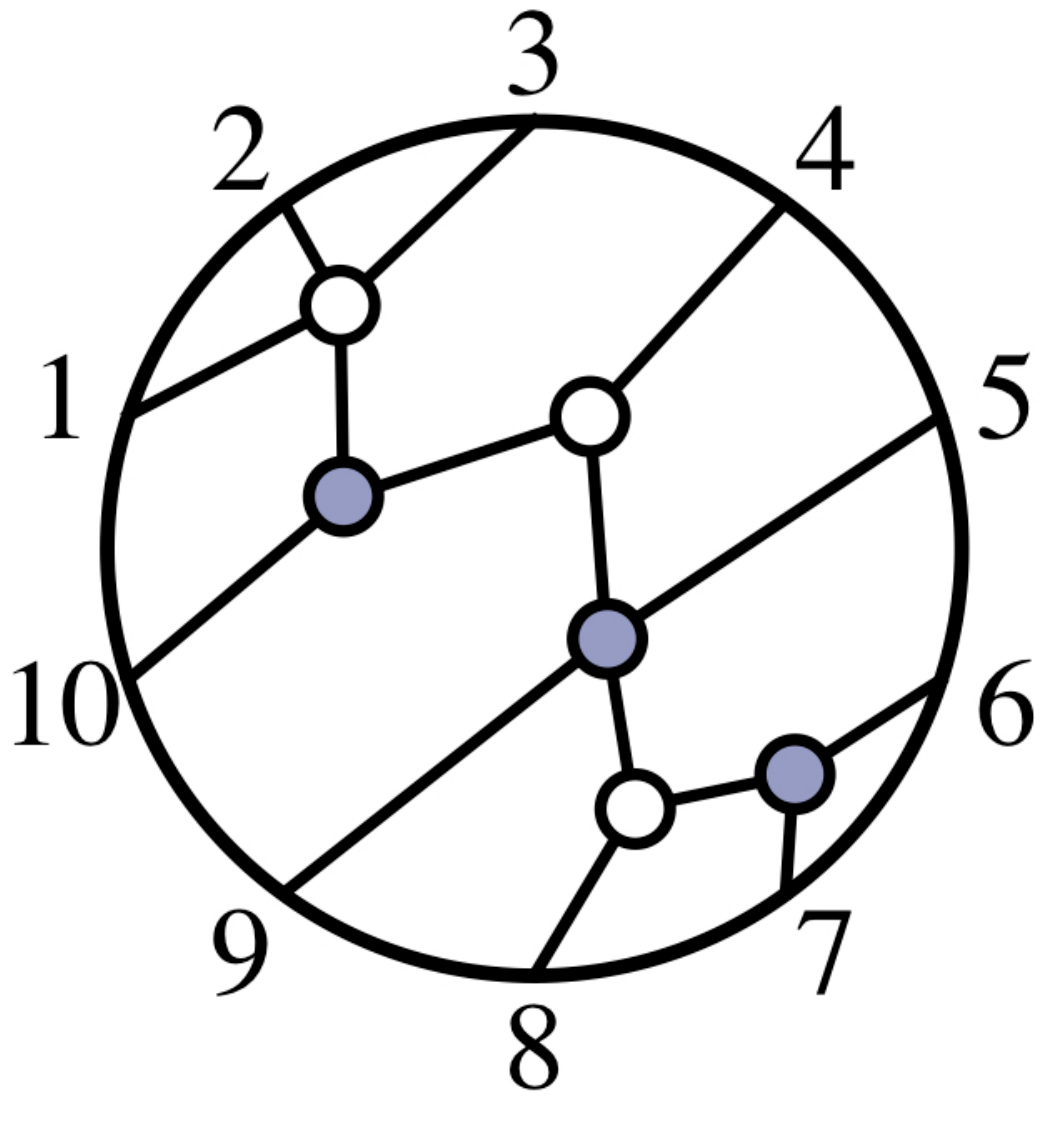}  
\newline \tiny\{2,3,10,9,11,8,16,15,17,14\}&
{\tiny
$\left(
\begin{array}{cccccccccc}
 1 & \alpha_9 & \alpha_8 & \alpha_6 & \alpha_6 \alpha_7 & 0 & 0 & 0 & 0 & 0 \\
 0 & 0 & 0 & 1 & \alpha_7 & 0 & 0 & 0 & 0 & -\alpha_1 \\
 0 & 0 & 0 & 0 & 1 & 0 & -\alpha_3 & -\alpha_3 \alpha_4 & 0 & 0 \\
 0 & 0 & 0 & 0 & 0 & 1 & \alpha_5 & 0 & 0 & 0 \\
 0 & 0 & 0 & 0 & 0 & 0 & 1 & \alpha_4 & \alpha_2 & 0 \\
\end{array}
\right)$}
\\
\hline

$\begin{array}{c} (3.5d) \\n=10\\k=5\\d=9 \end{array}$ &
\includegraphics[width=0.1\textwidth]{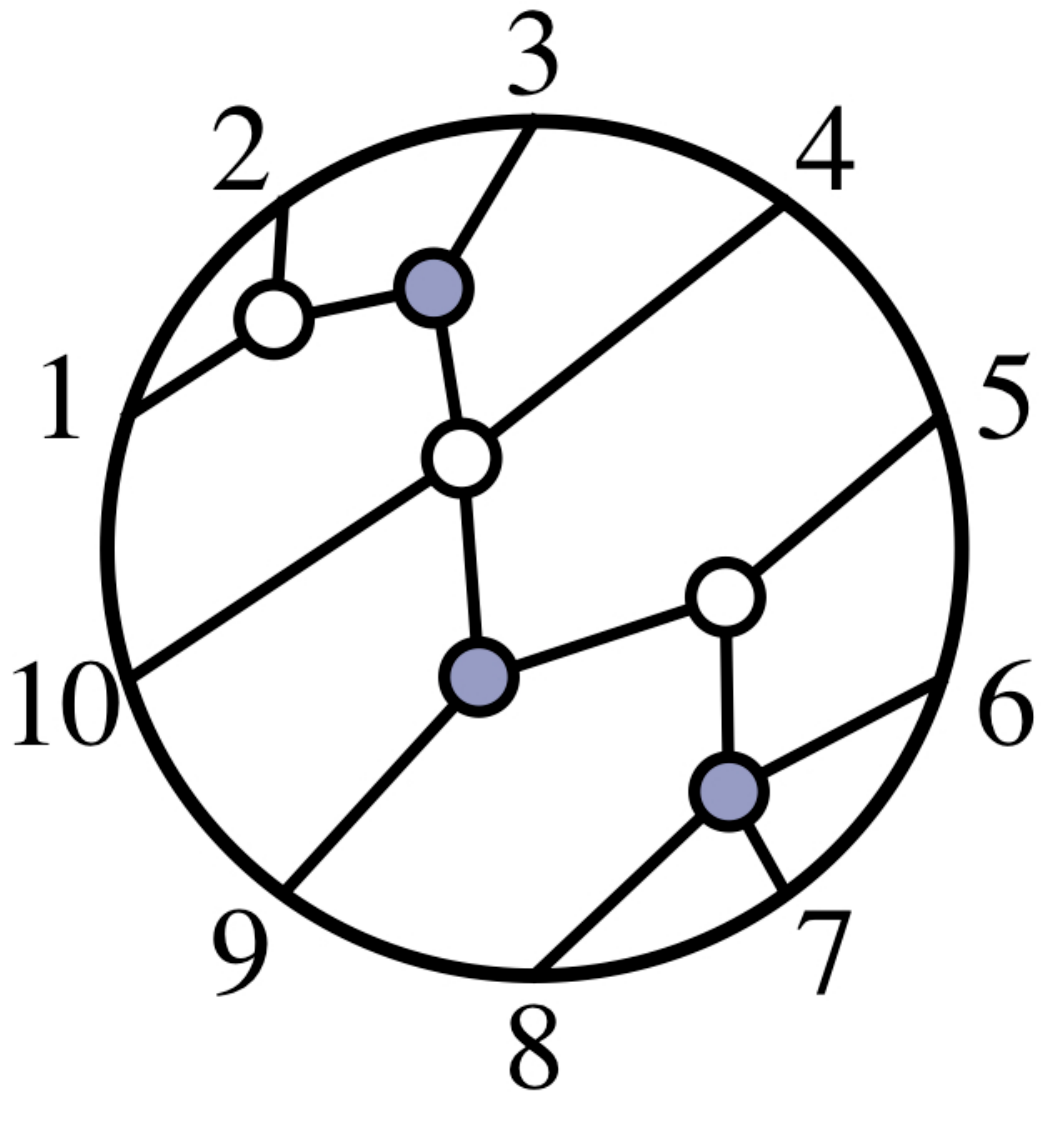}  
\newline\tiny\{2,4,11,9,8,10,16,17,15,13\} &
{\tiny
$\left(
\begin{array}{cccccccccc}
 1 & \alpha_9 & \alpha_8 & 0 & 0 & 0 & 0 & 0 & 0 & 0 \\
 0 & 0 & 1 & \alpha_7 & \alpha_5 & \alpha_5 \alpha_6 & 0 & 0 & 0 & -\alpha_1 \\
 0 & 0 & 0 & 0 & 1 & \alpha_6 & 0 & 0 & \alpha_2 & 0 \\
 0 & 0 & 0 & 0 & 0 & 1 & \alpha_4 & 0 & 0 & 0 \\
 0 & 0 & 0 & 0 & 0 & 0 & 1 & \alpha_3 & 0 & 0 \\
\end{array}
\right)$}
\\
\hline

$\begin{array}{c} (3.6a) \\n=6\\k=3\\d=7 \end{array}$ &
\includegraphics[width=0.1\textwidth]{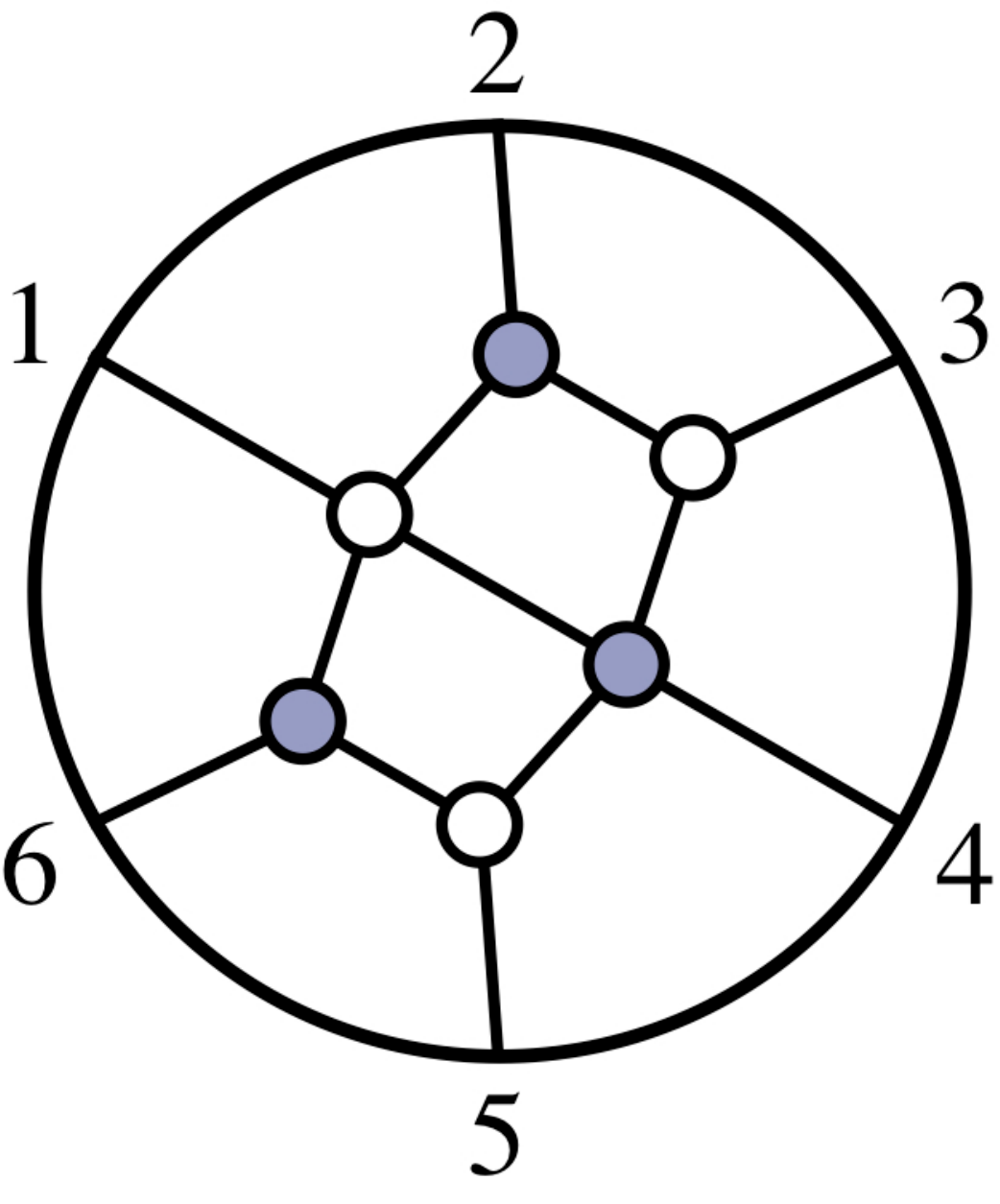}  
\newline\tiny\{3,5,6,8,7,10\}&
\tiny{$\left(
\begin{array}{cccccc}
 1 & \alpha_5+\alpha_7 & \alpha_5 \alpha_6 & -\alpha_2 & -\alpha_2 \alpha_3 & 0 \\
 0 & 1 & \alpha_6 & \alpha_4 & 0 & 0 \\
 0 & 0 & 0 & 1 & \alpha_3 & \alpha_1 \\
\end{array}
\right)$}
\\
\hline

$\begin{array}{c} (3.6b) \\n=8\\k=4\\d=8 \end{array}$ &
\includegraphics[width=0.1\textwidth]{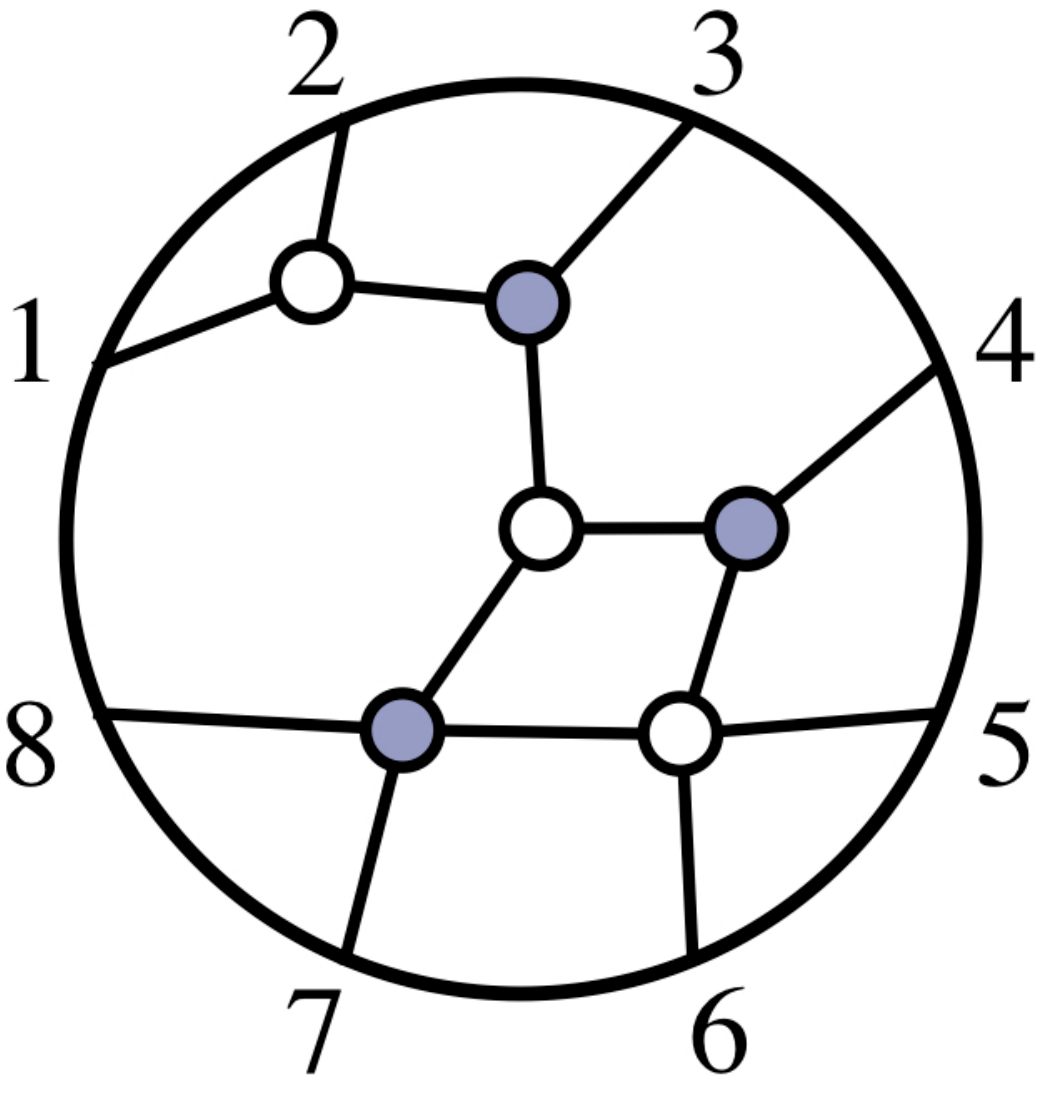}  
\newline{\tiny \{2,5,9,8,6,11,12,15\}}&
{\tiny
$\left(
\begin{array}{cccccccc}
 1 & \alpha_8 & \alpha_7 & 0 & 0 & 0 & 0 & 0 \\
 0 & 0 & 1 & \alpha_3+\alpha_6 & \alpha_3 \alpha_5 & \alpha_3 \alpha_4 & 0 & 0 \\
 0 & 0 & 0 & 1 & \alpha_5 & \alpha_4 & \alpha_2 & 0 \\
 0 & 0 & 0 & 0 & 0 & 0 & 1 & \alpha_1 \\
\end{array}
\right)$}
\\
\hline

$\begin{array}{c} (3.6c) \\n=8\\k=4\\d=8 \end{array}$ &
\includegraphics[width=0.1\textwidth]{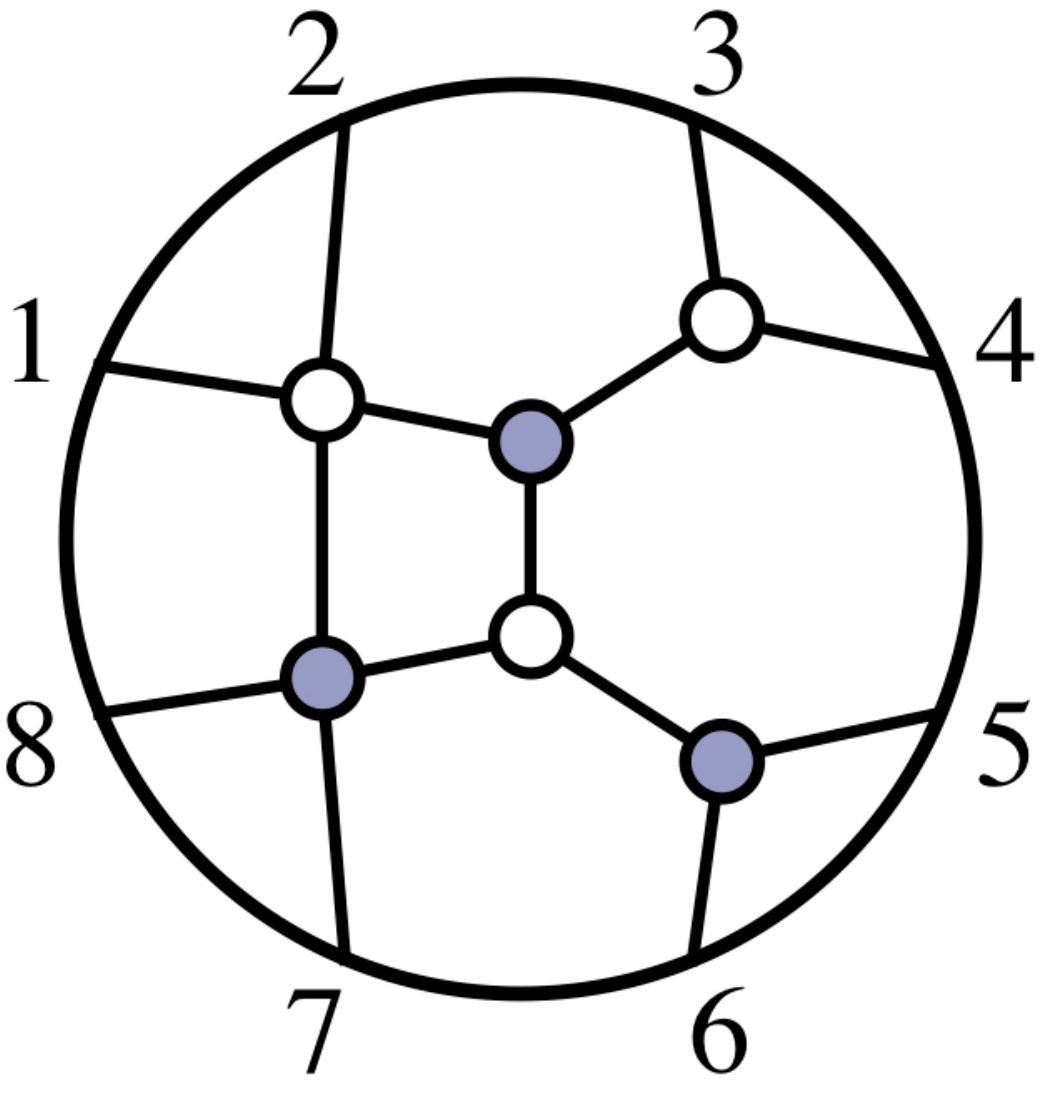}  
\newline{\tiny \{2,6,4,8,9,13,11,15\}}&
{\tiny
$\left(
\begin{array}{cccccccc}
 1 & \alpha_8 & \alpha_4+\alpha_6 & (\alpha_4+\alpha_6) \alpha_7 & \alpha_4 \alpha_5 & 0 & 0 & 0 \\
 0 & 0 & 1 & \alpha_7 & \alpha_5 & 0 & -\alpha_2 & 0 \\
 0 & 0 & 0 & 0 & 1 & \alpha_3 & 0 & 0 \\
 0 & 0 & 0 & 0 & 0 & 0 & 1 & \alpha_1 \\
\end{array}
\right)$}
\\
\hline

$\begin{array}{c} (3.6d) \\n=10\\k=5\\d=9 \end{array}$ &
\includegraphics[width=0.1\textwidth]{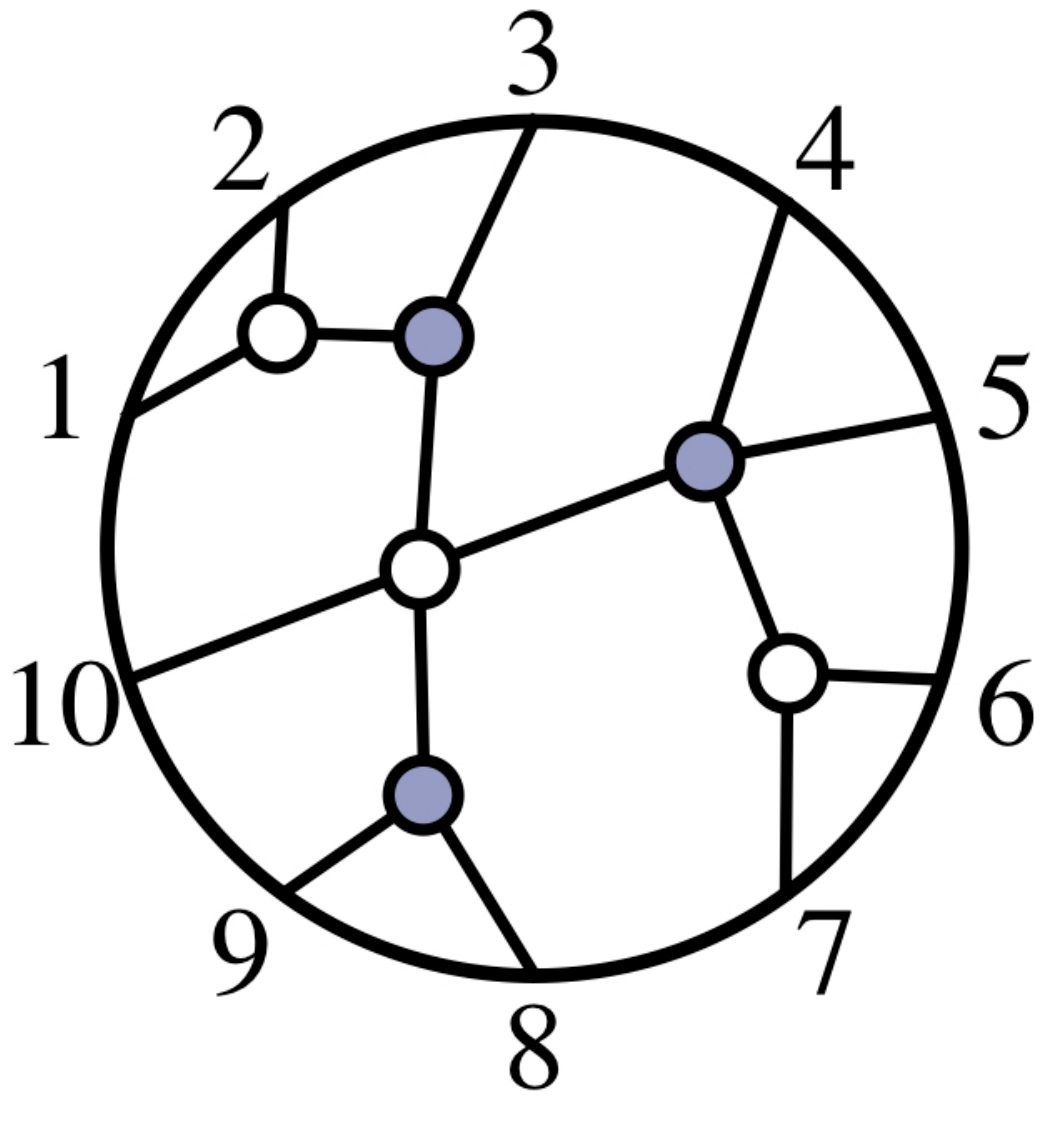}  
\newline{\tiny \{2,6,11,9,14,7,15,10,18,13\}}&
{\tiny
$\left(
\begin{array}{cccccccccc}
 1 & \alpha_9 & \alpha_8 & 0 & 0 & 0 & 0 & 0 & 0 & 0 \\
 0 & 0 & 1 & \alpha_7 & 0 & 0 & 0 & \alpha_3 & 0 & -\alpha_1 \\
 0 & 0 & 0 & 1 & \alpha_6 & 0 & 0 & 0 & 0 & 0 \\
 0 & 0 & 0 & 0 & 1 & \alpha_5 & \alpha_4 & 0 & 0 & 0 \\
 0 & 0 & 0 & 0 & 0 & 0 & 0 & 1 & \alpha_2 & 0 \\
\end{array}
\right)$}
\\
\hline

$\begin{array}{c} (3.6e) \\n=8\\k=4\\d=8 \end{array}$ &
\includegraphics[width=0.1\textwidth]{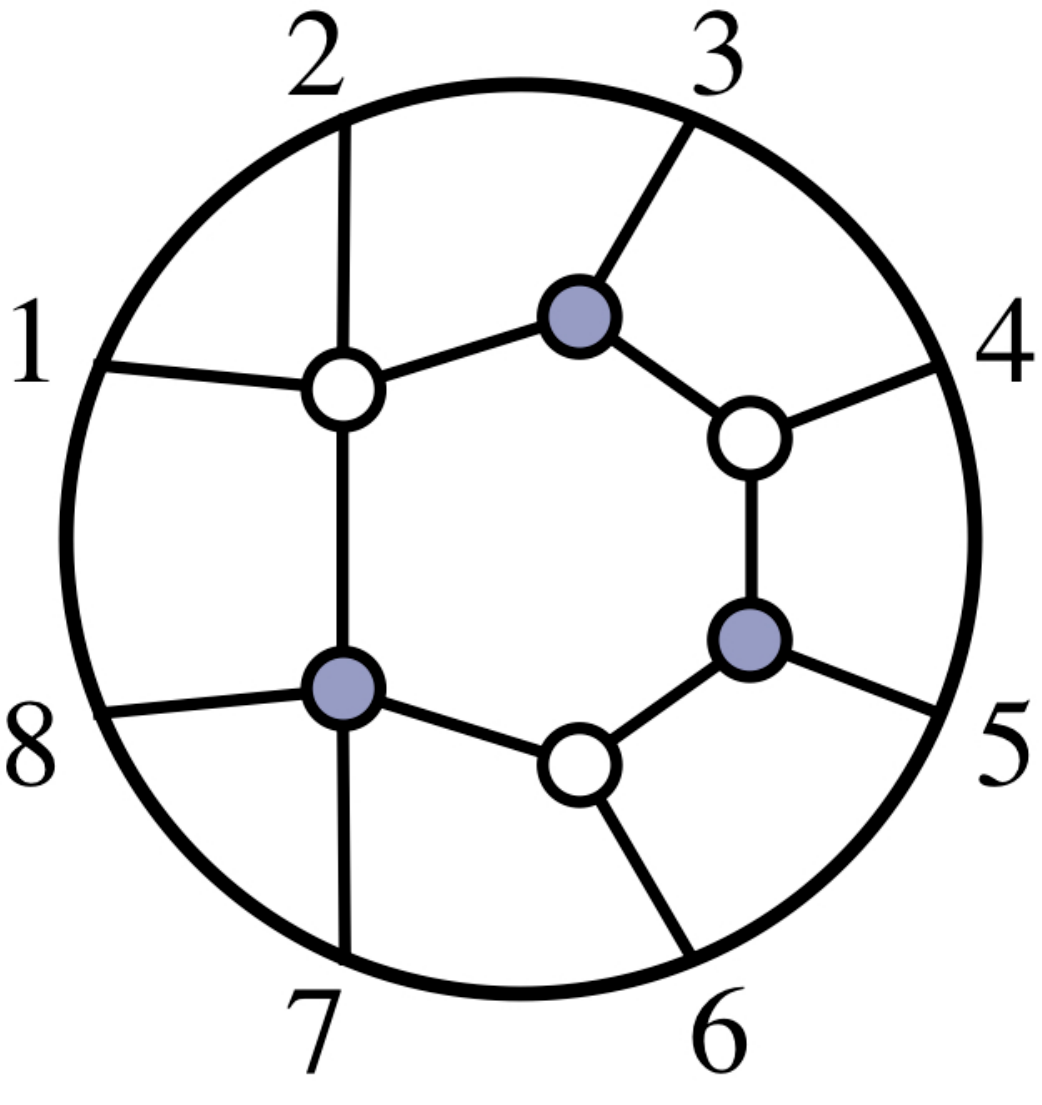}  
\newline\tiny\{2,4,8,6,11,9,13,15\} &
{\tiny
$\left(
\begin{array}{cccccccc}
 1 & \alpha_8 & \alpha_7 & 0 & -\alpha_3 & -\alpha_3 \alpha_4 & 0 & 0 \\
 0 & 0 & 1 & \alpha_6 & \alpha_5 & 0 & 0 & 0 \\
 0 & 0 & 0 & 0 & 1 & \alpha_4 & \alpha_2 & 0 \\
 0 & 0 & 0 & 0 & 0 & 0 & 1 & \alpha_1 \\
\end{array}
\right)$}
\\
\hline

$\begin{array}{c} (3.6f) \\n=10\\k=5\\d=9 \end{array}$ &
\includegraphics[width=0.1\textwidth]{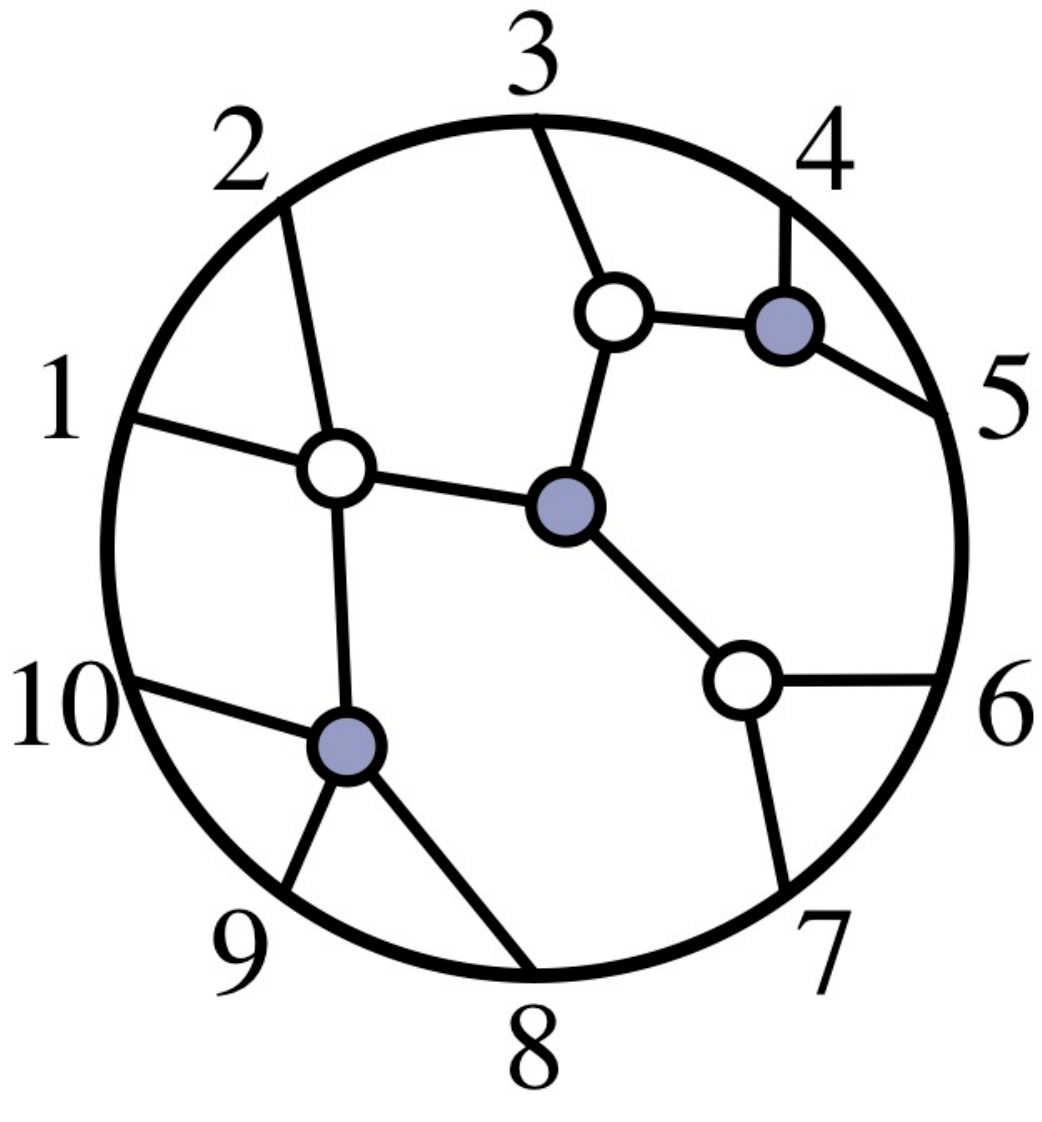}  
\newline\tiny\{2,6,5,10,14,7,13,11,18,19\} &
{\tiny
$\left(
\begin{array}{cccccccccc}
 1 & \alpha_9 & \alpha_7 & \alpha_7 \alpha_8 & 0 & 0 & 0 & \alpha_3 & 0 & 0 \\
 0 & 0 & 1 & \alpha_8 & 0 & -\alpha_5 & -\alpha_4 & 0 & 0 & 0 \\
 0 & 0 & 0 & 1 & \alpha_6 & 0 & 0 & 0 & 0 & 0 \\
 0 & 0 & 0 & 0 & 0 & 0 & 0 & 1 & \alpha_2 & 0 \\
 0 & 0 & 0 & 0 & 0 & 0 & 0 & 0 & 1 & \alpha_1 \\
\end{array}
\right)$}
\\
\hline

$\begin{array}{c} (3.6g) \\n=10\\k=5\\d=9 \end{array}$ &
\includegraphics[width=0.1\textwidth]{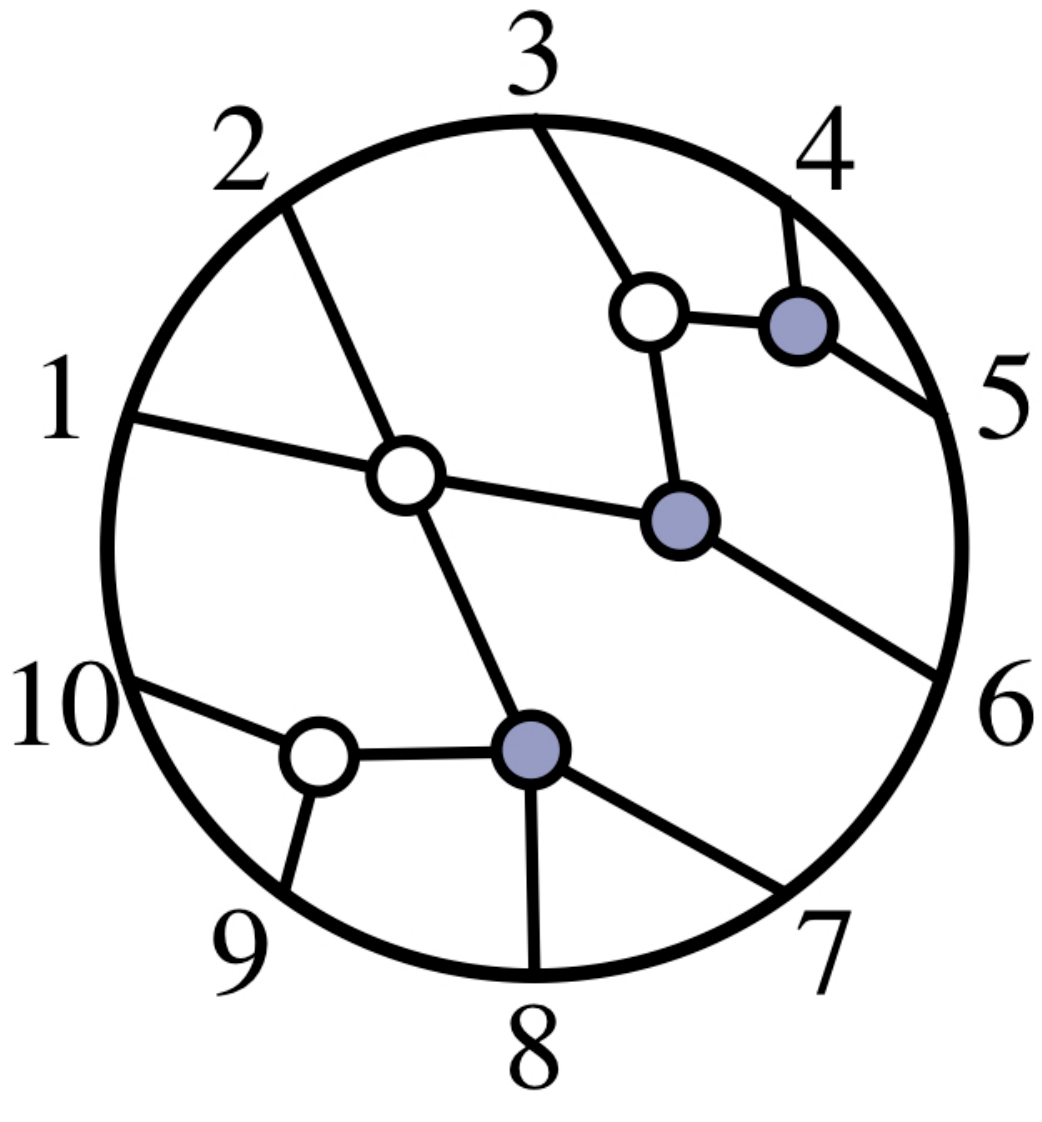}  
\newline\tiny \{2,6,5,9,14,13,11,17,10,18\}&
{\tiny
$\left(
\begin{array}{cccccccccc}
 1 & \alpha_9 & \alpha_7 & \alpha_7 \alpha_8 & 0 & 0 & \alpha_4 & 0 & 0 & 0 \\
 0 & 0 & 1 & \alpha_8 & 0 & -\alpha_5 & 0 & 0 & 0 & 0 \\
 0 & 0 & 0 & 1 & \alpha_6 & 0 & 0 & 0 & 0 & 0 \\
 0 & 0 & 0 & 0 & 0 & 0 & 1 & \alpha_3 & 0 & 0 \\
 0 & 0 & 0 & 0 & 0 & 0 & 0 & 1 & \alpha_2 & \alpha_1 \\
\end{array}
\right)$}
\\
\hline

$\begin{array}{c} (3.6h) \\n=10\\k=5\\d=9 \end{array}$ &
\includegraphics[width=0.1\textwidth]{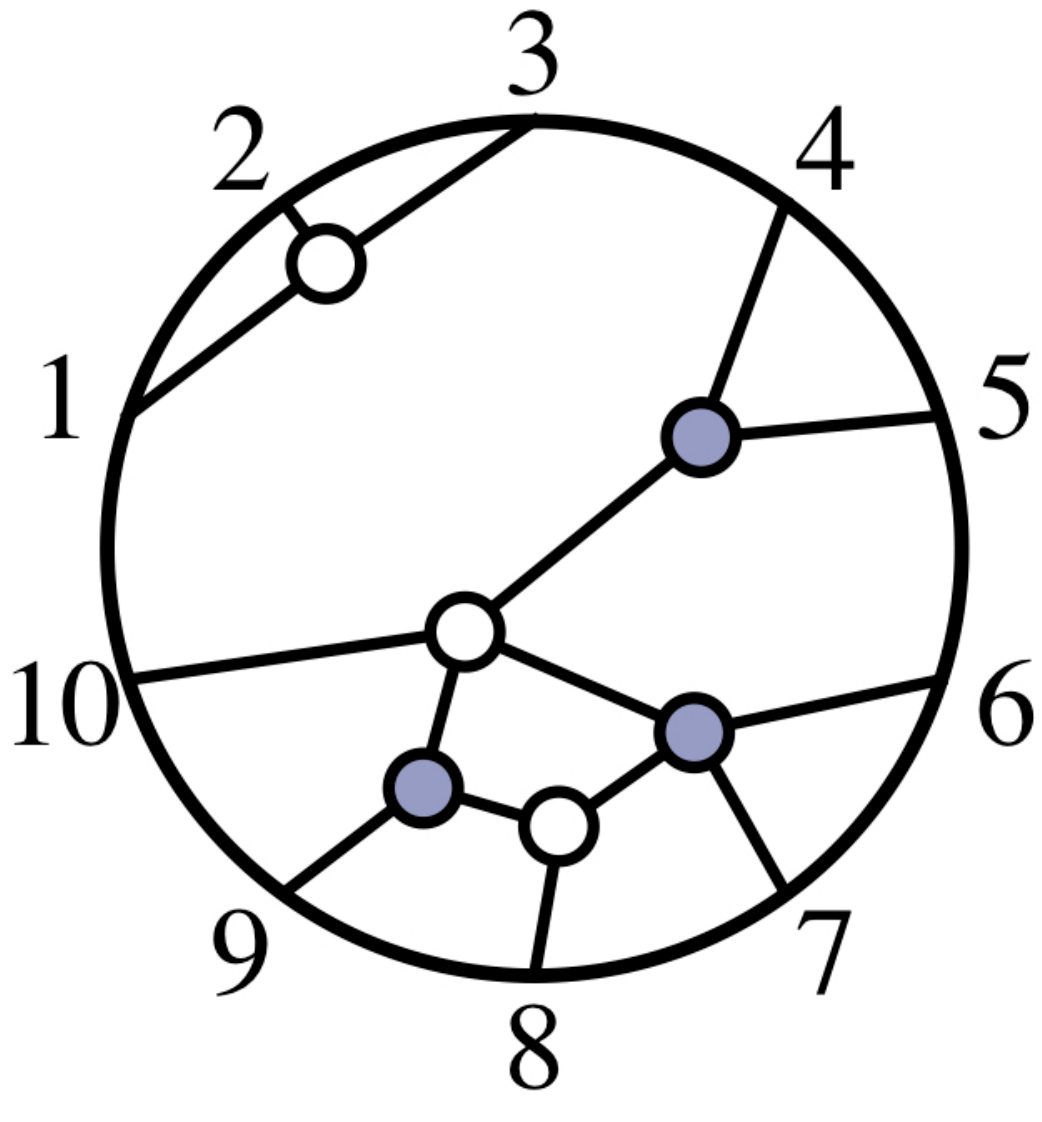}  
\newline \tiny\{2,3,11,8,14,9,16,10,17,15\}&
{\tiny
$\left(
\begin{array}{cccccccccc}
 1 & \alpha_9 & \alpha_8 & 0 & 0 & 0 & 0 & 0 & 0 & 0 \\
 0 & 0 & 0 & 1 & \alpha_7 & 0 & 0 & 0 & 0 & 0 \\
 0 & 0 & 0 & 0 & 1 & \alpha_6 & -\alpha_3 & -\alpha_3 \alpha_4 & 0 & \alpha_1 \\
 0 & 0 & 0 & 0 & 0 & 1 & \alpha_5 & 0 & 0 & 0 \\
 0 & 0 & 0 & 0 & 0 & 0 & 1 & \alpha_4 & \alpha_2 & 0 \\
\end{array}
\right)$}
\\
\hline

$\begin{array}{c} (3.6i) \\n=10\\k=5\\d=9 \end{array}$ &
\includegraphics[width=0.1\textwidth]{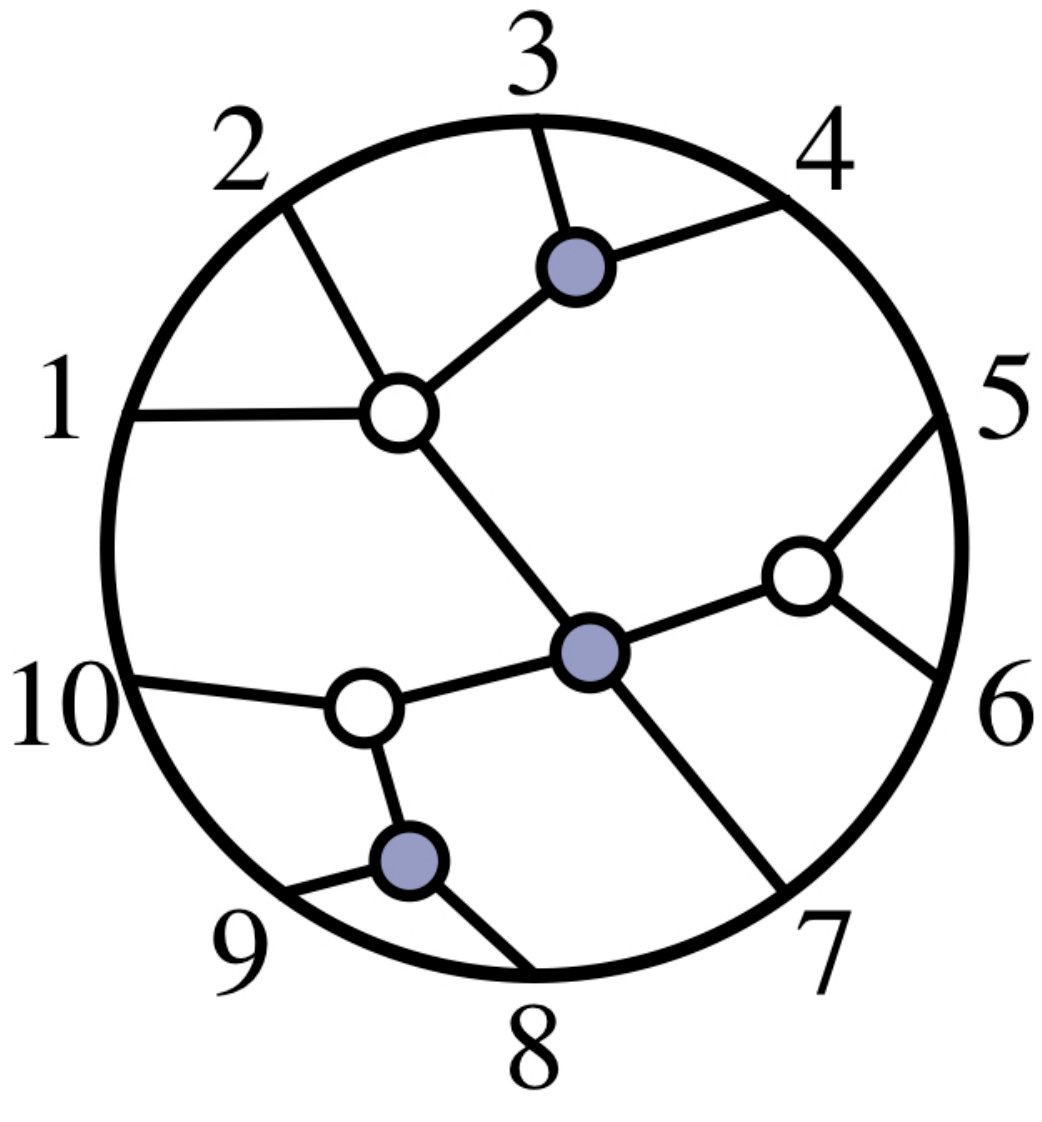}  
\newline \tiny\{2,4,9,13,6,11,15,10,18,17\}&
{\tiny
$\left(
\begin{array}{cccccccccc}
 1 & \alpha_9 & \alpha_8 & 0 & -\alpha_5 & -\alpha_5 \alpha_6 & 0 & 0 & 0 & 0 \\
 0 & 0 & 1 & \alpha_7 & 0 & 0 & 0 & 0 & 0 & 0 \\
 0 & 0 & 0 & 0 & 1 & \alpha_6 & \alpha_4 & 0 & 0 & 0 \\
 0 & 0 & 0 & 0 & 0 & 0 & 1 & \alpha_3 & 0 & -\alpha_1 \\
 0 & 0 & 0 & 0 & 0 & 0 & 0 & 1 & \alpha_2 & 0 \\
\end{array}
\right)$}
\\
\hline

$\begin{array}{c} (3.6j) \\n=8\\k=4\\d=8 \end{array}$ &
\includegraphics[width=0.1\textwidth]{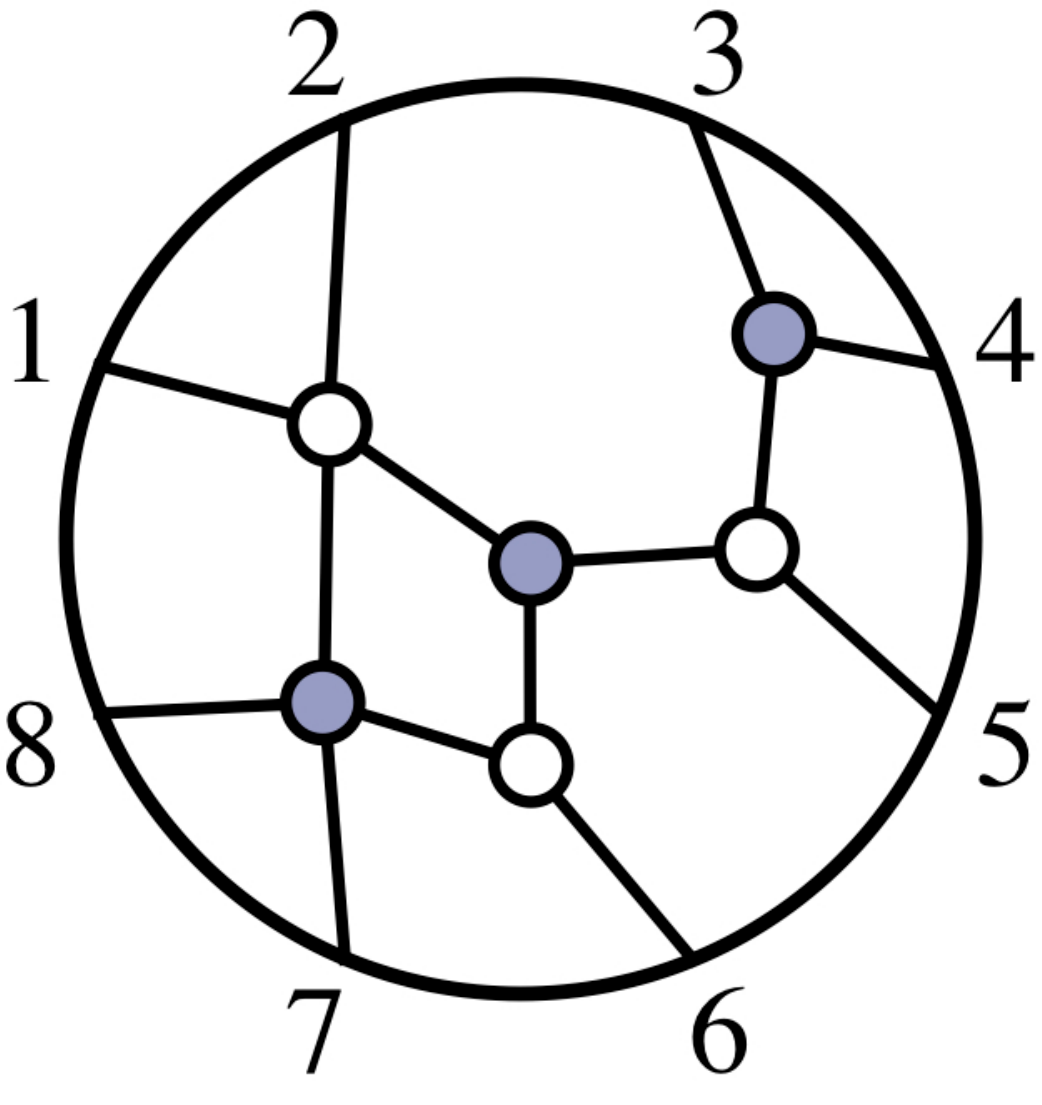}  
\newline\tiny\{2,6,5,11,8,9,12,15\} &
{\tiny
$\left(
\begin{array}{cccccccc}
 1 & \alpha_8 & 0 & -\alpha_3-\alpha_5 & (-\alpha_3-\alpha_5) \alpha_6 & -\alpha_3 \alpha_4 & 0 & 0 \\
 0 & 0 & 1 & \alpha_7 & 0 & 0 & 0 & 0 \\
 0 & 0 & 0 & 1 & \alpha_6 & \alpha_4 & \alpha_2 & 0 \\
 0 & 0 & 0 & 0 & 0 & 0 & 1 & \alpha_1 \\
\end{array}
\right)$}
\\
\hline

$\begin{array}{c} (3.6k) \\n=10\\k=5\\d=9 \end{array}$ &
\includegraphics[width=0.1\textwidth]{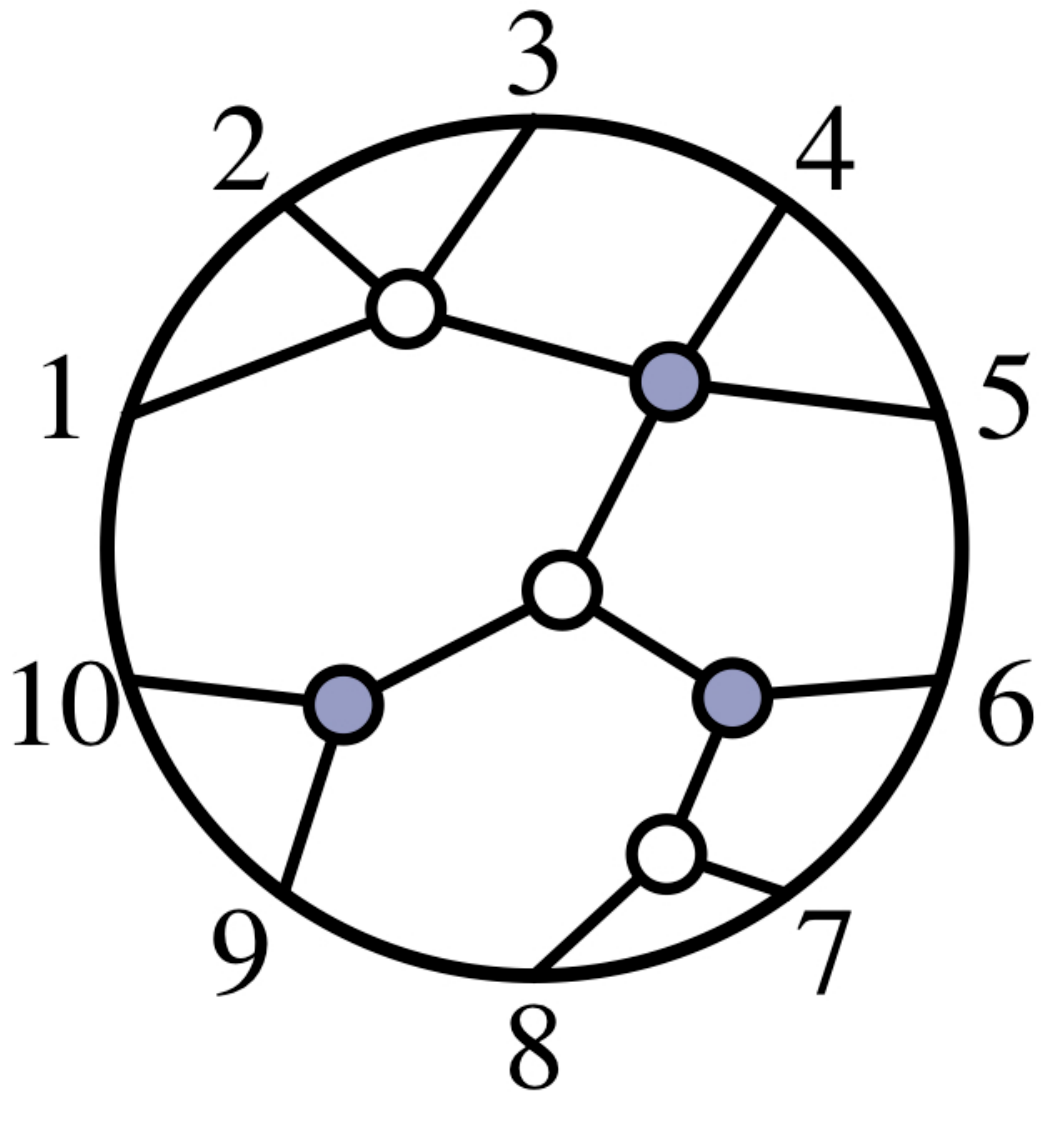}  
\newline\tiny\{2,3,7,11,14,10,8,16,15,19\} &
{\tiny
$\left(
\begin{array}{cccccccccc}
 1 & \alpha_9 & \alpha_8 & \alpha_7 & 0 & 0 & 0 & 0 & 0 & 0 \\
 0 & 0 & 0 & 1 & \alpha_6 & 0 & 0 & 0 & 0 & 0 \\
 0 & 0 & 0 & 0 & 1 & \alpha_5 & 0 & 0 & -\alpha_2 & 0 \\
 0 & 0 & 0 & 0 & 0 & 1 & \alpha_4 & \alpha_3 & 0 & 0 \\
 0 & 0 & 0 & 0 & 0 & 0 & 0 & 0 & 1 & \alpha_1 \\
\end{array}
\right)$}
\\
\hline

$\begin{array}{c} (3.13a) \\n=12\\k=6\\d=11 \end{array}$ &
\includegraphics[width=0.1\textwidth]{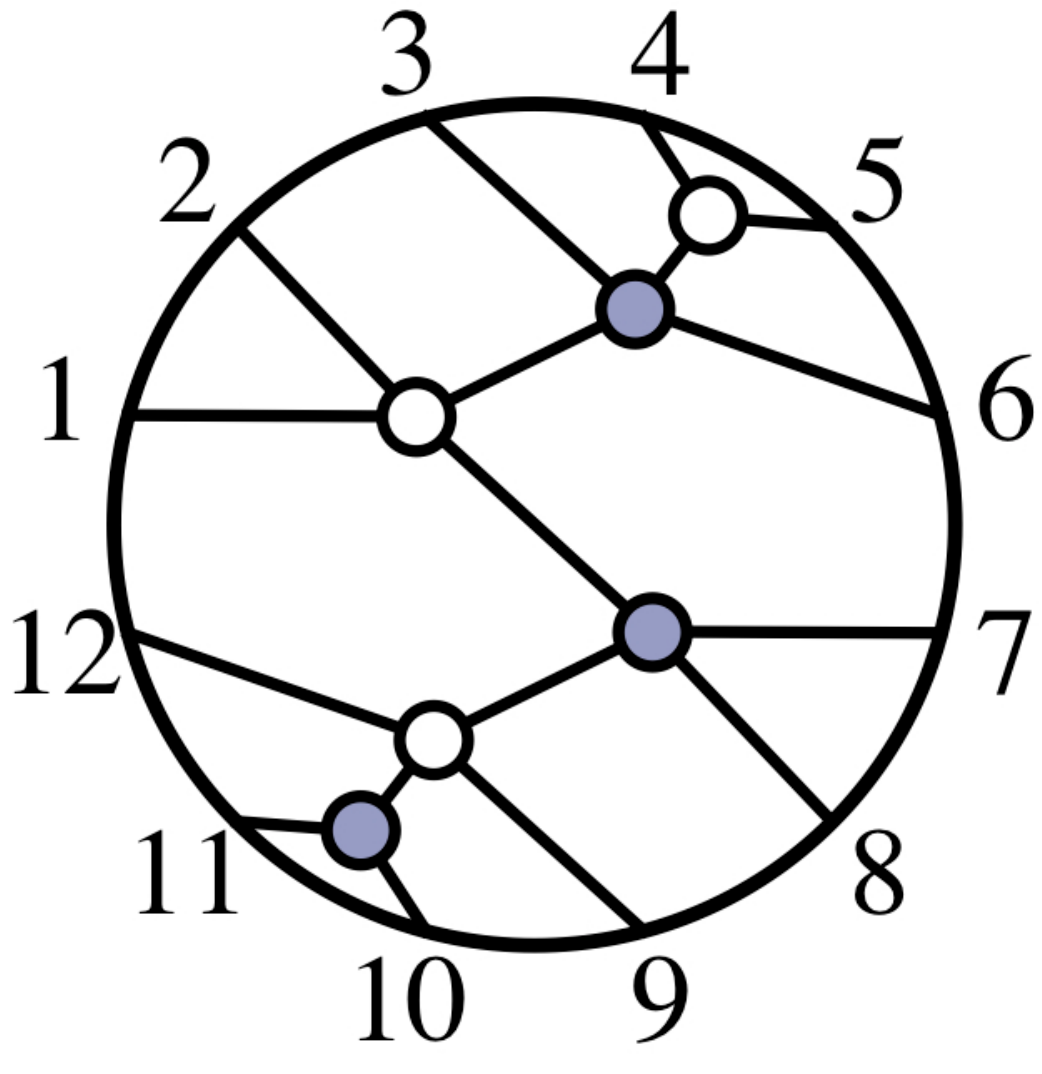}  
\newline\tiny \{2,6,9,5,15,16,13,19,11,12,22,20\}&
{\tiny
$\left(
\begin{array}{cccccccccccc}
 1 & \alpha_{11} & \alpha_{10} & 0 & 0 & 0 & \alpha_6 & 0 & 0 & 0 & 0 & 0 \\
 0 & 0 & 1 & \alpha_8 & \alpha_8 \alpha_9 & 0 & 0 & 0 & 0 & 0 & 0 & 0 \\
 0 & 0 & 0 & 1 & \alpha_9 & \alpha_7 & 0 & 0 & 0 & 0 & 0 & 0 \\
 0 & 0 & 0 & 0 & 0 & 0 & 1 & \alpha_5 & 0 & 0 & 0 & 0 \\
 0 & 0 & 0 & 0 & 0 & 0 & 0 & 1 & \alpha_4 & \alpha_3 & 0 & -\alpha_1 \\
 0 & 0 & 0 & 0 & 0 & 0 & 0 & 0 & 0 & 1 & \alpha_2 & 0 \\
\end{array}
\right)$}
\\
\hline

$\begin{array}{c} (3.13b) \\n=12\\k=6\\d=11 \end{array}$ &
\includegraphics[width=0.1\textwidth]{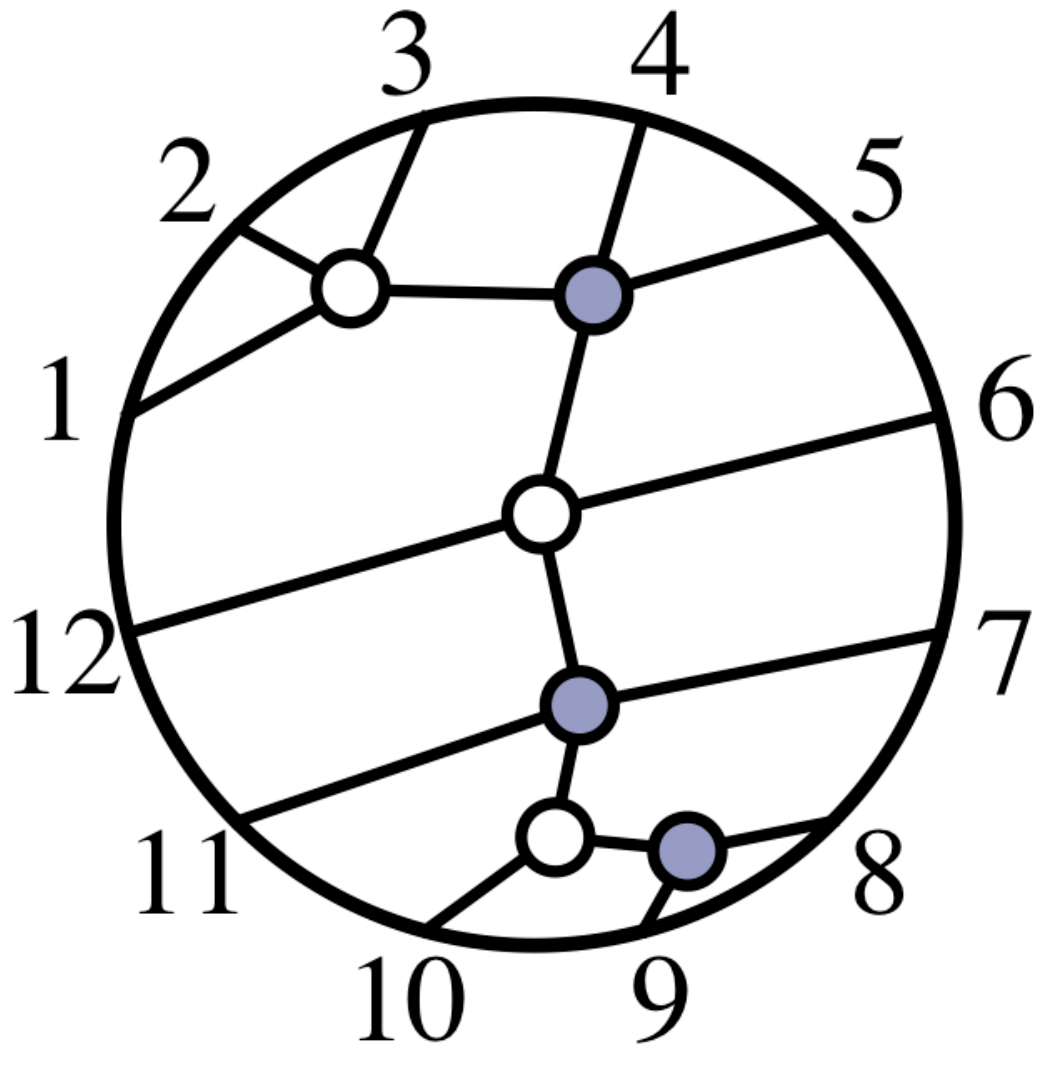}  
\newline\tiny \{2,3,6,13,16,11,12,10,20,19,21,17\}&
{\tiny
$\left(
\begin{array}{cccccccccccc}
 1 & \alpha_{11} & \alpha_{10} & \alpha_9 & 0 & 0 & 0 & 0 & 0 & 0 & 0 & 0 \\
 0 & 0 & 0 & 1 & \alpha_8 & 0 & 0 & 0 & 0 & 0 & 0 & 0 \\
 0 & 0 & 0 & 0 & 1 & \alpha_7 & \alpha_6 & 0 & 0 & 0 & 0 & -\alpha_1 \\
 0 & 0 & 0 & 0 & 0 & 0 & 1 & 0 & -\alpha_3 & -\alpha_3 \alpha_4 & 0 & 0 \\
 0 & 0 & 0 & 0 & 0 & 0 & 0 & 1 & \alpha_5 & 0 & 0 & 0 \\
 0 & 0 & 0 & 0 & 0 & 0 & 0 & 0 & 1 & \alpha_4 & \alpha_2 & 0 \\
\end{array}
\right)$}
\\
\hline

$\begin{array}{c} (3.13c) \\n=12\\k=6\\d=11 \end{array}$ &
\includegraphics[width=0.1\textwidth]{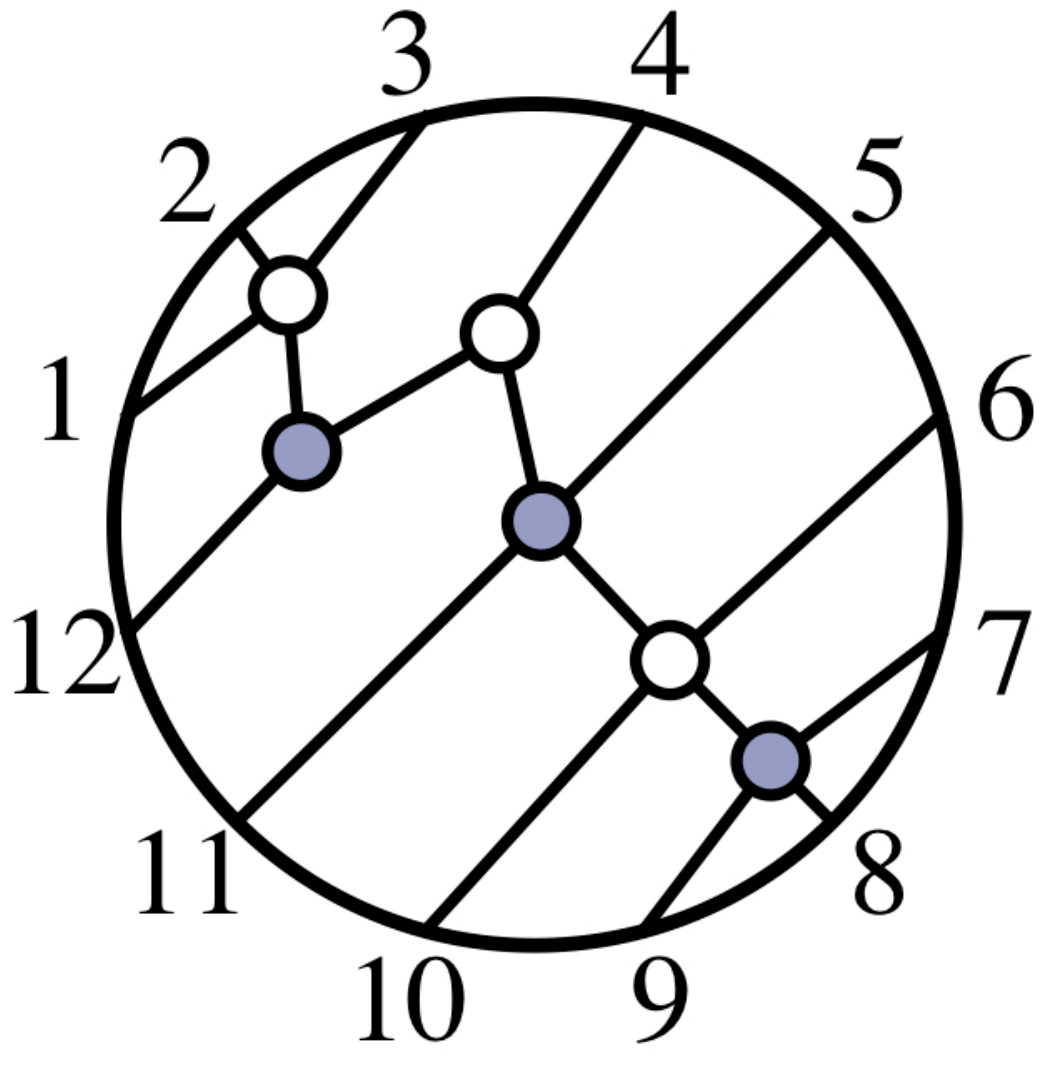}  
\newline\tiny\{2,3,12,11,13,9,10,19,20,17,18,16\} &
{\tiny
$\left(
\begin{array}{cccccccccccc}
 1 & \alpha_{11} & \alpha_{10} & \alpha_8 & \alpha_8 \alpha_9 & 0 & 0 & 0 & 0 & 0 & 0 & 0 \\
 0 & 0 & 0 & 1 & \alpha_9 & 0 & 0 & 0 & 0 & 0 & 0 & \alpha_1 \\
 0 & 0 & 0 & 0 & 1 & \alpha_3 & \alpha_3 \alpha_7 & 0 & 0 & \alpha_3 \alpha_4 & 0 & 0 \\
 0 & 0 & 0 & 0 & 0 & 1 & \alpha_7 & 0 & 0 & \alpha_4 & \alpha_2 & 0 \\
 0 & 0 & 0 & 0 & 0 & 0 & 1 & \alpha_6 & 0 & 0 & 0 & 0 \\
 0 & 0 & 0 & 0 & 0 & 0 & 0 & 1 & \alpha_5 & 0 & 0 & 0 \\
\end{array}
\right)$}
\\
\hline

$\begin{array}{c} (3.13d) \\n=12\\k=6\\d=11 \end{array}$ &
\includegraphics[width=0.1\textwidth]{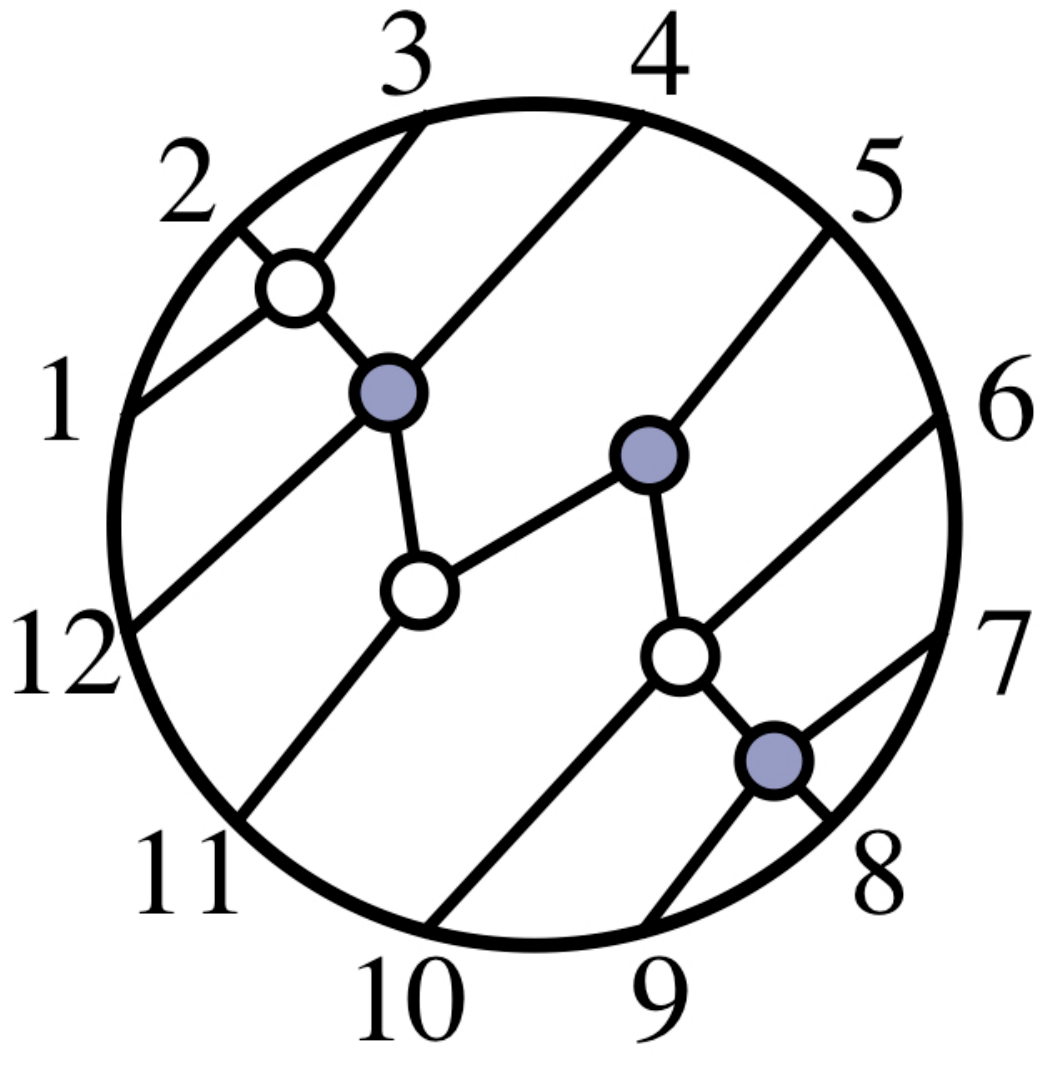}  
\newline\tiny\{2,3,12,13,11,9,10,19,20,17,16,18\} &
{\tiny
$\left(
\begin{array}{cccccccccccc}
 1 & \alpha_{11} & \alpha_{10} & \alpha_9 & 0 & 0 & 0 & 0 & 0 & 0 & 0 & 0 \\
 0 & 0 & 0 & 1 & 0 & -\alpha_2 & -\alpha_2 \alpha_8 & 0 & 0 & -\alpha_2 \alpha_5 & -\alpha_2 \alpha_3 & 0 \\
 0 & 0 & 0 & 0 & 1 & \alpha_4 & \alpha_4 \alpha_8 & 0 & 0 & \alpha_4 \alpha_5 & 0 & 0 \\
 0 & 0 & 0 & 0 & 0 & 1 & \alpha_8 & 0 & 0 & \alpha_5 & \alpha_3 & \alpha_1 \\
 0 & 0 & 0 & 0 & 0 & 0 & 1 & \alpha_7 & 0 & 0 & 0 & 0 \\
 0 & 0 & 0 & 0 & 0 & 0 & 0 & 1 & \alpha_6 & 0 & 0 & 0 \\
\end{array}
\right)$}
\\
\hline

$\begin{array}{c} (3.13e) \\n=12\\k=6\\d=11 \end{array}$ &
\includegraphics[width=0.1\textwidth]{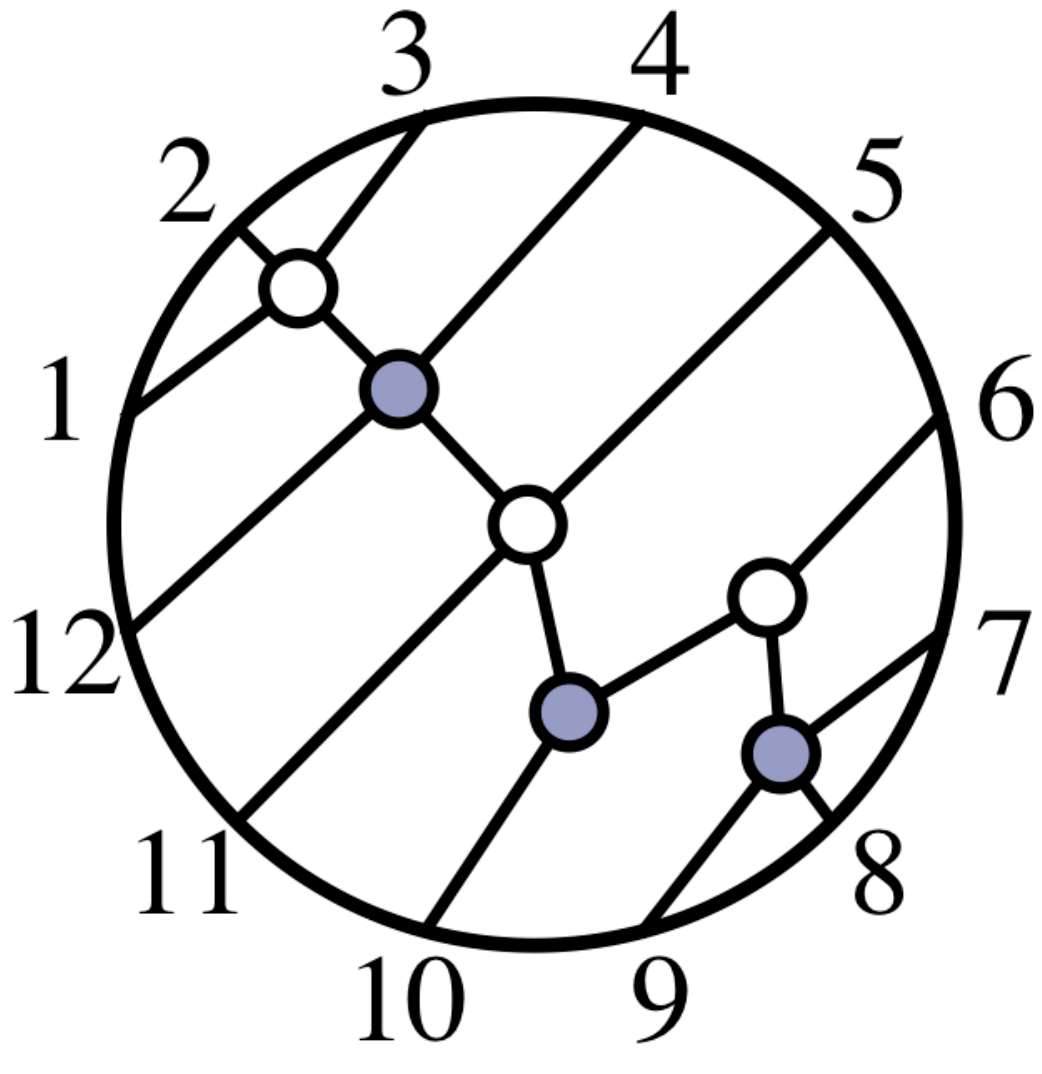}  
\newline\tiny\{2,3,12,13,10,9,11,19,20,18,16,17\} &
{\tiny
$\left(
\begin{array}{cccccccccccc}
 1 & \alpha_{11} & \alpha_{10} & \alpha_9 & 0 & 0 & 0 & 0 & 0 & 0 & 0 & 0 \\
 0 & 0 & 0 & 1 & \alpha_2 & \alpha_2 \alpha_7 & \alpha_2 \alpha_7 \alpha_8 & 0 & 0 & 0 & -\alpha_2 \alpha_3 & 0 \\
 0 & 0 & 0 & 0 & 1 & \alpha_7 & \alpha_7 \alpha_8 & 0 & 0 & 0 & -\alpha_3 & -\alpha_1 \\
 0 & 0 & 0 & 0 & 0 & 1 & \alpha_8 & 0 & 0 & \alpha_4 & 0 & 0 \\
 0 & 0 & 0 & 0 & 0 & 0 & 1 & \alpha_6 & 0 & 0 & 0 & 0 \\
 0 & 0 & 0 & 0 & 0 & 0 & 0 & 1 & \alpha_5 & 0 & 0 & 0 \\
\end{array}
\right)$}
\\
\hline

$\begin{array}{c} (3.13f) \\n=12\\k=6\\d=11 \end{array}$ &
\includegraphics[width=0.1\textwidth]{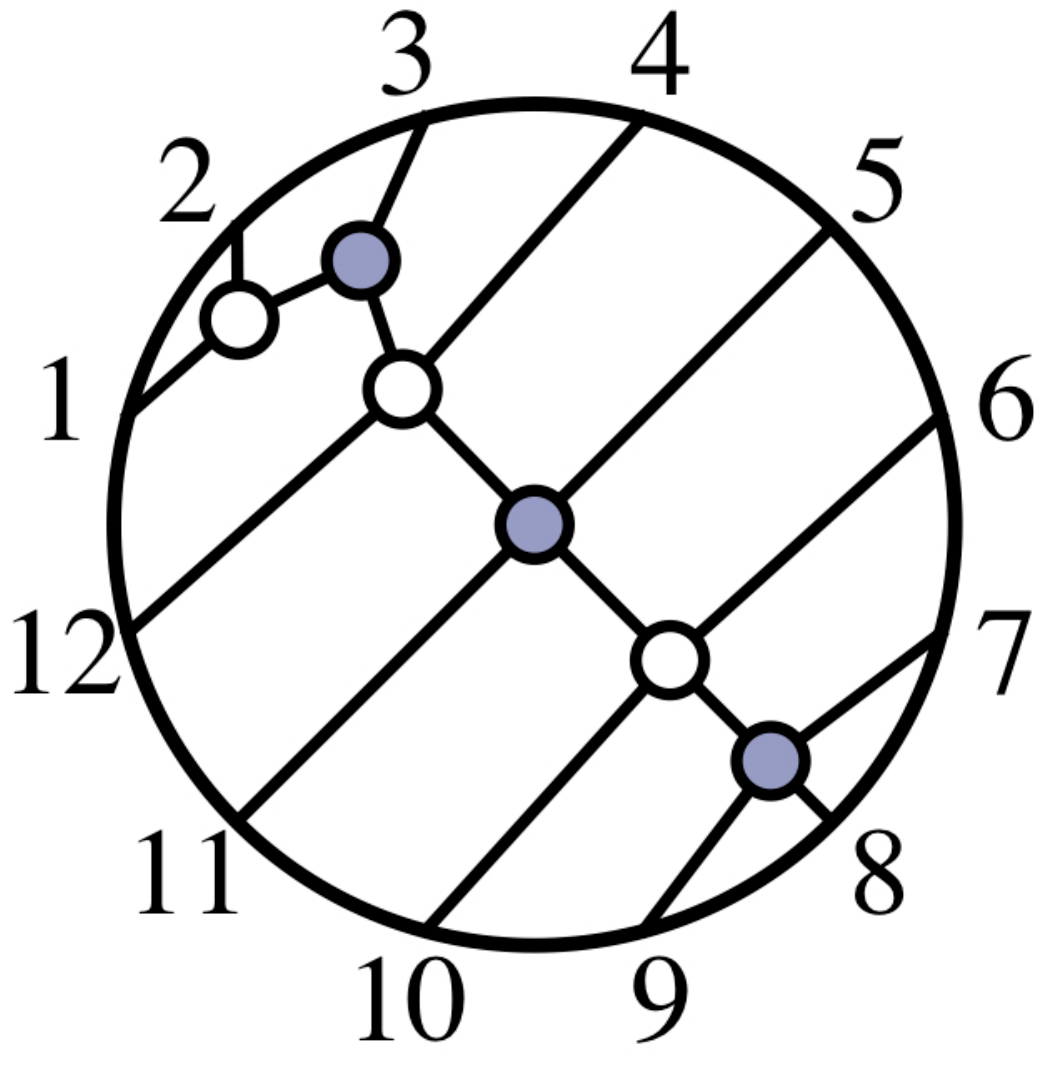}  
\newline\tiny\{2,4,13,11,12,9,10,19,20,17,18,15\} &
{\tiny
$\left(
\begin{array}{cccccccccccc}
 1 & \alpha_{11} & \alpha_{10} & 0 & 0 & 0 & 0 & 0 & 0 & 0 & 0 & 0 \\
 0 & 0 & 1 & \alpha_9 & \alpha_8 & 0 & 0 & 0 & 0 & 0 & 0 & \alpha_1 \\
 0 & 0 & 0 & 0 & 1 & \alpha_3 & \alpha_3 \alpha_7 & 0 & 0 & \alpha_3 \alpha_4 & 0 & 0 \\
 0 & 0 & 0 & 0 & 0 & 1 & \alpha_7 & 0 & 0 & \alpha_4 & \alpha_2 & 0 \\
 0 & 0 & 0 & 0 & 0 & 0 & 1 & \alpha_6 & 0 & 0 & 0 & 0 \\
 0 & 0 & 0 & 0 & 0 & 0 & 0 & 1 & \alpha_5 & 0 & 0 & 0 \\
\end{array}
\right)$}
\\
\hline

$\begin{array}{c} (3.13g) \\n=14\\k=7\\d=12 \end{array}$ &
\includegraphics[width=0.1\textwidth]{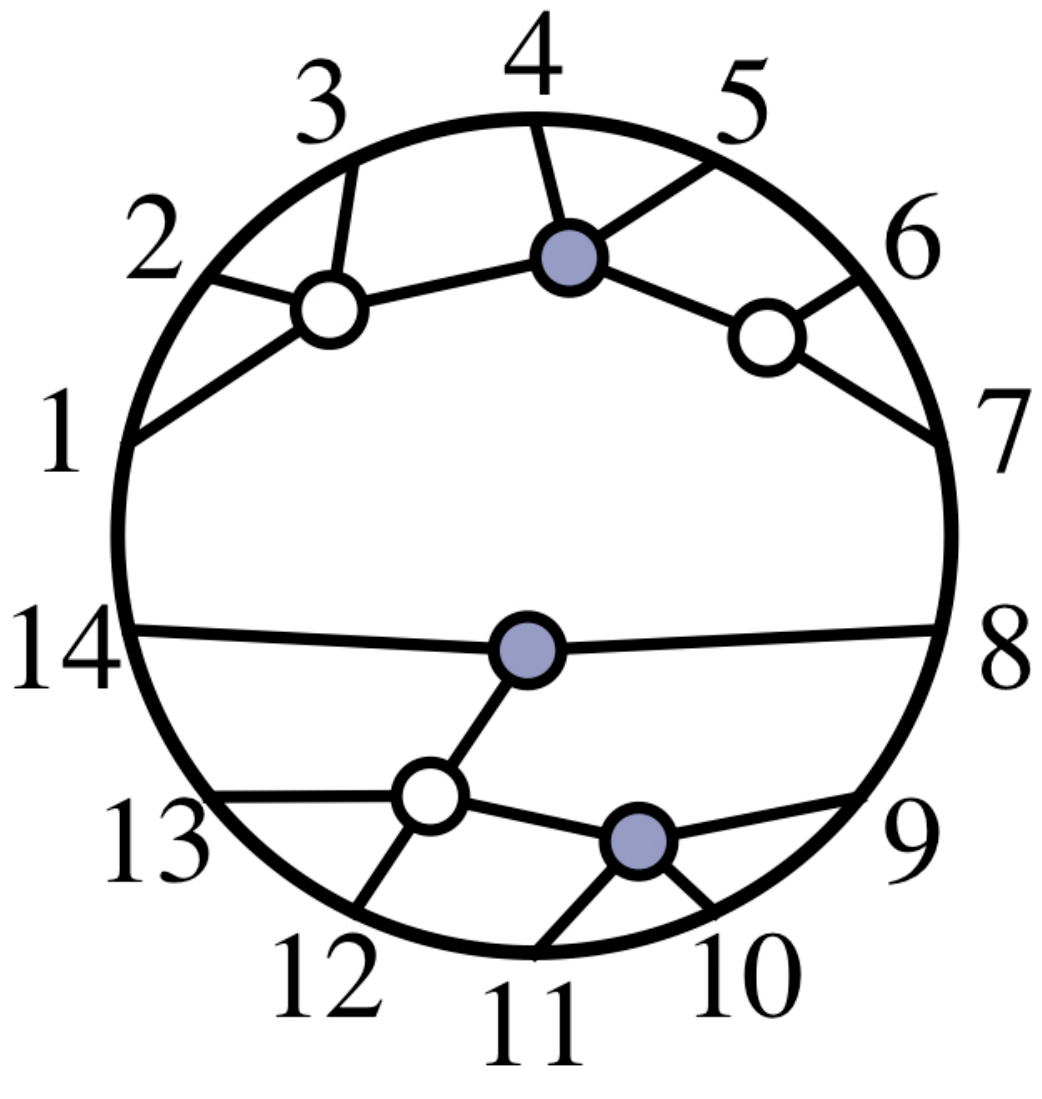}  
\newline\tiny\{2,3,6,15,18,7,19,14,12,23,24,13,22,25\} &
{\tiny
$\left(
\begin{array}{cccccccccccccc}
 1 & \alpha_{12} & \alpha_{11} & \alpha_{10} & 0 & 0 & 0 & 0 & 0 & 0 & 0 & 0 & 0 & 0 \\
 0 & 0 & 0 & 1 & \alpha_9 & 0 & 0 & 0 & 0 & 0 & 0 & 0 & 0 & 0 \\
 0 & 0 & 0 & 0 & 1 & \alpha_8 & \alpha_7 & 0 & 0 & 0 & 0 & 0 & 0 & 0 \\
 0 & 0 & 0 & 0 & 0 & 0 & 0 & 1 & 0 & 0 & \alpha_2 & \alpha_2 \alpha_4 & \alpha_2 \alpha_3 & 0 \\
 0 & 0 & 0 & 0 & 0 & 0 & 0 & 0 & 1 & \alpha_6 & 0 & 0 & 0 & 0 \\
 0 & 0 & 0 & 0 & 0 & 0 & 0 & 0 & 0 & 1 & \alpha_5 & 0 & 0 & 0 \\
 0 & 0 & 0 & 0 & 0 & 0 & 0 & 0 & 0 & 0 & 1 & \alpha_4 & \alpha_3 & \alpha_1 \\
\end{array}
\right)$}
\\
\hline

$\begin{array}{c} (3.13h) \\n=14\\k=7\\d=12 \end{array}$ &
\includegraphics[width=0.1\textwidth]{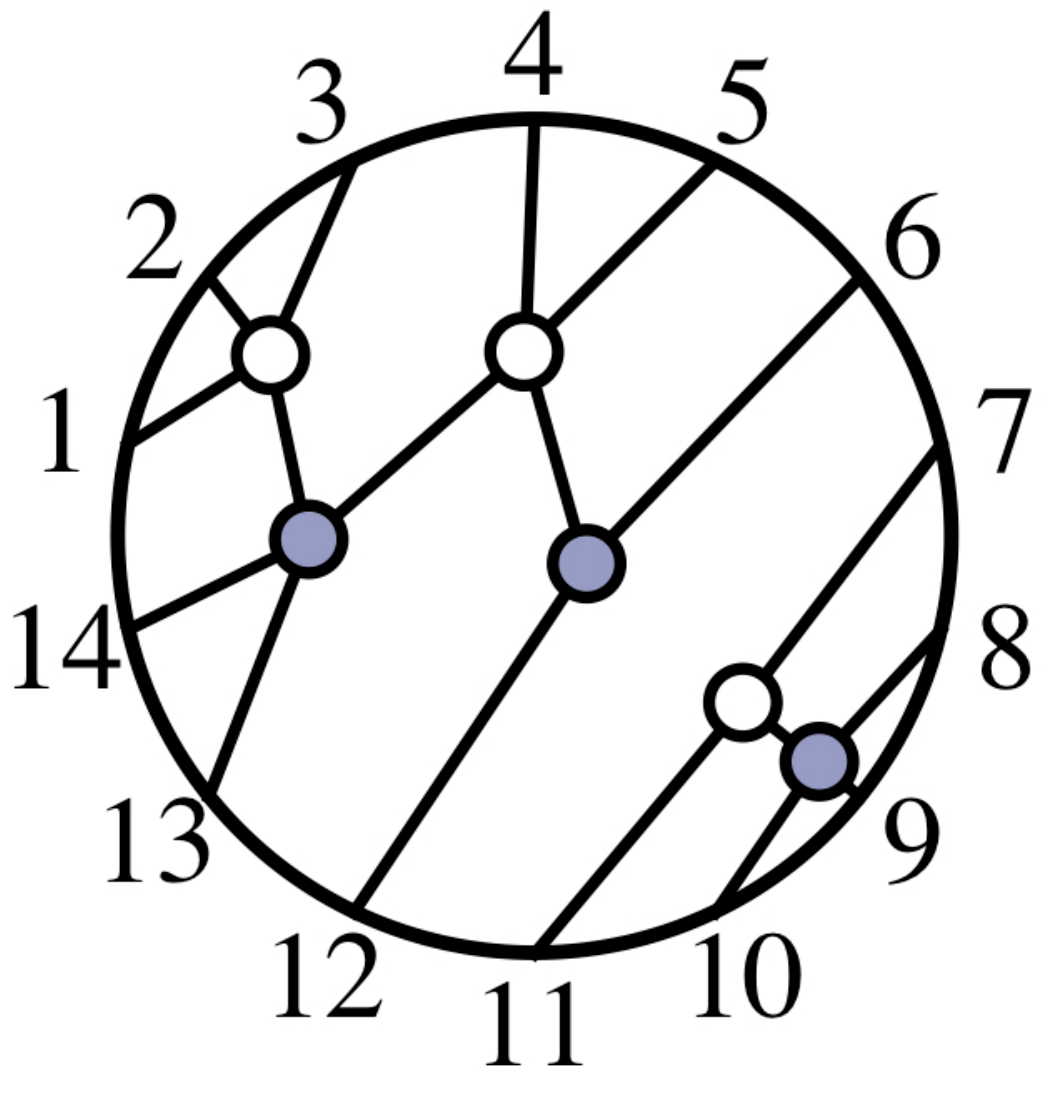}  
\newline\tiny \{2,3,14,5,12,15,10,11,22,23,21,20,18,27\}&
{\tiny
$\left(
\begin{array}{cccccccccccccc}
 1 & \alpha_{12} & \alpha_{11} & \alpha_8 & \alpha_8 \alpha_{10} & \alpha_8 \alpha_9 & 0 & 0 & 0 & 0 & 0 & 0 & 0 & 0 \\
 0 & 0 & 0 & 1 & \alpha_{10} & \alpha_9 & 0 & 0 & 0 & 0 & 0 & 0 & \alpha_2 & 0 \\
 0 & 0 & 0 & 0 & 0 & 1 & 0 & 0 & 0 & 0 & 0 & -\alpha_3 & 0 & 0 \\
 0 & 0 & 0 & 0 & 0 & 0 & 1 & \alpha_7 & 0 & 0 & \alpha_4 & 0 & 0 & 0 \\
 0 & 0 & 0 & 0 & 0 & 0 & 0 & 1 & \alpha_6 & 0 & 0 & 0 & 0 & 0 \\
 0 & 0 & 0 & 0 & 0 & 0 & 0 & 0 & 1 & \alpha_5 & 0 & 0 & 0 & 0 \\
 0 & 0 & 0 & 0 & 0 & 0 & 0 & 0 & 0 & 0 & 0 & 0 & 1 & \alpha_1 \\
\end{array}
\right)$}
\\
\hline

$\begin{array}{c} (3.13i) \\n=14\\k=7\\d=12 \end{array}$ &
\includegraphics[width=0.1\textwidth]{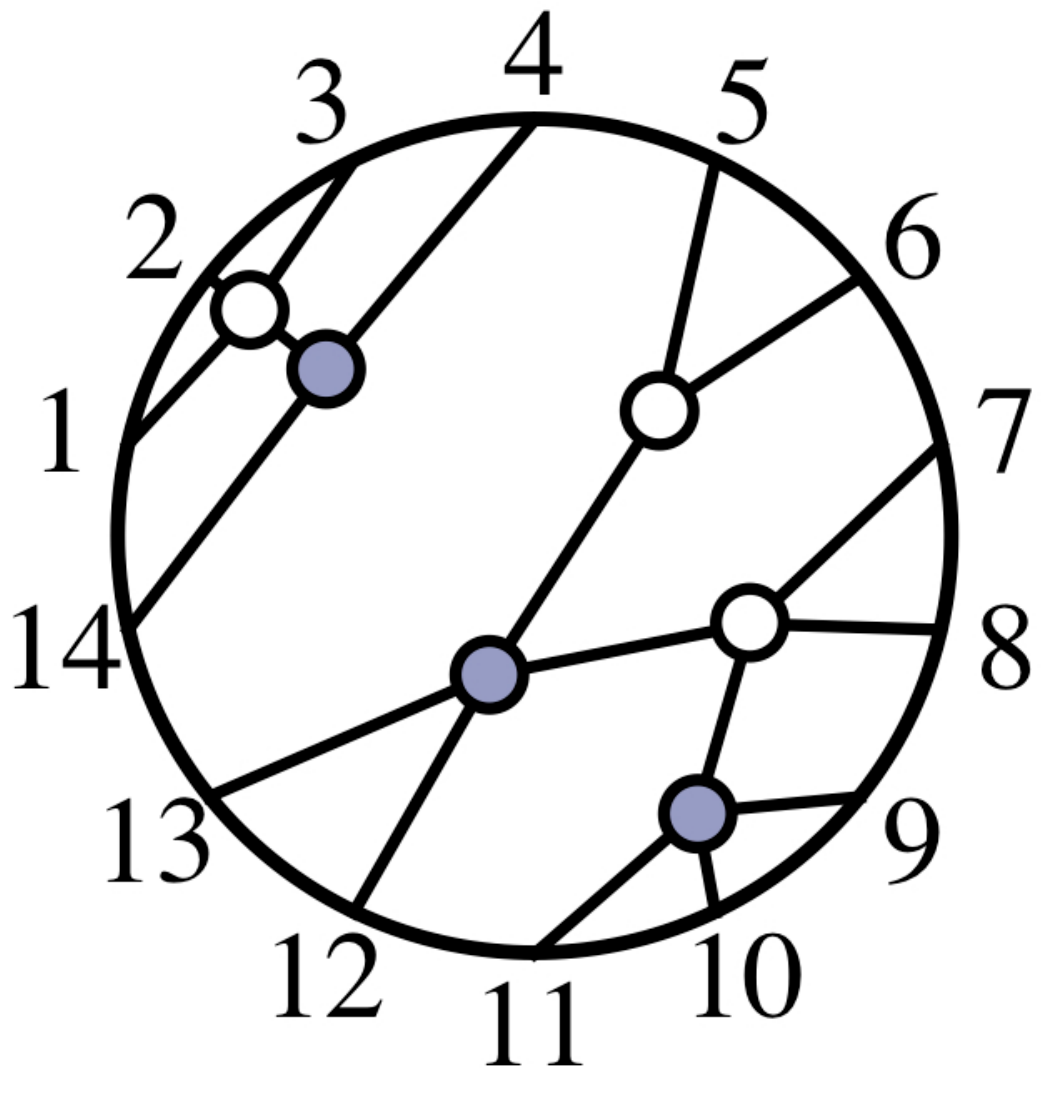}  
\newline\tiny\{2,3,14,15,6,13,8,11,19,23,24,21,26,18\} &
{\tiny
$\left(
\begin{array}{cccccccccccccc}
 1 & \alpha_{12} & \alpha_{11} & \alpha_{10} & 0 & 0 & 0 & 0 & 0 & 0 & 0 & 0 & 0 & 0 \\
 0 & 0 & 0 & 1 & 0 & 0 & 0 & 0 & 0 & 0 & 0 & 0 & 0 & -\alpha_1 \\
 0 & 0 & 0 & 0 & 1 & \alpha_9 & \alpha_6 & \alpha_6 \alpha_8 & \alpha_6 \alpha_7 & 0 & 0 & 0 & 0 & 0 \\
 0 & 0 & 0 & 0 & 0 & 0 & 1 & \alpha_8 & \alpha_7 & 0 & 0 & \alpha_3 & 0 & 0 \\
 0 & 0 & 0 & 0 & 0 & 0 & 0 & 0 & 1 & \alpha_5 & 0 & 0 & 0 & 0 \\
 0 & 0 & 0 & 0 & 0 & 0 & 0 & 0 & 0 & 1 & \alpha_4 & 0 & 0 & 0 \\
 0 & 0 & 0 & 0 & 0 & 0 & 0 & 0 & 0 & 0 & 0 & 1 & \alpha_2 & 0 \\
\end{array}
\right)$}
\\
\hline

$\begin{array}{c} (3.13j) \\n=14\\k=7\\d=12 \end{array}$ &
\includegraphics[width=0.1\textwidth]{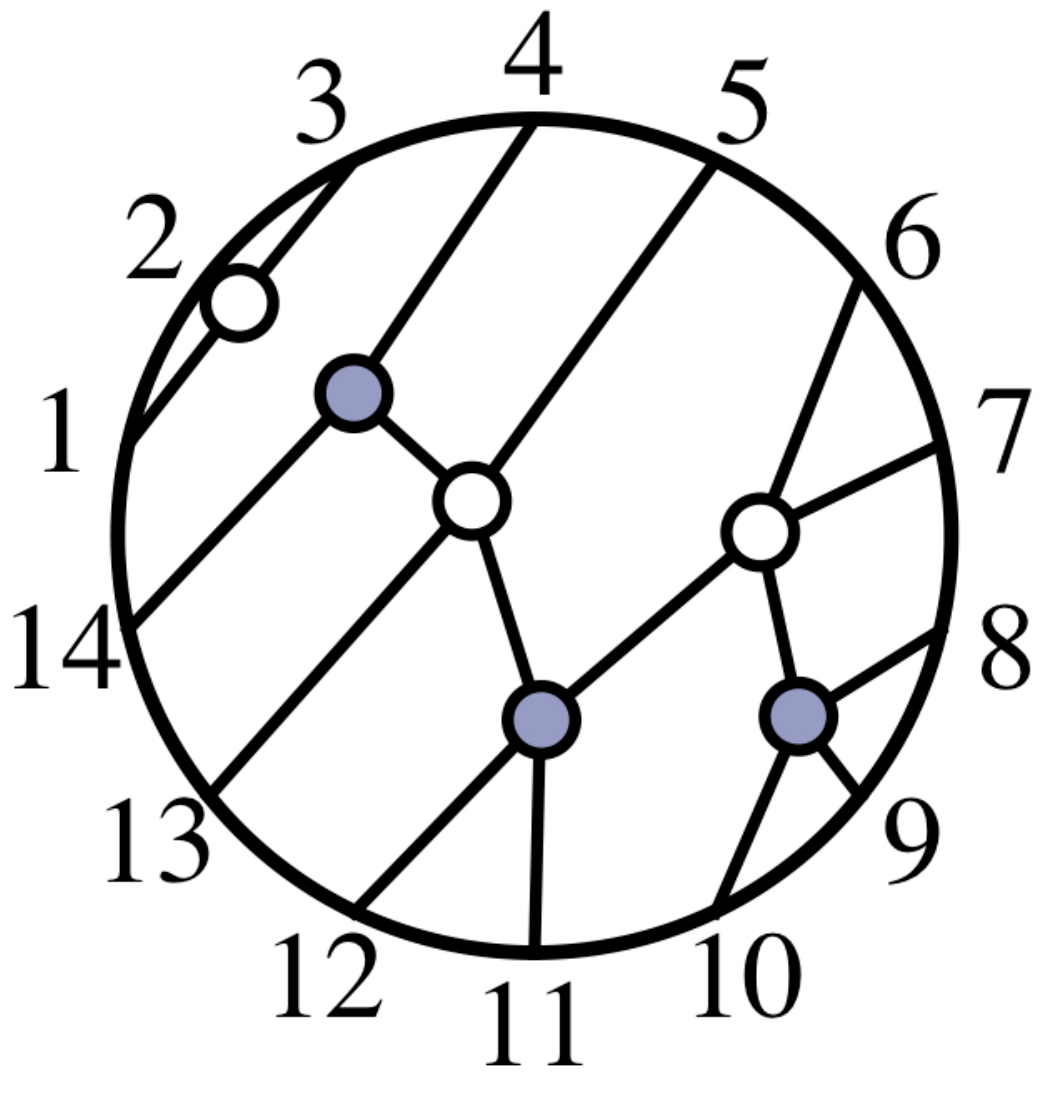}  
\newline\tiny \{2,3,15,14,12,7,10,13,22,23,20,25,18,19\}&
{\fontsize{4}{4.8}\selectfont
$\left(
\begin{array}{cccccccccccccc}
 1 & \alpha_{12} & \alpha_{11} & 0 & 0 & 0 & 0 & 0 & 0 & 0 & 0 & 0 & 0 & 0 \\
 0 & 0 & 0 & 1 & \alpha_2 & \alpha_2 \alpha_8 & \alpha_2 \alpha_8 \alpha_{10} & \alpha_2 \alpha_8 \alpha_9 & 0 & 0 & 0 & 0 & \alpha_2 \alpha_3 & 0 \\
 0 & 0 & 0 & 0 & 1 & \alpha_8 & \alpha_8 \alpha_{10} & \alpha_8 \alpha_9 & 0 & 0 & 0 & 0 & \alpha_3 & \alpha_1 \\
 0 & 0 & 0 & 0 & 0 & 1 & \alpha_{10} & \alpha_9 & 0 & 0 & \alpha_5 & 0 & 0 & 0 \\
 0 & 0 & 0 & 0 & 0 & 0 & 0 & 1 & \alpha_7 & 0 & 0 & 0 & 0 & 0 \\
 0 & 0 & 0 & 0 & 0 & 0 & 0 & 0 & 1 & \alpha_6 & 0 & 0 & 0 & 0 \\
 0 & 0 & 0 & 0 & 0 & 0 & 0 & 0 & 0 & 0 & 1 & \alpha_4 & 0 & 0 \\
\end{array}
\right)$}
\\
\hline

$\begin{array}{c} (3.13k) \\n=14\\k=7\\d=12 \end{array}$ &
\includegraphics[width=0.1\textwidth]{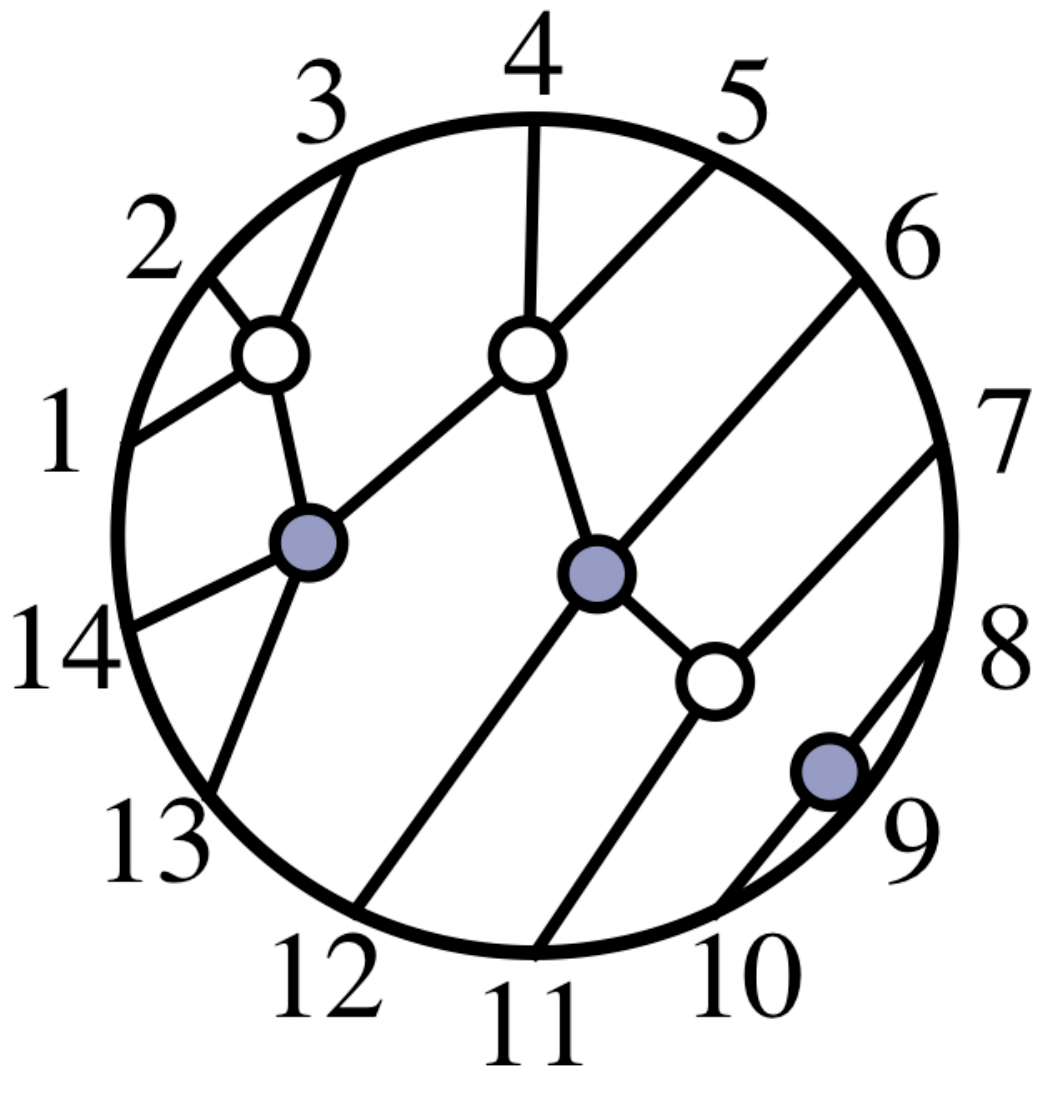}  
\newline\tiny\{2,3,14,5,12,15,11,10,22,23,20,21,18,27\} &
{\tiny
$\left(
\begin{array}{cccccccccccccc}
 1 & \alpha_{12} & \alpha_{11} & \alpha_8 & \alpha_8 \alpha_{10} & \alpha_8 \alpha_9 & 0 & 0 & 0 & 0 & 0 & 0 & 0 & 0 \\
 0 & 0 & 0 & 1 & \alpha_{10} & \alpha_9 & 0 & 0 & 0 & 0 & 0 & 0 & \alpha_2 & 0 \\
 0 & 0 & 0 & 0 & 0 & 1 & \alpha_4 & 0 & 0 & 0 & \alpha_4 \alpha_5 & 0 & 0 & 0 \\
 0 & 0 & 0 & 0 & 0 & 0 & 1 & 0 & 0 & 0 & \alpha_5 & \alpha_3 & 0 & 0 \\
 0 & 0 & 0 & 0 & 0 & 0 & 0 & 1 & \alpha_7 & 0 & 0 & 0 & 0 & 0 \\
 0 & 0 & 0 & 0 & 0 & 0 & 0 & 0 & 1 & \alpha_6 & 0 & 0 & 0 & 0 \\
 0 & 0 & 0 & 0 & 0 & 0 & 0 & 0 & 0 & 0 & 0 & 0 & 1 & \alpha_1 \\
\end{array}
\right)$}
\\
\hline

$\begin{array}{c} (3.13l) \\n=14\\k=7\\d=12 \end{array}$ &
\includegraphics[width=0.1\textwidth]{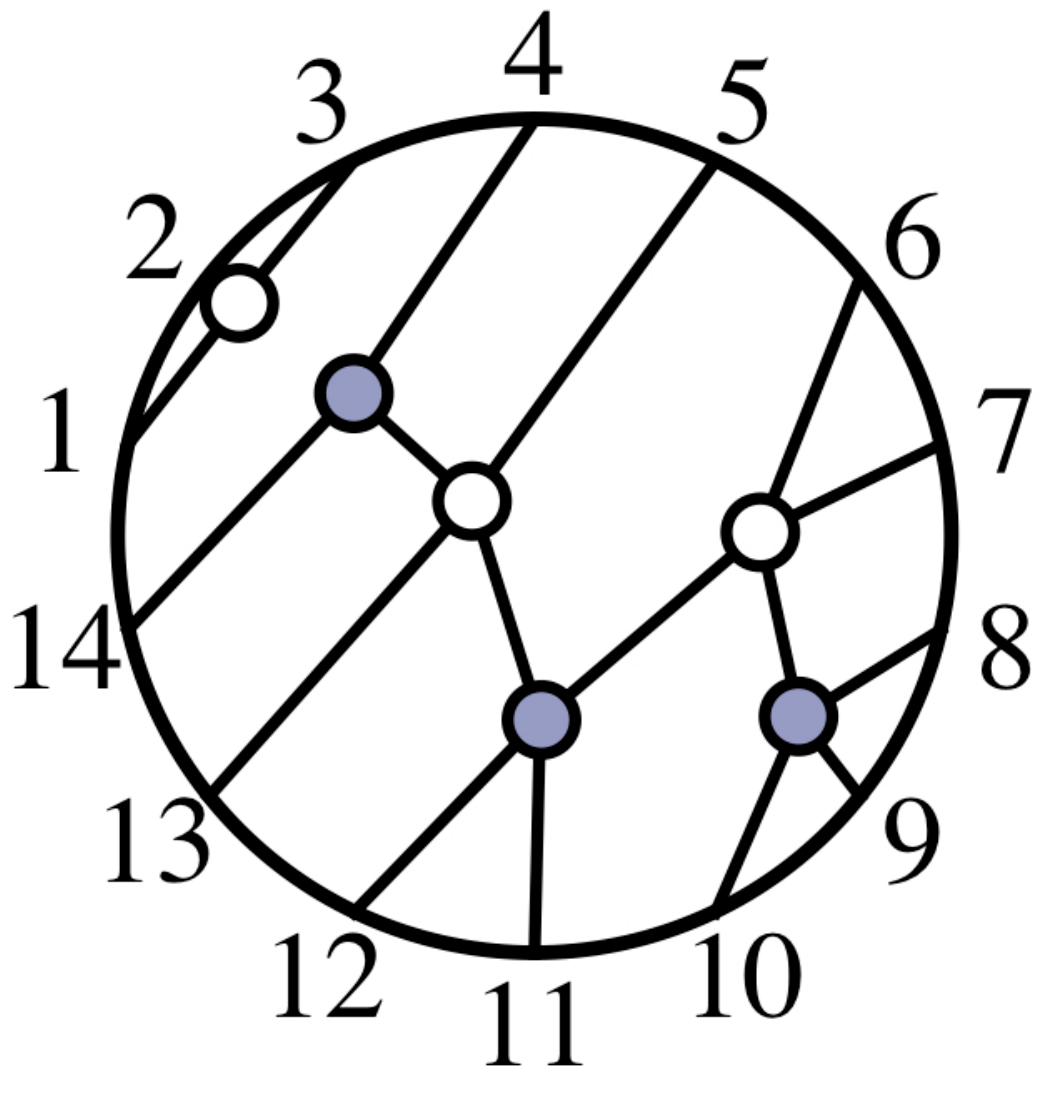}  
\newline\tiny\{2,3,15,14,12,7,10,13,22,23,20,25,18,19\} &
{\tiny
$\left(
\begin{array}{cccccccccccccc}
 1 & \alpha_{12} & \alpha_{11} & 0 & 0 & 0 & 0 & 0 & 0 & 0 & 0 & 0 & 0 & 0 \\
 0 & 0 & 0 & 1 & \alpha_2 & \alpha_2 \alpha_8 & \alpha_2 \alpha_8 \alpha_{10} & \alpha_2 \alpha_8 \alpha_9 & 0 & 0 & 0 & 0 & \alpha_2 \alpha_3 & 0 \\
 0 & 0 & 0 & 0 & 1 & \alpha_8 & \alpha_8 \alpha_{10} & \alpha_8 \alpha_9 & 0 & 0 & 0 & 0 & \alpha_3 & \alpha_1 \\
 0 & 0 & 0 & 0 & 0 & 1 & \alpha_{10} & \alpha_9 & 0 & 0 & \alpha_5 & 0 & 0 & 0 \\
 0 & 0 & 0 & 0 & 0 & 0 & 0 & 1 & \alpha_7 & 0 & 0 & 0 & 0 & 0 \\
 0 & 0 & 0 & 0 & 0 & 0 & 0 & 0 & 1 & \alpha_6 & 0 & 0 & 0 & 0 \\
 0 & 0 & 0 & 0 & 0 & 0 & 0 & 0 & 0 & 0 & 1 & \alpha_4 & 0 & 0 \\
\end{array}
\right)$}
\\
\hline

$\begin{array}{c} (3.14a) \\n=10\\k=5\\d=10 \end{array}$ &
\includegraphics[width=0.1\textwidth]{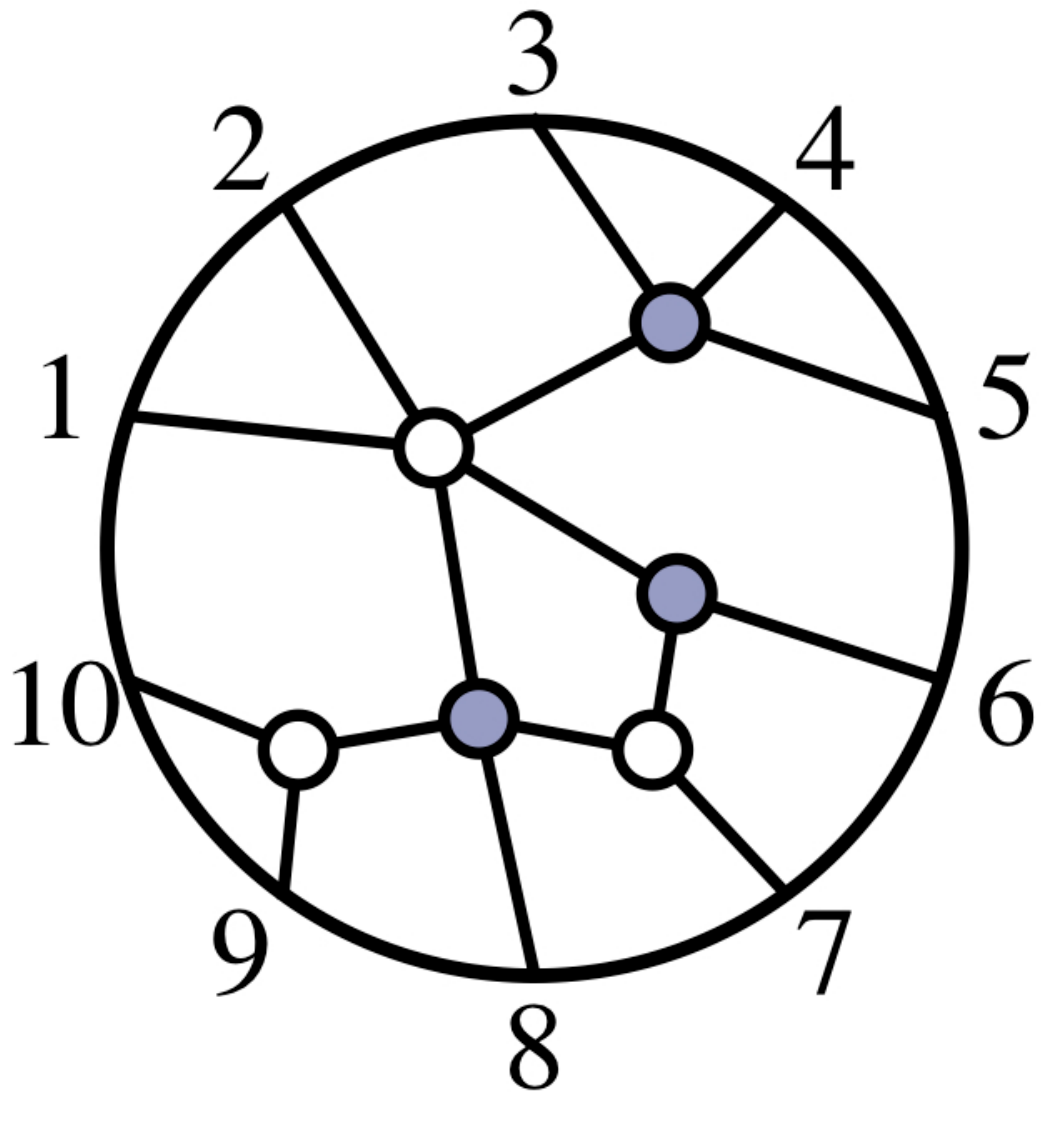}  
\newline\tiny\{2,5,7,13,14,9,11,16,10,18\} &
{\tiny
$\left(
\begin{array}{cccccccccc}
 1 & \alpha_{10} & \alpha_9 & 0 & 0 & \alpha_4+\alpha_6 & \alpha_4 \alpha_5 & 0 & 0 & 0 \\
 0 & 0 & 1 & \alpha_8 & 0 & 0 & 0 & 0 & 0 & 0 \\
 0 & 0 & 0 & 1 & \alpha_7 & 0 & 0 & 0 & 0 & 0 \\
 0 & 0 & 0 & 0 & 0 & 1 & \alpha_5 & \alpha_3 & 0 & 0 \\
 0 & 0 & 0 & 0 & 0 & 0 & 0 & 1 & \alpha_2 & \alpha_1 \\
\end{array}
\right)$}
\\
\hline

$\begin{array}{c} (3.14b) \\n=10\\k=5\\d=10 \end{array}$ &
\includegraphics[width=0.1\textwidth]{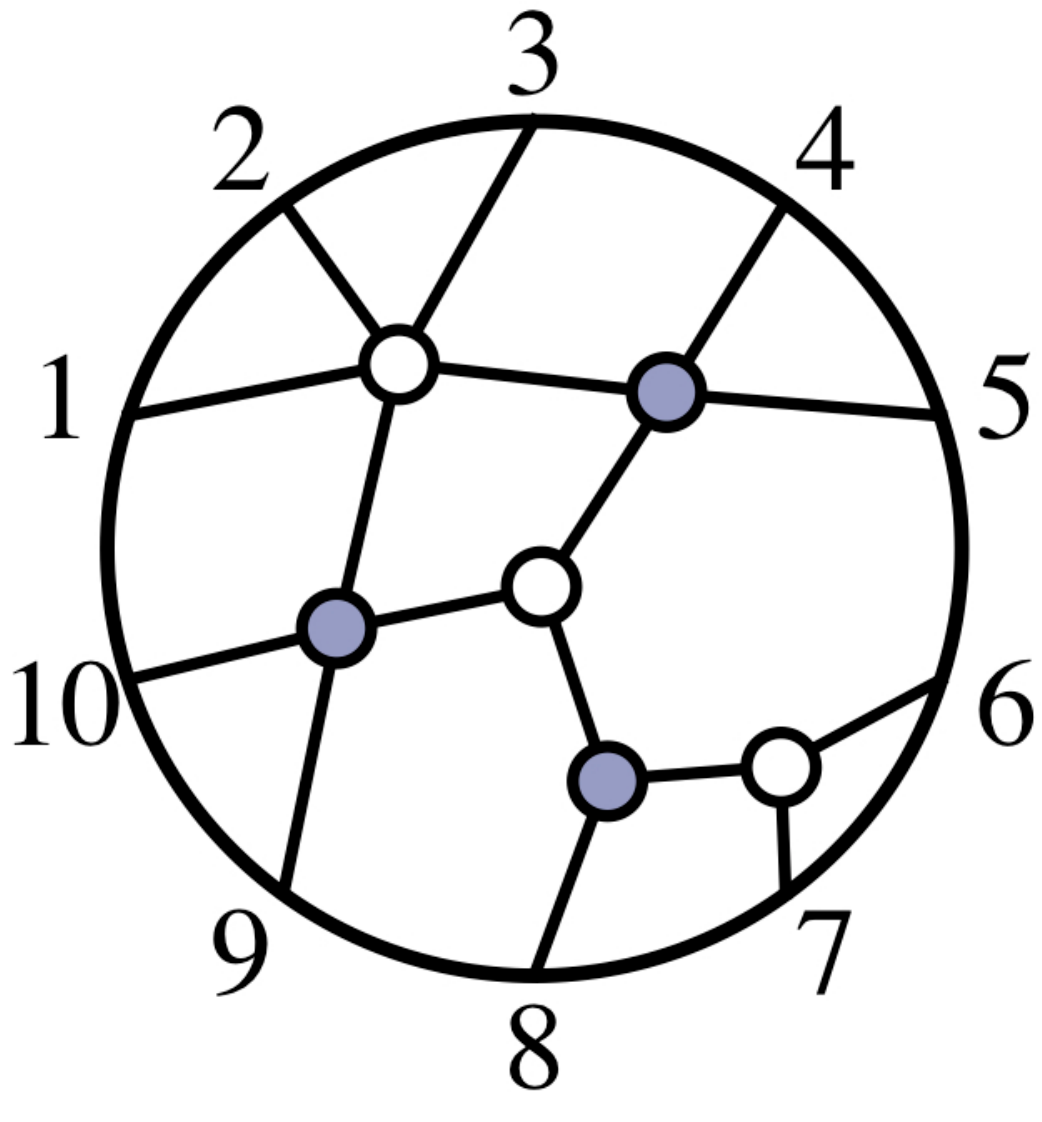}  
\newline\tiny\{2,3,8,10,14,7,11,16,15,19\} &
{\tiny
$\left(
\begin{array}{cccccccccc}
 1 & \alpha_{10} & \alpha_9 & \alpha_8 & -\alpha_4 & -\alpha_4 \alpha_5 & -\alpha_4 \alpha_5 \alpha_6 & 0 & 0 & 0 \\
 0 & 0 & 0 & 1 & \alpha_7 & 0 & 0 & 0 & 0 & 0 \\
 0 & 0 & 0 & 0 & 1 & \alpha_5 & \alpha_5 \alpha_6 & 0 & -\alpha_2 & 0 \\
 0 & 0 & 0 & 0 & 0 & 1 & \alpha_6 & \alpha_3 & 0 & 0 \\
 0 & 0 & 0 & 0 & 0 & 0 & 0 & 0 & 1 & \alpha_1 \\
\end{array}
\right)$}
\\
\hline

$\begin{array}{c} (3.14c) \\n=8\\k=4\\d=9 \end{array}$ &
\includegraphics[width=0.1\textwidth]{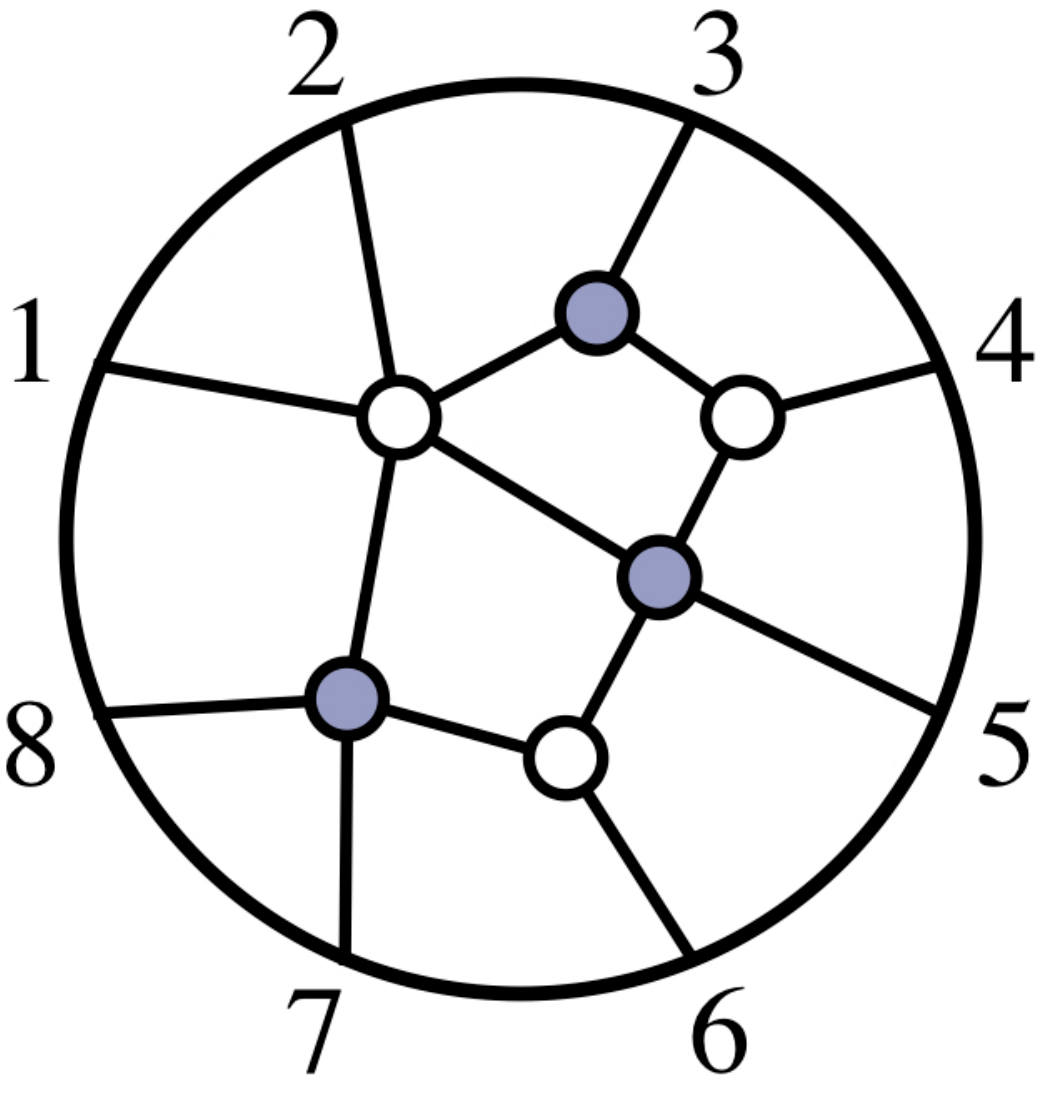}  
\newline\tiny\{2,4,6,8,11,9,13,15\} &
{\tiny
$\left(
\begin{array}{cccccccc}
 1 & \alpha_9 & \alpha_6+\alpha_8 & \alpha_6 \alpha_7 & -\alpha_3 & -\alpha_3 \alpha_4 & 0 & 0 \\
 0 & 0 & 1 & \alpha_7 & \alpha_5 & 0 & 0 & 0 \\
 0 & 0 & 0 & 0 & 1 & \alpha_4 & \alpha_2 & 0 \\
 0 & 0 & 0 & 0 & 0 & 0 & 1 & \alpha_1 \\
\end{array}
\right)$}
\\
\hline

$\begin{array}{c} (3.14d) \\n=8\\k=4\\d=9 \end{array}$ &
\includegraphics[width=0.1\textwidth]{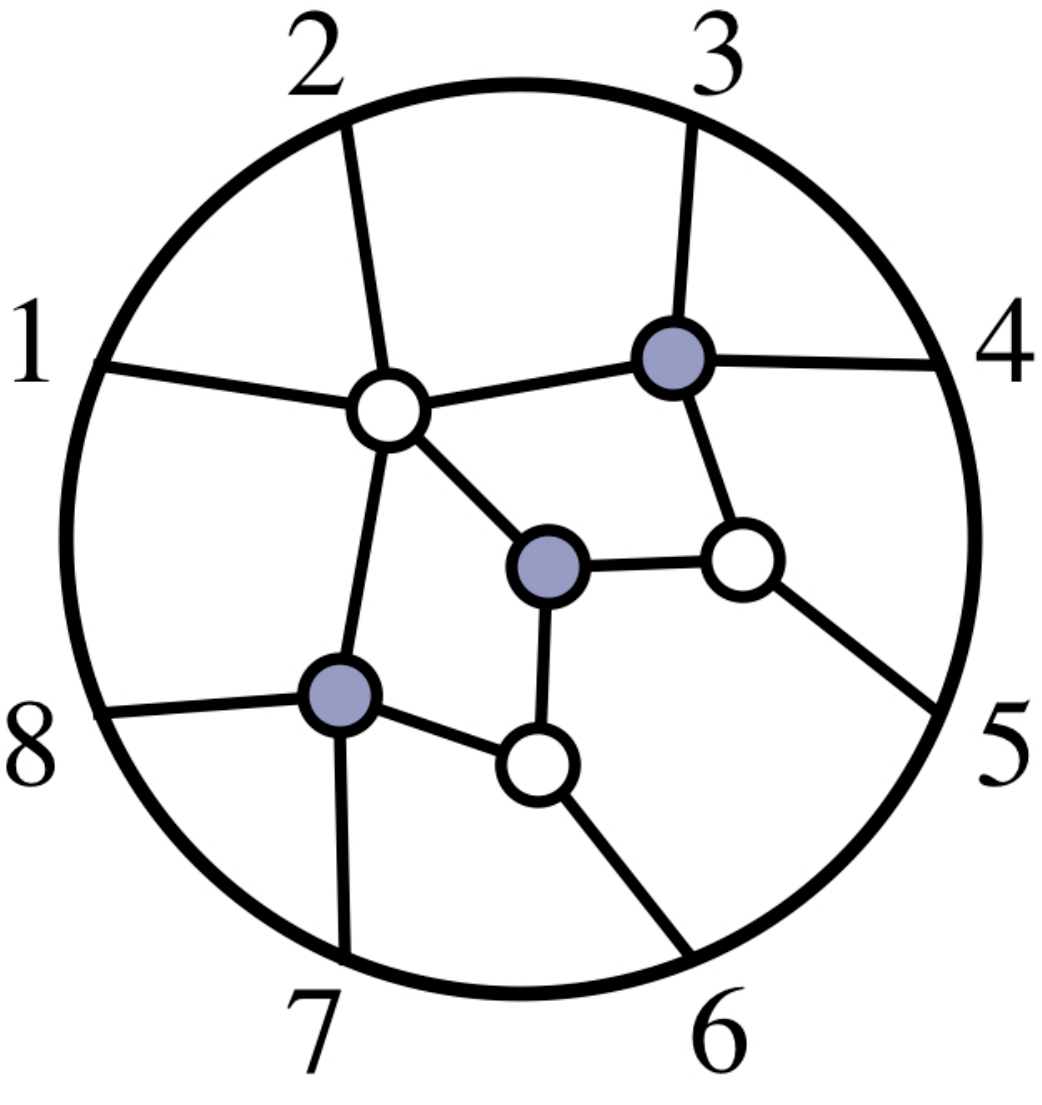}  
\newline\tiny\{2,5,6,11,8,9,12,15\} &
{\tiny
$\left(
\begin{array}{cccccccc}
 1 & \alpha_9 & \alpha_8 & -\alpha_3-\alpha_5 & (-\alpha_3-\alpha_5) \alpha_6 & -\alpha_3 \alpha_4 & 0 & 0 \\
 0 & 0 & 1 & \alpha_7 & 0 & 0 & 0 & 0 \\
 0 & 0 & 0 & 1 & \alpha_6 & \alpha_4 & \alpha_2 & 0 \\
 0 & 0 & 0 & 0 & 0 & 0 & 1 & \alpha_1 \\
\end{array}
\right)$}
\\
\hline

$\begin{array}{c} (3.14e) \\n=10\\k=5\\d=10 \end{array}$ &
\includegraphics[width=0.1\textwidth]{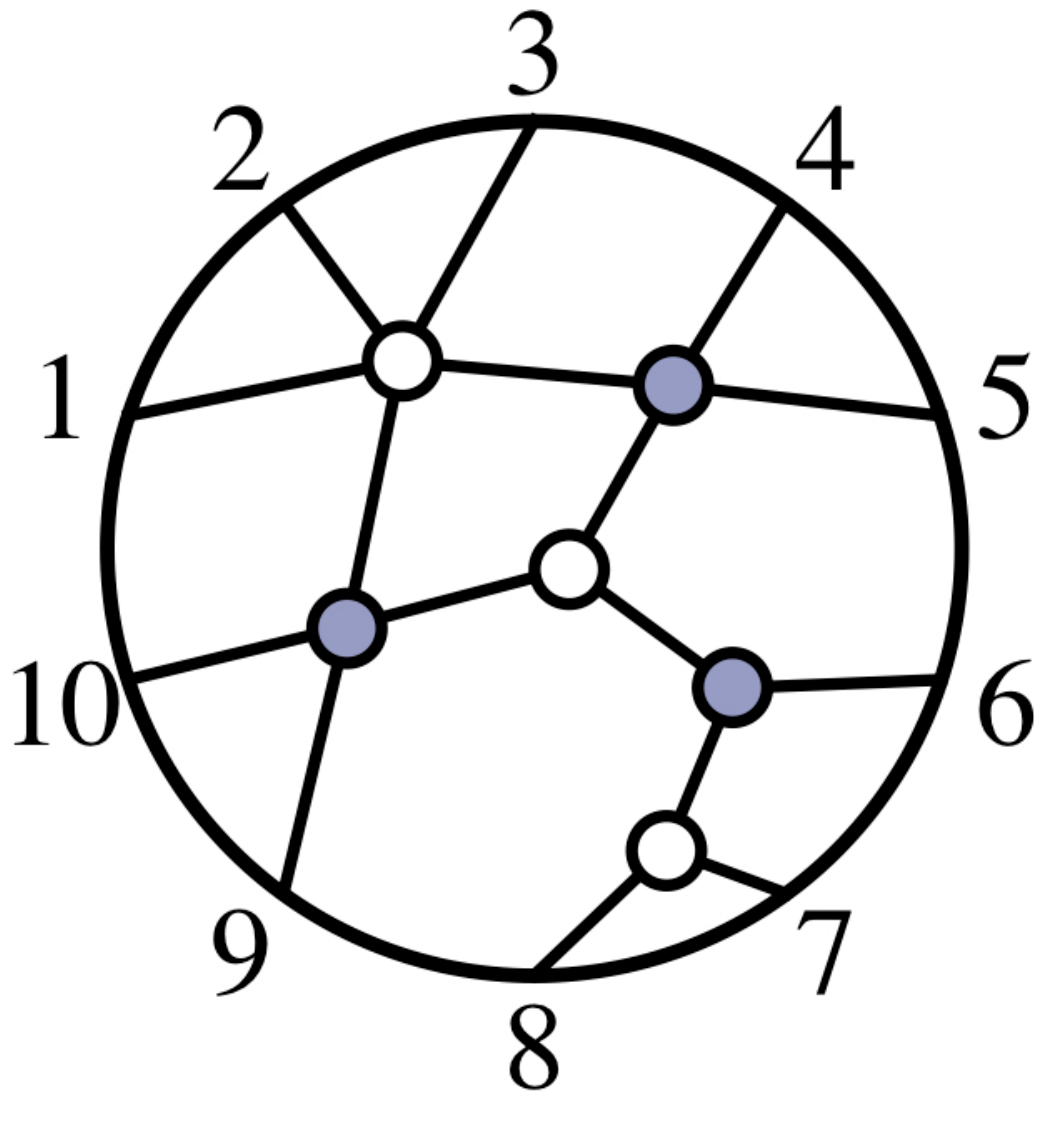}  
\newline\tiny\{2,3,7,10,14,11,8,16,15,19\} &
{\tiny
$\left(
\begin{array}{cccccccccc}
 1 & \alpha_{10} & \alpha_9 & \alpha_8 & -\alpha_5 & -\alpha_5 \alpha_6 & 0 & 0 & 0 & 0 \\
 0 & 0 & 0 & 1 & \alpha_7 & 0 & 0 & 0 & 0 & 0 \\
 0 & 0 & 0 & 0 & 1 & \alpha_6 & 0 & 0 & -\alpha_2 & 0 \\
 0 & 0 & 0 & 0 & 0 & 1 & \alpha_4 & \alpha_3 & 0 & 0 \\
 0 & 0 & 0 & 0 & 0 & 0 & 0 & 0 & 1 & \alpha_1 \\
\end{array}
\right)$}
\\
\hline

$\begin{array}{c} (3.14f) \\n=12\\k=6\\d=11 \end{array}$ &
\includegraphics[width=0.1\textwidth]{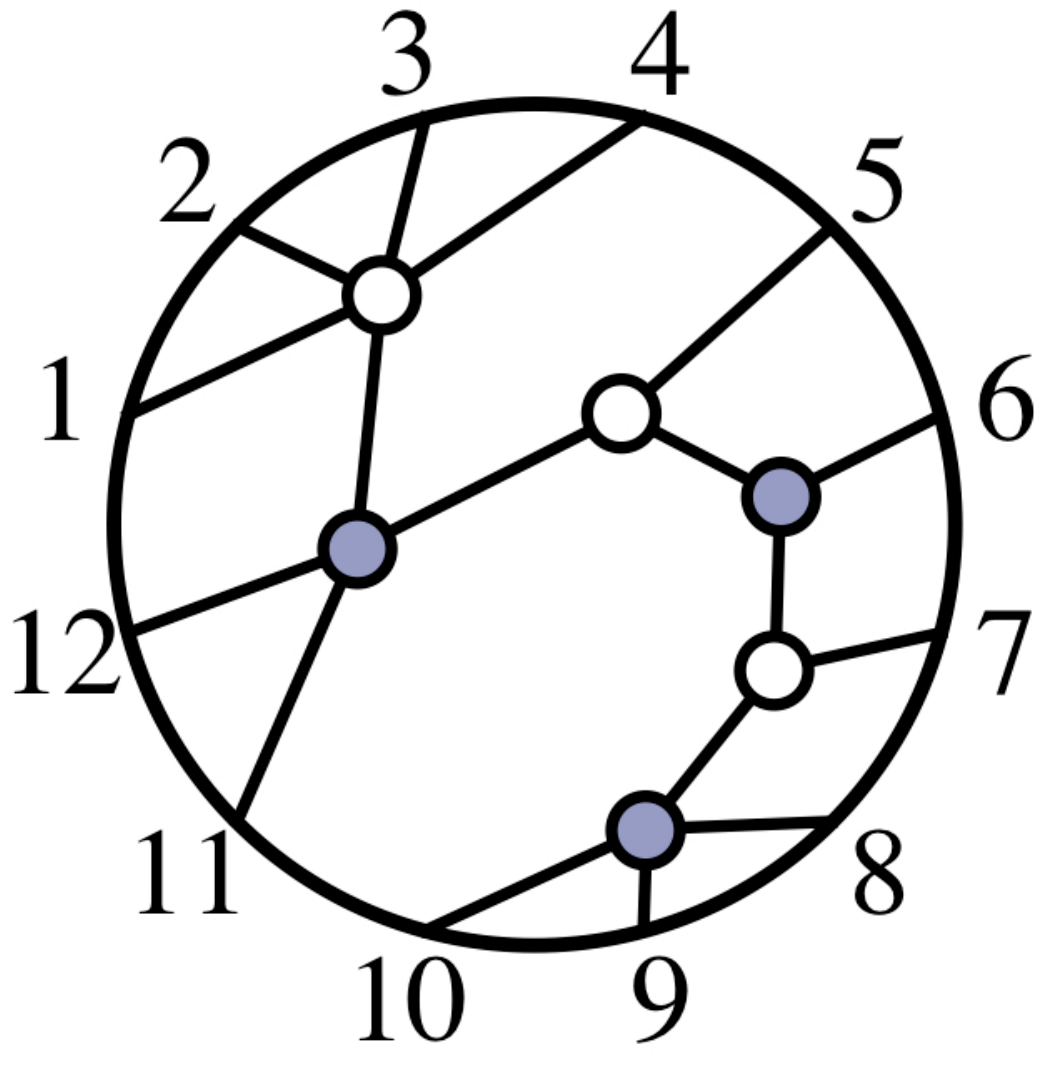}  
\newline\tiny\{2,3,4,12,7,13,10,18,20,21,17,23\} &
{\tiny
$\left(
\begin{array}{cccccccccccc}
 1 & \alpha_{11} & \alpha_{10} & \alpha_9 & \alpha_7 & \alpha_7 \alpha_8 & 0 & 0 & 0 & 0 & 0 & 0 \\
 0 & 0 & 0 & 0 & 1 & \alpha_8 & 0 & 0 & 0 & 0 & -\alpha_2 & 0 \\
 0 & 0 & 0 & 0 & 0 & 1 & \alpha_6 & \alpha_5 & 0 & 0 & 0 & 0 \\
 0 & 0 & 0 & 0 & 0 & 0 & 0 & 1 & \alpha_4 & 0 & 0 & 0 \\
 0 & 0 & 0 & 0 & 0 & 0 & 0 & 0 & 1 & \alpha_3 & 0 & 0 \\
 0 & 0 & 0 & 0 & 0 & 0 & 0 & 0 & 0 & 0 & 1 & \alpha_1 \\
\end{array}
\right)$}
\\
\hline

$\begin{array}{c} (3.14g) \\n=12\\k=6\\d=11 \end{array}$&
\includegraphics[width=0.1\textwidth]{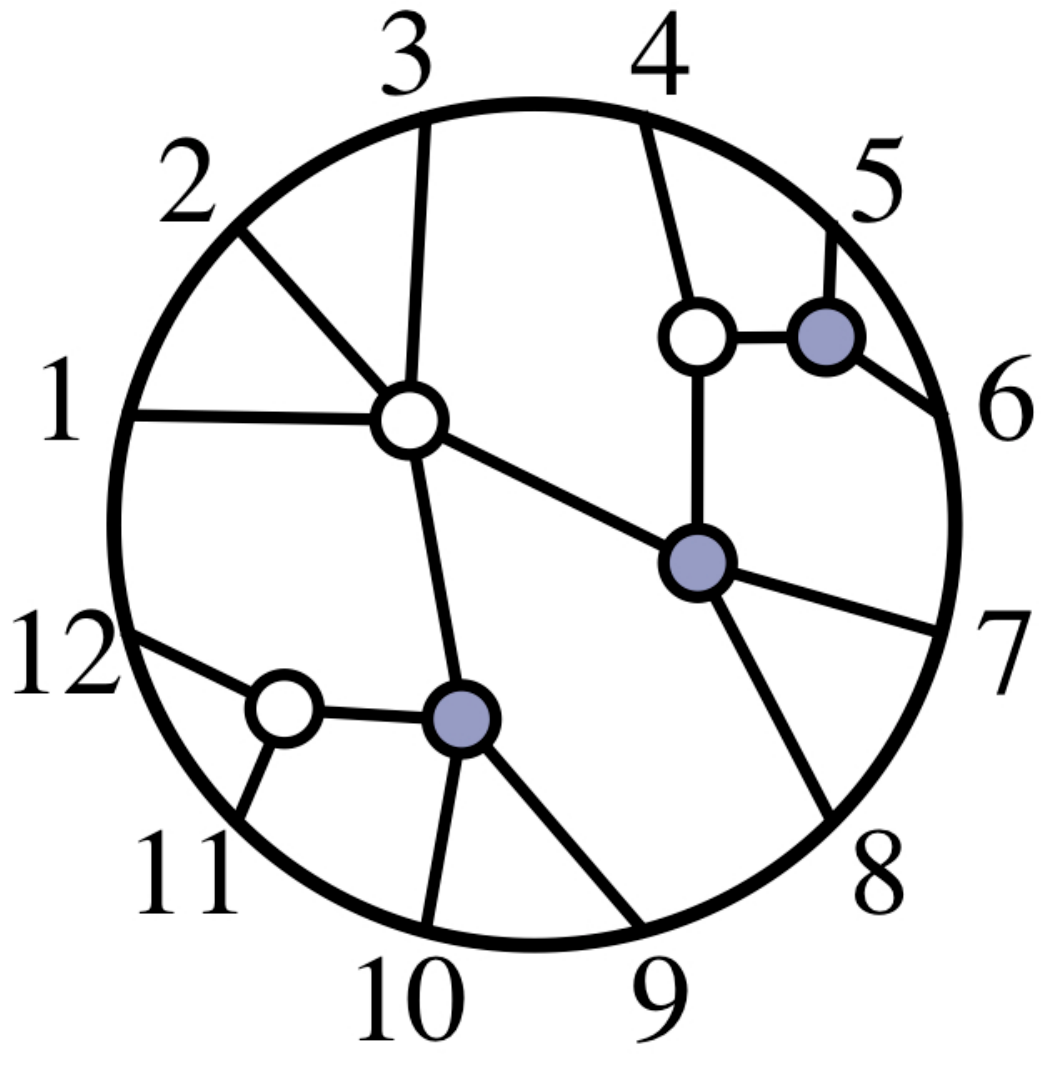}  
\newline\tiny\{2,3,8,6,11,17,16,19,13,21,12,22\} &
{\tiny
$\left(
\begin{array}{cccccccccccc}
 1 & \alpha_{11} & \alpha_{10} & \alpha_8 & \alpha_8 \alpha_9 & 0 & 0 & 0 & -\alpha_4 & 0 & 0 & 0 \\
 0 & 0 & 0 & 1 & \alpha_9 & 0 & -\alpha_6 & 0 & 0 & 0 & 0 & 0 \\
 0 & 0 & 0 & 0 & 1 & \alpha_7 & 0 & 0 & 0 & 0 & 0 & 0 \\
 0 & 0 & 0 & 0 & 0 & 0 & 1 & \alpha_5 & 0 & 0 & 0 & 0 \\
 0 & 0 & 0 & 0 & 0 & 0 & 0 & 0 & 1 & \alpha_3 & 0 & 0 \\
 0 & 0 & 0 & 0 & 0 & 0 & 0 & 0 & 0 & 1 & \alpha_2 & \alpha_1 \\
\end{array}
\right)$}
\\
\hline

$\begin{array}{c} (3.14h) \\n=12\\k=6\\d=11 \end{array}$ &
\includegraphics[width=0.1\textwidth]{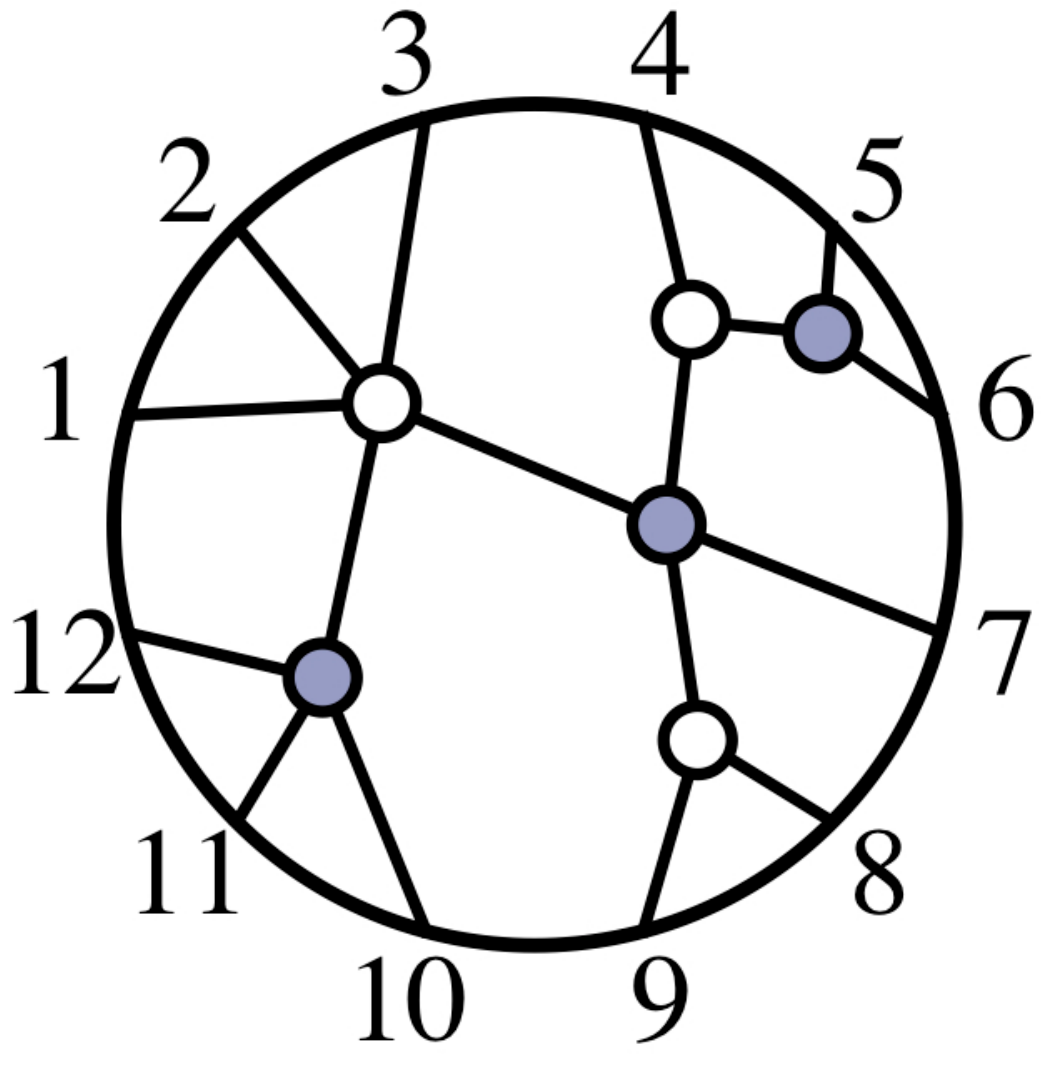}  
\newline\tiny\{2,3,8,6,12,17,16,9,19,13,22,23\} &
{\tiny
$\left(
\begin{array}{cccccccccccc}
 1 & \alpha_{11} & \alpha_{10} & \alpha_8 & \alpha_8 \alpha_9 & 0 & 0 & 0 & 0 & -\alpha_3 & 0 & 0 \\
 0 & 0 & 0 & 1 & \alpha_9 & 0 & -\alpha_6 & 0 & 0 & 0 & 0 & 0 \\
 0 & 0 & 0 & 0 & 1 & \alpha_7 & 0 & 0 & 0 & 0 & 0 & 0 \\
 0 & 0 & 0 & 0 & 0 & 0 & 1 & \alpha_5 & \alpha_4 & 0 & 0 & 0 \\
 0 & 0 & 0 & 0 & 0 & 0 & 0 & 0 & 0 & 1 & \alpha_2 & 0 \\
 0 & 0 & 0 & 0 & 0 & 0 & 0 & 0 & 0 & 0 & 1 & \alpha_1 \\
\end{array}
\right)$}
\\
\hline

$\begin{array}{c} (3.14i) \\n=10\\k=5\\d=10 \end{array}$&
\includegraphics[width=0.1\textwidth]{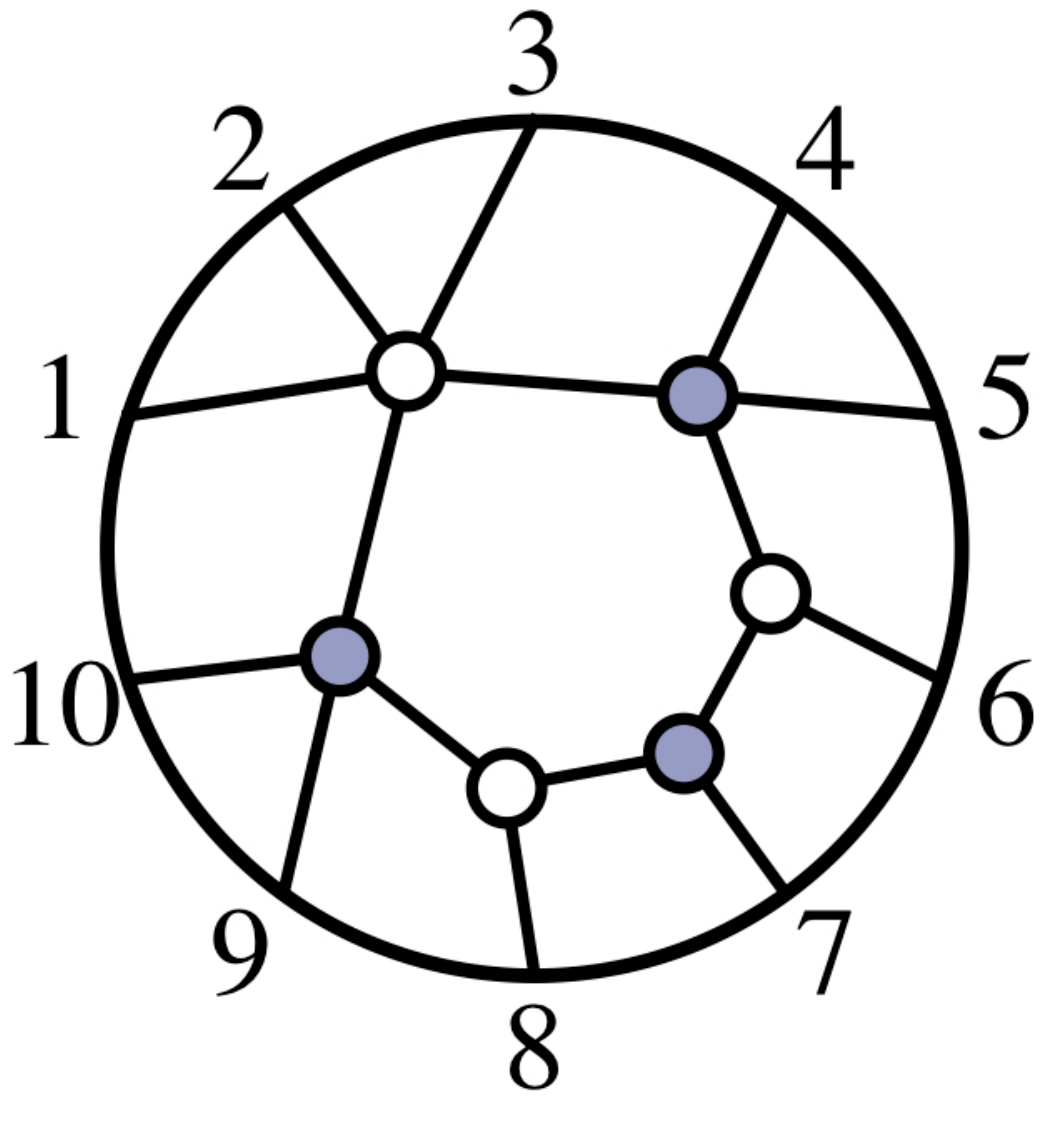}  
\newline\tiny\{2,3,6,10,14,8,15,11,17,19\} &
{\tiny
$\left(
\begin{array}{cccccccccc}
 1 & \alpha_{10} & \alpha_9 & \alpha_8 & 0 & 0 & \alpha_3 & \alpha_3 \alpha_4 & 0 & 0 \\
 0 & 0 & 0 & 1 & \alpha_7 & 0 & 0 & 0 & 0 & 0 \\
 0 & 0 & 0 & 0 & 1 & \alpha_6 & \alpha_5 & 0 & 0 & 0 \\
 0 & 0 & 0 & 0 & 0 & 0 & 1 & \alpha_4 & \alpha_2 & 0 \\
 0 & 0 & 0 & 0 & 0 & 0 & 0 & 0 & 1 & \alpha_1 \\
\end{array}
\right)$}
\\
\hline

$\begin{array}{c} (3.14j) \\n=10\\k=5\\d=10 \end{array}$ &
\includegraphics[width=0.1\textwidth]{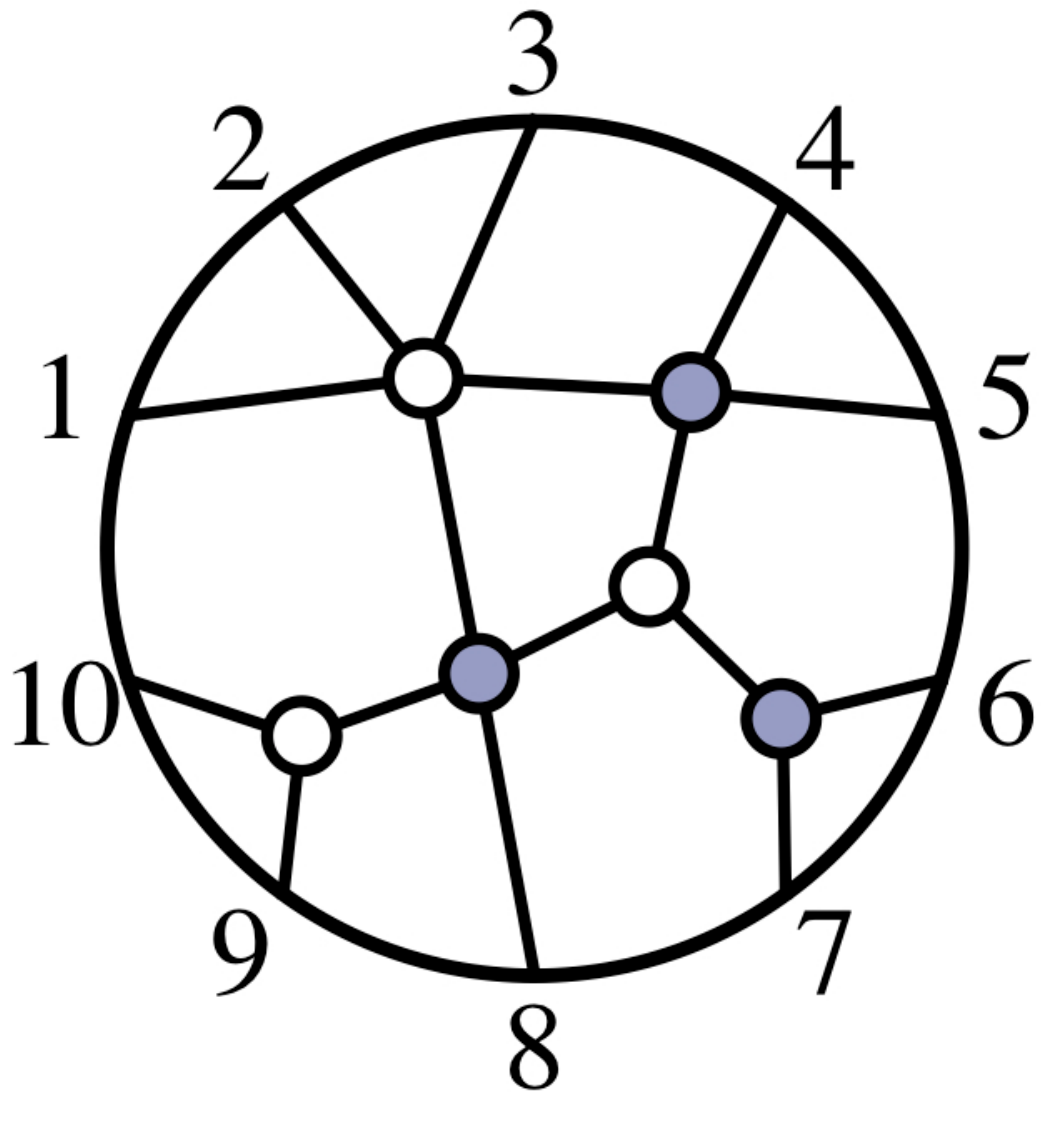}  
\newline\tiny\{2,3,7,9,14,11,16,15,10,18\} &
{\tiny
$\left(
\begin{array}{cccccccccc}
 1 & \alpha_{10} & \alpha_9 & \alpha_8 & -\alpha_5 & -\alpha_5 \alpha_6 & 0 & 0 & 0 & 0 \\
 0 & 0 & 0 & 1 & \alpha_7 & 0 & 0 & 0 & 0 & 0 \\
 0 & 0 & 0 & 0 & 1 & \alpha_6 & 0 & -\alpha_3 & 0 & 0 \\
 0 & 0 & 0 & 0 & 0 & 1 & \alpha_4 & 0 & 0 & 0 \\
 0 & 0 & 0 & 0 & 0 & 0 & 0 & 1 & \alpha_2 & \alpha_1 \\
\end{array}
\right)$}
\\
\hline

$\begin{array}{c} (3.14k) \\n=10\\k=5\\d=10 \end{array}$ &
\includegraphics[width=0.1\textwidth]{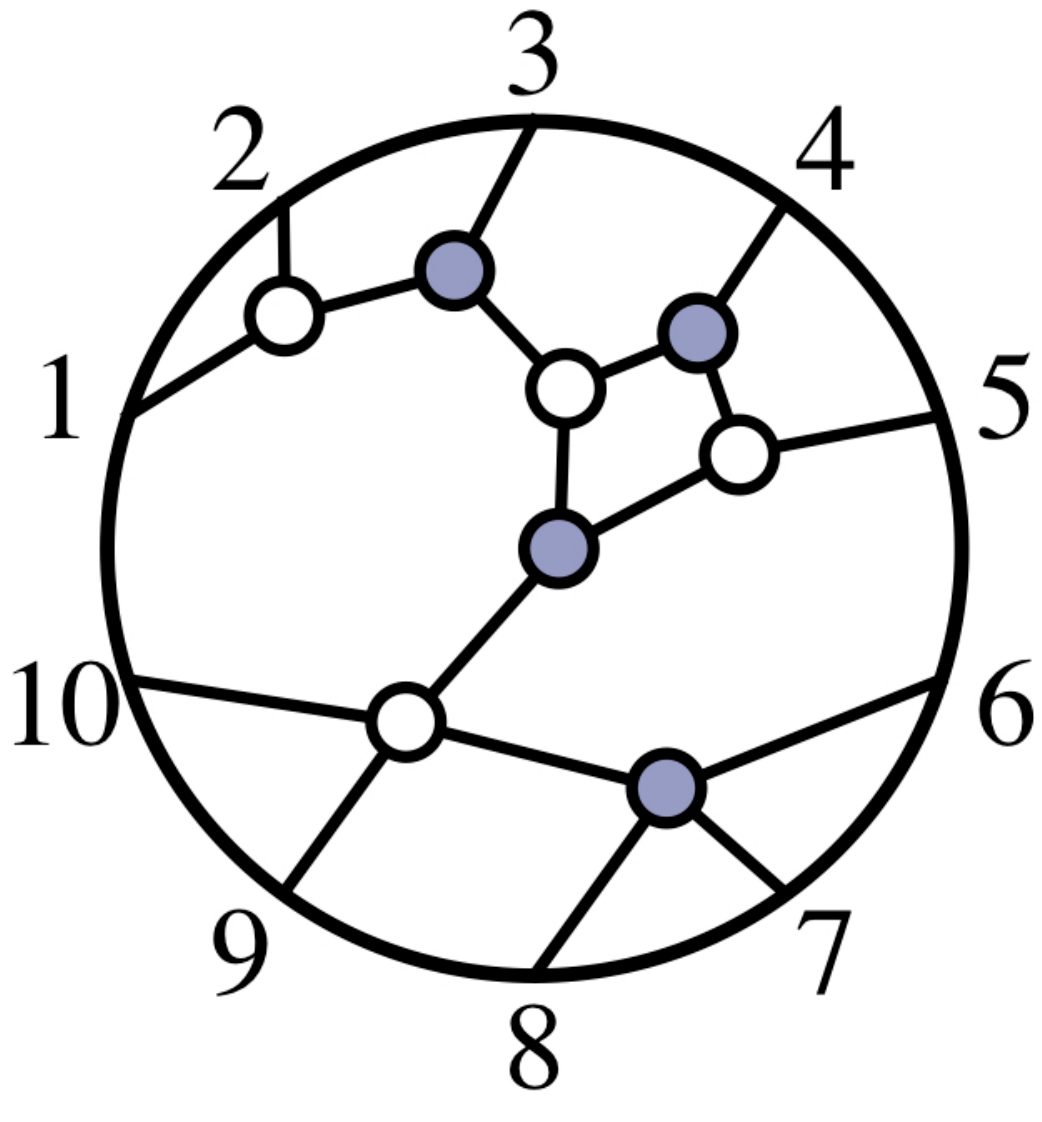}  
\newline\tiny\{2,5,11,8,13,9,16,17,10,14\} &
{\tiny
$\left(
\begin{array}{cccccccccc}
 1 & \alpha_{10} & \alpha_9 & 0 & 0 & 0 & 0 & 0 & 0 & 0 \\
 0 & 0 & 1 & \alpha_6+\alpha_8 & \alpha_6 \alpha_7 & 0 & 0 & 0 & 0 & 0 \\
 0 & 0 & 0 & 1 & \alpha_7 & \alpha_5 & 0 & 0 & \alpha_2 & \alpha_1 \\
 0 & 0 & 0 & 0 & 0 & 1 & \alpha_4 & 0 & 0 & 0 \\
 0 & 0 & 0 & 0 & 0 & 0 & 1 & \alpha_3 & 0 & 0 \\
\end{array}
\right)$}
\\
\hline

$\begin{array}{c} (3.14l) \\n=8\\k=4\\d=9 \end{array}$ &
\includegraphics[width=0.1\textwidth]{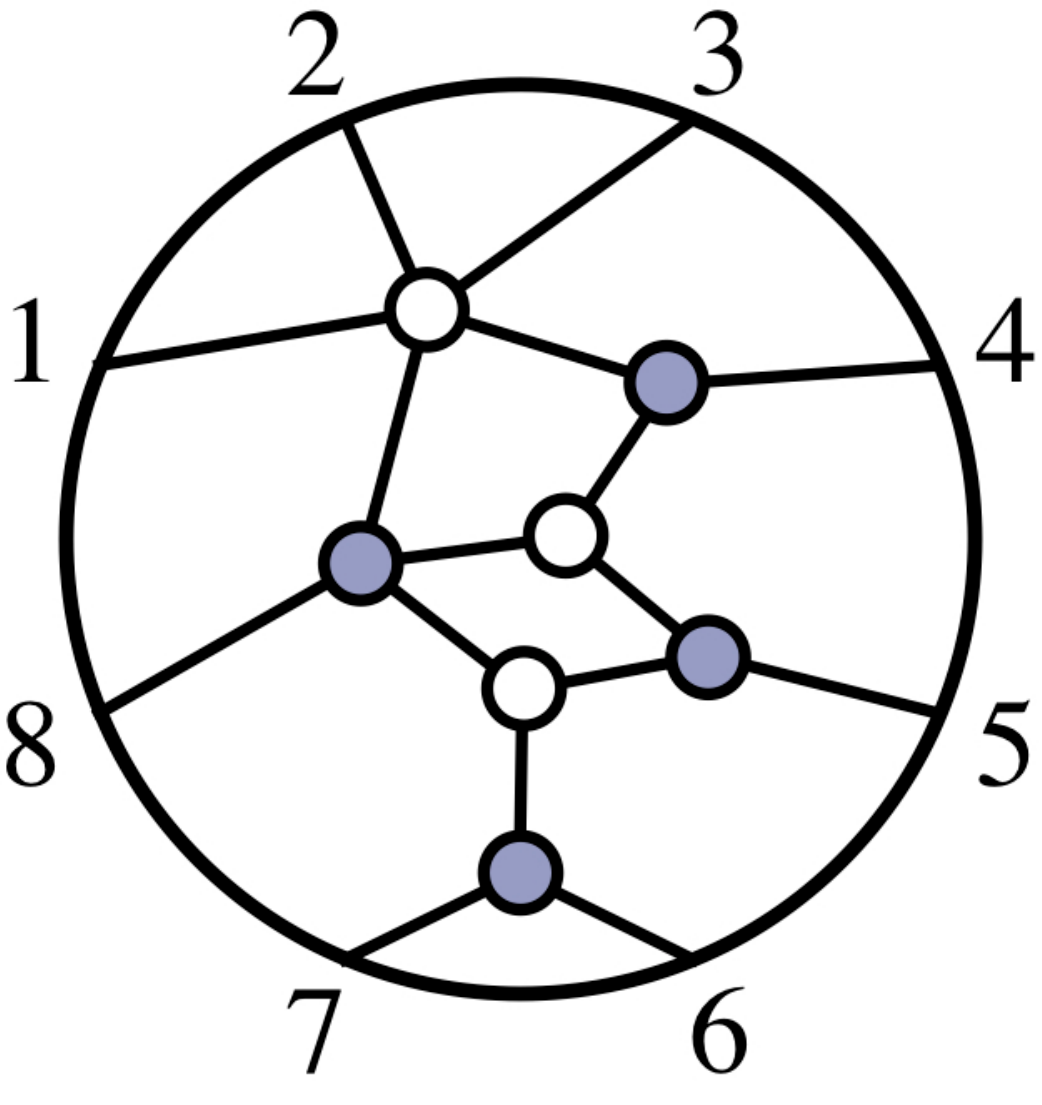}  
\newline\tiny\{2,3,7,8,9,12,14,13\} &
{\tiny
$\left(
\begin{array}{cccccccc}
 1 & \alpha_9 & \alpha_8 & \alpha_5+\alpha_7 & \alpha_5 \alpha_6 & 0 & 0 & 0 \\
 0 & 0 & 0 & 1 & \alpha_3+\alpha_6 & \alpha_3 \alpha_4 & 0 & 0 \\
 0 & 0 & 0 & 0 & 1 & \alpha_4 & 0 & -\alpha_1 \\
 0 & 0 & 0 & 0 & 0 & 1 & \alpha_2 & 0 \\
\end{array}
\right)$}
\\
\hline

$\begin{array}{c} (3.14m) \\n=8\\k=4\\d=9 \end{array}$ &
\includegraphics[width=0.1\textwidth]{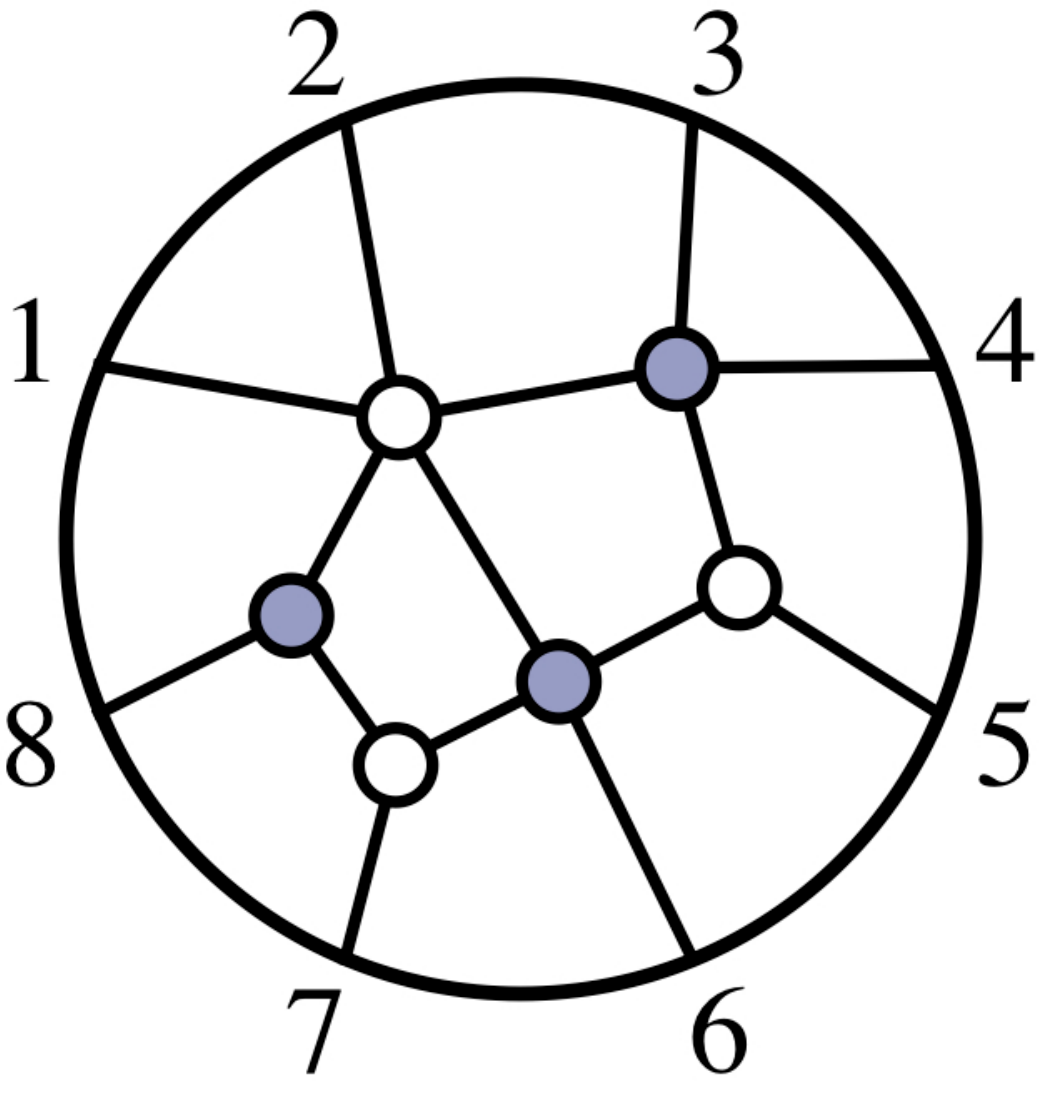}  
\newline\tiny\{2,5,7,11,8,12,9,14\} &
{\tiny
$\left(
\begin{array}{cccccccc}
 1 & \alpha_9 & \alpha_8 & -\alpha_5 & -\alpha_5 \alpha_6 & \alpha_2 & \alpha_2 \alpha_3 & 0 \\
 0 & 0 & 1 & \alpha_7 & 0 & 0 & 0 & 0 \\
 0 & 0 & 0 & 1 & \alpha_6 & \alpha_4 & 0 & 0 \\
 0 & 0 & 0 & 0 & 0 & 1 & \alpha_3 & \alpha_1 \\
\end{array}
\right)$}
\\
\hline

$\begin{array}{c} (3.26a) \\n=14\\k=7\\d=13 \end{array}$&
\includegraphics[width=0.1\textwidth]{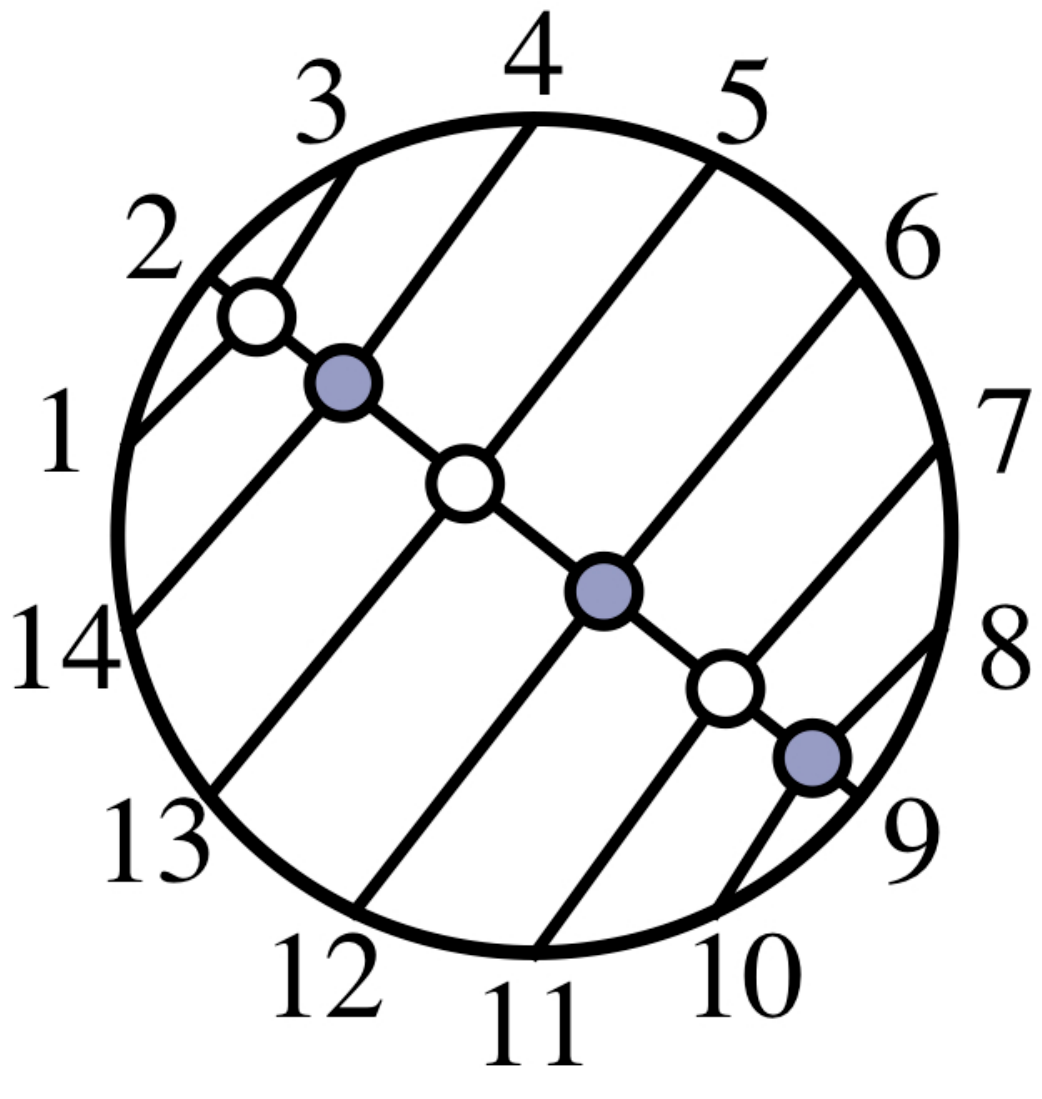}  
\newline\tiny\{2,3,14,15,12,13,10,11,22,23,20,21,18,19\} &
{\tiny
$\left(
\begin{array}{cccccccccccccc}
 1 & \alpha_{13} & \alpha_{12} & \alpha_{11} & 0 & 0 & 0 & 0 & 0 & 0 & 0 & 0 & 0 & 0 \\
 0 & 0 & 0 & 1 & \alpha_2 & \alpha_2 \alpha_{10} & 0 & 0 & 0 & 0 & 0 & 0 & \alpha_2 \alpha_3 & 0 \\
 0 & 0 & 0 & 0 & 1 & \alpha_{10} & 0 & 0 & 0 & 0 & 0 & 0 & \alpha_3 & \alpha_1 \\
 0 & 0 & 0 & 0 & 0 & 1 & \alpha_5 & \alpha_5 \alpha_9 & 0 & 0 & \alpha_5 \alpha_6 & 0 & 0 & 0 \\
 0 & 0 & 0 & 0 & 0 & 0 & 1 & \alpha_9 & 0 & 0 & \alpha_6 & \alpha_4 & 0 & 0 \\
 0 & 0 & 0 & 0 & 0 & 0 & 0 & 1 & \alpha_8 & 0 & 0 & 0 & 0 & 0 \\
 0 & 0 & 0 & 0 & 0 & 0 & 0 & 0 & 1 & \alpha_7 & 0 & 0 & 0 & 0 \\
\end{array}
\right)$}
\\
\hline

$\begin{array}{c} (3.27a) \\n=10\\k=5\\d=11 \end{array}$ &
\includegraphics[width=0.1\textwidth]{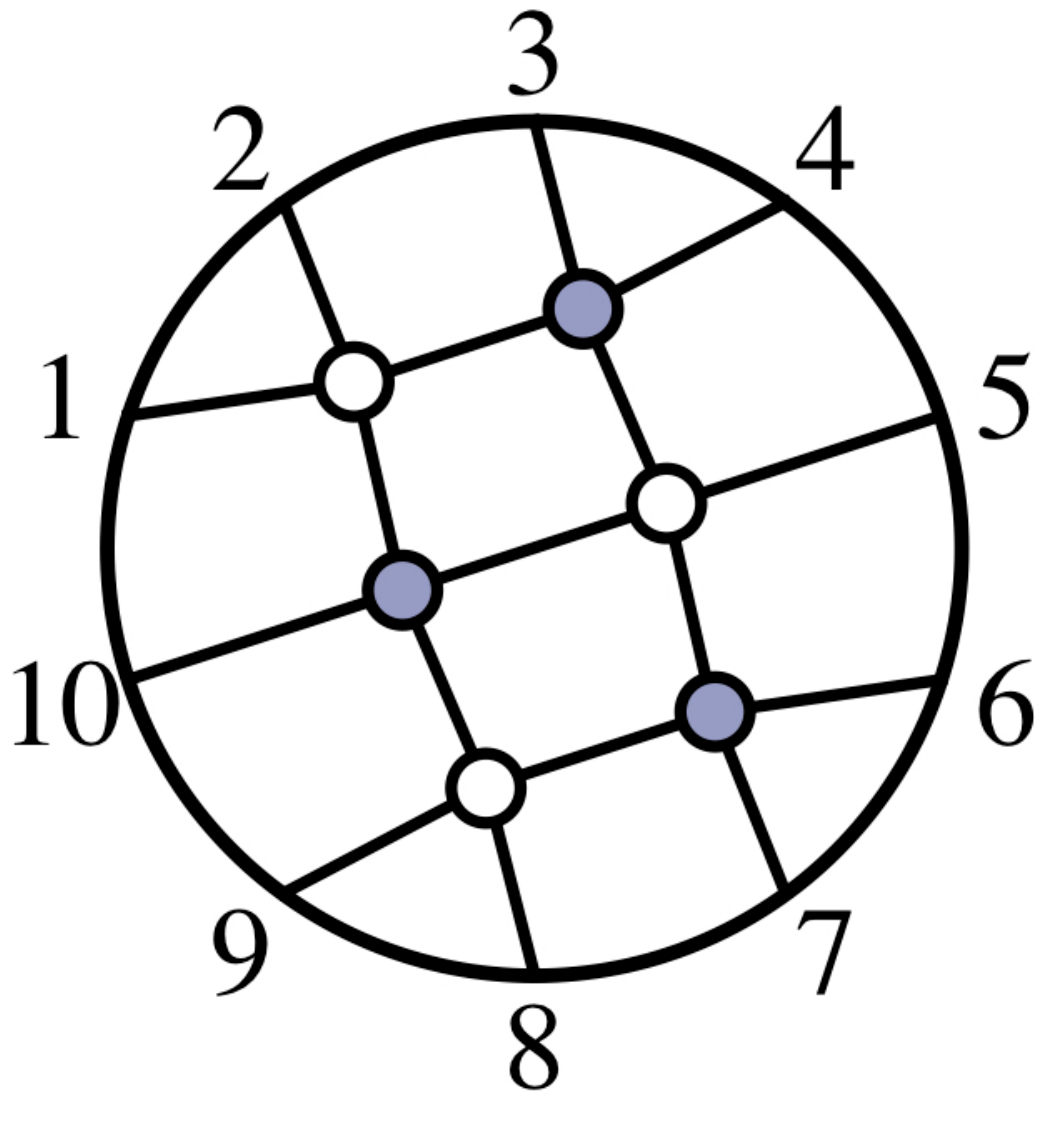}  
\newline\tiny\{2,5,10,13,8,11,16,9,14,17\} &
{\tiny
$\left(
\begin{array}{cccccccccc}
 1 & \alpha_{11} & \alpha_{10} & -\alpha_6 & -\alpha_6 \alpha_8 & -\alpha_6 \alpha_7 & 0 & 0 & 0 & 0 \\
 0 & 0 & 1 & \alpha_9 & 0 & 0 & 0 & 0 & 0 & 0 \\
 0 & 0 & 0 & 1 & \alpha_8 & \alpha_7 & -\alpha_2 & -\alpha_2 \alpha_4 & -\alpha_2 \alpha_3 & 0 \\
 0 & 0 & 0 & 0 & 0 & 1 & \alpha_5 & 0 & 0 & 0 \\
 0 & 0 & 0 & 0 & 0 & 0 & 1 & \alpha_4 & \alpha_3 & \alpha_1 \\
\end{array}
\right)$}
\\
\hline

$\begin{array}{c} (3.27b) \\n=14\\k=7\\d=13 \end{array}$ &
\includegraphics[width=0.1\textwidth]{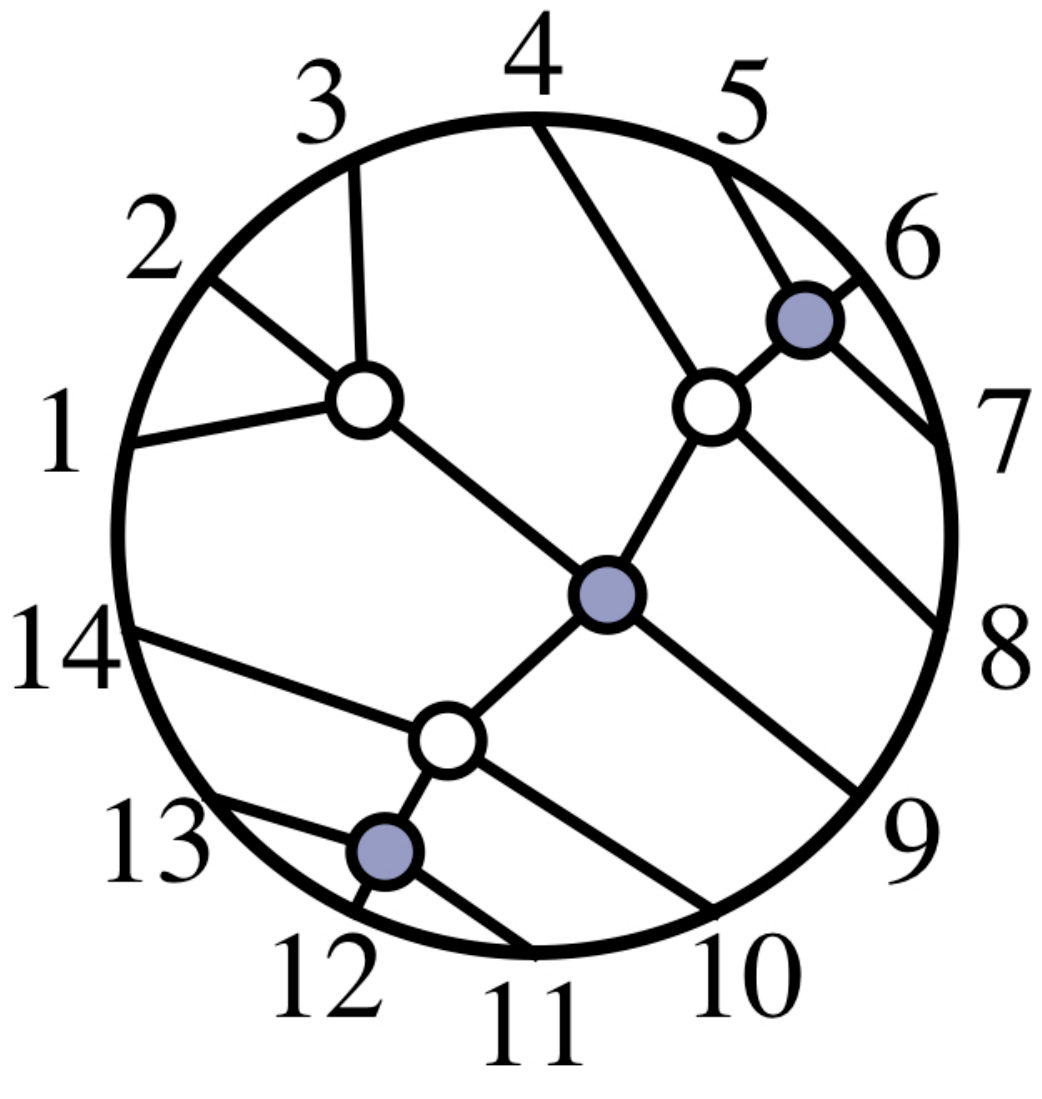}  
\newline\tiny\{2,3,10,7,8,19,20,15,18,13,14,25,26,23\} &
{\tiny
$\left(
\begin{array}{cccccccccccccc}
 1 & \alpha_{13} & \alpha_{12} & \alpha_7 & \alpha_7 \alpha_{11} & 0 & 0 & \alpha_7 \alpha_8 & 0 & 0 & 0 & 0 & 0 & 0 \\
 0 & 0 & 0 & 1 & \alpha_{11} & 0 & 0 & \alpha_8 & \alpha_6 & 0 & 0 & 0 & 0 & 0 \\
 0 & 0 & 0 & 0 & 1 & \alpha_{10} & 0 & 0 & 0 & 0 & 0 & 0 & 0 & 0 \\
 0 & 0 & 0 & 0 & 0 & 1 & \alpha_9 & 0 & 0 & 0 & 0 & 0 & 0 & 0 \\
 0 & 0 & 0 & 0 & 0 & 0 & 0 & 0 & 1 & \alpha_5 & \alpha_4 & 0 & 0 & \alpha_1 \\
 0 & 0 & 0 & 0 & 0 & 0 & 0 & 0 & 0 & 0 & 1 & \alpha_3 & 0 & 0 \\
 0 & 0 & 0 & 0 & 0 & 0 & 0 & 0 & 0 & 0 & 0 & 1 & \alpha_2 & 0 \\
\end{array}
\right)$}
\\
\hline

$\begin{array}{c} (3.27c) \\n=14\\k=7\\d=13 \end{array}$ &
\includegraphics[width=0.1\textwidth]{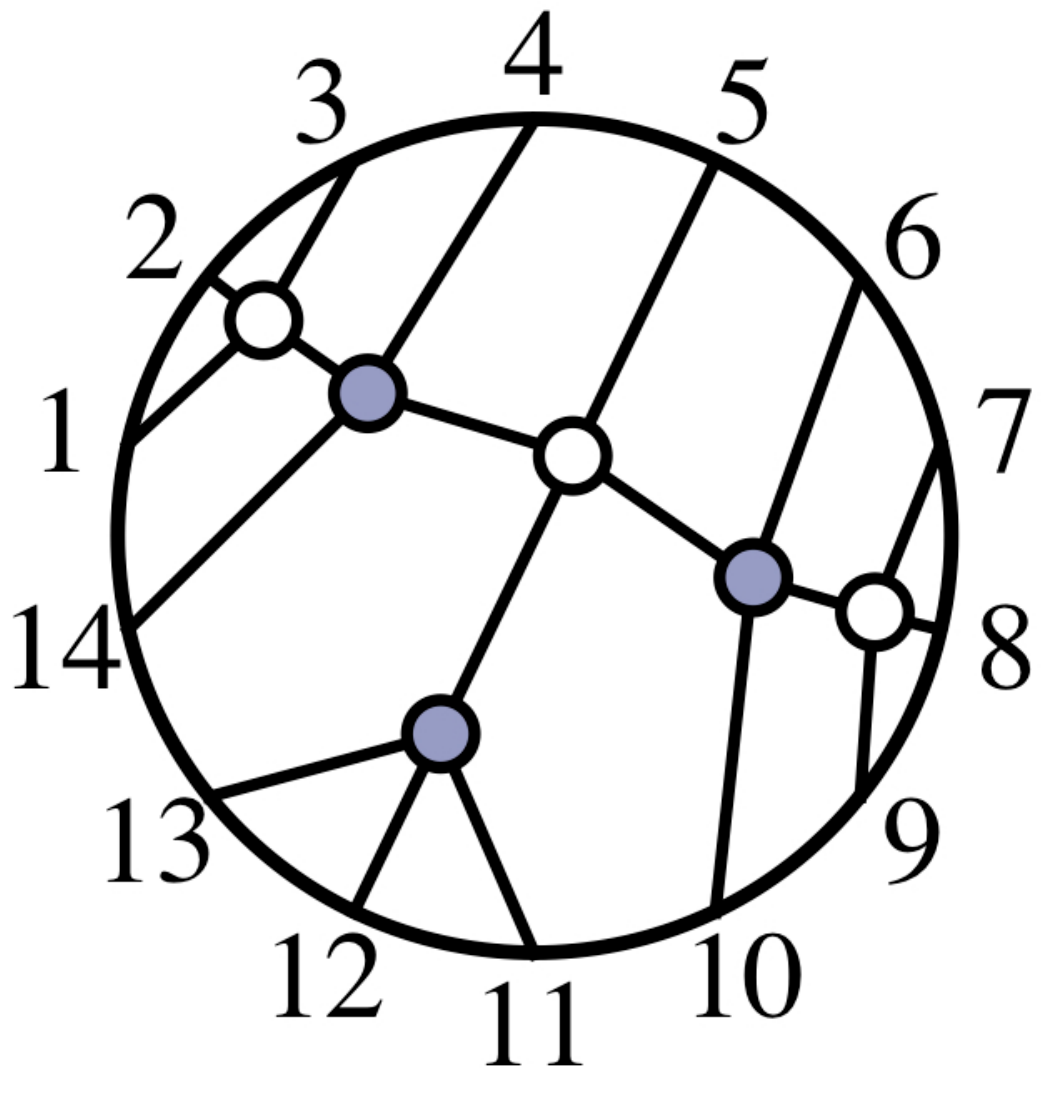}  
\newline\tiny\{2,3,14,15,10,13,8,9,20,21,18,25,26,19\} &
{\tiny
$\left(
\begin{array}{cccccccccccccc}
 1 & \alpha_{13} & \alpha_{12} & \alpha_{11} & 0 & 0 & 0 & 0 & 0 & 0 & 0 & 0 & 0 & 0 \\
 0 & 0 & 0 & 1 & \alpha_4 & \alpha_4 \alpha_{10} & 0 & 0 & 0 & 0 & \alpha_4 \alpha_5 & 0 & 0 & 0 \\
 0 & 0 & 0 & 0 & 1 & \alpha_{10} & 0 & 0 & 0 & 0 & \alpha_5 & 0 & 0 & \alpha_1 \\
 0 & 0 & 0 & 0 & 0 & 1 & \alpha_7 & \alpha_7 \alpha_9 & \alpha_7 \alpha_8 & 0 & 0 & 0 & 0 & 0 \\
 0 & 0 & 0 & 0 & 0 & 0 & 1 & \alpha_9 & \alpha_8 & \alpha_6 & 0 & 0 & 0 & 0 \\
 0 & 0 & 0 & 0 & 0 & 0 & 0 & 0 & 0 & 0 & 1 & \alpha_3 & 0 & 0 \\
 0 & 0 & 0 & 0 & 0 & 0 & 0 & 0 & 0 & 0 & 0 & 1 & \alpha_2 & 0 \\
\end{array}
\right)$}
\\
\hline

$\begin{array}{c} (3.27d) \\n=12\\k=6\\d=12 \end{array}$ &
\includegraphics[width=0.1\textwidth]{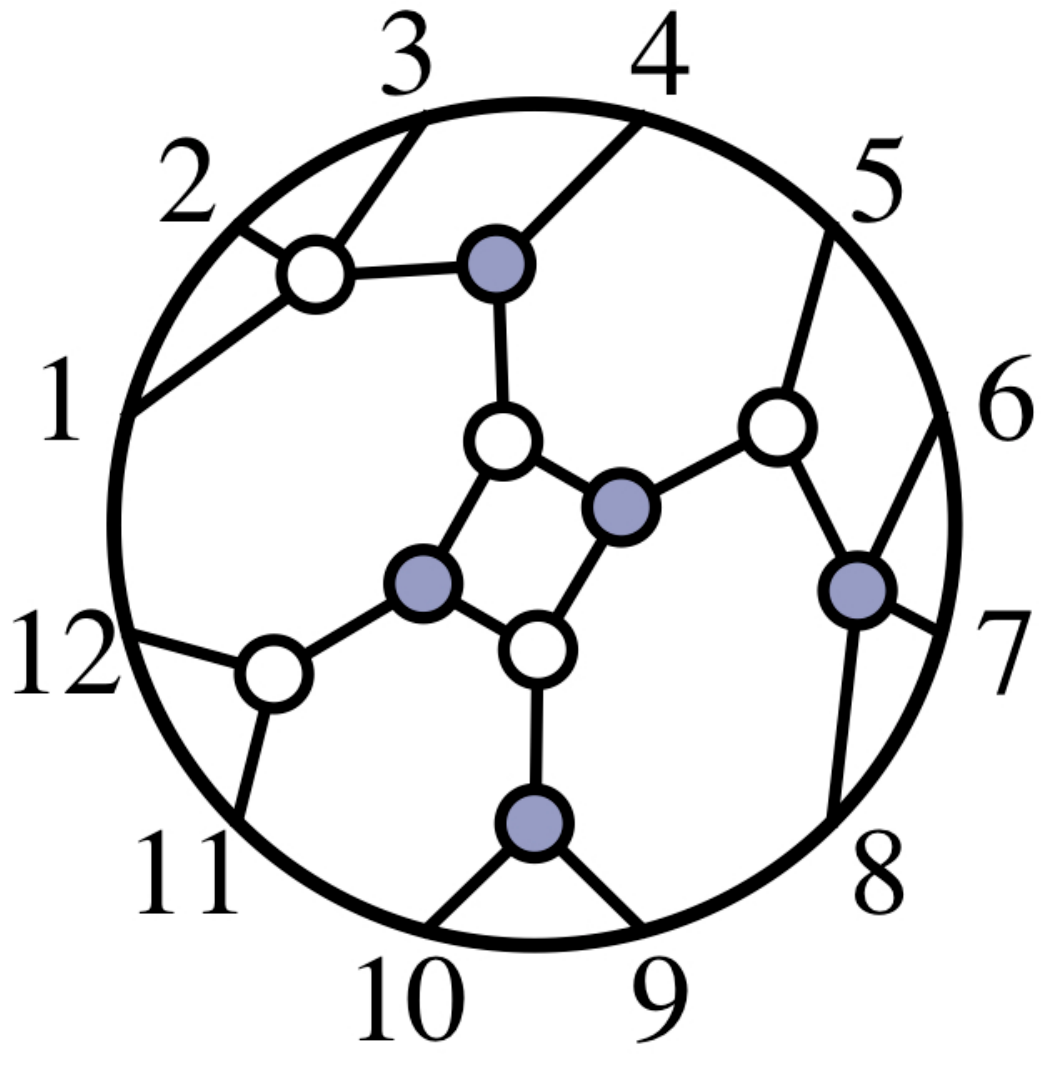}  
\newline\tiny\{2,3,10,13,8,11,18,19,16,21,12,17\} &
{\tiny
$\left(
\begin{array}{cccccccccccc}
 1 & \alpha_{12} & \alpha_{11} & \alpha_{10} & 0 & 0 & 0 & 0 & 0 & 0 & 0 & 0 \\
 0 & 0 & 0 & 1 & \alpha_4+\alpha_8 & (\alpha_4+\alpha_8) \alpha_9 & 0 & 0 & \alpha_4 \alpha_5 & 0 & 0 & 0 \\
 0 & 0 & 0 & 0 & 1 & \alpha_9 & 0 & 0 & \alpha_5 & 0 & -\alpha_2 & -\alpha_1 \\
 0 & 0 & 0 & 0 & 0 & 1 & \alpha_7 & 0 & 0 & 0 & 0 & 0 \\
 0 & 0 & 0 & 0 & 0 & 0 & 1 & \alpha_6 & 0 & 0 & 0 & 0 \\
 0 & 0 & 0 & 0 & 0 & 0 & 0 & 0 & 1 & \alpha_3 & 0 & 0 \\
\end{array}
\right)$}
\\
\hline

$\begin{array}{c} (3.28a) \\n=12\\k=6\\d=12 \end{array}$ &
\includegraphics[width=0.1\textwidth]{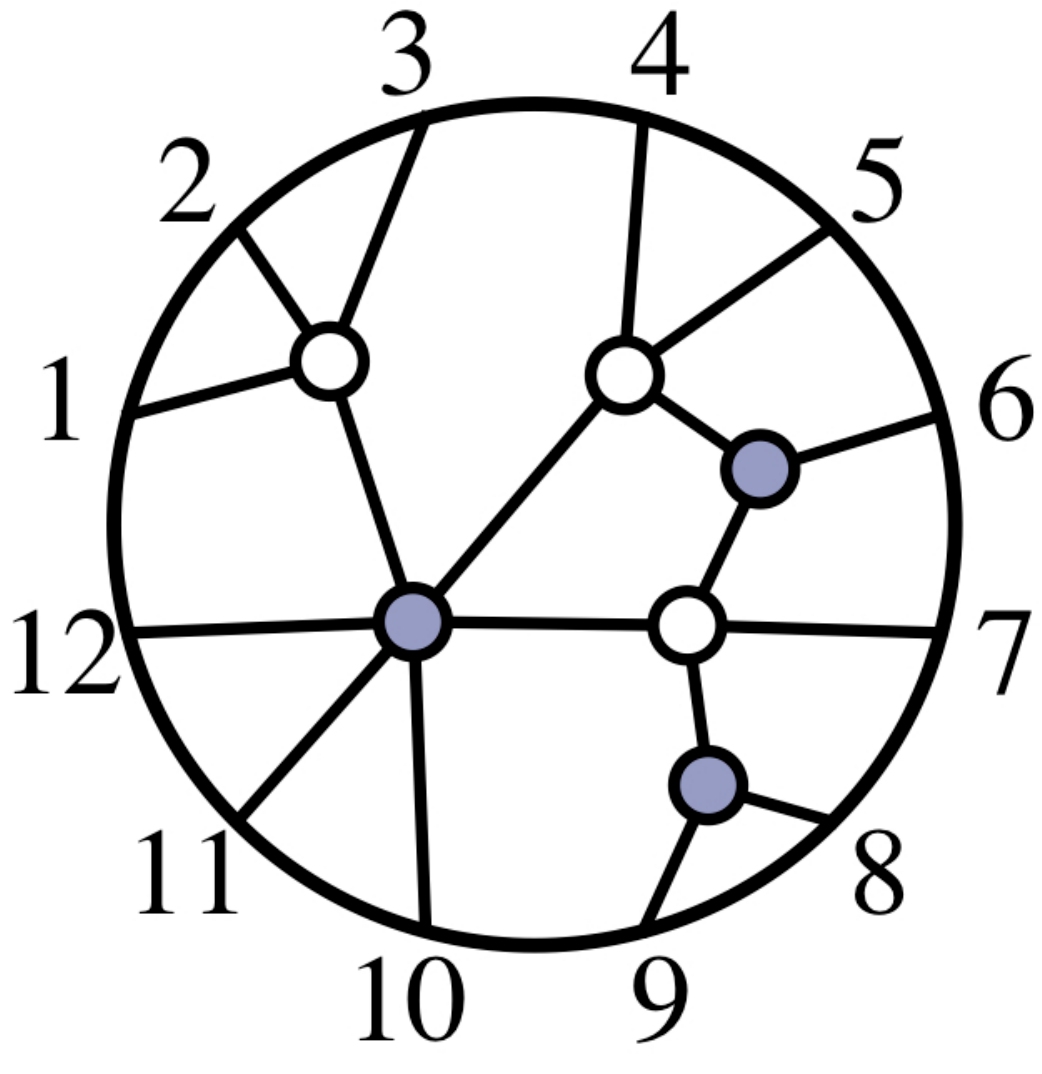}  
\newline\tiny\{2,3,12,5,7,13,9,16,20,18,22,23\} &
{\tiny
$\left(
\begin{array}{cccccccccccc}
 1 & \alpha_{12} & \alpha_{11} & \alpha_8 & \alpha_8 \alpha_{10} & \alpha_8 \alpha_9 & 0 & 0 & 0 & 0 & 0 & 0 \\
 0 & 0 & 0 & 1 & \alpha_{10} & \alpha_5+\alpha_9 & \alpha_5 \alpha_7 & \alpha_5 \alpha_6 & 0 & 0 & 0 & 0 \\
 0 & 0 & 0 & 0 & 0 & 1 & \alpha_7 & \alpha_6 & 0 & -\alpha_3 & 0 & 0 \\
 0 & 0 & 0 & 0 & 0 & 0 & 0 & 1 & \alpha_4 & 0 & 0 & 0 \\
 0 & 0 & 0 & 0 & 0 & 0 & 0 & 0 & 0 & 1 & \alpha_2 & 0 \\
 0 & 0 & 0 & 0 & 0 & 0 & 0 & 0 & 0 & 0 & 1 & \alpha_1 \\
\end{array}
\right)$}
\\
\hline

$\begin{array}{c} (3.28b) \\n=10\\k=5\\d=11 \end{array}$ &
\includegraphics[width=0.1\textwidth]{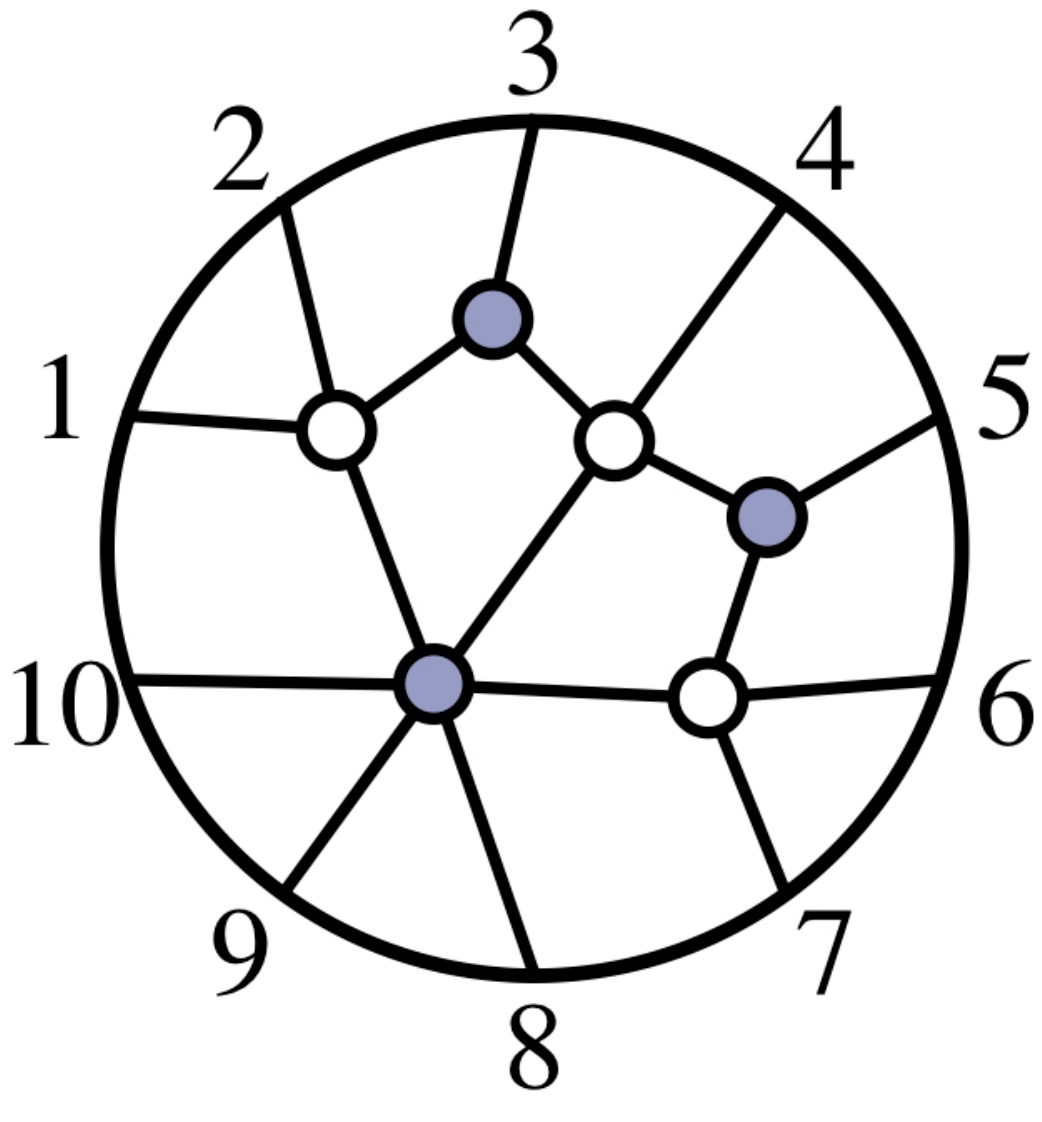}  
\newline\tiny\{2,4,10,6,11,7,13,15,18,19\} &
{\tiny
$\left(
\begin{array}{cccccccccc}
 1 & \alpha_{11} & \alpha_7+\alpha_{10} & \alpha_7 \alpha_9 & \alpha_7 \alpha_8 & 0 & 0 & 0 & 0 & 0 \\
 0 & 0 & 1 & \alpha_9 & \alpha_4+\alpha_8 & \alpha_4 \alpha_6 & \alpha_4 \alpha_5 & 0 & 0 & 0 \\
 0 & 0 & 0 & 0 & 1 & \alpha_6 & \alpha_5 & \alpha_3 & 0 & 0 \\
 0 & 0 & 0 & 0 & 0 & 0 & 0 & 1 & \alpha_2 & 0 \\
 0 & 0 & 0 & 0 & 0 & 0 & 0 & 0 & 1 & \alpha_1 \\
\end{array}
\right)$}
\\
\hline

$\begin{array}{c} (3.28c) \\n=12\\k=6\\d=12 \end{array}$ &
\includegraphics[width=0.1\textwidth]{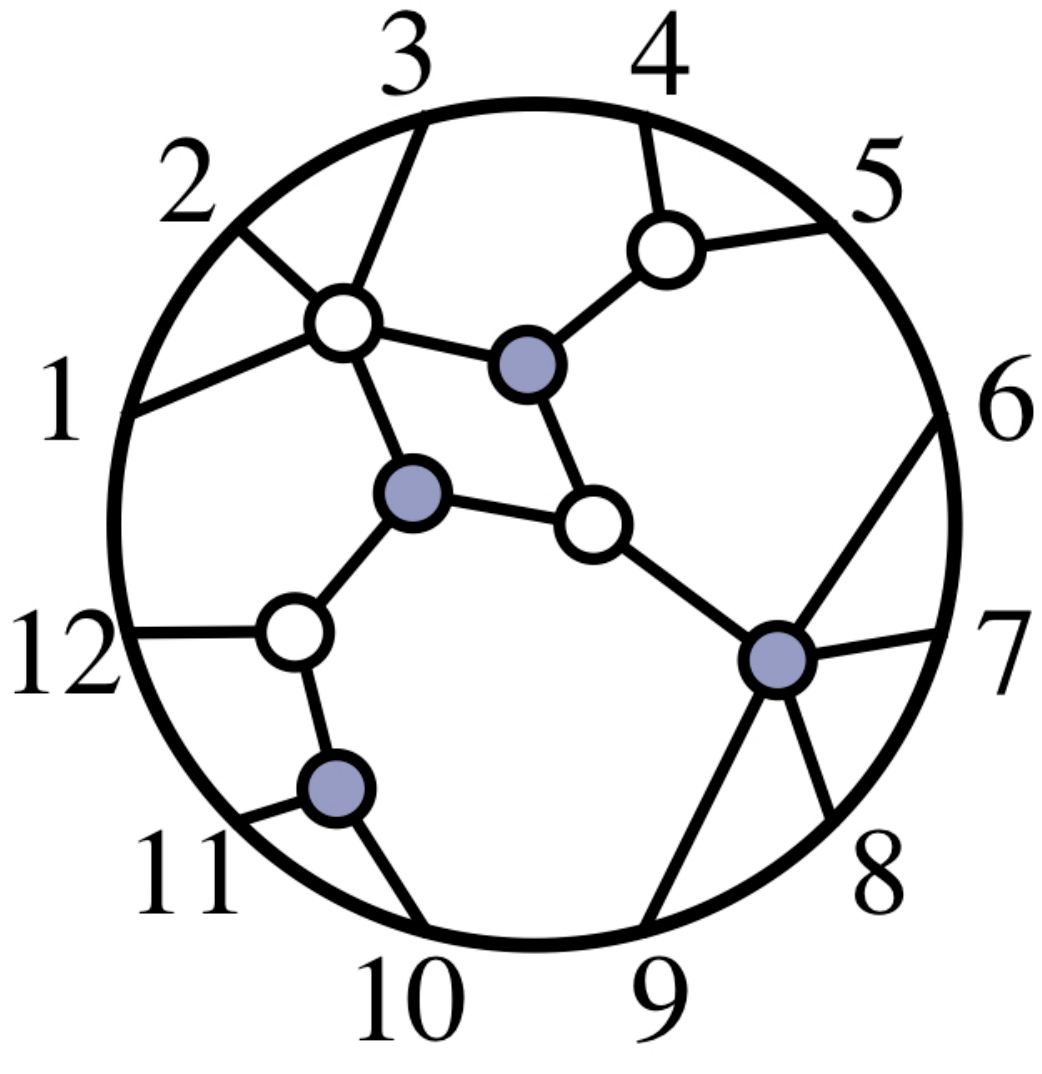}  
\newline\tiny\{2,3,9,5,11,13,18,19,20,12,22,16\} &
{\tiny
$\left(
\begin{array}{cccccccccccc}
 1 & \alpha_{12} & \alpha_{11} & \alpha_7+\alpha_9 & (\alpha_7+\alpha_9) \alpha_{10} & \alpha_7 \alpha_8 & 0 & 0 & 0 & 0 & 0 & 0 \\
 0 & 0 & 0 & 1 & \alpha_{10} & \alpha_8 & 0 & 0 & 0 & -\alpha_3 & 0 & \alpha_1 \\
 0 & 0 & 0 & 0 & 0 & 1 & \alpha_6 & 0 & 0 & 0 & 0 & 0 \\
 0 & 0 & 0 & 0 & 0 & 0 & 1 & \alpha_5 & 0 & 0 & 0 & 0 \\
 0 & 0 & 0 & 0 & 0 & 0 & 0 & 1 & \alpha_4 & 0 & 0 & 0 \\
 0 & 0 & 0 & 0 & 0 & 0 & 0 & 0 & 0 & 1 & \alpha_2 & 0 \\
\end{array}
\right)$}
\\
\hline

$\begin{array}{c} (3.28d) \\n=12\\k=6\\d=12 \end{array}$ &
\includegraphics[width=0.1\textwidth]{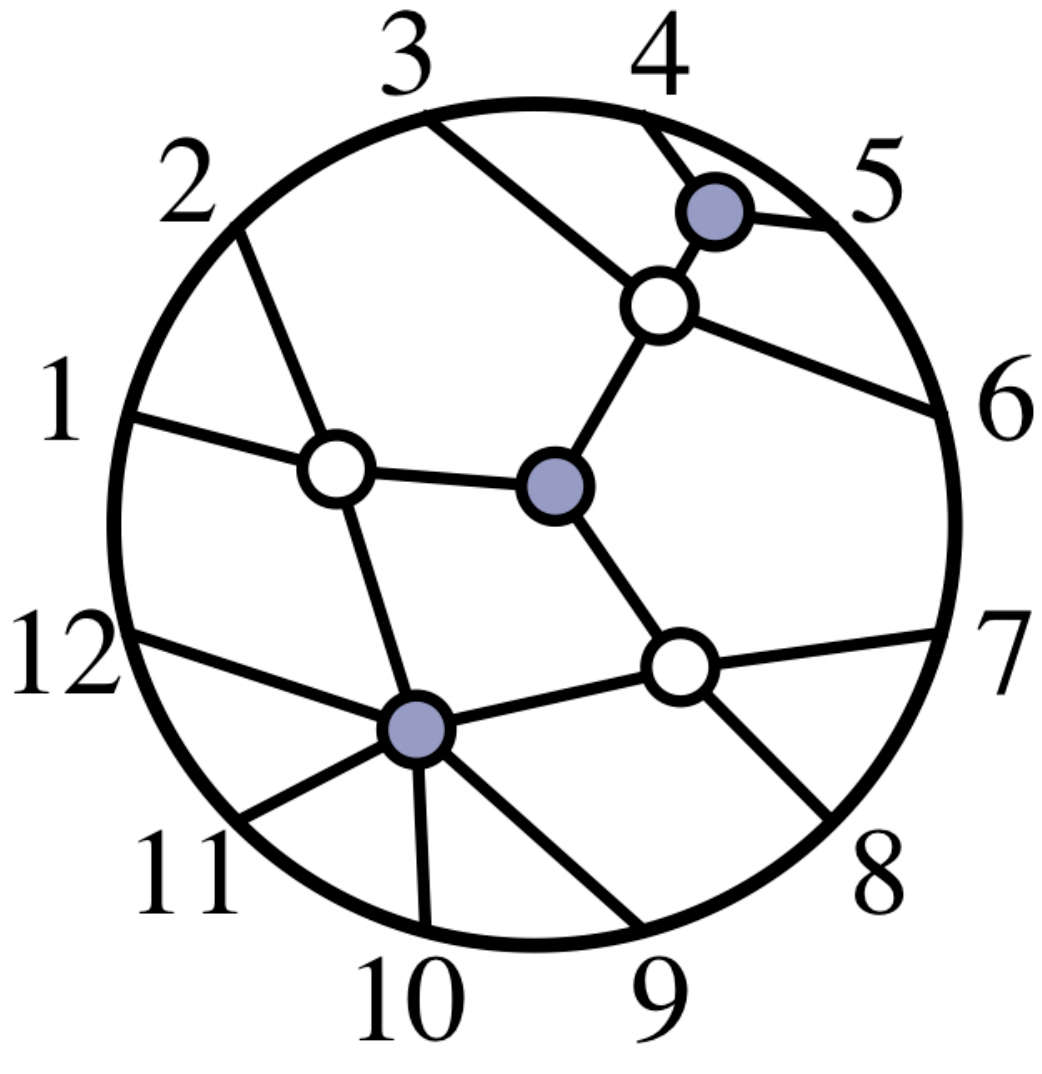}  
\newline\tiny\{2,7,5,6,16,12,8,13,15,21,22,23\} &
{\tiny
$\left(
\begin{array}{cccccccccccc}
 1 & \alpha_{12} & \alpha_5+\alpha_8 & (\alpha_5+\alpha_8) \alpha_{11} & 0 & -(\alpha_5+\alpha_8) \alpha_9 & -\alpha_5 \alpha_7 & -\alpha_5 \alpha_6 & 0 & 0 & 0 & 0 \\
 0 & 0 & 1 & \alpha_{11} & 0 & -\alpha_9 & -\alpha_7 & -\alpha_6 & -\alpha_4 & 0 & 0 & 0 \\
 0 & 0 & 0 & 1 & \alpha_{10} & 0 & 0 & 0 & 0 & 0 & 0 & 0 \\
 0 & 0 & 0 & 0 & 0 & 0 & 0 & 0 & 1 & \alpha_3 & 0 & 0 \\
 0 & 0 & 0 & 0 & 0 & 0 & 0 & 0 & 0 & 1 & \alpha_2 & 0 \\
 0 & 0 & 0 & 0 & 0 & 0 & 0 & 0 & 0 & 0 & 1 & \alpha_1 \\
\end{array}
\right)$}
\\
\hline

$\begin{array}{c} (3.28e) \\n=10\\k=5\\d=11 \end{array}$ &
\includegraphics[width=0.1\textwidth]{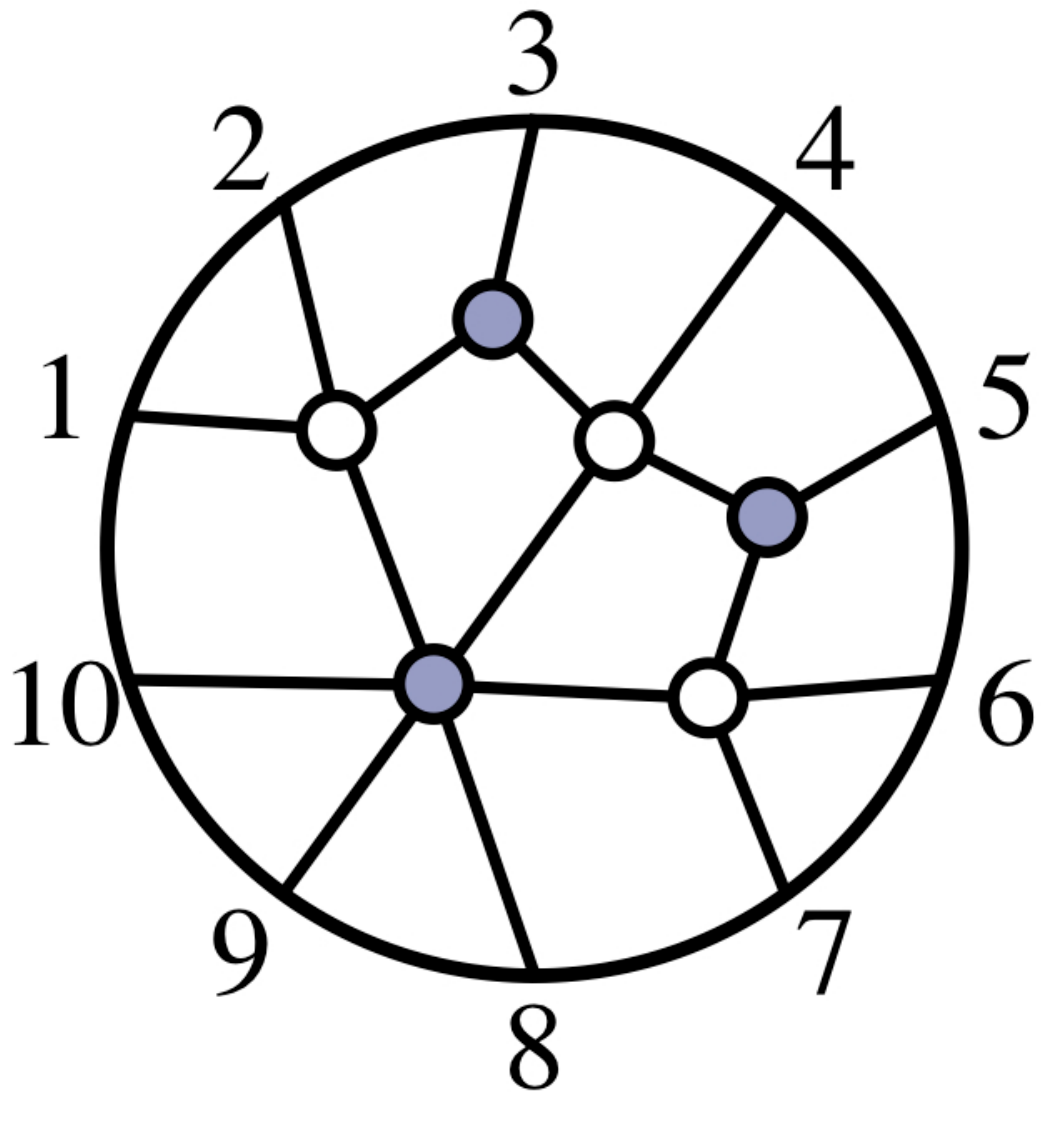}  
\newline\tiny\{2,4,10,6,11,7,13,15,18,19\} &
{\tiny
$\left(
\begin{array}{cccccccccc}
 1 & \alpha_{11} & \alpha_7+\alpha_{10} & \alpha_7 \alpha_9 & \alpha_7 \alpha_8 & 0 & 0 & 0 & 0 & 0 \\
 0 & 0 & 1 & \alpha_9 & \alpha_4+\alpha_8 & \alpha_4 \alpha_6 & \alpha_4 \alpha_5 & 0 & 0 & 0 \\
 0 & 0 & 0 & 0 & 1 & \alpha_6 & \alpha_5 & \alpha_3 & 0 & 0 \\
 0 & 0 & 0 & 0 & 0 & 0 & 0 & 1 & \alpha_2 & 0 \\
 0 & 0 & 0 & 0 & 0 & 0 & 0 & 0 & 1 & \alpha_1 \\
\end{array}
\right)$}
\\
\hline

$\begin{array}{c} (3.28f) \\n=14\\k=7\\d=13 \end{array}$ &
\includegraphics[width=0.1\textwidth]{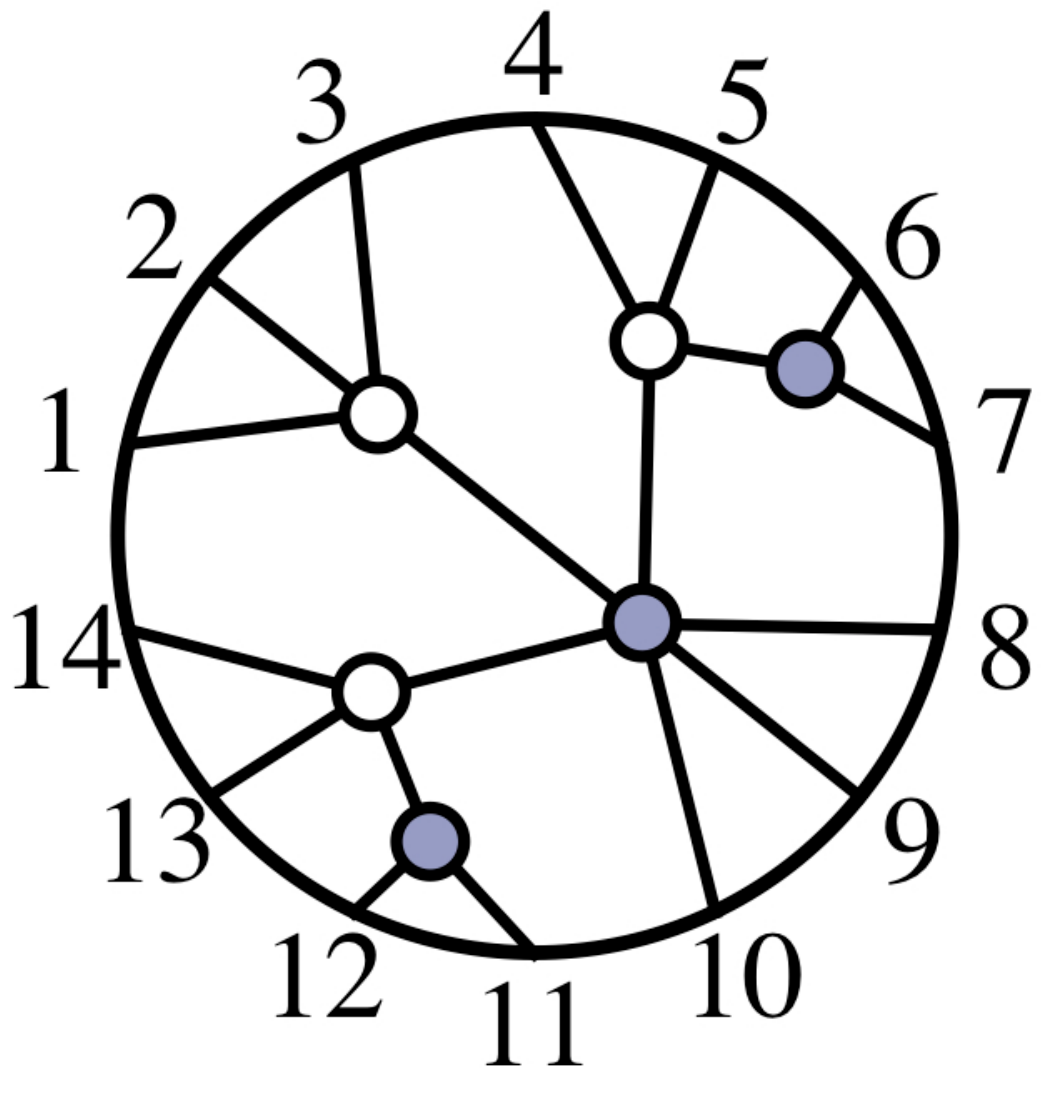}  
\newline\tiny\{2,3,12,5,7,15,20,18,22,23,13,25,14,24\} &
{\tiny
$\left(
\begin{array}{cccccccccccccc}
 1 & \alpha_{13} & \alpha_{12} & \alpha_9 & \alpha_9 \alpha_{11} & \alpha_9 \alpha_{10} & 0 & 0 & 0 & 0 & 0 & 0 & 0 & 0 \\
 0 & 0 & 0 & 1 & \alpha_{11} & \alpha_{10} & 0 & -\alpha_7 & 0 & 0 & 0 & 0 & 0 & 0 \\
 0 & 0 & 0 & 0 & 0 & 1 & \alpha_8 & 0 & 0 & 0 & 0 & 0 & 0 & 0 \\
 0 & 0 & 0 & 0 & 0 & 0 & 0 & 1 & \alpha_6 & 0 & 0 & 0 & 0 & 0 \\
 0 & 0 & 0 & 0 & 0 & 0 & 0 & 0 & 1 & \alpha_5 & 0 & 0 & 0 & 0 \\
 0 & 0 & 0 & 0 & 0 & 0 & 0 & 0 & 0 & 1 & \alpha_4 & 0 & -\alpha_2 & -\alpha_1 \\
 0 & 0 & 0 & 0 & 0 & 0 & 0 & 0 & 0 & 0 & 1 & \alpha_3 & 0 & 0 \\
\end{array}
\right)$}
\\
\hline

$\begin{array}{c} (3.28g) \\n=14\\k=7\\d=13 \end{array}$ &
\includegraphics[width=0.1\textwidth]{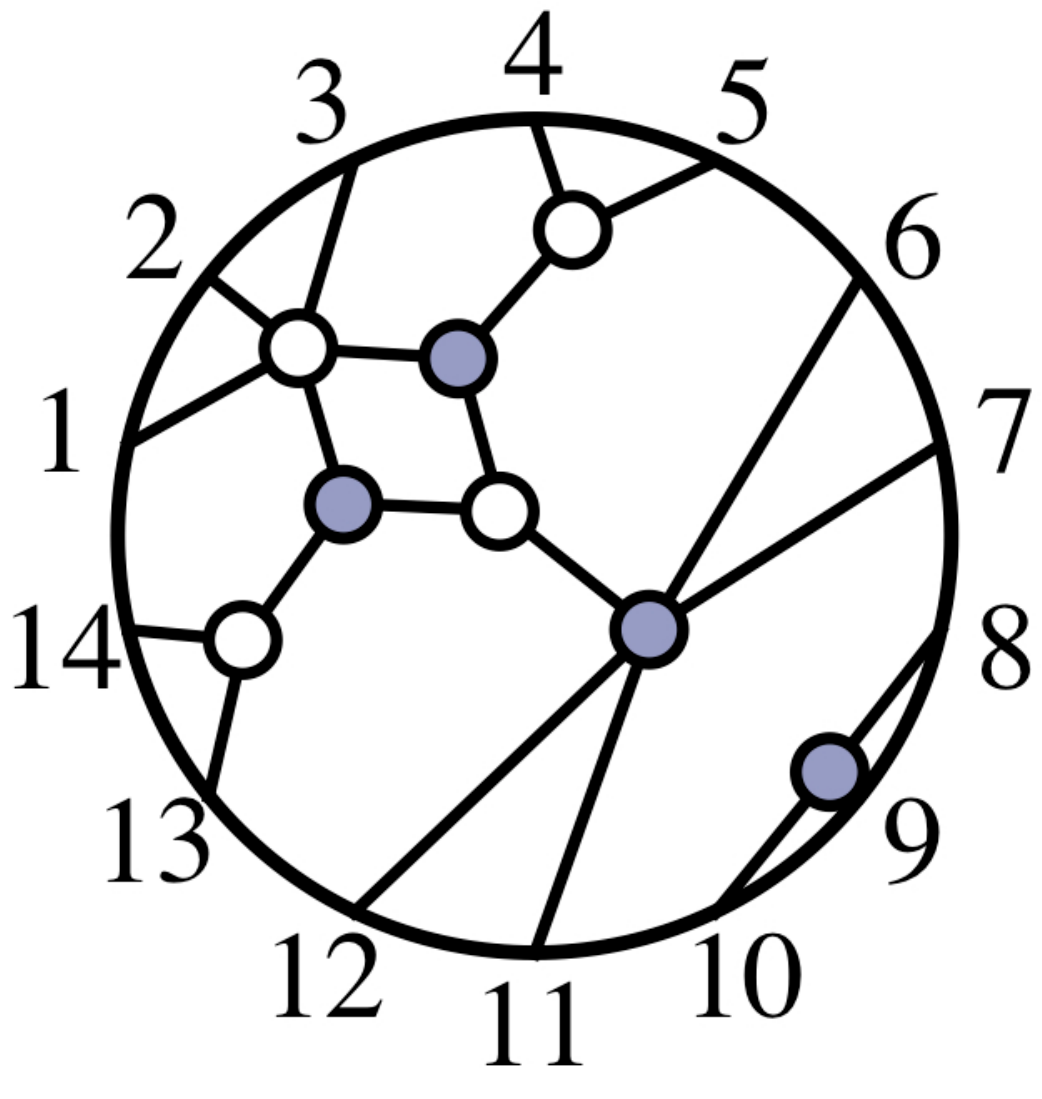}  
\newline\tiny\{2,3,12,5,13,15,20,10,22,23,21,25,14,18\} &
{\tiny
$\left(
\begin{array}{cccccccccccccc}
 1 & \alpha_{13} & \alpha_{12} & \alpha_8+\alpha_{10} & (\alpha_8+\alpha_{10}) \alpha_{11} & \alpha_8 \alpha_9 & 0 & 0 & 0 & 0 & 0 & 0 & 0 & 0 \\
 0 & 0 & 0 & 1 & \alpha_{11} & \alpha_9 & 0 & 0 & 0 & 0 & 0 & 0 & -\alpha_2 & -\alpha_1 \\
 0 & 0 & 0 & 0 & 0 & 1 & \alpha_7 & 0 & 0 & 0 & 0 & 0 & 0 & 0 \\
 0 & 0 & 0 & 0 & 0 & 0 & 1 & 0 & 0 & 0 & \alpha_4 & 0 & 0 & 0 \\
 0 & 0 & 0 & 0 & 0 & 0 & 0 & 1 & \alpha_6 & 0 & 0 & 0 & 0 & 0 \\
 0 & 0 & 0 & 0 & 0 & 0 & 0 & 0 & 1 & \alpha_5 & 0 & 0 & 0 & 0 \\
 0 & 0 & 0 & 0 & 0 & 0 & 0 & 0 & 0 & 0 & 1 & \alpha_3 & 0 & 0 \\
\end{array}
\right)$}
\\
\hline

$\begin{array}{c} (3.28h) \\n=12\\k=6\\d=12 \end{array}$&
\includegraphics[width=0.1\textwidth]{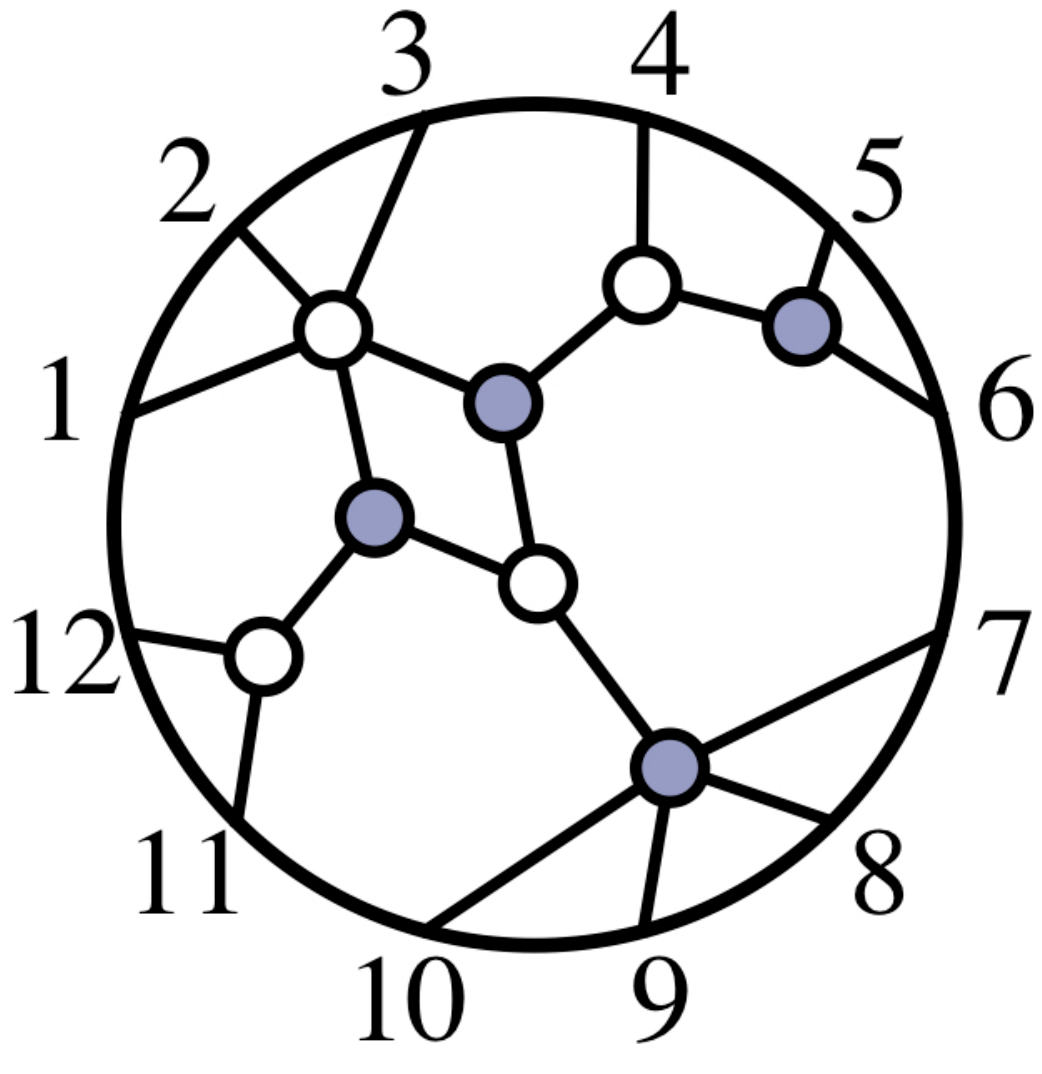}  
\newline\tiny\{2,3,10,6,11,17,13,19,20,21,12,16\} &
{\tiny
$\left(
\begin{array}{cccccccccccc}
 1 & \alpha_{12} & \alpha_{11} & \alpha_6+\alpha_9 & (\alpha_6+\alpha_9) \alpha_{10} & 0 & -\alpha_6 \alpha_7 & 0 & 0 & 0 & 0 & 0 \\
 0 & 0 & 0 & 1 & \alpha_{10} & 0 & -\alpha_7 & 0 & 0 & 0 & \alpha_2 & \alpha_1 \\
 0 & 0 & 0 & 0 & 1 & \alpha_8 & 0 & 0 & 0 & 0 & 0 & 0 \\
 0 & 0 & 0 & 0 & 0 & 0 & 1 & \alpha_5 & 0 & 0 & 0 & 0 \\
 0 & 0 & 0 & 0 & 0 & 0 & 0 & 1 & \alpha_4 & 0 & 0 & 0 \\
 0 & 0 & 0 & 0 & 0 & 0 & 0 & 0 & 1 & \alpha_3 & 0 & 0 \\
\end{array}
\right)$}
\\
\hline

$\begin{array}{c} (3.28i) \\n=12\\k=6\\d=12 \end{array}$ &
\includegraphics[width=0.1\textwidth]{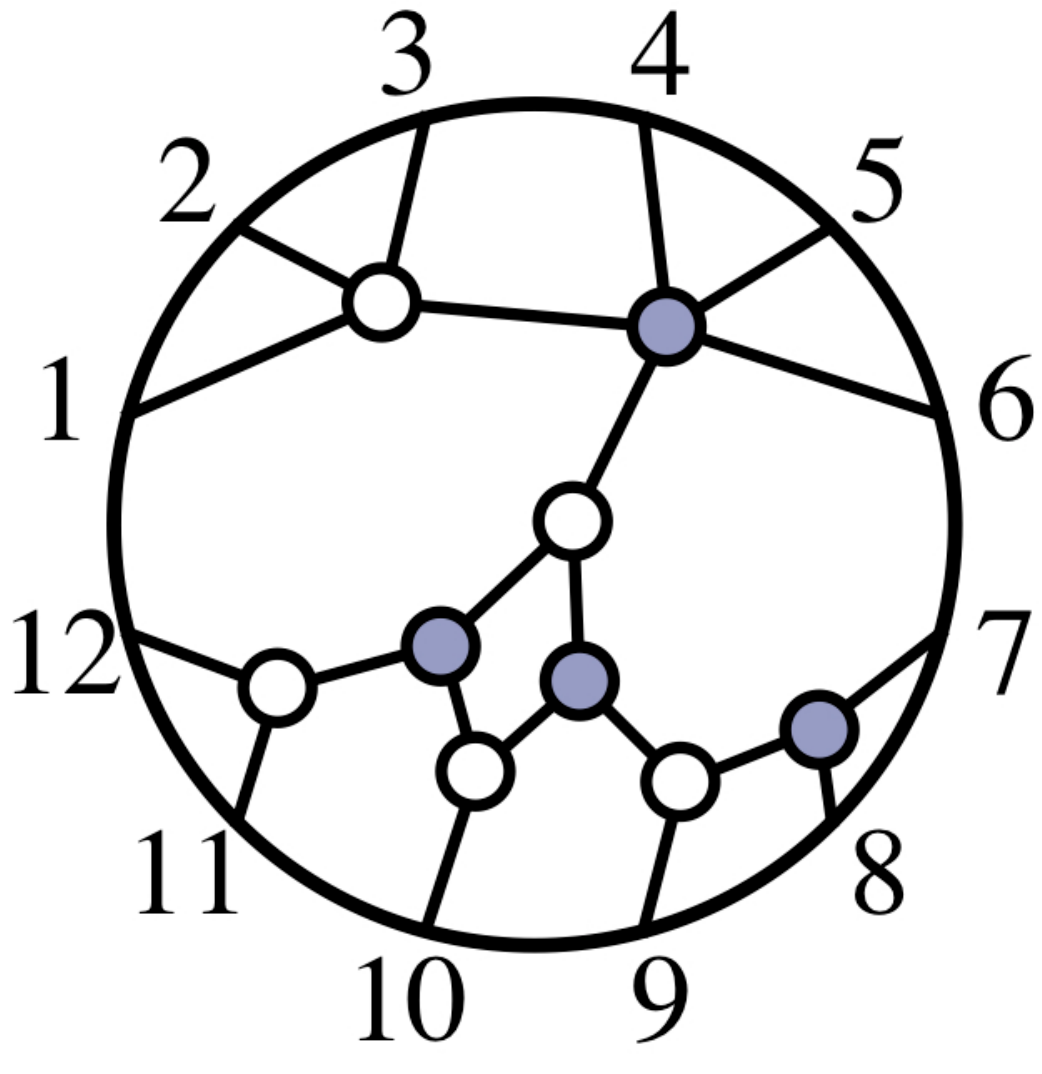}  
\newline\tiny\{2,3,10,13,16,17,9,19,11,18,12,20\} &
{\tiny
$\left(
\begin{array}{cccccccccccc}
 1 & \alpha_{12} & \alpha_{11} & \alpha_{10} & 0 & 0 & 0 & 0 & 0 & 0 & 0 & 0 \\
 0 & 0 & 0 & 1 & \alpha_9 & 0 & 0 & 0 & 0 & 0 & 0 & 0 \\
 0 & 0 & 0 & 0 & 1 & \alpha_8 & 0 & 0 & 0 & 0 & 0 & 0 \\
 0 & 0 & 0 & 0 & 0 & 1 & 0 & -\alpha_3-\alpha_5 & (-\alpha_3-\alpha_5) \alpha_6 & -\alpha_3 \alpha_4 & 0 & 0 \\
 0 & 0 & 0 & 0 & 0 & 0 & 1 & \alpha_7 & 0 & 0 & 0 & 0 \\
 0 & 0 & 0 & 0 & 0 & 0 & 0 & 1 & \alpha_6 & \alpha_4 & \alpha_2 & \alpha_1 \\
\end{array}
\right)$}
\\
\hline

\end{longtable}
\end{landscape}
\bibliographystyle{ieeetr}
\bibliography{grassman_bib}

\end{document}